\documentclass{SciPost}

\hypersetup{
    colorlinks,
    linkcolor={red!50!black},
    citecolor={blue!50!black},
    urlcolor={blue!80!black}
}

\usepackage[bitstream-charter]{mathdesign}
\DeclareSymbolFont{usualmathcal}{OMS}{cmsy}{m}{n}
\DeclareSymbolFontAlphabet{\mathcal}{usualmathcal}

\fancypagestyle{SPstyle}{
\fancyhf{}
\lhead{\colorbox {scipostblue}{\bf \color{white} ~Lecture Notes }}
\rhead{{\bf \color{scipostdeepblue} ~ }}

\fancyfoot[C]{\textbf{\thepage}}
}

\newcommand{\adag}{a^{\dagger}}

\newcommand\ket[1]{| #1\rangle}
\newcommand\kket[1]{| #1\rangle\rangle}
\newcommand\bra[1]{\langle #1|}
\newcommand\bbra[1]{\langle\langle #1|}

\newcommand\adaga{a^\dagger a}

\usepackage[framemethod=TikZ]{mdframed}

\newcounter{ex}[section]
\renewcommand{\theex}{\arabic{section}.\arabic{ex}}

\newcounter{examp}[section]
\renewcommand{\theexamp}{E\arabic{section}.\arabic{examp}}

\newenvironment{examp}[2][]{%
    \refstepcounter{examp}
\ifstrempty{#1}%
{\mdfsetup{%
    frametitle={%
        \tikz[baseline=(current bounding box.east),outer sep=0pt]
        \node[anchor=east,rectangle,fill=scipostblue]
        {\strut {\bf \color{white}Example~\theexamp}};}
    }%
}{\mdfsetup{%
    frametitle={%
        \tikz[baseline=(current bounding box.east),outer sep=0pt]
        \node[anchor=east,rectangle,fill=scipostblue]
        {\strut {\bf \color{white}Example~\theexamp:~#1}};}%
    }%
}%
\mdfsetup{%
innertopmargin=10pt,innerbottommargin=10pt,linecolor=scipostblue,%
    linewidth=2pt,topline=true,%
    frametitleaboveskip=\dimexpr-\ht\strutbox\relax%
}
\begin{mdframed}[]\relax}{%
\end{mdframed}}

\begin{document}

\pagestyle{SPstyle}

\begin{center}{\Large \textbf{\color{scipostdeepblue}{
Open quantum systems\\
}}}\end{center}

\begin{center}\textbf{
Erik Torrontegui\textsuperscript{1$\star$} and
Ricardo Puebla\textsuperscript{1$\dagger$}
}\end{center}

\begin{center}
{\bf 1} Department of Physics, Universidad Carlos III de Madrid, Avenida de la Universidad 30, 28911 Legan\'es, Madrid, Spain
\\[\baselineskip]
$\star$ \href{mailto:email1}{\small etorront@fis.uc3m.es}\,,\quad
$\dagger$ \href{mailto:email2}{\small rpuebla@fis.uc3m.es}
\end{center}

\section*{\color{scipostdeepblue}{Abstract}}
\textbf{\boldmath{%
Open quantum systems describe quantum systems that interact unavoidably with their environments. These notes introduce the physical origins of decoherence and dissipation in realistic settings, developing the theoretical framework of reduced dynamics and quantum master equations within the connection to thermodynamics in the quantum regime. Additionally, a set of relevant exercises are proposed to reinforce the acquisition of knowledge. The text equips readers with the tools to model and analyze open quantum dynamics.
}}

\vspace{\baselineskip}








\vspace{10pt}
\noindent\rule{\textwidth}{1pt}
\tableofcontents
\noindent\rule{\textwidth}{1pt}
\vspace{10pt}


\section{Introduction}\label{c:intro}

The theory of open quantum systems \cite{BRE02, Lidar20, Szankowski23} plays a central role in every area of quantum technologies. The central motivation to study this theoretical framework is straightforward: treating quantum systems as perfectly isolated is an idealization that never holds in reality. No physical system—quantum or classical—can be completely shielded from external perturbations or interactions with other degrees of freedom. This limitation is especially critical in quantum systems, where uniquely quantum features such as superposition and entanglement are notoriously fragile in the presence of environmental noise.  

A paradigmatic manifestation of this fragility is quantum decoherence—the gradual loss of quantum coherence or entanglement between registers or subsystems as time evolves from their initial preparation. But why does decoherence occur, and how does it unfold? Addressing these questions lies at the heart of open quantum system theory. The “environment” in this context refers to the collection of external degrees of freedom that interact—directly or indirectly—with the system of interest. Understanding how these environmental interactions degrade, modify, or sometimes even induce quantum properties is crucial not only from a fundamental standpoint—touching on deep issues such as the measurement problem, the foundations of quantum mechanics, and the quantum-to-classical transition—but also from a practical one.

Indeed, identifying the physical mechanisms behind decoherence can enable us to mitigate or even eliminate unwanted noise channels, extending the lifetime of quantum information. Moreover, a detailed description of an open system may justify treating it as an ideal closed system up to a certain time scale.  Conversely, in certain settings, carefully engineered system–environment interactions can be harnessed as a resource—for example, to extract information about the surrounding environment, or to realize novel sensing, control, and thermodynamic protocols. In this sense, mastering open quantum system dynamics is both a challenge and an opportunity for the future of quantum science and technology.

It should not be surprising that the framework of open quantum systems is of fundamental importance in virtually every field of modern quantum technologies. For example, early developments in quantum information processing, quantum computation, and quantum communication were often based on the idealization of perfectly isolated quantum systems. In reality, such isolation is impossible to achieve: every quantum system interacts, to some extent, with its surrounding environment. Consequently, assessing any genuine quantum advantage requires revisiting these protocols within a more realistic open-system setting, where environmental noise and losses are explicitly accounted for. Moreover, several other key areas—such as quantum thermodynamics (addressing the exchange of work and heat, the emergence of irreversibility, and the design of quantum heat engines) or quantum sensing, deal with quantum systems that are inherently open. In these settings, decoherence processes such as dephasing, relaxation, and dissipation are not merely detrimental side effects, but sometimes fundamental features that can be engineered or exploited. Understanding, modeling, and ultimately controlling these open-system dynamics is therefore indispensable for both fundamental research and the practical deployment of quantum technologies.

Follows a brief description of the contents covered in each of the Sections:
\begin{itemize}
    \item In Section~\ref{c:intro} we provide a review of the basics of quantum dynamics for isolated or closed systems. This includes the density matrix formalism, measurements, quantum channels, coherence and purity, and the concept of subsystems (i.e. partial trace). 
    \item In Section~\ref{c:OQS} we commence our journey into the open system dynamics. We will see how imposing reasonable mathematical conditions into the dynamical equation of an open quantum system leads into a generic form known as Gorini-Kossakowski-Sudarshan-Lindblad (GKSL) master equation. We will see relevant examples and introduce the powerful notation for the Liouvillian superoperator. 
    \item In Section~\ref{c:micro} we will derive the GKSL master equation from first principles, i.e. from a microscopic  model doing physically-motivated approximations to the dynamics. We will see how this general theory can be applied to two relevant scenarios: a harmonic oscillator exchanging excitations with an environment, and a spin undergoing pure dephasing noise. 
    \item In Section~\ref{c:thermo} we will introduce the basics of quantum thermodynamics, that is, how to define work and heat in a quantum system, the notion of entropy and finally some basics for quantum heat engines and refrigerators. 
    \item In Section~\ref{c:fr} we include more advanced and technical derivations related to different parts covered in the previous four main sections.
\end{itemize}

\subsection{Revisiting closed quantum systems} 
Let us briefly revisit the basics of closed quantum systems \cite{Sakurai}. The quantum state $\ket{\psi(t_0)}$ of a system, whose Hamiltonian is $H\in\mathbb{H}(\mathcal{H})$, where $\mathcal{H}$ is the Hilbert space of the system and $\mathbb{H}(\mathcal{H})$ denotes the subset of Hermitian operators acting on $\mathcal{H}$, i.e. $H\in\mathbb{H}(\mathcal{H}) \Leftrightarrow H^\dagger=H$.  The quantum state evolves in time according to the Schr\"odinger equation,
\begin{align}
    \frac{d}{dt}\ket{\psi(t)}=-i H \ket{\psi(t)},
\end{align}
where $\ket{\psi(t)}$ denotes the pure quantum state of the system at time $t$, fulfilling the normalization $\langle \psi(t)|\psi(t) \rangle=1, \forall t$. Note that we have set $\hbar=1$, which we will keep for convenience throughout the lecture notes. 

The Schr\"odinger equation implies that if a closed system finds itself in a pure quantum state at a time $t_0$, $\ket{\psi(t_0)}$, the system will remain in a pure state at later times as it undergoes a unitary time evolution dictated by the Hamiltonian $H$, i.e.
\begin{align}
    \ket{\psi(t)}=\mathcal{T}\left[e^{-i\int_{t_0}^t dt' H}\right] \ket{\psi(t_0)}. 
\end{align}
This can be written in a more compact manner by defining the time-evolution operator, $U(t,t_0)$, which evolves a state from time $t_0$ to time $t$, 
\begin{align}\label{eq:Udef}
    U(t_1,t_0)=\mathcal{T}\left[ e^{-i\int_{t_0}^{t_1} dt H}\right].
\end{align}
Here $\mathcal{T}$ denotes the time-ordering operator that orders products of time-dependent operators such that their time-arguments increase from
right to left:
\begin{align}
    \mathcal{T}\left[ U(t_1,t_0)U(t_2,t_1)\right]= U(t_2,t_1)U(t_1,t_0), \qquad t_0\leq t_1\leq t_2, 
\end{align}
and the time-evolution operator fulfills the composition property, that is,
\begin{align}
    U(t_2,t_1)U(t_1,t_0)=U(t_2,t_0).
\end{align}
Note that if the Hamiltonian $H$ is time-independent,  then the time-evolution operator adopts the well-known form
\begin{align}
    U(t_1,t_0)=e^{-i(t_1-t_0)H},
\end{align}
given that the Hamiltonian $H$ at different times commute with itself.

The time-evolution operator is unitary, i.e.
\begin{align}
    U(t_1,t_0)U^{\dagger}(t_1,t_0)=\mathbb{I},
\end{align}
where ${}^\dagger$ denotes the adjoint or conjugate transpose operation. 
This can be derived from Eq.~\eqref{eq:Udef}:
\begin{align}
    U^\dagger (t_1,t_0)&= \left[ \mathcal{T}\left[ e^{-i\int_{t_0}^{t_1} dt H}\right]\right]^{\dagger}=\left[\lim_{\delta t\rightarrow 0} e^{-i\delta t H(t_1)}e^{-i\delta t H(t_1-\delta t)}\cdots e^{-i\delta t H(t_0+\delta t)}e^{-i\delta t H(t_0)} \right]^\dagger\nonumber\\ &=\lim_{\delta t\rightarrow 0} e^{+i \delta t H^\dagger(t_0)}e^{+i \delta t H^\dagger(t_0+\delta t)}\cdots e^{+i \delta t H^\dagger(t_1-\delta t)}e^{+i \delta t H^\dagger(t_1)}\nonumber\\
    &= \lim_{\delta t\rightarrow 0} e^{+i \delta t H(t_0)}e^{+i \delta t H(t_0+\delta t)}\cdots e^{+i \delta t H(t_1-\delta t)}e^{+i \delta t H(t_1)}=\mathcal{T}^\dagger \left[ e^{+i\int_{t_0}^{t_1} dt H}\right].
\end{align}
Using the previous expression, one can verify $U(t_1,t_0)U^{\dagger}(t_1,t_0)=U^\dagger (t_1,t_0)U(t_1,t_0)=\mathbb{I}$. 

The time-evolution operator also obeys a dynamical equation. Given that, by definition, $\ket{\psi(t)}=U(t,t_0)\ket{\psi(t_0)}$, we can make use of the Schr\"odinger equation to find
\begin{align}\label{eq:ddtU}
    \frac{d}{dt} \left[U(t,t_0)\ket{\psi(t_0)}\right]=-i H(t) U(t,t_0)\ket{\psi(t_0)} \Rightarrow \frac{d}{dt} U(t,t_0)=-iH(t) U(t,t_0),
\end{align}
since this must be valid for any initial state $\ket{\psi(t_0)}$, and with the initial condition
\begin{align}
    U(t_0,t_0)=\mathbb{I}.
\end{align}

\subsubsection{Heisenberg picture}
The Schr\"odinger picture is the standard framework to compute the dynamics of a quantum system. Yet, moving to other pictures may prove  useful to extract dynamical features or to simplify its calculation, both in a closed or an open context, as we will see during the course.

In the Heisenberg picture, the states do not evolve in time, but operators do. Recall that this is just a mathematical tool as the physics we aim to describe must not change, i.e. the observed dynamics is independent of the mathematical framework used to describe or compute expectation values and observable properties. Let us use a subscript $H$ to denote the Heisenberg picture, and $A$ to denote a generic operator, possibly time-dependent in the Schr\"odinger picture, and $t_0$ the initial time. The expectation value of the operator $A(t)\in\mathbb{H}(\mathcal{H})$ reads as
\begin{align}
    \langle A(t)\rangle= \bra{\psi(t)}A(t)\ket{\psi(t)}=\bra{\psi(t_0)}U^\dagger(t,t_0)A(t)U(t,t_0)\ket{\psi(t_0)}=\bra{\psi(t_0)}A_H(t)\ket{\psi(t_0)},
\end{align}
where we have defined the operator $A(t)$ in the Heisenberg picture as
\begin{align}
    A_H(t)=U^\dagger(t,t_0)A(t) U(t,t_0).
\end{align}
Given that $\ket{\psi_H(t)}=\ket{\psi(t_0)}$ (by definition of the Heisenberg picture), the question is then what sort of dynamical equation should $A_H(t)$ follow. Taking time-derivative to $A_H(t)$, we find
\begin{align}
    \frac{d}{dt} A_H(t)&=\frac{d}{dt}\left[U^\dagger(t,t_0)A(t) U(t,t_0) \right]\nonumber\\
    &=iU^\dagger(t,t_0) H(t) A(t)U(t,t_0)-iU^\dagger(t,t_0)A(t)H(t)U(t,t_0)\nonumber\\&\qquad+U^\dagger(t,t_0)\left(\frac{\partial A(t)}{\partial t} \right)U(t,t_0)\nonumber\\
    &=iU^\dagger(t,t_0) H(t) U(t,t_0)U^\dagger(t,t_0) A(t)U(t,t_0)\nonumber\\&\qquad-iU^\dagger(t,t_0)A(t)U(t,t_0)U^\dagger(t,t_0)H(t)U(t,t_0)\nonumber\\&\qquad+U^\dagger(t,t_0)\left(\frac{\partial A(t)}{\partial t} \right)U(t,t_0)
\end{align}
where we have used Eq.~\eqref{eq:ddtU}, $\frac{d}{dt}U(t,t_0)=-iH(t)U(t,t_0)$. Since $A_H(t)=U^\dagger(t,t_0)A(t)U(t,t_0)$, we can write
\begin{align}
    \frac{d}{dt}A_H(t)=i[H_H(t),A_H(t)]+\left(\frac{\partial A(t)}{\partial t} \right)_H,
\end{align}
where $H_H(t)$ corresponds to the Hamiltonian in the Heisenberg picture. 
If the operator $A$, in the Schr\"odinger picture, does not depend explicitly on time, then the last term of the right-hand side of the equation vanishes.

\subsubsection{Interaction picture}\label{ss:int_pic}
The interaction picture is an intermediate framework where states and operators evolve in time. This may seem an unnecessary complication at first, but the interaction picture is extremely useful to compute quantum dynamics and perform approximations. For that, we consider that the Hamiltonian of a system in the Schr\"odinger picture can be split in two, 
\begin{align}
    H(t)=H_0+V(t),
\end{align}
where typically $V(t)$ refers to a possibly time-dependent interaction. The first step in the interaction picture is to define the quantum state as (subscript $I$ denotes interaction frame),
\begin{align}\label{eq:psiI}
\ket{\psi_I(t)}=U^\dagger_0(t,t_0)\ket{\psi(t)}=e^{i (t-t_0)H_0}\ket{\psi(t)},
\end{align}
where $\ket{\psi(t)}$ is the state in the Schr\"odinger picture at time $t$, and $U_0(t,t_0)$ refers to the time-evolution operator of $H_0$ alone. The operators are defined as
\begin{align}\label{eq:AI}
    A_I(t)=U_0^\dagger(t,t_0) A(t) U_0(t,t_0).
\end{align}
Both the state and operators coincide at the initial time $t_0$, since $U(t_0,t_0)=\mathbb{I}$. 
From the definitions in Eqs.~\eqref{eq:psiI} and~\eqref{eq:AI}, it follows that the expectation value of $A(t)$ matches with the one in the Schr\"odinger frame (as it should be to be a valid physical transformation):
\begin{align}
    \langle A_I(t)\rangle&= \bra{\psi_I(t)}A_I(t)\ket{\psi_I(t)}=\bra{\psi(t)}U_0(t,t_0) U_0^\dagger(t,t_0)A(t) U_0(t,t_0)U_0^\dagger(t,t_0) \ket{\psi(t)}\nonumber\\&=\bra{\psi(t)}A(t)\ket{\psi(t)}=\langle A(t)\rangle.
\end{align}
The Hamiltonian $H_0$ remains invariant in the interaction picture, $H_{0,I}=H_0$ since $U_0(t,t_0)$ commutes with $H_0$. However, in general $V(t)$ does not:
\begin{align}
    V_I(t)=U^\dagger_0(t,t_0)V(t)U_0(t,t_0).
\end{align}
In this manner, one can find the dynamical equation for $\ket{\psi_I(t)}$, 
\begin{align}
    \frac{d}{dt}\ket{\psi_I(t)}&=\frac{d}{dt}\left[ U_0^\dagger(t,t_0)\ket{\psi(t)}\right]=i H_0U^\dagger_0(t,t_0)\ket{\psi(t)}+U_0^\dagger(t,t_0)\frac{d}{dt}\ket{\psi(t)}\nonumber\\&=iH_0 U_0^\dagger(t,t_0)\ket{\psi(t)}-i U_0^\dagger(t,t_0)(H_0+V(t))\ket{\psi(t)}\nonumber\\
    &=iU_0^\dagger(t,t_0) H_0 U_0(t,t_0)\ket{\psi_I(t)}-iU_0^\dagger(t,t_0)(H_0+V(t))U_0(t,t_0)\ket{\psi_I(t)}\nonumber\\
    &=-iU_0^\dagger(t,t_0)V(t)U_0(t,t_0 \ket{\psi_I(t)}\nonumber\\&=-i V_I(t)\ket{\psi_I(t)}, \label{eq:psiI_dyn}
\end{align}
with initial condition $\ket{\psi_I(t_0)}=\ket{\psi(t_0)}$. 
Therefore, the state $\ket{\psi_I(t)}$ evolves according to a Schr\"odinger equation where $V_I(t)$ plays the role of the Hamiltonian of the system.  In addition, one can show, following similar steps as done to the Heisenberg picture, that operators evolve according to
\begin{align}
    \frac{d}{dt}A_I(t)=i[H_0,A_I(t)]+\left( \frac{\partial A(t)}{\partial t}\right)_I.
\end{align}
For an explicit relation among the Schr\"odinger and interaction pictures see Sec.~\ref{moremat:Int_Schro}.

\subsection{Brief survey on density matrix formalism}
If the system is described in terms of a  statistical ensemble \cite{BRE02}, i.e. by a mixture over a set of different pure states $\ket{\phi_n}$ with different probabilities $c_n$, then it is convenient to use the density matrix $\rho$ defined as 
\begin{align}\label{eq:rho}
    \rho=\sum_{n} c_n \ket{\phi_n}\bra{\phi_n},
\end{align}
where $\{c_n\}$ are positive real coefficients that add up to $1$ to ensure the normalization, 
\begin{align}
    {\rm Tr}[\rho]={\rm Tr}[\sum_n c_n \ket{\phi_n}\bra{\phi_n}]=\sum_n c_n {\rm Tr}[\ket{\phi_n}\bra{\phi_n}]=\sum_n c_n =1,
\end{align}
since the trace of the outer product, ket-bra, of a pure state is one, ${\rm Tr}[\ket{\phi_n}\bra{\phi_n}]=1$.   

The  matrix $\rho$ represents a valid quantum state if it is a non-negative complex Hermitian matrix with unit trace, $\rho^\dagger=\rho$, $\rho\geq 0$ and ${\rm Tr}[\rho]=1$. The spectral decomposition of $\rho$ can be written as
\begin{align}\label{eq:rho_spectral}
    \rho=\sum_{m=1}^{d_H} \alpha_m \ket{\varphi_m}\bra{\varphi_m},
\end{align}
with $d_H=\dim[\mathcal{H}]$ the dimension of the Hilbert space $\mathcal{H}$, where $\{\ket{\varphi_m}\}$ for an orthonormal basis of $\mathcal{H}$, i.e. $\langle \varphi_m|\varphi_{m'}\rangle=\delta_{m,m'}$. Therefore, the eigenvalues of $\rho$, which are the coefficients $\alpha_m$, can be associated to the probability of finding the state in the pure state $\ket{\varphi_m}$, so that they need to be real, positive and add up to $1$ (i.e. ${\rm Tr}[\rho]=1$). Since the eigenvalues are real, $\rho$ is Hermitian (i.e. $\rho^\dagger=\rho$), and since they represent probabilities, $\rho\geq 0$. More formally, the states $\rho$ fulfilling these conditions form a subset of the Hermitian operators on the total Hilbert space $\mathbb{H}(\mathcal{H})$ denoted by $\mathcal{D}(\mathcal{H})=\{ \rho\in\mathbb{H}(\mathcal{H})| \ {\rm Tr}[\rho]=1, \rho\geq 0\}$.

Note that the spectral decomposition does not match, in general, with the ensemble decomposition given in Eq.~\eqref{eq:rho}. In general, the ensemble decomposition is not unique: There are infinitely different mixtures, defined by the set of weights and states $\{c_n,\ket{\phi_n}\}$ that result in the same mixed state $\rho$ (see Box~\ref{examp:ensemble} for an example). Finally, we recall that any pure quantum state (any ket or wavevector) can be expressed as a density matrix, by the outer product $\rho=\ket{\psi}\bra{\psi}$. However, the converse is not true: Not all quantum states can be written as kets or wavevectors. The space of valid quantum states is larger than that spanned by pure quantum states.

\begin{examp}[Different ensembles, same density matrix]{examp:ensemble}\label{examp:ensemble}
Suppose two different single-qubit experiments. In the first one, denoted $A$, the qubit is prepared with $c_1=1/3$ in the state $\ket{0}$ and with $c_2=2/3$ in $\ket{1}$. Therefore, the density matrix reads
\begin{align}
    \rho_A=\frac{1}{3}\ket{0}\bra{0}+\frac{2}{3}\ket{1}\bra{1}=\begin{pmatrix} 1/3 & 0\\ 0 & 2/3        
    \end{pmatrix}.
\end{align}
    In the second experiment, $B$, the qubit is prepared in four different states, $\ket{+}$, $\ket{-}$, $\ket{1}$ and $\ket{0}$ with probabilities $c_+=1/4$, $c_-=1/4$, $c_1=1/12$ and $c_0=5/12$, respectively. The density matrix in this case reads as
    \begin{align}
        \rho_B=\frac{1}{4}\ket{+}\bra{+}+\frac{1}{4}\ket{-}\bra{-}+\frac{1}{12}\ket{0}\bra{0}+\frac{5}{12}\ket{1}\bra{1}=\begin{pmatrix} 1/3 & 0\\ 0 & 2/3        
    \end{pmatrix}.
    \end{align}
    Although the qubits are prepared in different ways, their average states $\rho_A$ and $\rho_B$ coincide. This means that in average, these two qubits yield equivalent measurement outcomes. Without further assumptions, no experiment would be able to distinguish them. 
\end{examp}

\subsubsection{Purity and completely mixed states}
An important quantity when dealing with mixed states is its purity. The purity $P$ is a measure of how close a quantum state $\rho\in\mathcal{D}(\mathcal{H})$ is to be a pure state. The quantity $P$ is a positive real value, $P: \mathcal{D}(\mathcal{H})\rightarrow \mathbb{R}$ and it is defined as
\begin{align}
    P\equiv {\rm Tr}[\rho^2].
\end{align}
The maximum value that the purity can take is $1$, meaning that $\rho$ is a pure state. If $\rho=\ket{\psi}\bra{\psi}$, which corresponds to Eq.~\eqref{eq:rho} where only one coefficient is $1$ and the rest zero, then
\begin{align}
    \rho^2=\rho \ \rho= \ket{\psi}\bra{\psi}\ \ket{\psi}\bra{\psi}=\ket{\psi}\bra{\psi}=\rho.
\end{align}
 Therefore, for $\rho$ pure, it follows
 \begin{align}
     P={\rm Tr}[\rho^2]={\rm Tr}[\rho]=1.
 \end{align}
This is a sufficient and necessary condition, i.e.
\begin{align}
    P=1 \Leftrightarrow \rho=\ket{\psi}\bra{\psi}, \ \rho\ {\rm is \ a \ pure \ state}.
\end{align}
If the state $\rho$ is pure, it means that the spectral decomposition in Eq.~\eqref{eq:rho_spectral} yields $\alpha_1=1$ and $\alpha_{m>1}=0$. 

On the contrary, one may ask the minimum value that the purity can take. What is the state $\rho$ that minimizes the purity? Using the general spectral decomposition of $\rho$, we can write
\begin{align}
\rho^2=\sum_{n=1}^{d_H}\sum_{m=1}^{d_H}\alpha_m \alpha_n\ket{\varphi_n}\bra{\varphi_n}\ \ket{\varphi_m}\bra{\varphi_m} = \sum_{n=1}^{d_H} \alpha_m \alpha_n \ket{\varphi_n} \delta_{n,m}\bra{\varphi_m}=\sum_{n=1}^{d_H}\alpha_n^2\ket{\varphi_n}\bra{\varphi_n}
\end{align}
so that
\begin{align}
    P={\rm Tr}[\rho^2]={\rm Tr}\left[ \sum_{n=1}^{d_H}\alpha_n^2\ket{\varphi_n}\bra{\varphi_n}\right]=\sum_{n=1}^{d_H}\alpha_n^2.
\end{align}
 When $\alpha_n=1/d_H, \forall n\in\{1,\ldots,d_H\}$~\footnote{This can be shown using the Cauchy-Schwarz inequality. Suppose two vectors in a $d_H$-dimensional space, such that $\ket{u}=\sum_{n=1}^{d_H} \alpha_n\ket{n}$ and $\ket{v}=\sum_{n=1}^{d_H}\ket{n}$. Then, the inequality reads $|\langle u|v\rangle|^2\leq \langle u|u\rangle \langle v |v\rangle$. Substituting the expressions, we find $(\sum_{n=1}^{d_H} \alpha_n)^2 \leq (\sum_{n=1}^{d_H})(\sum_{n=1}^{d_H}\alpha_n^2)\Rightarrow \sum_{n=1}^{d_H} \alpha_n^2\geq 1/d_H$, where we have already used $\sum_{n=1}^{d_H}\alpha_n=1$. Hence, the minimum value reads $\sum_{n=1}^{d_H}\alpha_n^2=1/d_H$.}. the minimum $P$ is obtained, constrained to the normalization $\sum_{n=1}^{d_H}\alpha_n=1$. This state is known as the completely or maximally mixed state or random state,
\begin{align}
    \rho_{cm}=\frac{1}{d_H}\sum_{n=1}^{d_H}\ket{\varphi_n}\bra{\varphi_n}=\frac{1}{d_H}\mathbb{I},
\end{align}
where we have used the resolution of the identity, $\mathbb{I}=\sum_{n}^{d_H}\ket{\varphi_n}\bra{\varphi_n}$. Therefore, the purity $P$ takes values in the range
\begin{align}
    \frac{1}{d_H}\leq P\leq 1.
\end{align}

The completely mixed state $\rho_{cm}$ is simply proportional to the identity. It is a relevant state and will appear multiple times throughout the notes. The $\rho_{cm}$ state is also known as random state. For a single qubit, $d_H=2$, the probability of finding the system $\rho_{cm}$  in any quantum state $\ket{\varphi}$ is $1/2$ (and therefore also $1/2$ for the orthogonal state $\ket{\varphi_\perp}$). We will come back to this in Box~\ref{examp:rho_random}.  In addition, we recall that there are infinitely different ensembles that result in $\rho_{cm}$. 

Just to stress it again, the purity of the maximally mixed or random state takes its minimum value
\begin{align}
    P={\rm Tr}[\rho_{cm}^2]=\sum_{n=1}^{d_H}\frac{1}{d_H^2}=\frac{1}{d_H}.
\end{align}
The completely mixed state is diagonal, whose entries are all equal; it is a balanced classical mixture of all possible states (or simply, same probability over a complete set of orthonormal states). 

\subsubsection{Quantum coherence}
The state $\rho\in\mathcal{D}(\mathcal{H})$ provides a matrix representation to describe quantum systems. As we have seen, the diagonal entries correspond to the probabilities and must add up to $1$ (normalization). The off-diagonal entries are known as the coherent contribution. Since the state $\rho$ can be expressed in different basis, it should be clear that the coherence of a state is a basis-dependent quantity. Indeed, any quantum state $\rho$ contains, by definition, no coherence in its eigenbasis. 

Take for example a qubit which is found with probability $3/4$ in the state $\ket{+}=\frac{1}{\sqrt{2}}(\ket{1}+\ket{0})$, and with probability $1/4$ in the $\ket{-}=\frac{1}{\sqrt{2}}(\ket{1}-\ket{0})$, then
\begin{align}
    \rho=\frac{3}{4}\ket{+}\bra{+}+\frac{1}{4}\ket{-}\bra{-}.
\end{align}
In the $\{\ket{+},\ket{-} \}$ basis, we can write
\begin{align}
    \rho=\begin{pmatrix} 3/4 & 0\\ 0 & 1/4       
    \end{pmatrix},
\end{align}
and therefore the off-diagonal terms are simply zero --no coherence, as $\rho$ is already written in its eigenbasis. However, if we express the state $\rho$ in the computational basis $\{\ket{1},\ket{0}\}$, then
\begin{align}\label{eq:examplerho}
    \rho=\begin{pmatrix} 1/2 & 1/4\\ 1/4 & 1/2   
    \end{pmatrix},
\end{align}
which reveals a non-zero coherence given that the entries in the off-diagonal amount to $1/4$.

Importantly, completely mixed states show zero coherence in any basis of $\mathcal{H}$. Since $\rho_{cm}\propto \mathbb{I}$, any change of basis renders the identity unaffected. Indeed, a change of basis corresponds to a unitary transformation, $U$. Let us denote $\tilde{\rho}_{cm}$ the completely mixed state in a different basis, then
\begin{align}
    \tilde{\rho}_{cm}=U\rho_{cm} U^\dagger=U \left(\frac{1}{d_H} \mathbb{I}\right) U^\dagger=\frac{1}{d_H}\mathbb{I}=\rho_{cm}.
\end{align}

A measure of this basis-dependent quantum coherence can be defined as~\cite{Baumgratz2014}
\begin{align}
    C(\rho)\equiv \sum_{\substack{n,m \\ n\neq m}} |\rho_{n,m}|,
\end{align}
where $\rho_{n,m}$ denote the $n,m$ matrix elements of $\rho$, $\rho_{n,m}=\bra{\varphi_n}\rho\ket{\varphi_m}$, so that $C:\mathcal{D}(\mathcal{H})\rightarrow \mathbb{R}$, and $\{\ket{\varphi_m}\}$ the states that form the orthonormal basis chosen to represent the state.  In the example given in Eq.~\eqref{eq:examplerho}, it follows that $C(\rho)=1/4+1/4=1/2$. It can be shown that the states that maximize the coherence in the basis $\{\ket{\varphi_n}\}$ (known as maximally coherent states) are pure, $\rho_{coh}=\ket{\psi_{coh}}\bra{\psi_{coh}}$, being
\begin{align}
    \ket{\psi_{coh}}=\frac{1}{\sqrt{d_H}}\sum_{n=1}^{d_H}\ket{\varphi_n},
\end{align}
so that $\rho_{coh}=\frac{1}{d_H}\sum_{n,m=1}\ket{\varphi_n}\bra{\varphi_m}$ and therefore
\begin{align}
    C(\rho_{coh})=\sum_{\substack{n,m \\ n\neq m}} |\rho_{coh,n,m}|=d_H-1,
\end{align}
meaning that the measure of quantum coherence $C(\rho)$ takes values in the range
\begin{align}
    0\leq C(\rho)\leq d_H-1.
\end{align}

\subsubsection{Single qubits}
If the system of interest is a qubit, i.e. a quantum system with $d_H=2$, is particularly simple and allows for a graphical representation to draw quantum states, i.e. the Bloch sphere. 

An arbitrary ($2\times 2$) Hermitian matrix has four real parameters and can be expanded in the basis of operators of $SU(2)$, i.e. $\{\mathbb{I},\sigma_x,\sigma_y,\sigma_z\}$ consisting of the identity and the three Pauli matrices. Since the Pauli matrices are traceless, the coefficient of $\mathbb{I}$ in the expansion of a density matrix $\rho$ must be $\frac{1}{2}$, in order to have ${\rm Tr}[\rho]=1$. Thus $\rho$ may be expressed as
\begin{equation}
    \label{rho2}
    \rho=\frac{1}{2}(\mathbb{I}+\vec v\cdot\sigma)=\frac{1}{2}\begin{pmatrix}1+v_z\ v_x-iv_y\\ v_x+iv_y\ 1-v_z\end{pmatrix},
\end{equation}
where $\vec v=(v_x,v_y,v_z)$ and $\vec\sigma=(\sigma_x,\sigma_y,\sigma_z)$. The vector $\vec v$ is called the Bloch vector for the density operator $\rho$.  Any real Bloch vector $\vec v$ leads into a Hermitian operator $\rho$ with ${\rm Tr}[\rho]=1$, but in order for $\rho$ to be a density operator it must also be non-negative (i.e. $\rho\geq 0$ or equivalently $\bra{\phi}\rho\ket{\phi}\geq 0 \ \forall \ket{\phi}$). Which Bloch vectors yield legitimate density operators? That is, what does the non-negative condition on $\rho$ translate to in terms of the Bloch vectors $\vec v$?

To answer this, let us compute the eigenvalues of $\rho$. The trace of a matrix is equal to the {\itshape sum} of its eigenvalues, and the determinant is equal to the {\itshape product} of its eigenvalues. We know that ${\rm Tr}[\rho]=1$, and we can calculate ${\rm det}[\rho]$ from \eqref{rho2}
\begin{equation}
    {\rm det}[\rho]=\frac{1}{4}(1-|v|^2)=\frac{1}{2}(1-|v|) \ \frac{1}{2}(1+|v|),
\end{equation}
where $|\vec v|=\sqrt{v_x^2+v_y^2+v_z^2}$. It follows that the two eigenvalues of $\rho$ are $\frac{1}{2}(1\pm |v|)$. For 
$\rho$ to be non-negative, its eigenvalues have to be non-negative, and so $|v|$ (the length of the Bloch vector) cannot exceed $1$, i.e. $|\vec{v}|\leq 1$. Therefore, a general qubit state will lie not just on the Bloch sphere, i.e. not only states with Bloch vectors of magnitude $1$ (pure states), but instead within the sphere of vectors of magnitude {\itshape less} than or {\itshape equal} to $1$.

\begin{figure}[t!]
\begin{center}
\includegraphics[width=6.5cm]{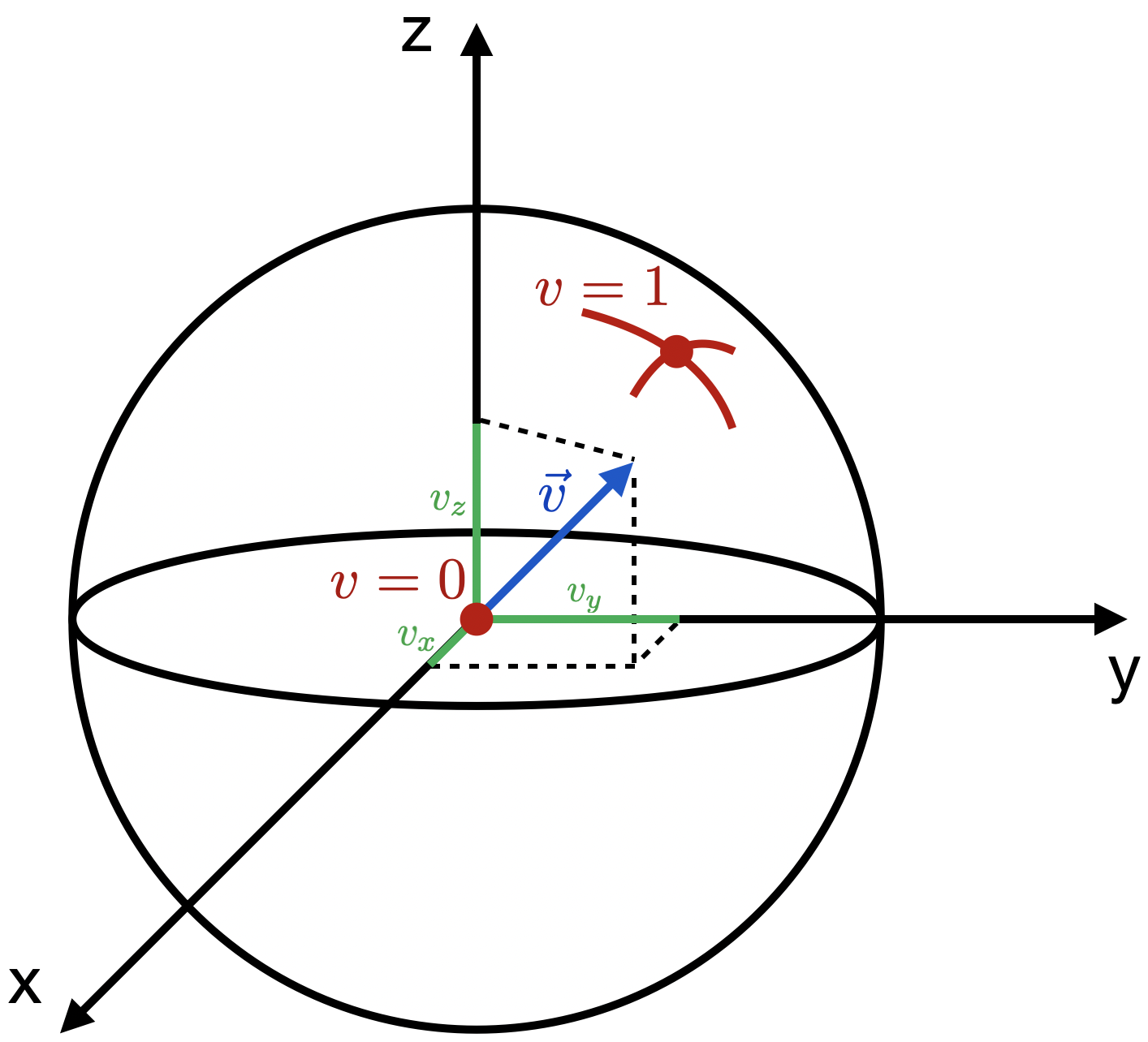}
\caption{Representation of a general qubit state $\rho$ on the Bloch sphere. The relation between the position of the state on the ball and the density matrix $\rho$ is established by the projections of the Bloch vector $\vec v$ on the Cartesian axis according to Eq. \eqref{rho2}. Pure states $\ket{\psi}$ ($|\vec v|=1$) are located on the surface while the completely mixed state $\rho_{cm}=\mathbb{I}/2$ ($|\vec v|=0$) is placed at the origin.}
\label{bloch-qubit}
\end{center}
\end{figure}
%

We can now visualize the convex set of ($2\times 2$) density matrices as a unit sphere in three-dimensional Euclidean space, see Fig.\ref{bloch-qubit}: the extrema points, which represent pure states, are the points on the boundary ($\vec v$ such that $|\vec v|=1$), i.e. on the surface of the sphere; the maximally mixed state $\mathbb{I}/2$ corresponds to $|v|=0$, i.e. the center of the sphere. In general, the length of the Bloch vector $\vec{v}$ can be thought of as the ``purity'' of a state. In particular, since the eigenvalues of the density matrix are $\frac{1}{2}(1\pm |v|)$, it follows that
\begin{align}
    {\rm Tr}[\rho^2]=\frac{1}{4}(1+|v|)^2+\frac{1}{4}(1-|v|)^2=\frac{1}{2}(1+|v|^2).
\end{align}
For higher dimensions, $d_H>2$,  the visualization becomes unfortunately much more complicated. 

\subsubsection{Measurements}

The measurement axiom of quantum mechanics states that, after a measurement, the state of a quantum system collapses according to the measured outcome. In this context, one typically  distinguishes between projective-valued measurements (PVMs) and positive-operator valued measurements (POVMs). 

\vspace{0.5cm}
\noindent\textbf{PVMs}
\vspace{0.2cm}

For that let us start with the basic notion of a measurement, known as projection-valued measurement (PVM). The expectation value an observable $A$, represented by a Hermitian operator $A$, i.e. $A\in\mathbb{H}(\mathcal{H})$,  when measuring a system in the state $\rho$ is
\begin{align}
    \langle A\rangle ={\rm Tr}[\rho A].
\end{align}
Since we are dealing with measurement outcomes, we can write
\begin{align}
    \langle A\rangle = \sum_n p_n a_n,
\end{align}
where $a_n$ are the eigenvalues of $A$ (possible outcomes) and $p_n$ the corresponding probability of recording the outcome $a_n$ when measuring it over $\rho$. In particular, writing $A$ via its spectral decomposition ($a_n\in \mathbb{R}$ since $A$ is an observable, i.e. Hermitian)
\begin{align}
    A=\sum_n a_n \ket{a_n}\bra{a_n},
\end{align}
with $\langle a_n|a_m \rangle=\delta_{n,m}$, then we can construct projectors for each of the eigenstates of $A$, i.e. $\Pi_n=\ket{a_n}\bra{a_n}$ so that $\Pi_n^2=\Pi_n$. In this manner, we can find the Born rule for density matrices: The  probability $p_n$ reads as
\begin{align}
    p_n=\langle \Pi_n\rangle={\rm Tr}[\rho \Pi_n]={\rm Tr}[\rho\ket{a_n}\bra{a_n}]=\bra{a_n}\rho\ket{a_n}. 
\end{align}

A state that has been measured, and the outcome $a_n$ has been recorded, will be updated according to
\begin{align}
    \rho\xrightarrow{{\rm measured \ outcome\ a_n}} \rho'=\Pi_n \rho \Pi_n^\dagger =\ket{a_n}\bra{a_n}\rho \ket{a_n}\bra{a_n} = p_n \ket{a_n}\bra{a_n}.
\end{align}
However, one must be careful since the previous expression is not a valid quantum state since, in general, it is not properly normalized (${\rm Tr}[\rho']\leq 1$) since $p_n\leq 1$ and $\ket{a_n}$ is a pure quantum state. Therefore, the valid post-measurement state conditioned to the outcome $a_n$ reads as
\begin{align}
    \rho\xrightarrow{{\rm measured \ outcome\ }a_n} \rho'=\frac{\Pi_n \rho \Pi_n^\dagger}{{\rm Tr}[\Pi_n \rho \Pi_n^\dagger]},
\end{align}
which is simply $\rho'=\ket{a_n}\bra{a_n}$. 
The previous expression denotes the irreversible change that a state $\rho$ suffers due to a selective measurement. 
The term ${\rm Tr}[\Pi_n \rho \Pi_n^\dagger]$, as we have seen above, corresponds to the probability $p_n$ that this outcome occurs (Born rule). This mathematical expression is the formal way to write that the state $\rho$ must be replaced by another according to the new information we have extracted from it. It turns out that the measurement problem of quantum mechanics is intimately linked with the open quantum dynamics that happens during the interaction between the system and the measurement apparatus! 

As a follow-up of the previous derivation is the scenario of non-selective measurements: What is the state $\rho'$ after a measurement if the outcome is not accessible or recorded (it cannot be retrieved)? This is equivalent, from a mathematical point of view, to the post-measurement state $\rho'$ averaged over all possible outcomes $a_n$.  In this case, the post-measurement state results from an incoherent superposition (i.e. a classical  mixture) of all possible outcomes:
\begin{align}
    \rho \xrightarrow{{\rm unknown \ outcome\ or \ averaged \ post-measurement}} \rho'=\sum_{n}\Pi_n \rho \Pi_n^\dagger=\sum_n p_n \ket{a_n}\bra{a_n}.
    \end{align}
The state $\rho'$ in this non-selective scenario is already normalized, given that $\sum_n p_n=1$ and $\{\ket{a_n}\}$ form an orthonormal basis of $\mathcal{H}$.

\begin{examp}[Completely mixed or random state]{examp:random}\label{examp:rho_random}
Let us show why $\rho_{cm}=\frac{1}{d_H}\mathbb{I}$ is known as random state. For simplicity, we consider a qubit $d_H=2$. In this manner, we can compute the expectation values of $\sigma_j$ with $j\in\{x,y,z\}$, 
\begin{align}
    \langle \sigma_j\rangle={\rm Tr}[\rho_{cm} \sigma_j]= \frac{1}{2}{\rm Tr}[\mathbb{I}\sigma_j]=0,
\end{align}
    since ${\rm Tr}[\sigma_j]=0$, the Pauli matrices are traceless. The average values are zero, meaning that with probability $1/2$ the outcomes $\pm 1$ are recorded. If we denote by $\ket{\alpha_+}$ and $\ket{\alpha_-}$ two orthogonal states and write a generic qubit observable as $A=\ket{\alpha_+}\bra{\alpha_+}-\ket{\alpha_-}\bra{\alpha_-}$, then 
    \begin{align}
        p_{\pm 1}={\rm Tr}[\rho_{cm} \Pi_\pm]=\frac{1}{2}{\rm Tr}[\mathbb{I}\ \ket{\alpha_\pm}\bra{\alpha_\pm}]=\frac{1}{2}{\rm Tr}[ \left( \ket{\alpha_+}\bra{\alpha_+}+\ket{\alpha_-}\bra{\alpha_-}\right)\ket{\alpha_\pm}\bra{\alpha_\pm}]=\frac{1}{2}.
    \end{align}
    That is, regardless of the observable $A$, the outcomes $\pm 1$ occur randomly with probability $1/2$.     
\end{examp}

\vspace{0.5cm}
\noindent\textbf{POVMs}
\vspace{0.2cm}

POVMs (Positive Operator-Valued Measures) describe the most general class of quantum measurements. While PVMs  are restricted to measurements represented by mutually orthogonal projectors, this constraint is not necessary in the general case. In fact, many realistic measurement processes do not correspond to projections onto orthogonal subspaces—due to noise, imperfect detectors, or deliberately generalized measurement schemes. POVMs therefore provide a more flexible and natural framework for describing quantum measurements in both theory and experiment.

Mathematically, a POVM is a set of positive semi-definite Hermitian matrices $\{M_n\}$ that add up to the identity. In other words, they form a set of operators that are non-negative and fulfill the completeness relation:
\begin{align}\label{eq:POVM_id}
    \sum_{n=1}^N M_n=\mathbb{I}.
\end{align}
In this manner, a POVM encompasses PVM when $\{M_n\}$ are the orthogonal projectors, $\sum_n \Pi_n=\mathbb{I}$. While the number of elements in a PVM is smaller or equal than the dimension of the Hilbert space $d_H$, lifting the orthogonal condition among different $M_n$ leads to the fact that the number of elements in a POVM can be actually larger than $d_H$.


If we denote by $n$ the recorded outcome associated to measuring $M_n$, then the probability of such outcome in a state $\rho$ reads as
\begin{align}
    p_n={\rm Tr}[\rho M_n].
\end{align}
Now, each of the positive semi-definite Hermitian operators $M_n$ can be further decomposed using the so-called Kraus operators, that is,
\begin{align}\label{eq:Kraus}
    M_n=K_n^\dagger K_n.
\end{align}
The $K_n$ are not necessarily Hermitian (self-adjoint) but $K_n^\dagger K_n$ must be, since $M_n^\dagger=M_n$. 
In this manner, the Born rule or probability to obtain the outcome $n$ from a POVM can be written as
\begin{align}
    p_n={\rm Tr}[\rho M_n]={\rm Tr}[\rho K_n^\dagger K_n]={\rm Tr}[K_n \rho K_n^\dagger],
\end{align}
and the post-measurement state, conditioned to the outcome $m$ (associated to $M_m$) reads as
\begin{align}
    \rho\xrightarrow{{\rm measured \ outcome \ m}} \frac{1}{p_m}K_m \rho K_m^\dagger,
\end{align}
where again the normalization $1/p_m$ is introduced so that the post-measurement state $\rho$ is a valid physical state (normalized).
In a similar fashion as for PVMs, the state $\rho$ after a POVM $\{M_n\}$ averaged over all possible outcomes, is given by
\begin{align}\label{eq:rho_av_meas}
    \rho_{av}=\sum_n K_n \rho K_n^\dagger.
\end{align}

\begin{examp}[A realistic single-photon detector]{examp:POVMexample}\label{examp:POVMexample}
Let us illustrate a POVM with an example: The measurement of a single photon with a realistic detection efficiency $0\leq \eta\leq 1$. For $\eta=1$, the detection is ideal, while for $\eta=0$ no measurement is performed. The measurement apparatus clicks if a photon has been detected, while it remains idle otherwise. This measurement can be described as a POVM with two operators, 
\begin{align}
    M_{click}&=\eta\ket{1}\bra{1},\\
    M_{no-click}&=\ket{0}\bra{0}+(1-\eta)\ket{1}\bra{1},
\end{align}
where $\ket{1}$ and $\ket{0}$ correspond to states containing $1$ or $0$ photons, respectively. Clearly, $M_{click}+M_{no-click}=\mathbb{I}$, so they satisfy the condition~\eqref{eq:POVM_id}, and are positive semi-definite Hermitian operators. For $\eta=1$ (ideal detection efficiency), the POVM reduces to a PVM as $M_{click}$ and $M_{no-click}$ become orthogonal projectors. The Kraus operators can be easily obtained as $M_{click}$ and $M_{no-click}$ are already diagonal in the computational basis, so that
\begin{align}
    K_{click}&=\sqrt{\eta}\ket{1}\bra{1},\\
    K_{no-click}&=\ket{0}\bra{0}+\sqrt{1-\eta}\ket{1}\bra{1},
\end{align}
which produce $K^\dagger_{click}K_{click}=M_{click}$, and $K^\dagger_{no-click}K_{no-click}=M_{no-click}$. 
Suppose a photon in a generic pure state of the form $\rho=\ket{\psi}\bra{\psi}$ with $\ket{\psi}=\alpha\ket{1}+\beta\ket{0}$ and $|\alpha|^2+|\beta|^2=1$. Then, if the detector clicks (assuming $\eta>0$), the post-measurement state becomes
\begin{align}
    \rho_{click}=\frac{1}{p_{click}}K_{click}\rho K^\dagger_{click}=\ket{1}\bra{1}.
\end{align}
This is expected given that a click can only take place if there is a photon (at least under the assumptions considered for this example). To the contrary, if there is no click, then
\begin{align}
    \rho_{no-click}&=\frac{1}{p_{no-click}}K_{no-click}\rho K^\dagger_{no-click}\nonumber\\&=\frac{1}{1-\eta|\alpha|^2}\left(|\alpha|^2(1-\eta)\ket{1}\bra{1}+|\beta|^2\ket{0}\bra{0}+\sqrt{1-\eta}\left[ \alpha \beta^* \ket{1}\bra{0}+\alpha^*\beta\ket{0}\bra{1}\right]\right).
\end{align}
Note that in general the state $\rho_{no-click}$ has a more involved expression than the one obtained for the ideal detector; setting  $\eta=1$, the state $\rho_{no-click}$ reduces to the expected $\ket{0}\bra{0}$ where the absence of click means that the state had no photons. 

In addition, one can compute the average state over many measurements. If $\eta=1$, one expects simply the mixture of having $0$ and $1$ photons, i.e. $\rho_{av}=|\alpha|^2\ket{1}\bra{1}+|\beta|^2\ket{0}\bra{0}$. For imperfect detectors $\eta<1$, however, the average state still keeps some coherence in the computational basis,
\begin{align}
    \rho_{av}&=K_{click}\rho K_{click}^\dagger+K_{no-click}\rho K_{no-click}^\dagger\nonumber\\&=|\alpha|^2\ket{1}\bra{1}+|\beta|^2\ket{0}\bra{0}+\sqrt{1-\eta}\left[a\beta^*\ket{1}\bra{0}+\alpha^*\beta\ket{0}\bra{1} \right].
\end{align}

\end{examp}

\subsubsection{Closed dynamics: von Neumann equation}

A closed system prepared in a pure quantum state evolves according to the Schr\"odinger equation, as we have already revisited above. Since the density matrix is a convex combination of pure states, one could compute the closed-system evolution of each of the states that form the ensemble $\{\ket{\phi_n}\}$ and then perform the ensemble at every time instant. That is, from
\begin{align}
    \ket{\phi_n(t)}=U(t,t_0)\ket{\phi_n(t_0)},
\end{align}
being $U(t,t_0)$ the time-evolution operator,  the density matrix at time $t_0$
\begin{align}
    \rho(t_0)=\sum_{n} c_n \ket{\phi_n(t_0)}\bra{\phi_n(t_0)},
\end{align}
evolves in time as 
\begin{align}
    \rho(t)&=\sum_n c_n \ket{\phi_n(t)}\bra{\phi_n(t)}=\sum_n c_n U(t,t_0)\ket{\phi_n(t_0)}\bra{\phi_n(t_0)}U^\dagger(t,t_0)\nonumber\\&=U(t,t_0)\left[\sum_{n}c_n \ket{\phi_n(t_0)}\bra{\phi_n(t_0)} \right]U^\dagger(t,t_0)=U(t,t_0)\rho(t_0)U^\dagger (t,t_0).
\end{align}
However, we can compute the dynamics of the density matrix alone. 
If we take the time derivative to this last expression we find
\begin{align}
    \frac{d}{dt}\rho(t)&=\frac{d}{dt} \left[U(t,t_0)\rho(t_0)U^\dagger (t,t_0) \right]\nonumber\\&=-i H(t)\left[U(t,t_0)\rho(t_0)U^\dagger (t,t_0) \right]+\left[U(t,t_0)\rho(t_0)U^\dagger (t,t_0)(+i H(t) \right]\nonumber\\&=-iH(t) \rho(t)+i \rho(t)H(t)=-i[H(t),\rho(t)],
\end{align}
where we have used again Eq.~\eqref{eq:ddtU}, $\frac{d}{dt}U(t,t_0)=-iH(t)U(t,t_0)$. 
In this manner, we have shown that the density matrix $\rho$ evolves according to  
\begin{align}
    \frac{d}{dt}\rho(t)=-i[H(t),\rho(t)],
\end{align}
which is known as the von Neumann equation \cite{von27}. This is equivalent to the Schr\"odinger equation but extended to deal with general quantum states that can be either pure or mixed, i.e. for any  density matrix.

\subsubsection{Quantum operations or channels}

\begin{figure}
    \centering
    \includegraphics[width=0.8\linewidth]{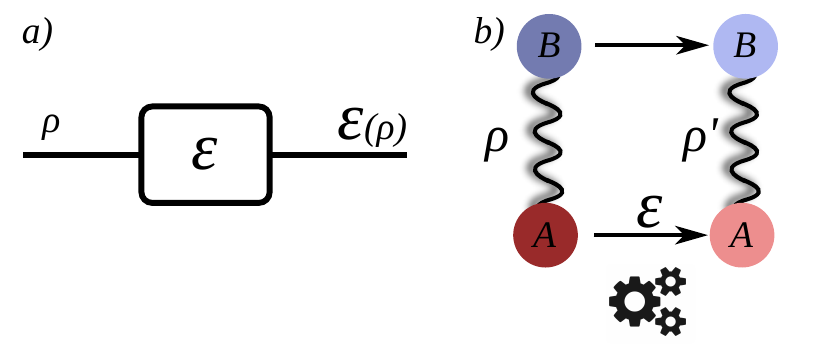}
    \caption{Quantum channels. The abstract representation of a quantum channel is shown in panel a), where an input state $\rho$ undergoes an operation $\mathcal{E}$ that results in the output state $\mathcal{E}(\rho)$. The panel b) shows the concept of complete positivity, that is, given a bipartite system whose global state $\rho$ is entangled, the action of a quantum channel $\mathcal{E}$ only on one of the parties must result in a physical state, i.e. $\mathcal{E}$ and $\mathcal{E}\otimes \mathbb{I}$ must be positive. This guarantees that the output state $\rho'=(\mathcal{E}\otimes \mathbb{I})(\rho)$ is physical for any input state $\rho$. }
    \label{fig:QChannel}
\end{figure}

Quantum states can be manipulated, i.e. they can undergo a transformation due to the action of an external agent, or simply by the dynamics dictated by the system. These quantum operations or channels can be described formally as linear maps \cite{Sudarshan}, defined as $\mathcal{E}: \mathcal{D}(\mathcal{H})\rightarrow \mathcal{D}(\mathcal{H})$. That is, $\mathcal{E}(\rho)$ represents a valid quantum operation or channel if it transforms a physical state $\rho\in\mathcal{D}(\mathcal{H})$ into another physical state $\mathcal{E}(\rho)\in\mathcal{D}(\mathcal{H})$. Although it might seem that these requirements are enough to define any channel or operations, they are actually insufficient. The reason is purely quantum mechanical due to the possibility of having entanglement between different systems. Suppose the following scenario: A qubit $A$ is entangled with another qubit $B$, and then one performs a quantum operation $\mathcal{E}$  only onto $A$, while $B$ remains idle (cf. Fig.~\ref{fig:QChannel}). It turns out that the previous requirements do not guarantee that the output state is physical. This extra condition is precisely what we need to impose to ensure physical operations. That is, $\mathcal{E}(\rho)$ must be  complete positivity, which refers to $\mathcal{E}\otimes \mathbb{I}$ being a positive operation, i.e. the quantum channel on $A$ plus doing nothing on $B$ must be a positive operation. A valid physical quantum channel or operation $\mathcal{E}(\rho)$ fulfills the following properties \cite{Nielsen}:
\begin{itemize}
    \item[i)] $\mathcal{E}(\rho)$ is a linear map: $\mathcal{E}(p_1\rho_1+p_2\rho_2)=p_1\mathcal{E}(\rho_1)+p_2\mathcal{E}(\rho_2)$. 
    \item[ii)] $\mathcal{E}(\rho)$ maps physical states into physical states: $\mathcal{E}: \mathcal{D}(\mathcal{H})\rightarrow \mathcal{D}(\mathcal{H})$. Therefore:
    \begin{itemize}
    \item[a)] Preserves Hermiticity: $\mathcal{E}(\rho)=\mathcal{E}(\rho)^\dagger$.
    \item[b)] Preserves trace: ${\rm Tr}[\mathcal{E}(\rho)]=1, \forall \rho\in \mathcal{D}(\mathcal{H})$.
    \item[c)] Preserves positivity: $\mathcal{E}(\rho)\geq 0, \forall \rho\in \mathcal{D}(\mathcal{H})$. 
    \end{itemize}
    \item[iii)] Complete positivity: $(\mathcal{E}\otimes \mathbb{I})(\rho)\geq 0, \forall \rho\in\mathcal{D}(\mathcal{H}\otimes \mathcal{H})$.  
\end{itemize}
The last condition can be checked by verifying that $(\mathcal{E}\otimes \mathbb{I})(\rho_{max-ent})$ is positive where $\rho_{max-ent}$ denotes a maximally entangled state in the bipartite system $\mathcal{H}\otimes \mathcal{H}$. 

Luckily, there is an important result, known as  Kraus' theorem \cite{kraus83}, that states that if a quantum channel can be written as an operator sum of the form \cite{Nielsen}
\begin{align}\label{eq:KrausQC}
    \mathcal{E}(\rho)=\sum_{n=1}^r K_n \rho K_n^\dagger,
\end{align}
where $\{K_n\}_{n=1,\ldots,r}$ are the Kraus operators satisfying the completeness relation
\begin{align}\label{eq:Kraus_trace}
    \sum_{n=1}^r K_n^\dagger K_n=\mathbb{I},
\end{align}
with $K_n^\dagger K_n\geq 0$ (non-negative), then $\mathcal{E}(\rho)$ is a complete positive and trace preserving  (CPTP) quantum channel. The smallest possible number $r$ is known as Kraus rank.  The linearity and Hermiticity conditions are clearly satisfied by any $\mathcal{E}(\rho)$ of the form in Eq.~\eqref{eq:KrausQC}. The trace preserving character follows from Eq.~\eqref{eq:Kraus_trace}. Moreover,  one can show that $\mathcal{E}(\rho)$ is CPTP $\iff$ $\mathcal{E}(\rho)=\sum_n^r K_n\rho K_n^\dagger$. Yet, verifying this statement is more involved; see Sec.~\ref{moremat:CPTP_Ep} if you are interested in the details of the proof. 

The expression in Eq.~\eqref{eq:KrausQC} should already be familiar:
\begin{itemize}
    \item Quantum channels can be viewed as the result of non-selective measurements, or equivalently, the result of an average over the different measurement outcomes, cf. Eq.~\eqref{eq:rho_av_meas}.
    \item Unitary time evolution is a simple quantum channel where  $\rho(t)=U(t,t_0)\rho(t_0)U^\dagger(t,t_0)$. Comparing this expression with Eq.~\eqref{eq:KrausQC}, we find $r=1$ and $K_1=U(t,t_0)$.  This should not be surprising as time evolution is a basic quantum operation.
    \item Quantum gates: Any quantum gate is described by a unitary $U$. Thus, similarly to unitary time evolution, $r=1$ with a unique Kraus operator $K_1=U$. 
\end{itemize}

\subsection{Towards open quantum systems: Subsystems}
We are now almost ready to begin our journey into open quantum systems, the focus of Sec.~\ref{c:OQS}. Before doing so, however, one final concept must be properly defined to streamline the derivations ahead: the physical notion of a subsystem and the mathematical tools required to describe it.
\begin{figure}
    \centering
    \includegraphics[width=0.6\linewidth]{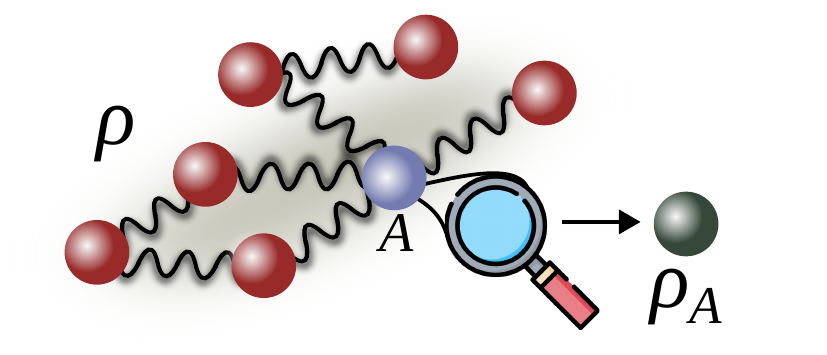}
    \caption{Graphical representation of a collection of subsystems, whose whole state is $\rho$. If we are interested solely on one of the subsystems, say $A$ (in blue), the reduced state $\rho_A$ consists in discarding the information about the subsystems we choose not to consider (in red). }
    \label{fig:sys_vs_sub}
\end{figure}
Let us consider a set of different subsystems $s_j$ with $j=1,\ldots, N$ being $N$ the total number of subsystems.  Each of these subsystems is made of a quantum register, whose Hilbert space is $\mathcal{H}_j$. Hence, any state of the total space $S$ lives in the tensor-product Hilbert space $\mathcal{H}=\otimes_{j=1}^N \mathcal{H}_j$. This is graphically sketched in Fig.~\ref{fig:sys_vs_sub}. Suppose that the initial state of the entire collection of subsystems is $\rho$. However, we have only access --or are interested in-- one or a few of these subsystems. How, then, should we describe the state of the relevant parts? The formal answer is given below, i.e. we need to trace out the degrees of freedom of the uninteresting subsystems. Physically, this amounts to discarding information about the parts we choose not to consider. At first glance, this may sound deceptively simple, reflecting our classical intuition. Yet in quantum mechanics, the act of tracing out subsystems can profoundly alter the states of those that remain.

\subsubsection{Partial trace}

The partial trace refers to tracing out one (or more) subsystems from the entire state. For simplicity, let us consider a bipartite case, with subsystems $A$ and $B$, so that the whole state is $\rho_{AB}\in\mathcal{D}(\mathcal{H}_A\otimes \mathcal{H}_B)$. Then, the reduced state of the subsystem $A$ is obtained by performing the partial trace over $B$, 
\begin{align}
    \rho_A={\rm Tr}_B[\rho_{AB}]=\sum_{n=1}^{d_B} \bra{n}_B \ \rho_{AB}\ \ket{n}_B,
\end{align}
where $\sum_{n=1}^{d_B}\ket{n}_B\bra{n}_B=\mathbb{I}_B$ is the closure or completeness relation in the Hilbert space of $B$ of dimension $d_B$. A crucial point here is that, if $A$ and $B$ are not entangled, then the partial trace over $B$ or $A$ leaves the state for each subsystem unaltered:
\begin{align}
    \rho_{AB}&=\sum_n \lambda_n \rho_n^A\otimes \rho_n^B \Rightarrow \rho_A={\rm Tr}_B[\rho_{AB}]\nonumber\\&=\sum_n \lambda_n \rho_n^A \otimes \sum_j \bra{j}_B \rho_n^B\ket{j}_B=\sum_n \lambda_n \rho_n^A {\rm Tr}_B[\rho_n^B]=\sum_n \lambda_n \rho_n^A,
\end{align}
and similarly for taking the trace over $B$,
\begin{align}
    \rho_{B}=\sum_n \lambda_n \rho_n^B.
\end{align}
As commented above, the physical interpretation of the partial trace over $B$ is that we ignore that particular system, i.e. we are just interested in describing the state locally for $A$. This is going to be crucial in the context of open quantum systems. For example, if we cannot access, measure or control $B$.   If there is entanglement between $A$ and $B$, then there is information that is lost when neglecting the other subsystem, and as a consequence $\rho_A$ will be different than when consider the whole state for $AB$. 

To remark this point, let us consider one of the Bell states, so that $\rho_{AB}=\ket{\phi^+}\bra{\phi^+}$ with $\ket{\phi^+}=\frac{1}{\sqrt{2}}(\ket{00}+\ket{11})$, 
then
\begin{align}\label{eq:rhoAB_Bell}
    \rho_{AB}=\frac{1}{2}\left(\ket{00}\bra{00}+\ket{00}\bra{11}+\ket{11}\bra{00}+\ket{11}\bra{11}\right) =\frac{1}{2}\begin{bmatrix} 1 & 0& 0& 1\\ 0 & 0 & 0& 0\\ 0 & 0 & 0& 0\\ 1& 0& 0& 1
    \end{bmatrix}.
\end{align}
Now, let us ask: How would the qubit $A$ look like if we simply ignore the qubit $B$? This is precisely what the partial trace does, i.e. we need to compute $\rho_A={\rm Tr}_B[\rho_{AB}]$, 
\begin{align}
\rho_A&={\rm Tr}_B[\rho_{AB}]=\bra{0}_B\rho_{AB}\ket{0}_B+\bra{1}_B\rho_{AB}\ket{1}_B\nonumber\\&=\frac{1}{2}\left( \ket{0}\bra{0}+\ket{1}\bra{1}\right)=\frac{1}{2}\mathbb{I}.
\end{align}
It turns out that $\rho_A=\mathbb{I}/2$ (and similarly, $\rho_B=\mathbb{I}/2$), so both reduced states become a maximally mixed state. Think about the implications of this result. We have started from a maximally entangled pure state $\ket{\phi^+}$ between qubits $A$ and $B$. But if we look locally to either of them, they are described by a classical  probabilistic mixture of the states $\ket{0}$ and $\ket{1}$, that is, a state with zero coherence and the lowest possible purity. We have not done anything on the qubits; this is just an information standpoint. How is this possible and what does it mean?

Suppose that Alice performs a measurement on her qubit $A$, but cannot access or measure qubit $B$ (in Bob's hands). Then, for any measurement that Alice performs on $A$, the outcomes would look completely random. That is, if $AB$ are maximally entangled and Alice is not able to know anything about $B$, then Alice's qubit is indistinguishable from a total random state, i.e. a maximally mixed state. Note that $\rho_{AB}$ (given in Eq.~\eqref{eq:rhoAB_Bell}) is very different from the product of their reduced states $\rho_A\otimes \rho_B$! This is indeed the very essence of entanglement: an entangled state cannot be individually described, they form an inseparable whole. Hence, discarding the information about the other party (Bob's qubit), Alice's qubit seem to be in a random state. Without any further considerations, this is indistinguishable from $\frac{1}{2}\mathbb{I}$. If we, however, can access Bob's information, then we will be able to observe the actual correlations of the outcomes and we will be able to certify that the state $\rho_{AB}$ was indeed a Bell state. 

In conclusion, this example illustrates two important points: 
\begin{itemize}
    \item[(i)] A mixed state for $A$ may be attributed to $A$ being in a pure and entangled state with another system, $B$, which we cannot control or access. This is very useful tool in quantum information, as it allows us to ascribed mixedness for system $A$ as a pure state on a larger, possible virtual system $AB$, including entanglement among them. This method is known as quantum state purification. 
    \item[(ii)] As we will see in the following section, this example should already showcase the complexity of open quantum systems due to information and entanglement. Suppose a closed system made of two interacting qubits $A$ and $B$. If we look only at qubit $A$, its state will be dramatically affected by the interaction with the uncontrollable qubit $B$.  In broad terms, this is the source of the complexity in open quantum systems.
\end{itemize}

\vspace{0.5cm}
\noindent\textbf{Operational recipe for two qubits}
\vspace{0.2cm}

It follows an operational recipe to compute the partial trace for bipartite qubit systems:
Considering the notation $\ket{c_a,c_b}$, then we can write the density operator $\rho_{AB}$ in term of coefficients $\rho_{c_a,c_b;c_a',c_b'}=\bra{c_a,c_b}\rho_{AB}\ket{c_a',c_b'}$, i.e.
\begin{align}
    \rho_{AB}=\begin{bmatrix} \rho_{00,00} & \rho_{00,01} & \rho_{00,10} & \rho_{00,11} \\  \rho_{01,00} & \rho_{01,01} & \rho_{01,10} & \rho_{01,11} \\
    \rho_{10,00} & \rho_{10,01} & \rho_{10,10} & \rho_{10,11} \\
    \rho_{11,00} & \rho_{11,01} & \rho_{11,10} & \rho_{11,11}         
    \end{bmatrix}.
\end{align}
Then, the partial trace over the subsystem $B$ consists in identifying terms with common index for $c_b=c_b'$ and sum for $c_b=0$ and $1$, and the location $c_a,c_a'$. Let us illustrate it with colors: 
\begin{align}
     \rho_{AB}=\begin{bmatrix} \textcolor{red}{\rho_{00,00}} & \rho_{00,01} & \textcolor{cyan}{\rho_{00,10}} & \rho_{00,11} \\  \rho_{01,00} & \textcolor{red}{\rho_{01,01}} & \rho_{01,10} & \textcolor{cyan}{\rho_{01,11}} \\
    \textcolor{orange}{\rho_{10,00}} & \rho_{10,01} & \textcolor{blue}{\rho_{10,10}} & \rho_{10,11} \\
    \rho_{11,00} & \textcolor{orange}{\rho_{11,01}} & \rho_{11,10} & \textcolor{blue}{\rho_{11,11}}       
    \end{bmatrix}.
\end{align}
Therefore, summing those elements we obtain the partial trace over the subsystem $B$:
\begin{align}
    \rho_A={\rm Tr}_B[\rho_{AB}]=\begin{bmatrix} \textcolor{red}{\rho_{00,00}+\rho_{01,01}} & \textcolor{cyan}{\rho_{00,10}+\rho_{01,11}} \\ 
    \textcolor{orange}{\rho_{10,00}+\rho_{11,01}} & \textcolor{blue}{\rho_{10,10}+\rho_{11,11}}\end{bmatrix}. 
\end{align}
Similarly, taking the trace over $A$, we need to identify terms with $c_a=c_a'$ and sum over $c_a=0$ and $1$ at place them at the location $c_b,c_b'$: 
\begin{align}
    \rho_{AB}=\begin{bmatrix} \textcolor{red}{\rho_{00,00}} & \textcolor{cyan}{\rho_{00,01}} & \rho_{00,10} & \rho_{00,11} \\  \textcolor{orange}{\rho_{01,00}} & \textcolor{blue}{\rho_{01,01}} & \rho_{01,10} & \rho_{01,11} \\
    \rho_{10,00} & \rho_{10,01} & \textcolor{red}{\rho_{10,10}} & \textcolor{cyan}{\rho_{10,11}} \\
    \rho_{11,00} & \rho_{11,01} & \textcolor{orange}{\rho_{11,10}} & \textcolor{blue}{\rho_{11,11}} 
    \end{bmatrix}.
\end{align}
That is, 
\begin{align}
    \rho_B={\rm Tr}_A[\rho_{AB}]=\begin{bmatrix} \textcolor{red}{\rho_{00,00}+\rho_{10,10}} & \textcolor{cyan}{\rho_{00,01}+\rho_{10,11}} \\ \textcolor{orange}{\rho_{01,00}+\rho_{11,10}} & \textcolor{blue}{\rho_{01,01}+\rho_{11,11}}
    \end{bmatrix}.
\end{align}
The states upon partial trace, $\rho_A$ and $\rho_B$, are also a valid states (obviously, provided $\rho_{AB}$ is a physical state), and is properly normalized. The normalization can be easily checked, ${\rm Tr}[\rho_{A}]=1$ and ${\rm Tr}[\rho_B]=1$.

\vspace{0.5cm}
\noindent\textbf{Useful identity: Operators on subsystems}
\vspace{0.2cm}

Finally, let us show a particular useful identity when dealing with partial and full traces. Suppose $X_{AB}$ to be an operator acting on the joined system $AB$, and the state is in a product form $\rho_{AB}=\rho_A\otimes\rho_B$, then
\begin{align}\label{eq:pt_1}
    {\rm Tr}_{AB}[X_{AB}(\rho_A\otimes\rho_B)]&={\rm Tr}_{A}[{\rm Tr}_B[X_{AB}(\rho_A\otimes \rho_B)]]={\rm Tr}_A[{\rm Tr}_B[X_{AB}(\mathbb{I}_A\otimes \rho_B)]\rho_A]\nonumber\\ &={\rm Tr}_A[\rho_A {\rm Tr}_B[X_{AB}(\mathbb{I}_A\otimes \rho_B)]]={\rm Tr}_A[\rho_A E].
\end{align}
Note that we have first split the trace on $A$ and then on $B$, and later we removed  the state $\rho_A$ from the trace over $B$. The advantage of this identity is that we can associate the result of tracing over $B$ as an operator acting on $A$. That is, $E={\rm Tr}_B[X_{AB}(\mathbb{I}_A\otimes \rho_B)]$ is an operator acting on $\mathcal{H}_A$.  This identity will be quite useful when dealing with system and environment in the following sections. 

Let us show how this can be done. By definition, the trace is
\begin{align}
    {\rm Tr}_{AB}[X_{AB}(\rho_A\otimes\rho_B)]=\sum_{i,n}\bra{i,n} X_{AB}(\rho_A\otimes\rho_B)\ket{i,n},
\end{align}
where $\ket{i,n}$ denotes $\ket{i}_A\otimes\ket{n}_B$, and form a complete orthonormal basis,  $\sum_{i}\ket{i}\bra{i}_A=\mathbb{I}_A$ and $\sum_{n}\ket{n}\bra{n}_B=\mathbb{I}_B$. We continue by analyzing how $(\rho_A\otimes \rho_B)$ acts on $\ket{j,n}$ employing the matrix elements,
\begin{align}
    (\rho_A\otimes \rho_B)\ket{j,n}=\sum_{k,m} (\rho_{A})_{kj}(\rho_B)_{mn}\ket{k,m}.
\end{align}
Therefore, 
\begin{align}
    \left({\rm Tr}_{B}[X_{AB}(\rho_A\otimes\rho_B)]\right)_{ij}&=\sum_n\bra{i,n} X_{AB} (\rho_A\otimes\rho_B)\ket{j,n}=\sum_n\bra{i,n} \sum_{k,m} X_{AB} (\rho_A)_{kj}(\rho_B)_{mn}\ket{k,m}\nonumber\\&=\sum_{k} \sum_{n,m} \bra{i,n}X_{AB}\ket{k,m}(\rho_A)_{kj}(\rho_B)_{mn}\nonumber\\&=\sum_k (\rho_A)_{kj}\sum_{n,m}\bra{i,n}X_{AB}\ket{k,m}(\rho_B)_{mn}\nonumber\\&=\sum_{k}(\rho_A)_{kj}\left({\rm Tr}_B[X_{AB}(\mathbb{I}\otimes \rho_B)] \right)_{ik}.
\end{align}
Expressing it in matrix form, we arrive to
\begin{align}\label{eq:prop_TrB}
  {\rm Tr}_{B}[X_{AB}(\rho_A\otimes\rho_B)]={\rm Tr}_B[X_{AB}(\mathbb{I}_A\otimes\rho_B)]\rho_A,
\end{align}
which, together with the cyclic property of the trace, completes the identity shown in Eq.~\eqref{eq:pt_1}.

\subsection{Problems to practice}

\vspace{0.5cm}
\noindent\textbf{Problem 1.1}
\vspace{0.2cm}

\noindent Suppose $U(t)$ to be a differentiable and invertible map dependent on the parameter $t$. Using the Jacobi's formula
    \begin{align}
        \frac{d}{dt} \det [U(t)]= \det[U(t)] {\rm Tr}\left[U^{-1}(t)\frac{dU(t)}{dt} \right],
        \nonumber
    \end{align}
    show that if $U(t)=e^{t X}$ with $t$ a scalar, then
    \begin{align}
        \det[e^{t X}]=e^{t {\rm Tr}[X]}.
        \nonumber
    \end{align}   

\vspace{0.5cm}
\noindent\textbf{Problem 1.2}
\vspace{0.2cm}
     
\noindent Suppose a system with a finite Hilbert space, $\dim \mathcal{H}_S=d$. The generators of the Lie algebra $\mathfrak{g}$ corresponding to $\textit{SU(d)}$ are $T_i$ for $i=1,\ldots d^2-1$, so that the unitary transformations in $\textit{SU}(d)$ can be expressed as $U(t)=e^{t X}$ with $X\in\mathfrak{g}$. Show that the generators must be traceless, i.e. ${\rm Tr}[T_i]=0,  \forall i=1,\ldots,d^2-1$.
     
\vspace{0.5cm}
\noindent\textbf{Problem 1.3}
\vspace{0.2cm}

\noindent Suppose a bosonic system with the usual creation and annihilation operators, $\adag$ and $a$ satisfying the commutation relation $[a,\adag]=\mathbb{I}$. Explain where is the catch in the following derivation:
    \begin{align}
        {\rm Tr}[\ [a,\adag]]&={\rm Tr}[\mathbb{I}] \Rightarrow
        {\rm Tr}[\ a\adag-\adaga]={\rm Tr}[\mathbb{I}] \Rightarrow
        {\rm Tr}[a\adag]-{\rm Tr}[\adaga]={\rm Tr}[\mathbb{I}] \Rightarrow\nonumber\\
        &\Rightarrow
        {\rm Tr}[\adaga]-{\rm Tr}[\adaga]={\rm Tr}[\mathbb{I}] \Rightarrow
        0=\infty
        \nonumber
    \end{align}

\vspace{0.5cm}
\noindent\textbf{Problem 1.4}
\vspace{0.2cm}

\noindent Consider a two-qubit system in which Alice holds one qubit and Bob holds the other. The combined state of both qubits is pure, and reads as 
    \begin{align}
        \ket{\psi}_{AB}=\frac{1}{\sqrt{2+|\lambda|^2}}(\lambda\ket{00}+\ket{10}+\ket{11}),
        \nonumber
    \end{align}
    for $\lambda\in\mathbb{C}$. 
    Answer:
    \begin{itemize}
        \item[(i)] Compute the density matrices for the combined and reduced systems of Alice and Bob, i.e. $\rho_{AB}$, $\rho_{A}={\rm Tr}_B[\rho_{AB}]$ and $\rho_B={\rm Tr}_A[\rho_{AB}]$.
        \item[(ii)] Compute the purity of Alice's qubit alone.
        \item[(iii)] From the previous result, for what a value(s) of $\lambda$ are the qubits not in an entangled state?
        \item[(iv)] If Alice performs a projective measurement on her qubit in the $Z$ basis, what is the average state (i.e. over all possible outcomes) of Bob's qubit? 
    \end{itemize}

\vspace{0.5cm}
\noindent\textbf{Problem 1.5}
\vspace{0.2cm}
    
\noindent Suppose two qubits $A$ and $B$. Initially, the state of the combined system is $\rho_A\otimes \rho_B$ with $\rho_B=\ket{k}\bra{k}$ pure. Then, they undergo an arbitrary unitary gate $U$, and after, the qubit $B$ suffers a sharp quantum measurement (projective) with orthogonal projectors $\Pi_{0}=\ket{0}\bra{0}$ and $\Pi_1=\ket{1}\bra{1}$. 
    Answer:
    \begin{itemize}
        \item[(i)] Show that, from the point of view of the qubit $A$ alone, this corresponds to a POVM $\{M_0,M_1\}$ with Kraus operators acting on $A$ given by $K_i=\bra{i}_B U\ket{k}_B$ such that $M_i=K_i^\dagger K_i$ for $i=0,1$ and $\sum_{i=0,1}K_i^\dagger K_i=\mathbb{I}_A$.
        \item[(ii)] Restrict now to $\rho_B=\ket{0}\bra{0}$ and $\rho_A$ a generic state, $\rho_A=\frac{1}{2}(1+\vec{v}\cdot\vec{\sigma})$ with Bloch vector $\vec{v}$ and $\vec{\sigma}=(\sigma_x,\sigma_y,\sigma_z)$. In addition, consider $U$ a controlled $R_y(\theta)$ gate (with $A$ the control and $B$ the target). Express the transformation of the average state after the measurements $\rho'_A$ as a quantum channel, i.e $\rho'_A=\mathcal{E}(\rho_A)$. What sort of operation is qubit $A$ suffering?
    \end{itemize}
\newpage
\section{Open Quantum Systems}\label{c:OQS}


\subsection{Understanding the challenge ahead}
\begin{figure}
    \centering
    \includegraphics[width=0.7\linewidth]{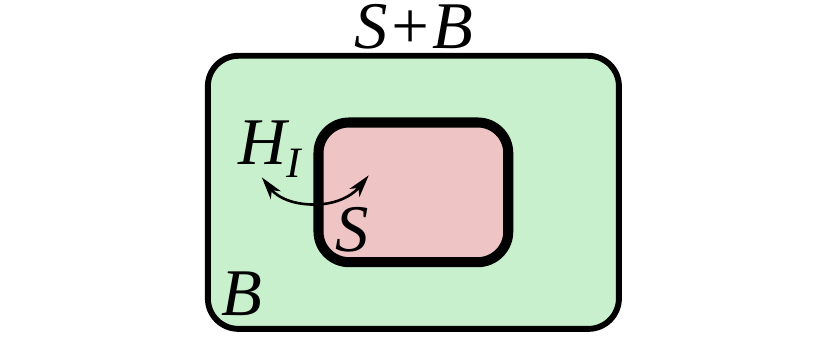}
    \caption{Schematic illustration of the open nature of a quantum system $S$ that interacts via $H_I$ with the degrees of freedom in $B$, that constitute the environment. The main goal of the theoretical framework of open quantum systems consists in finding a description of the dynamics that undergoes $S$ without explicitly including the degrees of freedom of $B$.}
    \label{fig:OQS_sketch}
\end{figure}

Let us consider a generic quantum system, that we denote by $S$, with Hilbert space $\mathcal{H}_S$ and state $\rho_S\in\mathcal{D}(\mathcal{H}_S)$. The Hamiltonian of the system $S$ alone is $H_S$. Yet, the system $S$ interacts with a collection of degrees of freedom $b_i$ that constitute the environment or bath $B$, see Fig.~\ref{fig:OQS_sketch}. The Hamiltonian of the environment is $H_B$ and its Hilbert space $\mathcal{H}_B$. Since the system and the environment interact, we can write without loss of generality the total Hamiltonian of $S+B$ as
\begin{align}\label{eq:H_SB}
    H=H_S+H_B+H_I,
\end{align}
with $H_I$ representing the $S$ and $B$ interaction in the Hilbert space $\mathcal{H}_S\otimes \mathcal{H}_B$ and either of the terms may be time dependent. Suppose that the initial state at time $t_0$ is $\rho(t_0)\in\mathcal{D}(\mathcal{H}_S\otimes \mathcal{H}_B)$, and that the environment $B$ accounts for all possible interacting degrees of freedom with the system of interest $S$. In addition, there are no other system that interacts with $B$, apart from $S$, that is not included in itself. In this manner, the dynamics of the combined systems $S+B$ follows the von Neumann equation because $S+B$ form a closed system,
\begin{align}
    \frac{d}{dt}\rho(t)=-i[H,\rho(t)],
\end{align}
with initial condition $\rho(t_0)$. If we are interested only on $S$, the state of the system at any time $t\geq t_0$ follows from tracing out $B$, 
\begin{align}\label{eq:rhoS_exact}
    \rho_S(t)={\rm Tr}_B[\rho(t)]={\rm Tr}_B[U(t,t_0)\rho(t_0)U^\dagger(t,t_0)].
\end{align}
This can be done also to the von Neumann equation equation, i.e.
\begin{align}\label{eq:rhoS_exact2}
    {\rm Tr}_B[\frac{d}{dt}\rho(t)]={\rm Tr}_B[-i[H,\rho(t)]] \Rightarrow \frac{d}{dt}\rho_S(t)=-i{\rm Tr}_B[\  [ H,\rho(t)]].
\end{align}

If the previous expressions can be explicitly computed, the open quantum dynamics of the system $S$ is already solved. However, solving Eq.~\eqref{eq:rhoS_exact} or~\eqref{eq:rhoS_exact2} is in the best case not feasible due to the large amount of degrees freedom that form the environment. For example, consider that $S$ is a single nucleus in a crystal lattice, which may significantly interact with hundreds of electrons and nuclei that are close to it. Then, we can estimate that the dimension of the Hilbert space of the bath amounts to $\dim \mathcal{H}_B=d_B\approx 2^{100}\sim 10^{30}$. This number is astronomically large: in general, storing the corresponding density matrix $\rho$ would be impossible, even if we could make use of all the digital memory available worldwide. 

In the worst case, solving Eq.~\eqref{eq:rhoS_exact} or~\eqref{eq:rhoS_exact2} is not even possible in principle. Why is that? Often, the microscopic Hamiltonian formulation of the environment and how it interacts with $S$ are unknown. That is, the terms $H_B$ and $H_I$ in Eq.~\eqref{eq:H_SB} are already problematic, let alone Eq.~\eqref{eq:rhoS_exact} or~\eqref{eq:rhoS_exact2}. Returning to our earlier example, such a formulation would require complete knowledge of the states, positions, and mutual interactions of hundreds of particles.   These challenges have strongly motivated the field of open quantum systems, where the goal is to develop approximations to Eq.~\eqref{eq:rhoS_exact2} that yield a tractable dynamical equation for the system $S$, or to construct phenomenological models that can successfully reproduce and predict experimental observations. 

In essence, we seek a dynamical equation for the system of interest, $S$, that incorporates the effect of its interactions with the environment, $B$, without explicitly including the degrees of freedom of $B$. This is precisely what it is written in Eq.~\eqref{eq:rhoS_exact2}. This expression can be  formulated in a compact form as
\begin{align}\label{eq:master_equation_gen}
    \frac{d}{dt}\rho_S(t)=\mathcal{L}(t)\{\rho_S(t)\},
\end{align}
where $\mathcal{L}(t)\{\cdot\}$ is an operator, possibly time dependent, acting solely on the Hilbert space of the system $\mathcal{H}_S$. Moreover, given that $\mathcal{L}(t)\{\cdot\}$ acts on operators, it is sometimes referred to as superoperator. In addition, since it defines the dynamics, we denote $\mathcal{L}(t)\{ \cdot\}$ as the Liouvillian superoperator\footnote{The term Liouvillian stems from classical mechanics, where the Liouvillian equation determines how the phase-space distribution evolves in time.}.  Finding the explicit form of $\mathcal{L}(t)\{\cdot\}$ is, of course, the key point we need to address.

\subsubsection{Recovering closed system dynamics}
Although we have not yet computed the operator $\mathcal{L}(t)\{ \cdot\}$, we know that it should reduce to the von Neumann equation in the limit when the system $S$ is perfectly isolated. 
In order to check this, we note that, if $S$ is closed, then the interaction between $S$ and $B$ vanishes, $H_I\rightarrow 0$. This leads to
\begin{align}
    {\rm Tr}_B[\frac{d}{dt}\rho(t)]=-i {\rm Tr}_B[ \ [H_S+H_B,\rho(t)]]=-i{\rm Tr}_B[\ [H_S\otimes \mathbb{I}_B,\rho(t)]+[\mathbb{I}_S\otimes H_B,\rho(t)]],
\end{align}
where we have used the fact that the local Hamiltonians for $S$ and $B$ act on different Hilbert spaces, i.e. $H_S\equiv H_S\otimes \mathbb{I}_B$ and similarly $H_B\equiv \mathbb{I}_S\otimes H_B$. Let us inspect both terms in the previous expression. Considering a generic state $\rho(t)$, the first term can be expanded as
\begin{align}
    {\rm Tr}_B [ \ [H_S\otimes \mathbb{I}_B,\rho(t)]]&={\rm Tr}_B [(H_S\otimes \mathbb{I}_B)\rho(t)]-{\rm Tr}_B[\rho(t)(H_S\otimes \mathbb{I}_B)]H_S  {\rm Tr}_B[\rho(t)]-{\rm Tr}_B[\rho(t)]H_S\nonumber\\&=H_S \rho_S(t)-\rho_S(t)H_S=[H_S,\rho_S(t)],
\end{align}
where we have used the property of the partial trace in Eq.~\eqref{eq:prop_TrB}, and used $\rho_S(t)={\rm Tr}_B[\rho(t)]$. For the second term, however, we have
\begin{align}
    {\rm Tr}_B [ \ [\mathbb{I}_A\otimes H_B,\rho(t)]]&={\rm Tr}_B [(\mathbb{I}_A\otimes H_B)\rho(t)]-{\rm Tr}_B[\rho(t)(\mathbb{I}_A\otimes H_B)]= 0,
\end{align}
because of the cyclic property of the trace. Note that the cyclic property only applies with respect to the system we are tracing over. Coming back to the dynamical equation, we finally arrive to
\begin{align}\label{eq:rhoS_closed}
    \frac{d}{dt}\rho_S(t)=-i[H_S,\rho_S(t)],
\end{align}
which is exactly what we  were looking for: if the system $S$ does not interact with the environment, then the dynamics of $S$ exactly reduces to the closed system dynamics dictated solely by $H_S$, with initial condition $\rho_S(t_0)={\rm Tr}_B[\rho(t)]$. A similar derivation can be found in Sec.~\ref{moremat:NonintB}. Summarizing, the Liouvillian superoperator satisfies
\begin{align}
    \lim_{H_I\rightarrow 0} \mathcal{L}(t)\{\rho_S(t)\}=-i[H_S,\rho_S(t)].
\end{align}
In the following we are going to see what is the  form for $\mathcal{L}(t)\{ \cdot\}$ when $H_I\neq 0$, under some reasonable assumptions.

\subsection{Reduced dynamics as a quantum channel}
For what follows, we will make a convenient and reasonable assumption: at some initial time $t_0$ we are able to prepare the system in a state $\rho_S(t_0)$ which is uncorrelated from the state of the environment, i.e. $\rho(t_0)=\rho_S(t_0)\otimes \rho_B(t_0)$, where $\rho_B(t_0)$ may correspond to a thermal equilibrium state. Then, Eq.~\eqref{eq:rhoS_exact} defines a map $V(t)$ between density matrices  $\mathcal{D}(\mathcal{H}_S)$ into itself, i.e.
\begin{align}\label{eq:rhoS_map}
    \rho_S(t_0)&\rightarrow \rho_S(t)=V(t,t_0)\{\rho_S(t_0)\}={\rm Tr}_B[U(t,t_0)(\rho_S(t_0)\otimes \rho_B(t_0))U^\dagger(t,t_0)],\nonumber\\
    V(t,t_0)&: {\mathcal D}(\mathcal{H}_S)\rightarrow {\mathcal D}(\mathcal{H}_S),
\end{align}
For a fixed $t$, this map can be seen as a quantum channel acting on $\rho_S(t_0)$ that dictates its state change at a future time $t\geq t_0$. For this reason, the $V(t,t_0)$ is known as a dynamical map. Considering $t$ to be a continuous varying parameter, $\{V(t,t_0)|t\geq t_0\}$ defines a family of dynamical maps where $V(t_0,t_0)$ is the identity map. 

From the point of view of the system $S$, it must be possible to completely characterized  the map $V(t,t_0)$ in terms of operators acting only on $\mathcal{H}_S$. For that reason, we develop further the expression in Eq.~\eqref{eq:rhoS_map} using the spectral decomposition of $\rho_B(t_0)$,
\begin{align}
    \rho_B(t_0)=\sum_n \lambda_n \ket{b_n}\bra{b_n},
\end{align}
where $\{\ket{b_n}\}$ forms an orthonormal basis of $\mathcal{H}_B$ and $\sum_n \lambda_n=1$ with $\lambda_n\geq 0, \forall n$. Plugging this into Eq.~\eqref{eq:rhoS_map}, we obtain
\begin{align}
    \rho_S(t)=V(t,t_0)\{\rho_S(t_0)\}&={\rm Tr}_B[U(t,t_0)(\rho_S(t_0)\otimes \sum_{m} \lambda_{m} \ket{b_{m}}\bra{b_{m}})U^\dagger(t,t_0)]\nonumber\\&=\sum_n \bra{b_n} \left(U(t,t_0)(\rho_S(t_0)\otimes \sum_{m} \lambda_{m} \ket{b_{m}}\bra{b_{m}})U^\dagger(t,t_0)  \right)\ket{b_n}\nonumber\\&=\sum_{n,m}\bra{b_n}U(t,t_0)\sqrt{\lambda_{m}}\ket{b_{m}}\rho_S(t_0) \bra{b_{m}} U^\dagger(t,t_0)\sqrt{\lambda_{m}}\ket{b_n}\nonumber\\ \label{eq:Wnm_terms}&=\sum_{n,m}W_{n,m}(t,t_0) \rho_S(t_0) W_{n,m}^\dagger(t,t_0),
\end{align}
where we have defined the operators
\begin{align}
    W_{n,m}(t,t_0)=\sqrt{\lambda_{m}}\bra{b_n}U(t,t_0)\ket{b_{m}}.
\end{align}
Note the similarity of Eq.~\eqref{eq:Wnm_terms} with a generic quantum channel, cf. Eq.~\eqref{eq:KrausQC}. Moreover, one can check that these operators $W_{n,m}(t,t_0)$ satisfy (this is left as an exercise)
\begin{align}
    \sum_{n,m}W_{n,m}^\dagger(t,t_0) W_{n,m}(t,t_0)=\mathbb{I}_S.
\end{align}
This latter condition ensures that the evolved state is properly normalized (as it was the case for a generic quantum channel)
\begin{align}
    {\rm Tr}_S[ V(t,t_0)\{ \rho_S(t_0)\}]=1, \ \forall t, \rho_S(t_0)\in\mathcal{D}(\mathcal{H}_S).
\end{align}
In this way, the dynamical map $V(t,t_0)$ is a completely-positive trace-preserving (CPTP) quantum operation. This is to be expected, since any unitary dynamics of the joint system $S+B$ necessarily yields physical states when restricted to the system $S$ alone. Although the previous derivations are useful, the operators $W_{n,m}(t,t_0)$ explicitly depend on the joint dynamics $U(t,t_0)$. In the following, we are going to see how this dynamical map can be further simplified, which leads to extra terms that are added to the von Neumann equation of the closed system, in terms of some operators acting on $\mathcal{H}_S$. These operators will be weighted by some  rate, which dictates their impact onto the dynamics of $S$.  

As a side remark, it is always instructive to check that the dynamical map reduces to the expected closed dynamics whenever the joint evolution can be decoupled (i.e. for non-interacting system and environment $H_I=0$). See Sec.~\ref{moremat:map_Hi0} for details.

\subsection{Tour de force: Markovian quantum master equation}
As noted above, the Eq.~\eqref{eq:Wnm_terms} provides a quantum channel description of the dynamical map. Yet, the operators that dictate the dynamics depend explicitly on the joint unitary evolution of the $S+B$ system. In order to derive a simplified dynamical equation for $S$ alone, i.e. to obtain an explicit expression for the Liouvillian superoperator $\mathcal{L}(t)\{\cdot \}$ we are going to make a further assumption: the evolution is Markovian and the generator time-independent. The Markovian property is crucial to define a quantum dynamical semigroup. A dynamical map $V(t)$ is said to fulfill the dynamical semigroup property if
\begin{align}
    V(t_2)V(t_1)=V(t_2+t_1).
\end{align}
For a time-dependent map, this should be translated into $V(t_2,t_1)V(t_1,t_0)=V(t_2,t_0)$. This property is very important and it is known as complete-positivity (CP) divisibility. 
In Sec.~\ref{c:micro} we will come back to this point, when we will derive from first principles the physical origin or requirements of such condition. For now, we simply assume that the dynamical map fulfills this requirement given that the environment degrees of freedom feature a characteristic time scale much shorter than those of the system of interest $S$. 

The dynamical semigroup or CP-divisibility property  allows us to express the dynamical map as the exponential of its generator, that we call $\mathcal{L}$, i.e. $V(t)=e^{\mathcal{L} t}$ so that 
\begin{align}
    \frac{d}{dt}\rho_S(t)=\mathcal{L}\{\rho_S(t)\}.
\end{align}
The superoperator $\mathcal{L}$ acts on the Liouville space: This Liouville space  is the set of operators $A\in\mathcal{H}_S$ such that ${\rm Tr}[A^\dagger A]$ is finite. For a dimension $\dim \mathcal{H}_S=d$, there are $d^2$ complex operators that form a complete orthonormal operators $F_i$, $i=1,2,\ldots, d^2$. The inner product (known as Hilbert-Schmidt)  in this space is defined as $(F_i,F_j)\equiv {\rm Tr}[F_i^\dagger F_j]$, which for these orthonormal operators reads as
\begin{align}
    (F_i,F_j)\equiv {\rm Tr}[F_i^\dagger F_j]=\delta_{i,j}.
\end{align}
Later, in Sec.~\ref{s:LiovSP}, we will use a different notation for the Hilbert-Schmidt inner  product in the Liouville space employing  double-kets and double-bras to denote vectorized operators, i.e. $(F_i,F_j)\equiv \langle \langle F_i| F_j\rangle \rangle$.\footnote{The distinct notation is kept intentionally. First, $(F_i,F_j)$ makes the equations lighter than double-kets or bras $\langle \langle F_i| F_j\rangle \rangle$. Second, this derivation has been traditionally  presented with this  notation rather than the one used for the Liouvillian superoperator in Sec.~\ref{s:LiovSP}. In any case, we emphasize that is just a notation issue: the inner product in this space is  ${\rm Tr}[A^\dagger B
]$ which is typically replaced by either of these  compact representations, i.e. $\langle \langle A|B\rangle \rangle \equiv (A,B)\equiv {\rm Tr}[A^\dagger B]$.}

For convenience, we set $F_{d^2}=\mathbb{I}_S/\sqrt{d}$, such that the other $d^2-1$ operators are traceless. Now, every operator in the Liouville space $A$ can be expressed as a linear combination of the orthonormal basis $F_i$,
\begin{align}
    A=\sum_{i}^{d^2}F_i (F_i,A).
\end{align}
Note that if we consider a single qubit, $d=2$, then $F_{1,2,3}$ are the usual Pauli matrices, while $F_4=\mathbb{I}/\sqrt{2}$. Making use of this relation, we can write
\begin{align}
    W_{n,m}(t)=\sum_{i=1}^{d^2} F_i(F_i,W_{n,m}(t)),
\end{align}
and therefore
\begin{align}
    V(t)\{ \rho_S(t_0)\}&=\sum_{n,m}W_{n,m}(t)\rho_S(t_0)W_{n,m}^\dagger(t)=\sum_{i,j}\sum_{n,m}F_i(F_i,W_{n,m}(t))\rho_S(t_0) F_j^\dagger(F_j,W_{n,m}(t))^*\nonumber\\&=\sum_{i,j}^{d^2}c_{i,j}(t) F_i\rho_S(t_0)F_j^\dagger,
\end{align}
with the coefficients $c_{i,j}(t)$ defined as
\begin{align}\label{eq:c_ij}
    c_{i,j}(t)=\sum_{n,m}(F_i,W_{n,m}(t))(F_j,W_{n,m}(t))^*.
\end{align}
Clearly the coefficient matrix is Hermitian, $c=c^\dagger$, but it can also be seen to be positive (cf. Sec.~\ref{moremat:C_positive}).

Up to this point, the derivation may appear to be little more than a mathematical detour, with the joint unitary evolution simply hidden inside the coefficients $c_{i,j}(t)$. One may then wonder: are we truly gaining any insight? The answer is yes, but unfortunately, to make this clear, we must first take a few additional steps. 

From the definition of the generator of the dynamical map, we can write (let us now drop the time dependence for $\rho_S$)
\begin{align}
    \frac{d}{dt}\rho_S=\mathcal{L}\{\rho_S\} \Rightarrow \mathcal{L}\{\rho_S\}=\lim_{\epsilon\rightarrow 0}\frac{1}{\epsilon}\left(V(\epsilon)\rho_S-\rho_S\right).
\end{align}
This means that
\begin{align}
&\mathcal{L}\{\rho_S\}=\lim_{\epsilon\rightarrow 0}\frac{1}{\epsilon}\left(\sum_{i,j}^{d^2}c_{i,j}(\epsilon)F_i\rho_SF_j^\dagger-\rho_S \right)=\nonumber\\&=\lim_{\epsilon\rightarrow 0}\left[\frac{c_{d^2,d^2}(\epsilon)-d}{d\epsilon}\rho_S+\frac{1}{\sqrt{d}}\sum_{i=1}^{d^2-1}\left(\frac{c_{i,d^2}(\epsilon)}{\epsilon}F_i\rho_S+\frac{c_{d^2,i}(\epsilon)}{\epsilon}\rho_S F_i^\dagger \right)+\sum_{i,j}^{d^2-1}\frac{c_{i,j}(\epsilon)}{\epsilon}F_i\rho_SF_j^\dagger\right],
\end{align}
from where we define different coefficients 
\begin{align}
    a_{i,j}&=\lim_{\epsilon\rightarrow 0}\frac{c_{i,j}(\epsilon)}{\epsilon}, \ i,j=1,\ldots,d^2-1\\
    a_{i,d^2}&=\lim_{\epsilon\rightarrow 0}\frac{c_{i,d^2}(\epsilon)}{\epsilon}, i=1,\ldots, d^2-1\\
    a_{d^2,d^2}&=\lim_{\epsilon\rightarrow 0}\frac{c_{d^2,d^2}(\epsilon)-d}{\epsilon},
\end{align}
as well as the following operators for convenience,
\begin{align}
    F&=\frac{1}{\sqrt{d}}\sum_{i=1}^{d^2-1}a_{i,d^2}F_i,\\
    G&=\frac{1}{2d}a_{d^2,d^2}\mathbb{I}_S+\frac{1}{2}(F+F^\dagger).
\end{align}
In this manner, the generator of the dynamical map can be written as
\begin{align}
    \mathcal{L}\{\rho_S\}=-\frac{1}{2}\left([F^\dagger-F,\rho_S]\right)+\{G,\rho_S\}+\sum_{i,j=1}^{d^2-1}a_{i,j}F_i\rho_S F_j^\dagger.
\end{align}
The first term can be regarded as an effective Hamiltonian part of the evolution (recall the von Neumann equation), 
\begin{align}
    H=-\frac{i}{2}(F^\dagger-F),
\end{align}
so that we arrive to
\begin{align}\label{eq:Lmap}
    \mathcal{L}\{\rho_S\}=-i[H,\rho_S]+\{G,\rho_S\}+\sum_{i,j=1}^{d^2-1}a_{i,j}F_i\rho_S F_j^\dagger.
\end{align}
This equation is not yet the most standard form because, in order to be a valid physical map, we need to impose ${\rm Tr}_S[\rho_S(t)]=1, \forall t$, meaning that ${\rm Tr}_S[\mathcal{L}\{\rho_S\}]=\frac{d}{dt}{\rm Tr}_S[\rho_S]=0$. Taking the trace to Eq.~\eqref{eq:Lmap},
\begin{align}
    {\rm Tr}_S[\mathcal{L}\{\rho_S\}]=0\Rightarrow {\rm Tr}_S\left[-i[H,\rho_S]+\{G,\rho_S\}+\sum_{i,j}^{d^2-1}a_{i,j}F_i \rho_SF_j^\dagger \right]=0.
\end{align}
The trace over the commutator simply vanishes. The second term can be worked out and find 
\begin{align}
    {\rm Tr}_S[\{G,\rho_S\}]={\rm Tr}_S[2G\rho_S].
\end{align}
For the third one, we can arrange terms using the cyclic property and express it as
\begin{align}
    {\rm Tr}_S\left[\sum_{i,j}^{d^2-1}a_{i,j} F_i \rho_S F_j^\dagger \right]={\rm Tr}_S\left[\sum_{i,j=1}^{d^2-1}a_{i,j}F_j^\dagger F_i \rho_S\right].
\end{align}
Therefore, we arrive to
\begin{align}
    {\rm Tr}_S[\mathcal{L}\{\rho_S\}]=0\Rightarrow {\rm Tr}_S\left[\left(2G+ \sum_{i,j=1}^{d^2-1}a_{i,j}F_j^\dagger F_i\right)\rho_S \right]=0,
    \end{align}
    so finally, the trace-preserving condition requires
    \begin{align}
        G=-\frac{1}{2}\sum_{i,j}^{d^2-1}a_{i,j}F_j^\dagger F_i.
    \end{align}
After all of these developments, we can give a general structure for a Markovian quantum master equation, 
\begin{align}\label{eq:MarkovMaster}
    \frac{d}{dt}\rho_S(t)=\mathcal{L}\{\rho_S\}=-i[H,\rho_S]+\sum_{i,j=1}^{d^2-1}a_{i,j}\left(F_i\rho_S F_j^\dagger-\frac{1}{2}\{F_j^\dagger F_i,\rho_S \} \right),
\end{align}
    recalling that the matrix $a_{i,j}$ is Hermitian and positive. 
\subsubsection{The Gorini-Kossakowski-Sudarshan-Lindblad equation}
Although the Markovian master equation in~\eqref{eq:MarkovMaster} is general, it is still not in the most standard form, known as Gorini-Kossakowski-Sudarshan-Lindblad (GKSL) master equation, or Lindblad for short \cite{GKS76,L76}. For that, we recall that the matrix $a_{i,j}$ is Hermitian and positive, and therefore it can be diagonalized $uau^\dagger$ and yield positive eigenvalues $\gamma_k\geq 0, {\rm for}\  k=1,\ldots,d^2-1$. This diagonalization transforms the operators $F_i$ according to the linear combination,
\begin{align}
    F_i=\sum_{k=1}^{d^2-1}u_{ki}A_k,
\end{align}
so that we can obtain the GKSL master equation in diagonal form:
\begin{align}\label{eq:GKSL}
    \frac{d}{dt}\rho_S(t)=\mathcal{L}\{\rho_S\}=-i[H,\rho_S]+\sum_{k=1}^{d^2-1}\gamma_k\left(A_k\rho_S A_k^\dagger-\frac{1}{2}\{A_k^\dagger A_k,\rho_S \} \right).
\end{align}
This master equation is arguably the most important result of this course. Recall that we started from the dynamical property and imposed the map to be CPTP. This result is known as Lindblad's theorem \cite{L76}: The generator of any CPTP quantum operation satisfying
the semigroup property must have the form of Eq.~\eqref{eq:GKSL}.

Let us explain more in detail each of the terms in the Markovian master equation, and some important issues:
\begin{itemize}
    \item The first term corresponds to the coherent or unitary evolution of the system $S$. 
    \item The second term refers to the non-unitary part of the evolution, where $\gamma_k$ is the corresponding rate at which the Lindblad operators $A_k$ affect the system. That is why it is often denoted as dissipator, i.e. \begin{align}D_k\{\rho\}=\gamma_k\left(A_k\rho A_k^\dagger-\frac{1}{2}\{A_k^\dagger A_k,\rho\}\right),\end{align} that remarks its non-unitary behavior. Therefore, $\mathcal{L}\{\rho\}=-i[H,\rho]+\sum_k D_k\{\rho\}$. 
    \item Each of these operators $A_k$ can be understood as the modes onto which $S$ can decay. If $\gamma_k=0, \forall k$, one recovers a fully coherent or unitary evolution. 
    \item Note that $H$ may not be exactly $H_S$ of the system $S$ alone. Why? The interaction of $S$ with $B$ does not only add extra terms to the dynamical equation, but it  may also introduce modifications into the coherent evolution, such as energy shifts. In general, the Hamiltonian of the system changes $H_S\rightarrow H$. At this stage, the argument may not be entirely convincing since we did not start from a physical model. Hopefully, this will become more clear in the microscopic derivation of the GKSL presented in Sec.~\ref{c:micro}.
    \item Recall that, while the $F_i$ were defined as the orthonormal operators of the Liouvillian space, $A_k$ stems from the diagonalization depending on the $a_{i,j}$ coefficient matrix. Now, since the GKSL master equation is invariant under the transformation $A_k\rightarrow A_k+a_k \mathbb{I}_{d^2}$ with $a_k\in\mathbb{C}$, it is always possible to choose $A_k$ as traceless Lindblad operators. See Sec.~\ref{moremat:GKSL_invariance} for details.
\end{itemize}

The Eq.~\eqref{eq:GKSL} gives the most general form of a Markovian evolution for a quantum system $S$ interacting with an environment $B$. In Sec.~\ref{c:micro} we will show how one recovers the GKSL master equation from a microscopic derivation, thus allowing us to explicitly obtain the rates $\gamma_k$ and the corresponding Lindblad operators $A_k$. If such microscopic description is not possible, the GKSL master equation can still be used in a   phenomenological manner: The  operators $A_k$ and their rates $\gamma_k$ can be fitted to reproduce experimental observations.  This second approach is routinely employed in many  quantum setups. However, one should bear in mind that there can be scenarios where the assumptions made to arrive to Eq.~\eqref{eq:GKSL} do not hold, thus hindering the applicability of the GKSL master equation to describe the evolution of your system of interest. This is the case, for example, if there is a backflow of information between system and environment, a characteristic feature of non-Markovian evolutions.

\subsubsection{Time-dependent GKSL equation}
In the previous derivations we have assumed a time-independent generator of the dynamics. This is not a prerequisite to derive Eq.~\eqref{eq:GKSL}, albeit it simplifies the calculations. Without entering into much details, one can show that, a dynamical map fulfilling the dynamical semigroup or CP-divisibility $V(t_2,t_1)V(t_1,t_0)=V(t_2,t_0)$, leads into a general form 
\begin{align}\label{eq:GKSL_t}
    \frac{d}{dt}\rho_S(t)=\mathcal{L}\{\rho_S\}=-i[H(t),\rho_S]+\sum_{k=1}^{d^2-1}\gamma_k(t)\left(A_k\rho_S A_k^\dagger-\frac{1}{2}\{A_k^\dagger A_k,\rho_S \} \right).
\end{align}
where $\gamma_k(t)\geq 0, \ \forall k, t$. That is, the evolution is still Markovian although the rates may be time dependent. A failure to meet the positivity of the rates implies that the map is not CP-divisible. For negative rates, the evolution or dynamical map cannot be divided as $V(t_2,t_0)=V(t_2,t_1)V(t_1,t_0)$ with CP dynamical maps $V(t_2,t_1)$ and $V(t_1,t_0)$, for some times $t_0\leq t_1\leq t_2$.  
This does not directly violate the physical validity of the evolved states $\rho_S(t)$. More specifically, $V(t_2,t_0)$ may still describe a physically sound dynamical evolution, but those sliced dynamical maps $V(t_2,t_1)$ and $V(t_1,t_0)$ do not. In other words, the negative rates  $\gamma_k(t_s)<0$ at some time $t_s$ for some $k$ mode can lead into valid physical states at later times $t\geq 0$. If this happens, the evolution is physical but no longer Markovian! The physical evolution from $t_0$ to $t$ keeps memory about its history, and therefore, the dynamical map from $t_1>t_0$ to $t_2$ does not guarantee a final physical state for any initial state $\rho(t_1)$.


\subsubsection{Adjoint GKSL master equation}
The GKSL master equation is written for the evolution of the quantum states $\rho(t)$. In analogy to the Heisenberg picture, which we reviewed in Sec.~\ref{c:intro}, it is possible to derive a  dynamical equation for operators, including the impact of decoherence processes. In this framework or picture, the state $\rho(t_0)$ remains constant, while operators evolve. 

Let us first define the adjoint dynamical map. As we did for closed systems, we start from  the expectation value of an operator $Q$, which can be written as
\begin{align}
    \langle Q(t)\rangle={\rm Tr}[Q\rho(t)]={\rm Tr}[QV(t,t_0)\{\rho(t_0)\}]={\rm Tr}[V^\dagger(t,t_0)\{Q\}\rho(t_0)].
\end{align}
Hence, $V^\dagger(t,t_0)$ is the adjoint map acting on the operator $Q$, so that $Q_H(t)=V^\dagger(t,t_0)\{Q\}$ refers to the operator in the Heisenberg picture at time $t$. The initial condition must be $Q_H(t_0)=Q$. Taking the time-derivative, we get
\begin{align}
    \frac{d}{dt}\langle Q(t)\rangle \equiv \frac{d}{dt}{\rm Tr}[Q_H(t)\rho(t_0)]={\rm Tr}[\left(\frac{d}{dt}Q_H(t)\right)\rho(t_0)].
    \end{align}
    On the other hand, the expectation value must be equal in the Schr\"odinger picture, so the time-derivative also reads as
    \begin{align}
        \frac{d}{dt}\langle Q(t)\rangle \equiv{\rm Tr}[Q\dot{\rho}(t)]={\rm Tr}[Q\mathcal{L}\{\rho(t)\}]={\rm Tr}[\mathcal{L}^\dagger\{ Q\}\rho(t)]={\rm Tr}[V^\dagger(t,t_0)\mathcal{L}^\dagger\{Q\}\rho(t_0)].
\end{align}
Equating both expressions we finally find the general expression for the adjoint master equation:
\begin{align}
    \frac{d}{dt}Q_H(t)=V^\dagger(t,t_0)\{ \mathcal{L}^\dagger\{Q\}\}.
\end{align}
This results in an intricate expression. It involves the generator and the map from initial to the evolved time $t$. So, in general, it is not enough the knowledge of the generator, one also needs the full dynamical adjoint map from initial to the time of interest. However, if the generator is time independent, $V^\dagger(t,t_0)$ commutes with $\mathcal{L}^\dagger$ so that we can write a much simpler and nicer time-local differential equation,
\begin{align}
    \frac{d}{dt}Q_H(t)=\mathcal{L}^\dagger\{Q_H(t)\}.
\end{align}

From an operational point of view, we need to arrange the terms of the GKSL master equation to find an explicit expression for the adjoint master equation $\mathcal{L}^\dagger\{Q\}$. In essence we need to isolate the state $\rho$ at the right from $\frac{d}{dt}{\rm Tr}[Q\rho(t)]$. This only requires the cyclic (and linear) trace property. For the first term in Eq.~\eqref{eq:GKSL} we have
\begin{align}
{\rm Tr}[Q(-i[H,\rho])]=-i{\rm Tr}[QH\rho-Q\rho H]=-i{\rm Tr}[QH\rho-HQ\rho]=i{\rm Tr}[[H,Q]\rho].
\end{align}
The second term results in
\begin{align}
    {\rm Tr}[Q A_k \rho A_k^\dagger]={\rm Tr}[A_k^\dagger Q A_k \rho],
\end{align}
and finally, for the third we only need to see that 
\begin{align}
    {\rm Tr}[Q \rho A_k^\dagger A_k]={\rm Tr}[A_k^\dagger A_k Q \rho].
\end{align}
In this case (time-independent generator),  the adjoint GKSL master equation adopts the form
\begin{align}\label{eq:adjoint_GKSL}
    \frac{d}{dt}Q_H(t)=\mathcal{L}^\dagger\{ Q_H(t\}=i[H,Q_H(t)]+\sum_k \gamma_k \left(A_k^\dagger Q_H(t) A_k-\frac{1}{2}\{A_k^\dagger A_k,Q_H(t)\} \right).
\end{align}
This is very similar to the normal GKSL master equation, but with the difference of the minus sign for the coherent part (check also the Heisenberg dynamical equation in closed systems) and the ordering for the first term of the dissipator. 

\vspace{0.5cm}
\noindent\textbf{Trace preserving implies unital adjoint master equation}
\vspace{0.2cm}

A quantum channel $\mathcal{E}$ is unital if it preserves the identity operator, $\mathcal{E}(\mathbb{I})=\mathbb{I}$. In the context of dynamical maps, this means that $V(t,t_0)\{\mathbb{I}\}=\mathbb{I}, \forall t\geq t_0$, or equivalently, its generator must fulfill  $\mathcal{L}\{\mathbb{I}\}=0$.

If the dynamical map preserves the trace, the adjoint master equation must be unital. That is, $\mathbb{I}_H(t)=\mathbb{I}, \forall t$ since ${\rm Tr}[\mathbb{I}_H(t)\rho(t_0)]={\rm Tr}[\mathbb{I}\rho(t)]=1$, and therefore $\mathcal{L}^\dagger\{\mathbb{I}\}=0$ must hold. Note, however, that the GKSL master equation in the Schr\"odinger picture is, in general, not unital. 

Substituting $Q_H\rightarrow \mathbb{I}$ in the adjoint GKSL master equation in~\eqref{eq:adjoint_GKSL} one can easily show that, indeed, it is unital: $\mathcal{L}^\dagger\{ \mathbb{I}\}=0$.

\subsubsection{Quantum regression theorem}
In many experimental settings, it is possible to  determine the frequency spectrum of certain observables, which are intimately related to multi-time correlation functions. These quantities, therefore,  offer insight of the system beyond single time expectation values. The quantum regression theorem \cite{Lax63} is an important tool in computing these time correlation functions, which is derived assuming a GKSL Markovian master equation. This theorem states that if the dynamics is governed by a GKSL Markovian master equation, the expectation values of multi-time correlation functions follow the same dynamics as the simple one-time expectations values but with modified initial conditions. Here we will just provide an operational recipe that follows from the main result of the quantum regression theorem. 

Suppose two operators $Q,P$, the two-time correlation function can be written as (for simplicity, we fix $t_0=0$)
\begin{align}
    C(t+\tau,t)=\langle Q(t+\tau)P(t)\rangle={\rm Tr}[Q(t+\tau)P(t)\rho(0)],
\end{align}
for $t,\tau\geq 0$, and with $Q(t+\tau)$ and $P(t)$ denoting the Heisenberg operator at times $t+\tau$ and $t$, respectively. The quantum regression theorem states that, if $Q(t)$ evolves with generator $\mathcal{L}^\dagger\{Q(t)\}$, so does $Q(t+\tau)P(t)$ but with a modified initial condition. Let us see this in detail for a time-independent GKSL master equation. Then,
\begin{align}
    C(t+\tau,t)={\rm Tr}[Q(t+\tau)P(t)\rho(0)]={\rm Tr}[e^{\tau\mathcal{L}^\dagger}Q e^{t\mathcal{L}^\dagger}P \rho(0)]={\rm Tr}[Qe^{\tau\mathcal{L}}\{P\rho(t)\}],
\end{align}
where we have made explicit the action of the map $V(\tau)=e^{\tau\mathcal{L}}$ on the operator $P\rho(t)$. From an operational point of view, the two-time correlation  function $C(t+\tau,t)$ requires:
\begin{itemize}
    \item Evolve the initial state from $\rho(0)$ to $\rho(t)$
    \item Compute the operator (possibly unnormalized) $P\rho(t)$ and evolve it in time by another $\tau$, to get $\rho_P(t+\tau)=e^{\tau\mathcal{L}}\{P\rho(t)\}$.
    \item Compute the expectation value of interest as $C(t+\tau,t)={\rm Tr}[Q\rho_P(t+\tau)]$.
\end{itemize}

In many experiments, the two-time correlation function is computed at the equilibrium state, that is, $C(\tau)=\lim_{t\rightarrow \infty} C(t+\tau,t)$. In this manner, one has to replace $\rho(t)$ by the corresponding long-time equilibrium state or steady state $\rho_{ss}$, i.e.
\begin{align}
    C(\tau)={\rm Tr}[Q e^{\tau \mathcal{L}}{P\rho_{ss}}].
\end{align}
This is a powerful result, but recall that it is based on a Markovian time-independent generator $\mathcal{L}$. 
The typical example to underline the importance of the quantum regression theorem consists in computing the emission properties of a quantum emitter. Indeed, the spectrum of the emitted radiation $S(\omega)$ is proportional to the Fourier transform of the two-time correlation function $C(\tau)=\langle \sigma^+(\tau)\sigma^-\rho_{ss}\rangle$. Although we skip the details here, note that under the previous assumptions, the spectrum can be computed analytically.   

\subsection{Liouvillian Superoperator}\label{s:LiovSP}
As we have seen, the GKSL equation is the most general form for a Markovian quantum dynamics. In doing so, we have talked about the Liouvillian $\mathcal{L}\{ \rho\}$, which is commonly known as Liouvillian \textit{superoperator} because it acts on operators, rather than on states as the standard operators. As we mentioned above, in the absence of interactions with the environment, one recovers the von Neumann equation, $\lim_{H_I\rightarrow 0}\mathcal{L}\{\rho\}=-i[H,\rho]$, but in general for a Markovian evolution one has the GKSL master equation, cf.~\eqref{eq:GKSL}. Now, how do we operate or handle this dynamical equation? In certain cases, the differential equation in matrix form may be exactly solvable. If not, one can still use numerical integration methods, such as Runge-Kutta, to obtain reliable numerical solutions. 
However, even for simple problems, this matrix formulation makes it difficult to gain insight into the dynamics. For that, it is useful to rewrite the Liouvillian superoperator acting on a vectorized density matrix (sometimes called the Fock-Liouville space):
\begin{align}
    \frac{d}{dt}\rho_S(t)=\mathcal{L}\{\rho_S(t)  \} \ \Rightarrow \  \frac{d}{dt}\kket{\rho_S(t)}=\mathbb{L}\kket{\rho_S(t)},
\end{align}
where the double-ket $\kket{\rho}$ denotes the column-stacked or column-ordered density matrix, i.e.
\begin{align}
    \rho=\begin{pmatrix}  \rho_{00} & \rho_{01} & \dots  & \rho_{0d}\\
    \rho_{10} & \rho_{11} & \dots & \rho_{1d} \\
    \vdots & \vdots & \ddots & \vdots\\
    \rho_{d0} & \dots  & \dots &\rho_{dd} 
    \end{pmatrix} \ \Rightarrow \ \kket{\rho}=\begin{pmatrix} \rho_{00} \\ \rho_{10} \\ \vdots \\ \rho_{d0} \\ \rho_{01}\\ \rho_{11}\\ \vdots \\ \rho_{dd}
    \end{pmatrix}.
\end{align}
Doing this, one can see that any superoperator of the form $AXB$ can be expressed as\footnote{The core of this transformation boils down to notice that $\kket{\ket{n}\bra{m}}=\ket{m}\otimes \ket{n}$. For example, say $E=\ket{0}\bra{1}=\begin{pmatrix} 0 & 1\\0 & 0\end{pmatrix}$, so that $\kket{E}=\begin{pmatrix} 0 & 0 & 1 & 0\end{pmatrix}^T$. This is equivalent to $\kket{E}=\ket{1}\otimes \ket{0}$. Then, $A\ket{n}\bra{m}B=(A\ket{n})(B^T\ket{m})^T$, and its vectorized form reads $\kket{A\ket{n}\bra{m}B}=(B^T\ket{m})\otimes (A\ket{n})=(B^T\otimes A)(\ket{m}\otimes \ket{n})$. Summing over all $m,n$ with the corresponding matrix element one recovers  Eq.~\eqref{eq:AXB_vec}. }
\begin{align}\label{eq:AXB_vec}
    \kket{A X B} = (B^T\otimes A)\kket{X},
\end{align}
where the superscript ${}^T$ refers to the transpose matrix operation. This vectorization is known as the Choi-Jamio{\l}kowski isomorphism \cite{Choi75, Jam72}.

Since the Liouvillian only contains terms like $A\rho B$, this allows us to give a systematic recipe to construct the Liouvillian superoperator acting, that we denote as $\mathbb{L}$, which acts on the vectorized density matrix, $\kket{\rho}$. Note that by doing so, the vectorized Liouvillian superoperator increases its dimension: For a system of dimension $d$, $\mathbb{L}$ is a complex-valued matrix of dimension $d^2\times d^2$. See Box~\ref{examp:super_master} for an example of how one transforms each of the terms of Eq.~\eqref{eq:GKSL} into the vectorized space.

\begin{examp}[GKSL master equation in vectorized form]{examp:super_master}\label{examp:super_master}
Let us see how each of the terms in the GKSL Markovian master equation transform in order to construct the Liouvillian superoperator $\mathbb{L}$. For that, we use the trick or mathematical identity given in Eq.~\eqref{eq:AXB_vec}. Expressing Eq.~\eqref{eq:GKSL} with identity matrices to match the expression $\kket{AXB}=(B^T\otimes A)\kket{X}$, 
\begin{align}
    \mathbb{I}\frac{d}{dt}\rho(t) \mathbb{I}=-i(H\rho(t)\mathbb{I}-\mathbb{I}\rho(t) H)+\sum_k \gamma_k\left(A_k \rho(t) A_k^\dagger-\frac{1}{2}A_k^\dagger A_k \rho(t) \mathbb{I}-\frac{1}{2}\mathbb{I}\rho(t) A_k^\dagger A_k \right)
\end{align}
so that 
\begin{align}
    \frac{d}{dt}\kket{\rho(t)}=\mathbb{L}\kket{\rho(t)},
\end{align}
with
\begin{align}
    \mathbb{L}=-i(\mathbb{I}\otimes H -H^T\otimes \mathbb{I})+\sum_k \gamma_k\left([A_k^\dagger]^T\otimes A_k-\frac{1}{2}(\mathbb{I}\otimes A_k^\dagger A_k)-\frac{1}{2}((A_k^\dagger A_k)^T\otimes \mathbb{I}) \right).
\end{align}
This can be written in a slightly more compact form  noting that $A^\dagger=(A^T)^*$, $(A^T)^T=A$ and $(A^*)^*=A$, 
\begin{align}
    \mathbb{L}=-i(\mathbb{I}\otimes H -H^T\otimes \mathbb{I})+\sum_k \gamma_k\left(A_k^*\otimes A_k-\frac{1}{2}(\mathbb{I}\otimes A_k^\dagger A_k)-\frac{1}{2}(A_k^T A_k^*\otimes \mathbb{I}) \right).
\end{align}
\end{examp}

It might seem that we have not gained much, but just increased the dimension of the problem we want to solve. To make the advantage explicit, recall that we know how to solve a differential equation of the form $\dot{v}=A v$. Indeed, if the Liouvillian $\mathbb{L}$ is time independent, we can directly write its solution as
\begin{align}
    \kket{\rho(t)}=e^{(t-t_0)\mathbb{L}}\kket{\rho(t_0)},
\end{align}
which is extremely useful in many relevant cases. For a time-dependent Liouvillian, we need to integrate over time keeping its causal or time order. This is similar to the unitary time evolution operator $U(t,t_0)$ we reviewed in Sec.~\ref{c:intro}. In this manner, we can write
\begin{align}
    \kket{\rho(t)}=\mathcal{T}e^{\int_{t_0}^t dt'\mathbb{L}(t')}\kket{\rho(t_0)}.
\end{align}
where $\mathcal{T}$ denotes the time ordering operator. In essence, the vectorization trick allows us to transform the master equation into a sort of Schr\"odinger-like equation, but obviously recalling the differences: the evolution is not unitary, unless $\gamma_k=0, \forall k$, and we are dealing with mixed states. The exponential of the Liouvillian superoperator $\mathbb{L}$ is enough to find the state at any time $t$. This method can be further developed to gain insight into the non-unitary dynamics.

The Liouvillian matrix $\mathbb{L}$ is, in general, a complex  non-Hermitian $d^2\times d^2$ matrix. Therefore, it is not always possible to find a diagonal form like $\mathbb{L}=VDV^{-1}$. If this were the case, the matrix $\mathbb{L}$ is said to be defective; it cannot be diagonalized\footnote{In order to check whether a matrix $A$ is defective, one starts by computing its eigenvalues, ${\rm det}(A-\lambda\mathbb{I})=0$. If at least of the eigenvalues has multiplicity $m>1$, but the eigenvectors associated to it span a vector space of dimension $n<m$, then $A$ is defective. That is, there are only $n<m$ linearly-independent eigenvectors associated to that eigenvalue.}.
So let us first consider that $\mathbb{L}$ can be diagonalized, and then in Sec.~\ref{ss:Jordan} we will discuss the most general case and its implications.  In Sec.~\ref{ss:SVD} we will briefly discuss how a singular value decomposition can also be used to compute steady states.

\subsubsection{Nondefective Liouvillian}
Suppose $\mathbb{L}$ is diagonalizable. Then, one can write 
\begin{align}\label{eq:LiovDiag}
    \mathbb{L}=V D V^{-1},
\end{align}
where $D$ is a diagonal matrix whose entries are the eigenvalues $\{ \lambda_\alpha\}$ of $\mathbb{L}$. In this case, one can find left and right eigenvectors, 
with the corresponding set of complex eigenvalues  $\{\lambda_\alpha\}$, 
\begin{align}    \mathbb{L}\kket{\tilde{r}_\alpha}&=\lambda_\alpha \kket{\tilde{r}_\alpha},\\
\bbra{\tilde{l}_\alpha}\mathbb{L}&=\lambda_\alpha\bbra{\tilde{l}_\alpha}.
\end{align}
Note that $\kket{\tilde{r}_\alpha}$ and $\kket{\tilde{l}_\alpha}$ are column vectors of dimension $d^2$. In addition, if $\bbra{\tilde{l}_\alpha}$ is a left eigenvector of $\mathbb{L}$, then $(\bbra{\tilde{l}_\alpha})^T=(\kket{\tilde{l}_\alpha})^*$ is a right eigenvector of $\mathbb{L}^T$. It is convenient to normalize the eigenvectors according to
\begin{align}
    \kket{r_\alpha}&=\frac{1}{\sqrt{\langle \langle \tilde{l}_\alpha | \tilde{r}_\alpha \rangle \rangle}}\kket{\tilde{r}_\alpha},\\
    \kket{l_\alpha}&=\frac{1}{\sqrt{\langle \langle \tilde{l}_\alpha | \tilde{r}_\alpha \rangle \rangle}}\kket{\tilde{l}_\alpha},
\end{align}
so that they obey the bi-orthonormal condition,
\begin{align}
    \langle \langle l_\alpha | r_\beta \rangle \rangle =\delta_{\alpha,\beta},
\end{align}
and fulfill the completeness relation
\begin{align}
    \sum_{\alpha=1}^{d^2} \kket{r_\alpha}\bbra{l_\alpha}=\mathbb{I}.
\end{align}
The right eigenvectors $\kket{r_\alpha}$ are the columns of $V$, while the row vectors of $V^{-1}$ are the left eigenvectors $\bbra{l_\alpha}$\footnote{Recall that for a Hermitian matrix $A$, the diagonalization results in $A=VDV^{\dagger}$, since $A=A^\dagger$. Therefore, in this special case, it is not necessary to distinguish between left and right eigenvectors: Each left eigenvector is simply the conjugate transpose of the corresponding right eigenvector.}. As a further remark: be careful when using a numerical or symbolic software to diagonalize $\mathbb{L}$. They typically output the matrix of eigenvectors $U$, which not necessarily corresponds to $V$ defined in Eq.~\eqref{eq:LiovDiag}, e.g. the output may refer to $V^{-1}$ rather than $V$, or with different ordering.

The diagonalization  of $\mathbb{L}$ is good news to compute the dynamics: we can expand the initial state $\kket{\rho(t_0)}$ in the eigenbasis and write the exact solution at any time $t\geq t_0$ as
\begin{align}
    \kket{\rho(t)}&=e^{(t-t_0) \mathbb{L}}\kket{\rho(t_0)}= e^{(t-t_0) \mathbb{L}}\left(\sum_{\alpha=1}^{d^2}\kket{r_\alpha}\bbra{l_\alpha}\right) \kket{\rho(t_0)}\nonumber\\ \label{eq:rhot_vec} &= \sum_{\alpha=1}^{d^2}\langle \langle l_\alpha | \rho(t_0) \rangle \rangle \kket{r_\alpha}e^{\lambda_\alpha (t-t_0)}=\sum_{\alpha=1}^{d^2} c_\alpha(t_0) \kket{r_\alpha}e^{\lambda_\alpha (t-t_0)},
\end{align}
where $c_\alpha(t_0)=\langle \langle l_\alpha|\rho(t_0)\rangle\rangle$ are the expansion coefficients of the initial state into the left eigenvectors.  
Note how this expression is very similar to a unitary evolution (cf. Sec.~\ref{moremat:H_vs_L}). 

If the Liouvillian is time dependent one can proceed in a similar fashion as for time-dependent Hamiltonian evolution: discretize the time in small steps and perform a Trotterization of the time-evolution operator.

\vspace{0.5cm}
\noindent\textbf{Left eigenvectors and conserved quantities: Trace preservation}
\vspace{0.2cm}

Interestingly, the left eigenvectors $\bbra{l_\alpha}$ with zero eigenvalue are related to conserved quantities. This can be seen, for example with the trace, i.e. ${\rm Tr}[\dot{\rho}]=0$. As we have seen, the GKSL master equation preserves the norm of the state, and therefore, the structure of the Liouvillian matrix $\mathbb{L}$ is such that one left eigenvector must correspond to the vectorized identity matrix, whose eigenvalue must be zero. That is,
\begin{align}
    \langle \langle \mathbb{I} | \rho(t_0) \rangle \rangle= \sum_{k=1}^{d}\rho_{k,k}(t_0)={\rm Tr}[\rho(t_0)]=1,
\end{align}
by definition of a physical valid state $\rho(t_0)$. Since the trace cannot change in time, ${\rm Tr}[\dot{\rho}(t)]=0$, or similarly,
\begin{align}
    {\rm Tr}[\dot{\rho}(t)]=0\Rightarrow {\rm Tr}[\dot{\rho}(t)]=\langle \langle \mathbb{I} | \dot{\rho}(t) \rangle \rangle=\bbra{\mathbb{I}}\mathbb{L}\kket{\rho(t)}=0 \Rightarrow \bbra{\mathbb{I}}\mathbb{L}=0,
\end{align}
where the last implication follows from the fact that the equality must hold for any state $\rho(t)$. Hence,  at least one eigenvalue is equal to zero, $\lambda_1=0$, whose left eigenvector is the vectorized identity matrix.\footnote{Note that when there are two or more eigenvalues equal to zero, it is possible that their associated left eigenvectors do not correspond directly to the vectorized identity matrix. In this case, due to the existence of extra conserved quantities, it must be possible to find a linear combination of them that yields  $\bbra{\mathbb{I}}$. } 

In general, if there a left eigenvector with zero eigenvalue, it corresponds to a conserved quantity. Suppose $\bbra{l_k}\mathbb{L}=0$ for some $k$. Then, this means that
\begin{align}
    \bbra{l_k}\mathbb{L}\kket{\rho(t)}=0 \Rightarrow \langle \langle l_k| \dot{\rho}(t)\rangle \rangle=0 \Rightarrow \frac{d}{dt}{\rm Tr}[l_k^\dagger\rho(t)]=0,
\end{align}
so that ${\rm Tr}[l_k^\dagger \rho(t)]$ remains constant throughout the evolution; the expectation value of the operator $l_k^\dagger$ is constant in time. Recall that $\bbra{l_k}=\left(\kket{l_k} \right)^\dagger$. We will present some examples to make this notation more clear. 

\vspace{0.5cm}
\noindent\textbf{Steady states and decay modes}
\vspace{0.2cm}

Let us inspect further Eq.~\eqref{eq:rhot_vec}. Although we have not said anything about the spectrum of $\mathbb{L}$, the evolution in Eq.~\eqref{eq:rhot_vec} should make clear that,  in order to ensure the physical validity of the evolved states, i.e. that the GKSL is a CPTP dynamical map, the spectrum $\{\lambda_\alpha\}$ of $\mathbb{L}$ must be such
\begin{align}
    {\rm Re}[\lambda_\alpha]\leq 0, \forall \alpha.
\end{align}
It is easy to see that if there is some eigenvalue with ${\rm Re}[\lambda_\alpha]>0$,  the long-time evolution will make the norm of the state to grow indefinitely, violating the trace preserving character of the dynamical map. Thus, all eigenvalues must fulfill ${\rm Re}[\lambda_\alpha]\leq 0$. 

Furthermore, we can distinguish between the kernel of $\mathbb{L}$ (i.e. the space of right eigenvectors with zero eigenvalues) from the rest, those right eigenvectors with associated eigenvalues with negative real part. Note that due to the trace preserving character of $\mathbb{L}$, we know that there is at least one eigenvalue equal to zero. That is, $d_K=\dim\{ {\rm ker}(\mathbb{L})\}\geq 1$, or equivalently ${\rm rank}(\mathbb{L})\leq d^2-1$.

The right eigenvectors of the kernel define the possible steady state solutions: 
\begin{align}
    \frac{d}{dt}\kket{\rho_{ss}}=0 \Rightarrow \mathbb{L}\kket{\rho_{ss}}=0.
\end{align}
That is, if $\kket{r_\beta}\in {\rm ker}(\mathbb{L})$, then $\mathbb{L}\kket{r_{\beta}}=0$, and therefore, any linear combination of these right-eigenvectors of the kernel results in a valid steady state solution, $\kket{\rho_{ss}}=\sum_\beta c_\beta \kket{r_\beta}$, provided ${\rm Tr}[\rho_{ss}]=1$, or similarly, $\langle \langle \mathbb{I}|\rho_{ss}\rangle \rangle=1$. If $d_K=1$, then the steady state is unique. Otherwise, $d_K>1$, the steady state depends on the initial configuration, i.e. on the coefficients $c_\beta(t_0)$. 

An alternative way to show the steady state consists in taking the long-time limit to Eq.~\eqref{eq:rhot_vec}, which results in
\begin{align}
    \lim_{t\rightarrow \infty} \kket{\rho(t)}=\lim_{t\rightarrow \infty} \sum_{\alpha=1}^{d^2} c_\alpha(t_0)e^{\lambda_\alpha (t-t_0)} \kket{r_\alpha}=\sum_{\beta\in {\rm ker}(\mathbb{L})}c_\beta(t_0)\kket{r_\beta}.
\end{align}
All the right-eigenvectors with ${\rm Re}[\lambda_\alpha]<0$ decay in time, and in the long-time limit, only those in the kernel survive. Note how the magnitude of ${\rm Re}[\lambda_\alpha]$ of the decay modes also determines the rate at which they are washed out.  Therefore, we can conclude that the eigenvalues of a non-defective Liouvillian superoperator $\mathbb{L}$, together with its left and right eigenvectors completely determine the dynamical evolution. They fix the steady or equilibrium state, as well as the rate at which other modes of the system decay with time. 

Before moving on, let us list the  powerful results we have obtained:
\begin{itemize}
    \item[(i)] Any trace-preserving dynamical map must have an associated generator $\mathbb{L}$ with a zero eigenvalue, whose left eigenvector is the vectorized identity matrix, and the right eigenvector is the steady state.
    \item[(ii)] If the zero eigenvalue is degenerate, there are extra conserved quantities and the steady state is not unique.
    \item[(iii)] The spectrum of $\mathbb{L}$ is such all eigenvalues have non-positive real part, ${\rm Re}[\lambda_\alpha]\leq 0$. 
\end{itemize}
These findings are illustrated in Box~\ref{examp:LiovSOP_qubit} for a particular example of a qubit with driving and dephasing noise. 

\begin{figure}[h]
    \centering
        \includegraphics[width=0.9\linewidth]{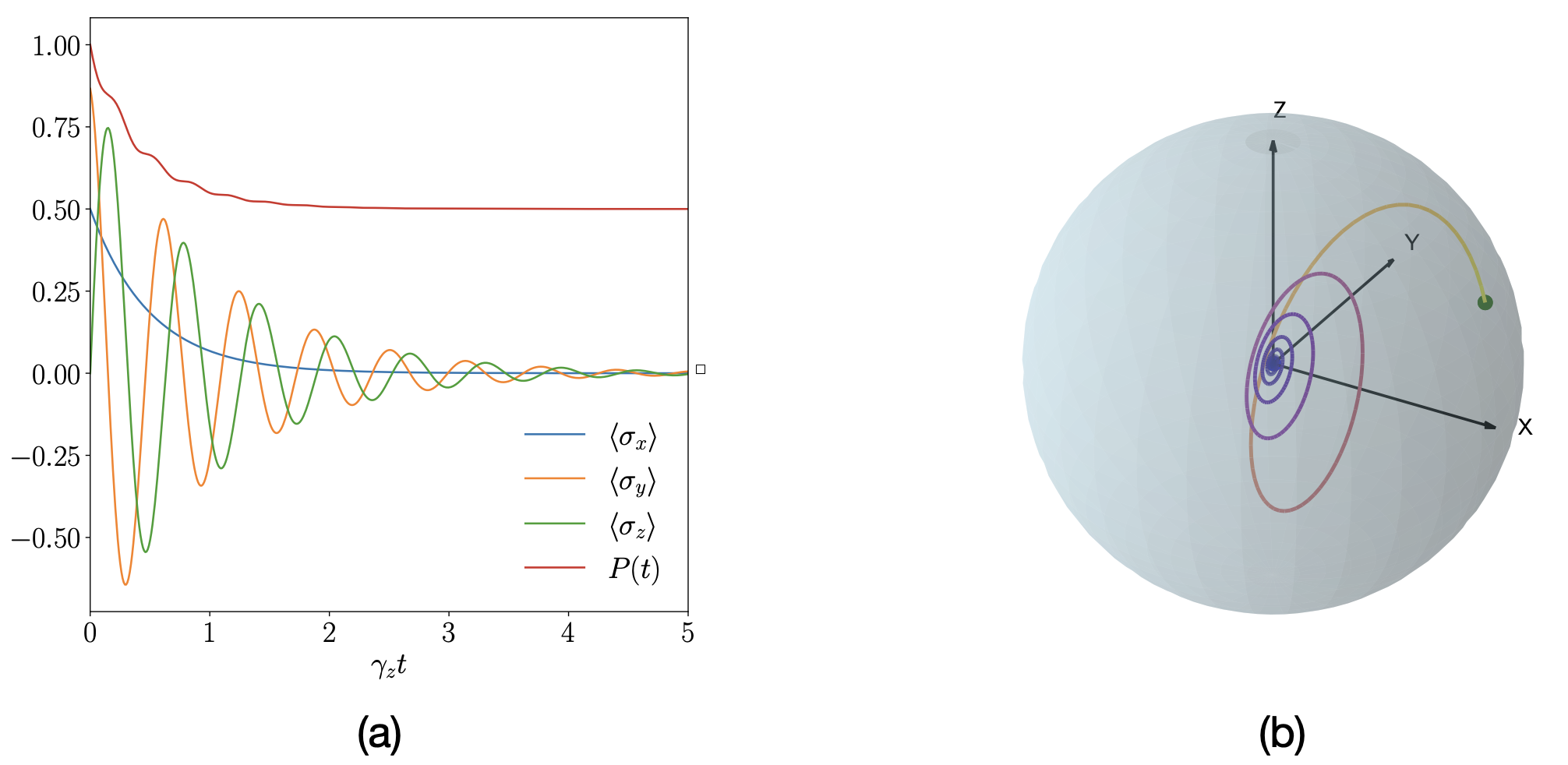}
    \caption{Dynamics of a qubit undergoing driving and dephasing, initially in a pure state $\rho(0)=\frac{1}{2}\begin{pmatrix}1 & e^{-i\pi/3} \\ e^{i\pi/3} & 1\end{pmatrix}$, and with parameters $\omega=10\gamma_z$. (a) Dynamics of the Bloch vector components, as well as the purity $P(t)={\rm Tr}[\rho^2(t)]$. In the long time limit, $\rho_{ss}=\mathbb{I}/2$. (b) Trajectory of the qubit state in the Bloch sphere.}
    \label{fig:examp_Liovqubit}
\end{figure}

\begin{examp}[Qubit Liouvillian superoperator]{examp:LiovSOP_qubit}\label{examp:LiovSOP_qubit}
Consider a single qubit whose Hamiltonian is $H=\omega\sigma_x/2$ and it undergoes Markovian dephasing at constant rate $\gamma_z$, whose Lindblad operator is $\sigma_z$. 
\begin{align}
    \frac{d}{dt}\rho(t)=-i[\frac{\omega}{2}\sigma_x,\rho(t)]+\gamma_z\left(\sigma_z\rho(t)\sigma_z -\rho(t)\right),
\end{align}
where we used the fact $\sigma_z^\dagger=\sigma_z$ and $\sigma_z\sigma_z=\mathbb{I}$. Using the Choi-Jamio{\l}kowski isomorphism (i.e., column-stacked vectorization), the master equation is transformed into $\frac{d}{dt}\kket{\rho(t)}=\mathbb{L}\kket{\rho(t)}$ $\mathbb{L}$ given by
\begin{align}
    \mathbb{L}=\frac{1}{2}\begin{pmatrix} 0 &-i\omega & i\omega & 0\\
    -i\omega & -4\gamma_z& 0 & i\omega\\
    i\omega & 0 & -4\gamma_z&-i\omega\\ 0& i\omega & -i\omega & 0\end{pmatrix}.
\end{align}
$\mathbb{L}$ is diagonalizable if $\omega\neq \pm \gamma_z$. The defective case will be discussed in Box~\ref{examp:LiovSOP_qubit_defective}. 
In this manner,  $\mathbb{L}=VDV^{-1}$ with $D={\rm diag}(0,-2\gamma_z,-\gamma_z+\sqrt{\gamma_z^2-\omega^2},-\gamma_z-\sqrt{\gamma_z^2-\omega^2})$ and 
\begin{align}
    V=\begin{pmatrix} \frac{1}{2} & 0 & -\frac{1}{4}\left(1+\frac{\gamma_z}{\Delta} \right) & -\frac{1}{4}\left(-1+\frac{\gamma_z}{\Delta} \right)\\
    0 & \frac{1}{2}& \frac{i\omega}{4\Delta} & \frac{-i\omega}{4\Delta}\\
    0 & \frac{1}{2} & \frac{-i\omega}{4\Delta} & \frac{i\omega}{4\Delta}\\
    \frac{1}{2} & 0 & \frac{1}{4}\left(1+\frac{\gamma_z}{\Delta} \right) & \frac{1}{4}\left(1-\frac{\gamma_z}{\Delta} \right)        
    \end{pmatrix},
\end{align}
with $\Delta=\sqrt{\gamma_z^2-\omega^2}$. The columns of $V$ are the right eigenvectors $\kket{r_\alpha}$. One can verify (by doing the inverse, and retrieving the left eigenvectors $\bbra{l_\alpha}$) that the bi-orthonormality and completeness relations are satisfied. This is best done, of course, via symbolic or numerical software. Notice in this example that: \begin{itemize}
    \item The steady state is unique (only one zero eigenvalue) iff $\omega\neq 0$. The steady state is given by the unvectorized right eigenvector $\kket{r_0}$, which is the maximally mixed state, $r_0=\mathbb{I}/2$.
    \item The left eigenvector $\bbra{l_0}$ corresponds to the vectorized identity $\bbra{l_0}=\bbra{\mathbb{I}}$; a requisite for normalized states at all times. 
    \item The eigenvalues are real for $\gamma_z\geq \omega$. This means that oscillatory behavior (underdamped regime) takes place only when $\gamma_z<\omega$, while overdamped for $\gamma_z\geq \omega$. Physically, the overdamped regime corresponds to a weak driving, so the dynamics are mostly governed by dephasing. 
\end{itemize}  
Doing this calculation, we obtain exactly the evolution of the state $\kket{\rho(t)}$ for any given initial state, $\kket{\rho(0)}$ using Eq.~\eqref{eq:rhot_vec}. This can be done analytically.  The results for the evolution of the Bloch-vector components, together with the purity $P(t)={\rm Tr}[\rho^2(t)]$ are plotted in Fig.~\ref{fig:examp_Liovqubit} for a particular pure initial state $\rho(0)$, and parameters $\omega=10\gamma_z$ (underdamped regime).

If $\omega=0$, then there are two zero eigenvalues. The state steady is no longer unique. One can check that $\rho_{s1}=\ket{0}\bra{0}$ and $\rho_{s2}=\ket{1}\bra{1}$ are valid steady states, i.e. $\mathcal{L}\{\rho_{sj}\}=0$. Hence, any state of the form $\rho_{ss}=p\rho_{s1}+(1-p)\rho_{s2}=p\ket{0}\bra{0}+(1-p)\ket{1}\bra{1}$ is also a valid steady state.

\end{examp}

\subsubsection{Singular value decomposition: Steady states}\label{ss:SVD}
Regardless of whether $\mathbb{L}$ is diagonalizable or defective, one can always do singular value decomposition: 
\begin{align}
\mathbb{L}=U \Sigma V^*,
\end{align}
where $\Sigma$ is a diagonal matrix with non-negative real values corresponding to the square root of the eigenvalues of $\mathbb{L}^\dagger\mathbb{L}$, and $U$ and $V$ are unitary matrices. As we have seen above, at least one eigenvalue must be zero (because the dynamical map is trace preserving). The kernel of $\mathbb{L}$ is translated into zero singular values, and therefore, all possible steady state solutions stem from linear combinations of the right singular vectors with zero singular value. That is, if one is solely interested in the steady state solution of the dynamics, then it is enough to compute the singular value decomposition. In the following, we discuss how the dynamics can be obtained for a defective $\mathbb{L}$ matrix.

\subsubsection{Defective Liouvillian: Jordan normal form}\label{ss:Jordan}
If $\mathbb{L}$ is defective, it cannot be diagonalized.   In that case, it is possible to find an almost-diagonal form, known as Jordan normal form, according to
\begin{align}
    \mathbb{L}=VJV^{-1},
\end{align}
where $V$ is an invertible matrix and $J$ a block-diagonal matrix, $J=\oplus_{\alpha=1}\oplus_{\beta=1}^{d(\alpha)} J_{\alpha,\beta}$, with $d(\alpha)$ the number of Jordan blocks for eigenvalue $\lambda_\alpha$ and each $J_{\alpha,\beta}$ can have dimension $n_{\alpha,\beta}$, 
\begin{align}
    J_{\alpha,\beta}=\begin{pmatrix} \lambda_\alpha & 1            & 0            & \cdots & 0 \\
0              & \lambda_\alpha & 1          & \ddots & \vdots \\
0              & 0            & \lambda_\alpha & \ddots & 0 \\
\vdots         & \ddots       & \ddots       & \ddots & 1 \\
0              & \cdots       & 0            & 0      & \lambda_\alpha
    \end{pmatrix}.
\end{align}
This can be written as $J_{\alpha,\beta}=\lambda_\alpha \mathbb{I}+T$, with $T$ the matrix containing $1$'s in the first upper diagonal. Note that $T^{n_{\alpha,\beta}}=0$, being the $n_{\alpha,\beta}$ the size of this block, which is related to the multiplicities of the eigenvalues. In addition, we remark that is it possible to have different Jordan blocks for the same eigenvalue.

Although the eigenvalues for $\mathbb{L}$ can be obtained in the standard manner, ${\rm det}(\mathbb{L}-\lambda\mathbb{I})=0$, being defective means that there are not enough linearly independent eigenvectors to span the whole space. How can we proceed then? We define generalized eigenvectors. For each Jordan block  with size $n_{\alpha,\beta}$, we have the chain
\begin{align}
    (\mathbb{L}-\lambda_\alpha\mathbb{I})\kket{r_{\alpha,\beta}^{(0)}}=0, \  (\mathbb{L}-\lambda_\alpha\mathbb{I})\kket{r_{\alpha,\beta}^{(1)}}=\kket{r_{\alpha,\beta}^{(0)}}, \dots, \mathbb{L}-\lambda_\alpha\mathbb{I})\kket{r_{\alpha,\beta}^{(n_{\alpha,\beta}-1)}}=\kket{r_{\alpha,\beta}^{((n_{\alpha,\beta}-2)}},
\end{align}
and similarly for left generalized eigenvectors. In this manner, within each Jordan block we can retrieve the bi-orthogonality condition
\begin{align}
    \langle \langle l_{\alpha,\beta}^{(i)}| r_{\alpha,\beta}^{(j)}\rangle \rangle=\delta_{i,j}.
\end{align}

\begin{figure}[t!]
    \centering
    \includegraphics[width=0.5\linewidth]{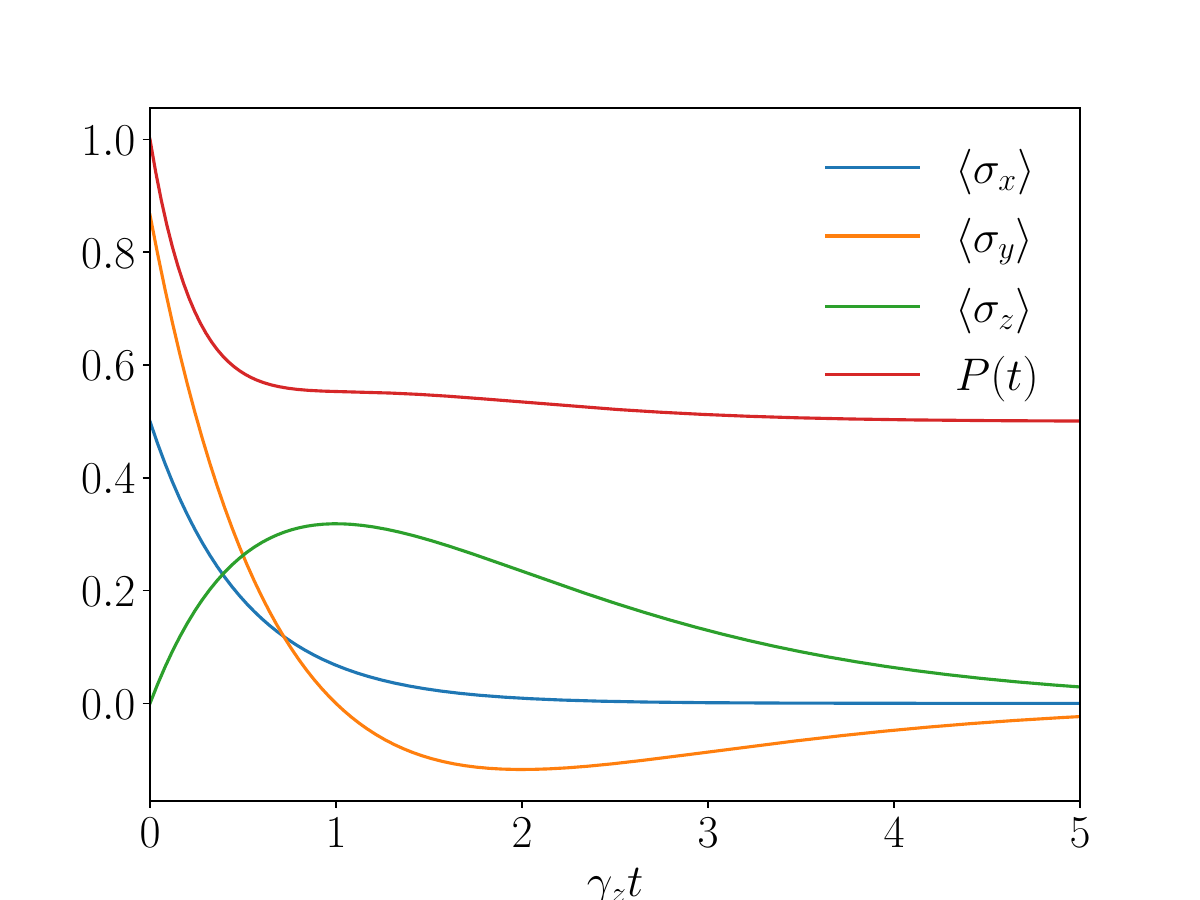}
    \caption{Dynamics of a qubit under driving and dephasing with a defective Liouvillian, $\gamma_z=\omega$. This gives rise to transient behavior where certain observables or properties scale as $t^ne^{-|\lambda_\alpha|t}$. In this simple case, appears as $v_z(t)\sim t e^{-\gamma_z t}$. }
    \label{fig:examp_Liovqubit_def}
\end{figure}

Now, we note that the matrix exponential of a Jordan block can be written as
\begin{align}
    e^{t J_{\alpha,\beta}}=e^{t \lambda_\alpha \mathbb{I}}e^{t T}=e^{t \lambda_\alpha} \sum_{n=0}^{n_{\alpha,\beta}-1}\frac{t^n}{n!}T^{n}=e^{t\lambda_\alpha}\sum_{n=0}^{n_{\alpha,\beta}-1}\frac{t^n}{n!}\sum_{m=0}^{n_{\alpha,\beta}-1-n} \kket{r_{\alpha,\beta}^{(m+n)}}\bbra{l_{\alpha,\beta}^{(m)}}.
\end{align}
So therefore, plugging this into $\kket{\rho(t)}=e^{t\mathbb{L}}\kket{\rho(0)}$ we find
\begin{align}\label{eq:rhot_vec_DEF}
\kket{\rho(t)}=\sum_{\alpha=1}e^{\lambda_\alpha t}\sum_{\beta=1}^{d(\alpha)}\sum_{n=0}^{n_{\alpha,\beta}-1}\frac{t^n}{n!}\sum_{m=0}^{n_{\alpha,\beta}-1-n} \kket{r_{\alpha,\beta}^{(m+n)}}\bbra{l_{\alpha,\beta}^{(m)}} \ \kket{\rho(0)}.
\end{align}
The main difference with respect to Eq.~\eqref{eq:rhot_vec} is the emergence of transient non-exponential behavior due to terms like $t^n e^{\lambda_\alpha t}$. That is, Markovian evolutions allow for $t^ne^{\lambda t}$ behavior with ${\rm Re}[\lambda]<0$. In any case, we can notice that:
\begin{itemize}
    \item[(i)] The spectrum of a defective Liouvillian must be again ${\rm Re}[\lambda_\alpha]\leq 0, \forall \alpha$. 
    \item There must be at least one eigenvalue $\lambda_\alpha=0$ (again, it defines the steady state(s)).
    \item[(ii)] The eigenvalue $\lambda_\alpha=0$ does not admit a Jordan block structure. That is, the blocks associated to $\lambda_\alpha=0$ must be scalars. The reason is again physically motivated: If $\lambda_\alpha=0$ had a Jordan-block structure with $n_{\alpha,\beta}>1$, then there would be terms in Eq.~\eqref{eq:rhot_vec_DEF} growing  polynomially in time, $t^n$, with $n\geq 1$, thus hindering the physical validity of the evolved states as $t\rightarrow\infty$. 
\end{itemize}

In Box~\ref{examp:LiovSOP_qubit_defective} we discuss a particular example of a defective Liouvillian (following the same physical problem of Box~\ref{examp:LiovSOP_qubit}). 

\begin{examp}[Defective Liouvillian for a qubit]{examp:LiovSOP_qubit_defective}\label{examp:LiovSOP_qubit_defective}
Let us consider the same physical problem of Box~\ref{examp:LiovSOP_qubit}, i.e. a qubit with driving $H=\omega\sigma_x/2$ and dephasing at constant rate $\gamma_z$. For $\omega=\gamma_z$ the Liouvillian is not diagonalizable. According to $\mathbb{L}=VJV^{-1}$, we find the Jordan normal form
\begin{align}
    J=\begin{pmatrix}0 & 0 & 0 & 0 \\
0 & -2\gamma_z & 0 & 0 \\
0 & 0 & -\gamma_z & 1 \\
0 & 0 & 0 & -\gamma_z
    \end{pmatrix}, 
\end{align}
Note the Jordan block for eigenvalue $-\gamma_z$. 
Again, the steady state is unique, $\rho_{ss}=\mathbb{I}/2$. One can then proceed as in Eq.~\eqref{eq:rhot_vec_DEF} and retrieve the exact state $\rho(t)$ at any time. Taking a generic initial state $\rho(0)$ with Bloch vector $\vec{v}=(v_x(0),v_y(0),v_z(0))$, Eq.~\eqref{eq:rhot_vec_DEF} allows us to find
\begin{align}
    v_x(t)&=e^{-2\gamma_z t}v_x(0),\\
    v_y(t)&=e^{-\gamma_z t}\left( v_y(0)-\gamma_z t (v_y(0)+ v_z(0))\right),\\ \label{eq:vz_def}
    v_z(t)&=e^{-\gamma_z t}\left(v_z(0)+\gamma_z t(v_y(0)+v_z(0) \right).
\end{align}
Note how the expectation values $\langle \sigma_{y,z}\rangle$ acquire a linear $t$ dependence at short times, a feature that reflects the defective nature of the underlying Liouvillian. This is plotted in Fig.~\ref{fig:examp_Liovqubit_def} for the same initial state as in Fig.~\ref{fig:examp_Liovqubit}, but with $\gamma_z=\omega$\footnote{For physical systems where the frequency is similar but not equal to the dephasing rate,  $\gamma_z\approx \omega$, the behavior is similar to the one shown in Fig.~\ref{fig:examp_Liovqubit_def} albeit the Liouvillian is no longer defective. That is, the dynamical behavior does not change sharply with $\gamma_z-\omega$, but rather in a smooth manner. A calculation for the general case leads, for example, to  $\langle \sigma_z\rangle=v_z(t)=e^{-\gamma_z t}\left( v_z(0)\cosh(t\Delta)+(\gamma_zv_z(0)+\omega v_y(0))\sinh(t\Delta)/\Delta\right)$ with $\Delta=\sqrt{\gamma_z^2-\omega^2}$. For $|\omega-\gamma_z|\ll 1$, the first order of the expansion corresponds to the Eq.~\eqref{eq:vz_def}. }. 

\end{examp}

\subsection{Problems to practice}

\noindent\textbf{Problem 2.1}

\noindent Verify that the dynamical equation for the system alone reduces to the von Neumann equation for a non-interacting scenario  $H=H_S\otimes \mathbb{I}_B+\mathbb{I}_S\otimes H_B$ and a product state $\rho(t)=\rho_S(t)\otimes \rho_B(t)$. That is, show that
\begin{align}
    {\rm Tr}_B[\frac{d}{dt}\rho(t)]=-i{\rm Tr}_B[[H,\rho_S(t)\otimes \rho_B(t)]] \Rightarrow \frac{d}{dt}\rho_S(t)=-i[H_S,\rho_S(t)].
    \nonumber
\end{align}

\vspace{0.5cm}
\noindent\textbf{Problem 2.2}
\vspace{0.2cm}

\noindent For an initial product state, $\rho(t_0)=\rho_S(t)\otimes \rho_B(t_0)$ with $\rho_B(t_0)=\sum_m \lambda_m \ket{b_m}\bra{b_m}$, the dynamical map $V(t,t_0)$ can be expressed as
\begin{align}
    V(t,t_0)\{ \rho_S(t_0)\}=\sum_{n,m}W_{n,m}(t,t_0)\rho_S(t_0)W_{n,m}^\dagger(t,t_0),
    \nonumber
\end{align}
with
$W_{n,m}(t,t_0)=\sqrt{\lambda_m}\bra{b_m}U(t,t_0)\ket{b_n}$,
being the $U(t,t_0)$ the unitary time evolution operator of the joint system $S+B$. 
Verify that these operators $W_{n,m}(t,t_0)$ satisfy 
\begin{align}
    \sum_{n,m} W_{n,m}^\dagger(t,t_0)W_{n,m}(t,t_0)=\mathbb{I}_S.
    \nonumber
\end{align}

\vspace{0.5cm}
\noindent\textbf{Problem 2.3}
\vspace{0.2cm}

\noindent Consider a two qubit system, $A$ and $B$. The initial pure state is $\ket{\psi(0)}_{AB}=\ket{00}$ and $H_A=H_B=\mathbb{I}_2$, while they interact according to $H_I=J \sigma_x\otimes \sigma_x$ with $J>0$. 
\begin{itemize}
    \item[(i)] Find the reduced state of the qubit $A$ at all times, $\rho_A(t)$.
    \item[(ii)] Explain what happens at times $Jt=\pi/4+n\pi/2$ with $n\in\mathbb{Z}$.
    \item[(iii)] Suppose we want to describe the dynamics of $A$ via a Lindblad master equation of the form
    \begin{align}
        \frac{d}{dt}\rho_A(t)=\gamma_x(t)\left(\sigma_x\rho_A(t)\sigma_x^\dagger-\frac{1}{2}\{\sigma_x^\dagger\sigma_x,\rho_A(t) \}\right).
        \nonumber
    \end{align}
    Knowing the form of the state $\rho_A(t)$ at all times obtained in \textit{i)}, determine the corresponding time-depending rate $\gamma_x(t)$. Is this a Markovian master equation? 
\end{itemize} 
\textit{Hint:} $e^{ -i t \sigma_x\otimes \sigma_x}=\cos t \ \mathbb{I}_4-i\sin t \  \sigma_x\otimes\sigma_x$. 

\vspace{0.5cm}
\noindent\textbf{Problem 2.4}
\vspace{0.2cm}

\noindent Consider a single qubit initially in the state $\ket{\psi(0)}=\ket{+}$ with system Hamiltonian $H_S=\omega \sigma_z/2$ which evolves under the Lindblad equation with a single jump operator $A_1=\sigma_z$ and rate $\gamma_z>0$. 
\begin{itemize}
    \item[(i)] Write down the dynamical equation for each of the components of the Bloch vector.
    \item[(ii)] Obtain the solution and draw/sketch its trajectory in the Bloch sphere.
    \item[(iii)] What is the characteristic time $T_2$ in terms of $\gamma_z$ for this dephasing process? 
\end{itemize}

\vspace{0.5cm}
\noindent\textbf{Problem 2.5}
\vspace{0.2cm}

\noindent Suppose a qubit with $H_S=\omega \sigma_z/2$ that undergoes spontaneous emission and absorption, corresponding to Lindblad operators $\sigma^-$ and $\sigma^+$, with rates $\gamma_-$ and $\gamma_+$, respectively.
\begin{itemize}
    \item[(i)] What relation should $\gamma_-$ and $\gamma_+$ satisfy so that the Lindblad equation ensures thermalization, i.e. the steady state of the Lindblad master equation corresponds to the thermal equilibrium state $\rho_{eq}$ at some temperature $T$?
\begin{align}
    \rho_{eq}=\frac{e^{-H_S/kT}}{{\rm Tr}[e^{-H_S/kT}]}=\frac{1}{2\cosh(\omega/(2k_bT))}\begin{pmatrix}e^{-\omega/(2k_bT)} & 0\\ 0 & e^{\omega/(2k_bT)} \end{pmatrix},
    \nonumber
\end{align}
with $k_b$ the Boltzmann constant.
\item[(ii)] If the previous relation holds for $\gamma_{\pm}$ but the rates are transformed according to $\gamma_{\pm}'=a \gamma_{\pm}$ with $a>0$, how does the steady state modify under $\gamma_{\pm}'$? What is the impact on the dynamics depending on $a$? Explain it, no need for calculations.
\end{itemize}


\newpage
\section{Microscopic Models}\label{c:micro}

In Sec.~\ref{c:OQS}, we examined how imposing certain mathematical conditions on the dynamical map (dynamical semigroup)  describing the evolution of an open quantum system leads to the Gorini –Kossakowski–Sudarshan–Lindblad (GKSL) master equation. The main assumptions were Markovian dynamics, an initial product state between the system and its environment, and a completely positive and trace-preserving (CPTP) map. Although we have already discussed examples illustrating open-system dynamics, these were so far introduced in a phenomenological manner. In those cases, the dissipative dynamics were constructed by assuming some values for the corresponding rates and jump operators that define the non-unitary part of the GKSL master equation. While this approach is valuable in many practical contexts, it remains conceptually unsatisfactory, as we have not yet derived the dynamical equation from first principles. This is precisely the goal of the present section.

\subsection{Microscopic origin of the Lindblad master equation}
We start again by writing the Hamiltonian of a systems $S$ interacting with an environment $B$ as
\begin{align}
    H=H_S+H_B+H_I.
\end{align}
Moving to the interaction picture (cf. Sec.~\ref{ss:int_pic}) with respect to $H_S+H_B$, we get that the new Hamiltonian is $H_I^I(t)=U^{\dagger}_{S+B}(t)H_I U_{S+B}(t)$ with the time-evolution operator of the local terms of the system and environment $U_{S+B}(t)=\mathcal{T}e^{-i\int_{0}^t dt'(H_S(t')+H_B(t'))}$. This is a convenient frame to describe the dynamics since the joint state at time $t$ can be written as (standard von Neumann equation)
\begin{align}\label{eq:rhoI_1}
    \frac{d}{dt}\rho^I(t)=-i[H_I^I(t),\rho^I(t)],
\end{align}
 where the superscript $I$ denotes operators in the interaction picture. In order to ease the notation, we will drop the superscript in the following derivation. At the end, we will come back to the Schr\"odinger picture and write the resulting master equation in the standard (Schr\"odinger) frame. 

 The Eq.~\eqref{eq:rhoI_1} is the exact description of the joint state. Performing a formal integration we obtain the von Neumann equation in integral form (assuming initial time $t_0=0$ for simplicity)
 \begin{align}
     \rho(t)=\rho(0)-i\int_{0}^tds [H_I(s),\rho(s)].
 \end{align}
This expression can now be used to replace $\rho(t)$ in Eq.~\eqref{eq:rhoI_1}, which yields
\begin{align}
    \frac{d}{dt}\rho(t)=-i[H_I(t),\rho(0)]-\int_{0}^t ds [H_I(t),[H_I(s),\rho(s)]].
\end{align}
Take now the trace over the environment, ${\rm Tr}_B[\dots]$ in the previous expression to obtain the dynamical equation for the system of interest, $\rho_S(t)$. In addition, one typically considers that ${\rm Tr}_B[H_I(t),\rho(0)]=0$, as it is the case in many relevant scenarios. Otherwise, one would simply carry on an additional term which could be included in the original $H_S$. Therefore, we can write
\begin{align}\label{eq:rhoI_2}
    \frac{d}{dt}\rho_S(t)=-\int_{0}^t ds {\rm Tr}_B \left[ [H_I(t),[H_I(s),\rho(s)]]\right].
\end{align}
The expression in Eq.~\eqref{eq:rhoI_2} is so far  general and exact. The task now consists in performing a series of approximations to Eq.~\eqref{eq:rhoI_2} so that we can bring it into the form of a GKSL master equation \cite{BRE02}. 

\subsubsection{Born-Markov approximation}

The first step is the so-called Born approximation. Note that Eq.~\eqref{eq:rhoI_2} depends explicitly on the joint state of the system and environment $\rho(t)$. If we assume that that interaction between the system and the environment is weak, the state of the environment will be negligibly affected by $H_I$. Hence, we can approximate the joint state $\rho(t)$ as
\begin{align}
    \rho(t)\approx \rho_S(t)\otimes \rho_B.
\end{align}
The weak coupling and Born approximations are indeed strong assumptions, although physically motivated. In this manner, we can rewrite Eq.~\eqref{eq:rhoI_2} as\footnote{Keep in mind that this is an approximation, but we do not carry on the approximate symbol to ease the notation.}
\begin{align}\label{eq:rhoI_3}
    \frac{d}{dt}\rho_S(t)=-\int_0^{t}ds {\rm Tr}_B \left[ [H_I(t),[H_I(s),\rho_S(s)\otimes \rho_B]]\right].
\end{align}
This is still a complicated integro-differential equation, which is not time-local. In order to continue with the derivation, we perform now the Markov approximation. 

In essence, the Markov approximation consists in replacing $\rho_S(s)$ by $\rho_S(t)$. That is, the evolution of the state from time $t$ to $t+\delta t$ depends only on the state $\rho_S(t)$, and not on the previous history from $0$ to $t$. This is exactly the property of a Markovian evolution:  memoryless dynamics. This leads into
\begin{align}\label{eq:rhoI_4}
    \frac{d}{dt}\rho_S(t)=-\int_0^{t}ds {\rm Tr}_B \left[ [H_I(t),[H_I(s),\rho_S(t)\otimes \rho_B]]\right].
\end{align}
Although this step renders a  time-local equation (in $\rho_S(t)$) it is not yet a Markovian master equation: the evolution still depends on the initial preparation time, so it does not fulfill the dynamical semigroup property. For that, we perform another step within the Markov approximation: Remove the dependence on the initial state by integrating over all infinite past times. This is best done with a change of variables, from $s$ to $t-s$, so that\footnote{In this step we express a function defined as $A(t)=\int_{0}^t ds f(s)$ in terms of $s'=t-s$, so that $A(t)=\int_{t}^0 (-ds') f(t-s')=\int_{0}^tds'f(t-s')=\int_0^t ds f(t-s)$ since $s'$ was the integration variable it can be renamed again as $s$. The Markov approximation is then complete when doing $A(t)\approx\int_0^\infty ds f(t-s)$. This last step is a good approximation provided the integrand vanishes sufficiently fast at large time differences. }
\begin{align}\label{eq:rhoI_5}
    \frac{d}{dt}\rho_S(t)=-\int_0^{\infty}ds {\rm Tr}_B \left[ [H_I(t),[H_I(t-s),\rho_S(t)\otimes \rho_B]]\right].
\end{align}
This approximation is justified  as long as the time scale in which the system changes, $\tau_R$, is much longer that the relaxation time of environment, $\tau_B$. That is, the approximation is good as long as the integrand vanishes sufficiently fast for $s\gg \tau_B$. Therefore, our dynamical description will not be able to account for changes in times comparable or smaller than $\tau_B$, i.e. it is a  master equation on a coarse-grained time axis.

\subsubsection{Secular or rotating-wave approximation}
The approximations made to bring the exact Eq.~\eqref{eq:rhoI_1} into Eq.~\eqref{eq:rhoI_5} are commonly known as Born-Markov approximations. They are, however, not enough to ensure that the resulting master equation is in a GKSL form. For that we still need to perform the third step: the secular or rotating-wave approximation. Let us see in detail how we can proceed from Eq.~\eqref{eq:rhoI_5} to arrive to the structure of Eq.~\eqref{eq:GKSL}. 

If we expand the commutators in Eq.~\eqref{eq:rhoI_5} we find
\begin{align}
    [H_I(t),[H_I(t-s),\rho_S(t)\otimes \rho_B]]&=H_I(t)H_I(t-s)\rho_S(t)\otimes \rho_B-H_I(t)\rho_S(t)\otimes \rho_B H_I(t-s)\nonumber\\
    &-H_I(t-s)\rho_S(t)\otimes \rho_B H_I(t)+\rho_S(t)\otimes \rho_B H_I(t-s)H_I(t).
\end{align}
Now, taking into account that all the operators are Hermitian, we can rewrite Eq.~\eqref{eq:rhoI_5} as
\begin{align}\label{eq:rhoI_6}
    \frac{d}{dt}\rho_S(t)=\int_0^\infty ds {\rm Tr}_B \left[ H_I(t-s)\rho_S(t)\otimes \rho_B H_I(t)-H_I(t)H_I(t-s)\rho_S(t)\otimes \rho_B +{\rm H.c.}\right],
\end{align}
where ${\rm H.c.}$ denotes the Hermitian conjugate of the two terms.

Now we must expand the integrand of Eq.~\eqref{eq:rhoI_6} to gain insight. For that, let us consider the most general form for the interaction Hamiltonian $H_I$, which can be written as
\begin{align}
  H_I=\sum_\alpha A_\alpha \otimes B_\alpha,
\end{align}
where $A_\alpha$ and $B_\alpha$ are operators acting on the Hilbert space of the system $\mathcal{H}_S$ and environment $\mathcal{H}_B$, respectively. Note that the strength of the interaction is absorbed inside the operators. 

In order to develop further the master equation, we recall that any generic operator can be decomposed in the  eigenbasis of the Hamilonian we are using as reference, i.e. using $H_S=\sum_{\varepsilon} \varepsilon \ket{\varepsilon}\bra{\varepsilon}$,  we can define the operator
\begin{align}
A_\alpha(\omega)=\sum_{\varepsilon'-\varepsilon=\omega}\ket{\varepsilon}\bra{\varepsilon} A_\alpha  \ket{\varepsilon'}\bra{\varepsilon'},
\end{align}
where the sum runs over pairs of eigenstates whose energy differs by $\omega$. This is a convenient way to express the interaction operators $A_\alpha(t)$. First, note that
\begin{align}\label{eq:Aw_def}
\sum_\omega A_\alpha(\omega)=\sum_{\varepsilon,\varepsilon'}\ket{\varepsilon}\bra{\varepsilon} A_\alpha  \ket{\varepsilon'}\bra{\varepsilon'}= A_\alpha,
  \end{align}
using the completeness relation, $\sum_{\varepsilon} \ket{\varepsilon}\bra{\varepsilon}=\mathbb{I}$. These operators fulfill also $A_\alpha^\dagger(\omega)=A_\alpha(-\omega)$. One can check that $[H_S,A_\alpha(\omega)]=-\omega A_\alpha(\omega)$, which enables the following closed expression in the interaction picture (see Sec.~\ref{moremat:Aw_int} for details),


\begin{align}
A_\alpha(t)=e^{i H_S t}A_\alpha e^{-i H_S t} =e^{i H_S t}\left(\sum_\omega A_\alpha(\omega)\right) e^{-i H_S t}=\sum_\omega e^{-i \omega t} A_\alpha(\omega),
\end{align}
while $B_\alpha(t)=e^{i H_B t} B_\alpha e^{-i H_B t}$, so that
\begin{align}\label{eq:HI_t}
  H_I(t)=\sum_{\alpha,\omega}e^{-i \omega t} A_\alpha(\omega)\otimes B_\alpha(t).
  \end{align}
In addition, note that $H_I(t)$ can be written in terms of the adjoint operators $A_\alpha^\dagger$ and $B_\alpha^\dagger$ since $H_I(t)=H_I^\dagger(t)$, i.e.
\begin{align}\label{eq:HI_t2}
    H_I(t)=\sum_{\alpha,\omega}e^{i\omega t} A_\alpha^\dagger(\omega)\otimes B_\alpha^\dagger(t).
\end{align}

We can now use these expressions (cf. Eq.~\eqref{eq:HI_t}-\eqref{eq:HI_t2}) to expand Eq.~\eqref{eq:rhoI_6}. Each of the terms in Eq.~\eqref{eq:rhoI_6} can be written as
\begin{align}
    {\rm Tr}_B&[H_I(t-s)\rho_S(t)\otimes \rho_B H_I(t)]=\nonumber\\
    &=\left(\sum_{\beta,\omega}e^{-i \omega (t-s)}A_\beta(\omega)\right) \rho_S(t) \left( \sum_{\alpha,\omega'} e^{i \omega' t} A^\dagger_\alpha(\omega')\right) {\rm Tr_B}\left[\sum_{\alpha,\beta}B_\beta(t-s)\rho_B B_\alpha^\dagger(t)\right]=\nonumber\\
    &=\sum_{\alpha,\beta}\sum_{\omega,\omega'} e^{-i(\omega-\omega')t}A_\beta(\omega)\rho_S(t)A_\alpha^\dagger(\omega') {\rm Tr}_B\left[e^{i\omega s} B_\alpha^\dagger(t)B_\beta(t-s)\rho_B\right]\\
    {\rm Tr}_B&[H_I(t)H_I(t-s)\rho_S(t)\otimes \rho_B ]=\nonumber\\
    &=\left(\sum_{\alpha,\omega'}e^{i \omega' t}A_\alpha^\dagger(\omega')\right) \left( \sum_{\beta,\omega} e^{-i \omega (t-s)} A_\beta(\omega)\right)\rho_S(t) {\rm Tr_B}\left[\sum_{\alpha,\beta}B_\alpha^\dagger(t)B_\beta(t-s)\rho_B\right]\nonumber\\&=\sum_{\alpha,\beta}\sum_{\omega,\omega'} e^{-i(\omega-\omega')t}A_\alpha^\dagger(\omega')A_\beta(\omega)\rho_S(t) {\rm Tr}_B\left[e^{i\omega s} B_\alpha^\dagger(t)B_\beta(t-s)\rho_B\right].
\end{align}
We introduce now a one-sided Fourier transform of the reservoir or environment correlation function,
\begin{align}
    \Gamma_{\alpha,\beta}(\omega)\equiv \int_0^\infty ds e^{i\omega s} C_{\alpha,\beta}(t,t-s),%
\end{align}
where the two-time correlation function reads as
\begin{align}
    C_{\alpha,\beta}(t,t-s)={\rm Tr}_B[B_\alpha^\dagger(t)B_\beta(t-s)\rho_B].
\end{align}
 In this way, we can rewrite Eq.~\eqref{eq:rhoI_6} as
\begin{align}\label{eq:rhoI_7}
\frac{d}{dt}\rho_S(t)=\sum_{\alpha,\beta}\sum_{\omega,\omega'}e^{-i(\omega-\omega')t} \Gamma_{\alpha,\beta}(\omega)\left( A_\beta(\omega)\rho_S(t)A_\alpha^\dagger(\omega')-A_\alpha^\dagger(\omega')A_\beta(\omega)\rho_S(t)\right) +{\rm H.c.}
\end{align}
Recall that the previous expression is equivalent to Eq.~\eqref{eq:rhoI_5}, but written in a convenient way. That is, we have not yet done any further approximation besides Born-Markov. However, it is important to remark that with this new development, the Markov approximation is justified as long as $C_{\alpha,\beta}(t,t-s)$ decay sufficiently fast in time (so that pushing the integration limit to $t\rightarrow \infty$ results in  a reasonable approximation). 

Now we can come back to the purpose of this derivation: the secular or rotating-wave approximation. This consists in keeping all the terms that are resonant (only those for $\omega=\omega')$, while discarding the off-resonant ones ($\omega\neq \omega'$).  This is justified as long as the coupling between the system and environment is weak or small, and the relaxation time of the system $\tau_R$ is much shorter than the characteristic time scale of the coherent system dynamics $\tau_S$. Indeed, the characteristic time $\tau_S\sim |\omega-\omega'|^{-1}$ for $\omega\neq \omega'$.  In that case, the terms that go with $\omega\neq \omega'$ will oscillate very rapidly with a small amplitude (weak coupling). These oscillating terms are negligible compared to the resonant ones. In addition, we typically consider an equilibrium state of the environment, i.e. $[H_B,\rho_B]=0$. In that case, the correlation function $C_{\alpha,\beta}(t,t-s)$ is homogeneous (and so is $\Gamma_{\alpha,\beta}(\omega)$), i.e. it does not depend on the time $t$ but only on the time difference $s$, $C_{\alpha,\beta}(s)\equiv C_{\alpha,\beta}(t,t-s)={\rm Tr}_B[B_\alpha^\dagger(t) B_\beta(t-s)\rho_B]={\rm Tr}_B[B_\alpha^\dagger(s)B_\beta(0)\rho_B]$\footnote{Intuitively, the condition $[H_B,\rho_B]=0$ means that the state of the environment is stationary, and therefore the two-time correlation function does not dependent on the exact time $t$ but only on the time difference $s$. Formally, this can be shown as follows. The operators at time $t$ and $t-s$ can be written as $B_\alpha^\dagger(t)=e^{i H_B t}B_\alpha^\dagger e^{-i H_B t}$, and $B_\beta(t-s)=e^{iH_B(t-s)}B_\beta e^{-iH_B(t-s)}$. Hence, ${\rm Tr}_B[B_\alpha^\dagger(t)B_\beta(t-s)\rho_B]={\rm Tr}_B[e^{i H_B t}B_\alpha^\dagger e^{-i H_B t}e^{i H_B(t-s)}B_\beta e^{-i H_B (t-s)}\rho_B]$. Now, since $H_B$ and $\rho_B$ commute, $e^{-i(t-s)H_B}\rho_B=\rho_B e^{-i(t-s)H_B}$. Then, using the cyclic property of the trace, we move $e^{-iH_B(t-s)}$ to the first place, ${\rm Tr}_B[e^{-iH_B(t-s)}e^{i H_B t }B_\alpha^\dagger e^{-i H_B s}B_\beta \rho_B]={\rm Tr}_B[e^{i H_B s}B_\alpha^\dagger e^{-i H_B s}B_\beta\rho_B]={\rm Tr}_B[B^\dagger_\alpha(s)B_\beta(0)\rho_B]$. }. Therefore, performing the secular or rotating-wave  approximation to Eq.~\eqref{eq:rhoI_6} we arrive to
\begin{align}\label{eq:rhoI_8}
    \frac{d}{dt}\rho_S(t)=\sum_\omega \sum_{\alpha,\beta}\Gamma_{\alpha,\beta}(\omega)\left(A_\beta(\omega)\rho_S(t)A_\alpha^\dagger(\omega)-A^\dagger_\alpha(\omega)A_\beta(\omega)\rho_S(t) \right)+{\rm H.c.}
\end{align}
We can split the Fourier transforms of the environment correlation functions as (see Sec.~\ref{moremat:Gammas} for details)
\begin{align}
    \Gamma_{\alpha,\beta}(\omega)=\frac{1}{2}\gamma_{\alpha,\beta}(\omega)+iS_{\alpha,\beta}(\omega),
\end{align}
so that
\begin{align}\label{eq:Sab}
    S_{\alpha,\beta}(\omega)&=\frac{1}{2i}\left(\Gamma_{\alpha,\beta}(\omega)-\Gamma_{\beta,\alpha}^*(\omega)\right), \\ \label{eq:gab}
    \gamma_{\alpha,\beta}(\omega)&=\Gamma_{\alpha,\beta}(\omega)+\Gamma_{\beta,\alpha}^*(\omega).
\end{align}

Let us now expand Eq.~\eqref{eq:rhoI_8} using the real and imaginary parts of $\Gamma_{\alpha,\beta}(\omega)$ to see the resulting form (for convenience we drop here the dependence on $\omega$)
\begin{align}
    \frac{d}{dt}\rho_S(t)&=\sum_{\omega}\sum_{\alpha,\beta}\left(\frac{1}{2}\gamma_{\alpha,\beta}+iS_{\alpha,\beta} \right)\left( A_\beta\rho_S(t) A_\alpha^\dagger-A_\alpha^\dagger A_\beta \rho_S(t)\right)\nonumber\\&+\sum_{\omega}\sum_{\alpha,\beta}\left(\frac{1}{2}\gamma_{\alpha,\beta}^*-iS_{\alpha,\beta}^* \right)\left( A_\alpha\rho_S(t) A_\beta^\dagger-\rho_S(t) A_\beta^\dagger A_\alpha \right).
\end{align}
We use now the fact that $S_{\alpha,\beta}^*(\omega)=S_{\beta,\alpha}(\omega)$ and $\gamma_{\alpha,\beta}^*(\omega)=\gamma_{\beta,\alpha}(\omega)$, and swap  the indices in the second line to be able to group them with the sum of the first line, i.e.
\begin{align}
    \frac{d}{dt}\rho_S(t)&=\sum_{\omega}\sum_{\alpha,\beta}\left(\frac{1}{2}\gamma_{\alpha,\beta}+iS_{\alpha,\beta} \right)\left( A_\beta\rho_S(t) A_\alpha^\dagger-A_\alpha^\dagger A_\beta \rho_S(t)\right)\nonumber\\&+\sum_{\omega}\sum_{\alpha,\beta}\left(\frac{1}{2}\gamma_{\alpha,\beta}-iS_{\alpha,\beta} \right)\left( A_\beta\rho_S(t) A_\alpha^\dagger-\rho_S(t) A_\alpha^\dagger A_\beta \right).\label{eq:rhoI_8}
\end{align}
We can distinguish two parts in Eq.~\eqref{eq:rhoI_8}. The first one proportional to $S_{\alpha,\beta}$ which can be cast as a commutator with $\rho_S(t)$, i.e.
\begin{align}
    -i\sum_{\omega}\sum_{\alpha,\beta}S_{\alpha,\beta}\left(A_\alpha^\dagger(\omega) A_\beta(\omega) \rho_S(t)-\rho_S(t) A_\alpha^\dagger(\omega) A_\beta(\omega)\right)=-i[H_{\rm LS},\rho_S(t)],
\end{align}
where we have defined a Hamiltonian $H_{\rm LS}=\sum_{\omega}\sum_{\alpha,\beta}S_{\alpha,\beta}(\omega)A^\dagger_\alpha(\omega)A_\beta(\omega)$.  Note that, by construction, $H_{\rm LS}=H_{\rm LS}^\dagger$. The second part that survives the sum in Eq.~\eqref{eq:rhoI_8} goes with $\gamma_{\alpha,\beta}$ which we define as the dissipator $D[\rho_S(t)]$
\begin{align}
    D[\rho_S(t)]\equiv \sum_{\omega}\sum_{\alpha,\beta} \gamma_{\alpha,\beta}(\omega) \left(A_\beta(\omega) \rho_S(t) A_\alpha^\dagger(\omega)-\frac{1}{2}\{A^\dagger_\alpha(\omega) A_\beta(\omega),\rho_S(t) \}\right).
\end{align}
This differential equation~\eqref{eq:rhoI_8} has already a recognizable form (a GKSL master equation but not yet in diagonal form, cf. Eq.~\eqref{eq:MarkovMaster}), which can be written as
\begin{align}
    \frac{d}{dt}\rho_S(t)=-i[H_{\rm LS},\rho_S(t)]+D[\rho_S(t)].
\end{align}
Before the last inspection of the resulting master equation, let us recall that we were working in the interaction picture. If we come back to the Schr\"odinger frame, we simply add the coherent evolution $H_S$ and obtain
\begin{align}\label{eq:rhoI_9}
    \frac{d}{dt}\rho_S(t)=&-i[H_S+H_{\rm LS},\rho_S(t)]+\nonumber \\&\quad+\sum_{\omega}\sum_{\alpha,\beta} \gamma_{\alpha,\beta}(\omega) \left(A_\beta(\omega) \rho_S(t) A_\alpha^\dagger(\omega)-\frac{1}{2}\{A^\dagger_\alpha(\omega) A_\beta(\omega),\rho_S(t) \}\right).
\end{align}

It is interesting to observe that the interaction with an environment leads into coherent and incoherent dynamics. The coherent part stems from the interaction as a form of renormalization of the system Hamiltonian (changing the energies of the system), $H_S\rightarrow H_S+H_{\rm LS}$, which is known as Lamb shift. 
The second part has the structure of the non-unitary dynamics of a Markov GKSL master equation, although not yet in diagonal form. However, in order to claim that the dynamics described by Eq.~\eqref{eq:rhoI_9} is indeed Markovian we need to ensure that $\gamma_{\alpha,\beta}(\omega)$ is a Hermitian and positive semi-definite matrix $\gamma_{\alpha,\beta}(\omega)\geq 0$ (similar to the general arguments we outlined in Sec.~\ref{c:OQS}, but now from a physical and microscopic perspective). This is indeed shown in Sec.~\ref{moremat:gab_pos}.

Let us review all the approximations we have made to derive a GKSL master equation:
\begin{itemize}
    \item The first approximation arises from the weak-coupling assumption. This motivates the condition $\rho(t)\approx \rho_S(t)\otimes \rho_B$, known as Born approximation.
    \item The second approximation is the Markov assumption, in which the master equation is i) made local in time by replacing the reduced density matrix $\rho_S(s)$ at past time by its form at the present time $\rho_S(t)$, and the integral is extended to infinite past times. This requires the integrand to vanish sufficiently fast for long times, and a characteristic time of the system $\tau_R$ much longer than the environment time $\tau_R\gg \tau_B$. These two approximation are typically known as Born-Markov approximations. 
    \item The third step is known as secular or rotating-wave approximation. Here the off-resonant terms $\omega\neq \omega'$ are neglected in favor of the resonant non-oscillating part. This is justified as long as the relaxation time of the system $\tau_R$ is much shorter than the characteristic time scale of the coherent part $\tau_S\gg \tau_R$. Note that these rapidly oscillating terms are accompanied by a weak coupling. This last approximation is essential to ensure that the master equation acquires a GKSL form. 
\end{itemize}

 In the following we provide two illustrative examples of relevant physical models, namely, a quantum harmonic oscillator and a qubit interacting with a continuum of bosonic modes. 

 \subsubsection{Breakdown of approximations}
 What happens if any of the three approximations or assumptions break down? In that case, the Markovian master equation derived under these conditions is no longer valid and fails to accurately capture the dynamics of the open quantum system. Describing such scenario (possibly, but not necessarily non-Markovian dynamics) is, however, a challenging task with no general analytical solution. Nonetheless, significant progress has been made through various approaches that relax or refine the standard approximations, allowing one to account for strong system–environment coupling, memory effects, and initial system–bath correlations, see for example Refs.~\cite{Prior2010,IlesSmith2016,Nazir2018,Strathearn2018,Tamascelli2018,Tanimura2020,Mascherpa2020,GonzalezBallestero2024,Settimo2025}.

 Just as an example, consider a system $S$ that interacts with the environment $B$. Suppose that the interaction is weak for a subset of modes of the environment, $B_w$,  but strong for the rest, $B_s$, such that $B_s\cup B_w=B$. In this case, treating on an equal footing $B_s$ and $B_w$ challenges the validity of the standard open system dynamics for $S$. A common strategy consists in enlarge the system by including the strongly-coupled modes $B_s$ into the extended system $S'=S\cup B_s$. In this way, the resulting dynamics approximating $S'$ as a system undergoing Markovian interactions with $B_w$ will correctly  capture  correlations between $S$ and $B_s$ or memory effects. This, however, can be done at the cost of increasing the effective size of the system of interest. In any case, approaches that go beyond Born-Markov and weak coupling approximations are outside of the scope of this course.

\subsection{Thermalization of a harmonic oscillator}
A typical example of the previous general derivation consists in a quantum harmonic oscillator interacting with a large, typically infinite, collection of bosonic modes that form the environment. This is indeed a simple model, but very illustrative to understand the microscopic origin of dissipative quantum dynamics. In particular, let us consider the three terms of the total Hamiltonian,
\begin{align}
    H_S&=\omega_0 a^\dagger a,\\
    H_B&=\sum_{k}\nu_k b_k^\dagger b_k,\\
    H_I&=\sum_k (g_k a^\dagger b_k+g_k^*ab_k^\dagger).
\end{align}
That is, the quantum harmonic oscillator of interest has a frequency $\omega_0$ and interacts exchanging excitations with bosonic modes at frequency $\nu_k$ of the environment at a rate $g_k$, which in general can be complex. Note that the environment is treated as a discrete set of modes. Later, we will take the  continuum limit, $H_B=\sum_k\nu_k b_k^\dagger b_k\rightarrow H_B=\int d\nu \ \nu \ b_\nu^\dagger b_\nu. $

Now, if we want to apply the previous prescription to arrive to a GKSL master equation we need to ensure:
\begin{itemize}
    \item At the beginning the initial state is in a product form, $\rho(0)=\rho_S(0)\otimes \rho_B$
    \item The coupling is weak so that $\rho_B$ remains unaffected throughout the evolution (Born)
    \item The response time of the environment is short compared to the characteristic time scale of the system (Markov)
    \item The intrinsic time of the system is longer than its relaxation time (rotating-wave approximation)
\end{itemize}
Under these conditions, we can workout the approximate Markov dynamics. Let us also assume a thermal environment, i.e. the state $\rho_B$ is a thermal equilibrium state (Gibbs state),
\begin{align}
    \rho_B=\frac{e^{-H_B/(k_b T)}}{{\rm Tr}[e^{- H_B/(k_b T)}]},
\end{align}
at temperature $T$ (where $k_b$ is the Boltzmann constant). 
For each of the modes, we have
\begin{align}
    \rho_{B,k}=\frac{1}{\overline{n}_k+1}\sum_{n=0}^{\infty} \left(\frac{\overline{n}_k}{\overline{n}_k+1}\right)^n\ket{n}\bra{n}_k,
\end{align}
with $\overline{n}_k=\langle b_k^\dagger b_k\rangle={\rm Tr}_B[b_k^\dagger b_k \rho_{B,k}]=(e^{\nu_k/(k_b T)}-1)^{-1}$ the average number of bosonic excitations in mode $k$, with frequency $\nu_k$. Therefore, $\rho_B=\bigotimes_k  \rho_{B,k}$. 

The interaction Hamiltonian can be expressed as
\begin{align}
    H_I&=A_1\otimes B_1+A_2\otimes B_2,\\
    A_1&=a^\dagger, \quad B_1=\sum_k g_k b_k, \quad 
    A_2=a, \quad  B_2=\sum_k g_k^*b_k^\dagger.
\end{align}
Therefore, we need to compute four different two-time correlation functions (or simply three, exploiting the symmetry $C_{1,2}(s)=C_{2,1}^*(-s)$). In particular,
\begin{align}
C_{1,1}(s)&={\rm Tr}_B[B_1^\dagger(s) B_1(0) \rho_B]={\rm Tr}_B[\left(\sum_{k} g_k^* e^{i\nu_k s}b_k^\dagger \right)\left(\sum_{k'} g_{k'} b_{k'}  \right)\rho_B]=\sum_{k} |g_k|^2 \overline{n}_k e^{i\nu_k s},\\
C_{1,2}(s)&={\rm Tr}_B[B_1^\dagger(s) B_2(0) \rho_B]={\rm Tr}_B[\left(\sum_{k} g_k^* e^{i\nu_k s}b_k^\dagger \right)\left(\sum_{k'} g_{k'}^*b_{k'}^\dagger  \right)\rho_B]=0,\\
C_{2,2}(s)&={\rm Tr}_B[B_2^\dagger(s) B_2(0) \rho_B]={\rm Tr}_B[\left(\sum_{k} g_k e^{-i\nu_k s}b_k \right)\left(\sum_{k'} g_{k'}^* b_{k'}^\dagger \right)\rho_B]=\sum_{k} |g_k|^2 (\overline{n}_k+1) e^{-i\nu_k s}, 
\end{align}
where ${\rm Tr}_B[b_k^\dagger b_{k'} \rho_{B,k}]=\overline{n}_k \delta_{k,k'}$ and $b_kb_k^\dagger=b_k^\dagger b_k +1$ so that ${\rm Tr}_B[b_k b_{k'}^\dagger \rho_{B,k}]=(\overline{n}_k+1)\delta_{k,k'}$. In addition, note that ${\rm Tr}_B[b_k b_{k'} \rho_B]=0$ as well as ${\rm Tr}_B[b_k^\dagger b_{k'}^\dagger \rho_B]=0$. Now, since $A_1$ and $A_2$ are particularly simple, we can write $A_1(t)=e^{i \omega_0 a^\dagger a t}a^\dagger e^{-i \omega_0 a^\dagger a t}=e^{i \omega_0 t}a^\dagger$, and similarly for $A_2(t)=e^{-i\omega_0 t}a$ without needing to worry about the frequency decomposition (which we did above for a general and physically motivated derivation). This means that
\begin{align}
    \Gamma_{1,1}(\omega_0)=\int_0^\infty ds e^{-i\omega_0 s} C_{1,1}(s)=\int_0^\infty ds \sum_k |g_k|^2 \overline{n}_k e^{-i(\omega_0-\nu_k)s}.
\end{align}
Note that the factor $e^{-i\omega_0 s}$ comes from $A_1(t-s)=e^{i\omega_0(t-s)} a^\dagger$. The term $e^{i\omega_0 t}$ will cancel $e^{-i\omega_0 t}$ coming from $A_1^\dagger(t)$ while $e^{-i\omega_0 s}$ enters into $\Gamma_{1,1}(\omega_0)$. 
In order to correctly integrate the previous diverging integral we introduce a parameter $\eta>0$ to regularize the integrand (and then take the limit to $\eta\rightarrow 0^+$),
\begin{align}
    \Gamma_{1,1}(\omega_0)&=\lim_{\eta\rightarrow 0^+} \int_{0}^\infty ds \sum_k e^{-i(\omega_0-\nu_k)s-\eta s}|g_k|^2\overline{n}_k=\lim_{\eta\rightarrow 0^+}\sum_{k}|g_k|^2\overline{n}_k\frac{1}{\eta+i(\omega_0-\nu_k)}\nonumber\\&=\lim_{\eta\rightarrow 0^+}\sum_{k}|g_k|^2\overline{n}_k\frac{\eta-i(\omega_0-\nu_k)}{\eta^2+(\omega_0-\nu_k)^2}.
\end{align}
Now we use the definition of the Dirac delta in limit form, $\pi \delta(x-a)=\lim_{\eta\rightarrow 0^+}\frac{\eta}{\eta^2+(x-a)^2}$. In this manner, the rate $\gamma_{1,1}(\omega_0)$, which is two times the real part of $\Gamma_{1,1}(\omega_0)$ reads as
\begin{align}
    \gamma_{1,1}(\omega_0)=2{\rm Re}[\Gamma_{1,1}(\omega_0)]=2\pi \sum_k |g_k|^2 \overline{n}_k \delta(\omega_0-\nu_k).
\end{align}
The imaginary part corresponds to $S_{1,1}(\omega_0)={\rm Im}[\Gamma_{1,1}(\omega_0)]$, 
\begin{align}\label{eq:S11}
    S_{1,1}(\omega_0)=-\lim_{\eta\rightarrow 0^+}\sum_{k}|g_k|^2 \overline{n}_k\frac{\omega_0-\nu_k}{\eta^2+(\omega_0-\nu_k)^2}=-{\rm P.V}\left(\sum_k |g_k|^2 \overline{n}_k\frac{1}{(\omega_0-\nu_k)} \right),
\end{align}
where ${\rm P.V.}$ denotes the Cauchy principal value (due to the divergence at $\omega_0=\nu_k$). Similarly, we can find $\gamma_{2,2}(\omega_0)$ and $S_{2,2}(\omega_0)$ (the other two terms are zero since $C_{1,2}(s)=C_{2,1}(s)=0$),
\begin{align}
    \gamma_{2,2}(\omega_0)&=2\pi \sum_{k}|g_k|^2 (\overline{n}_k+1)\delta(\omega_0-\nu_k),\\ \label{eq:S22}
    S_{2,2}(\omega_0)&={\rm P.V.}\left(\sum_k |g_k|^2 (\overline{n}_k+1)\frac{1}{(\omega_0-\nu_k)} \right).
\end{align}
Let us focus first on the noise rates $\gamma_{1,1}(\omega_0)$ and $\gamma_{2,2}(\omega_0)$ while the Lamb shift is discussed more in detail in Sec.~\ref{moremat:lamb}. The resulting rates take into account the strength of the interaction ($|g_k|^2$) for each of the modes $\nu_k$. This motivates the definition of the so-called spectral density $J(\omega)$: This quantity accounts for the strength of the interaction at a given frequency $\omega$. For a set of discrete modes, it is defined as
\begin{align}
    J(\omega)=\sum_{k} |g_k|^2 \delta(\omega-\nu_k),
\end{align}
however, as we will see below, in the continuum limit this can be a well-behaved and smooth function. The spectral density can also be expressed as $J(\omega)=|g(\omega)|^2D(\omega)$, i.e. as the product of the squared coupling system-environment $|g(\omega)|^2$ times the density of states at frequency $\omega$.

Using the spectral density, we can rewrite
\begin{align}
    \gamma_{1,1}(\omega_0)&=2\pi J(\omega_0) \overline{n}(\omega_0),\\
     \gamma_{2,2}(\omega_0)&=2\pi J(\omega_0) (\overline{n}(\omega_0)+1),
\end{align}
where $\overline{n}(\omega)=(e^{\omega/(k_bT)}-1)^{-1}$ denotes the amount of thermal excitations at frequency $\omega$ and temperature $T$. The rate $\gamma_{1,1}(\omega_0)$ is associated with the system operator $A_1=a^\dagger$, so it describes incoherent spontaneous absorption of excitations from the environment into the system. To the contrary, $\gamma_{2,2}(\omega_0)$ has an associated operator $A_2=a$, which accounts for spontaneous emission of excitations from the system into the environment. For this reason, one typically defines the noise rates as $\gamma_{\downarrow}\equiv \gamma_{2,2}(\omega_0)$ and $\gamma_{\uparrow}\equiv \gamma_{1,1}(\omega_0)$.  

This allows us to write the resulting GKSL master equation as (cf. Eq.~\eqref{eq:rhoI_9})
\begin{align}\label{eq:MasterEq_HO}
    \frac{d}{dt}\rho_S(t)=-i[(\Delta+ \omega_0)a^\dagger a,\rho_S(t)]&+\gamma_{\uparrow}(\omega_0) \left(a^\dagger \rho_S(t) a-\frac{1}{2}\{a a^\dagger,\rho_S(t) \}\right)+\nonumber\\&+\gamma_\downarrow(\omega_0) \left(a \rho_S(t) a^\dagger-\frac{1}{2}\{a^\dagger a,\rho_S(t) \}\right),
\end{align}
where $\Delta$ denotes the Lamb energy shift (see Sec.~\ref{moremat:lamb}). Thanks to $C_{1,2}(s)=C_{2,1}(s)=0$, the master equation is already in diagonal form (cf. Eq.~\eqref{eq:GKSL}) where the non-negativity of the rates $\gamma_{\downarrow},\gamma_{\uparrow}$ is evident.

It is interesting to analyze the zero temperature limit, $T\rightarrow 0$. In this scenario, $\overline{n}(\omega_0)\rightarrow 0$, and therefore spontaneous absorption is suppressed $\gamma_{\uparrow}=0$: There are no excitations in the environment, and therefore it cannot excite the system. Yet, $\gamma_{\downarrow}=2\pi J(\omega_0)$, so the system is still able to emit excitations into the environment. In the long time limit, for $T\rightarrow 0$, the system losses all the excitations (i.e. it dissipates all its energy) into the environment. In general,  since the rates fulfill 
\begin{align}
\frac{\gamma_{\downarrow}}{\gamma_{\uparrow}}=\frac{2\pi J(\omega_0)(\overline{n}(\omega_0)+1)}{2\pi J(\omega_0)\overline{n}(\omega_0)}=\frac{(e^{\omega/(k_b T)}-1)^{-1}+1}{(e^{\omega/(k_b T)}-1)^{-1}}=e^{\omega_0/(k_b T)},
\end{align}
the harmonic oscillator equilibrates into a thermal state at the same temperature as the environment (Sec.~\ref{moremat:steadystate_ho}
 for further details). This condition is known as  detailed balanced \cite{boltzmann1964lectures, tolman1938principles}: The probability of incoherent transitions that go from an energy level $n$ to $n-1$ must be proportional to the transitions from $n-1$ to $n$, where the proportional constant is the Boltzmann factor, $e^{(\omega_n-\omega_{n-1})/(k_b T)}$ being $\omega_{n}-\omega_{n-1}$ the energy difference among the levels.

\subsection{Spin-boson model}
Another interesting situation takes place when the system of interest is a single qubit or a spin-$\frac{1}{2}$ particle interacting with an infinite number of bosonic degrees of freedom. If the interaction is such that it exchanges excitations, we fall back to the case considered above but simply substituting $a^\dagger$ by the spin raising operator $\sigma^+$, and similarly for $a$ and $\sigma^-$. For that reason, we consider a different interaction, namely, a longitudinal coupling that just shifts the energy levels of the spin, 
\begin{align}
    H=\frac{\Omega}{2}\sigma_z+\sum_{k}\nu_k b_k^\dagger b_k +\sum_k \sigma_z(g_kb_k+g_k^*b_k^\dagger).
\end{align}
This model can be solved exactly since the interaction commutes with the local energy term of the system. However, we will just follow the standard steps to describe the open system dynamics under the Born-Markov and rotating-wave approximations. 

In this case, the interaction can be written as $H_I=A_1\otimes B_1$, such that
\begin{align}
    A_1=\sigma_z,\quad B_1=\sum_k (g_k b_k+g_k^* b_k^\dagger).
\end{align}
The Fourier transform of the two-time correlation function reads as
\begin{align}
    \Gamma&=\int_0^\infty ds {\rm Tr}_B[B_1^\dagger(s)B_1(0)\rho_B]\nonumber\\&=\int_0^\infty ds {\rm Tr}_B[\left(\sum_k g_k e^{-i\nu_k s}b_k+g_k^* e^{i\nu_k t}b_k^\dagger \right)\left(\sum_{k'} g_{k'} b_{k'}+g_{k'}^* b_{k'}^\dagger\right)\rho_B]\nonumber\\
    &=\int_0^\infty ds \sum_k |g_k|^2(2\cos(\nu_k s)\overline{n}_k+e^{-i\nu_k}).
\end{align}
Note that since $A_1(t)=\sigma_z$ (it is time independent since it commutes with $H_S$), there is no term $e^{i\omega s}$ in the correlation function. Doing the same trick as in the harmonic oscillator, we introduce a parameter $\eta$ to renormalize the integral,
\begin{align}
    \Gamma=\lim_{\eta\rightarrow 0^+} \int_0^\infty ds \sum_k |g_k|^2(2\cos(\nu_k s)\overline{n}_k+e^{-i\nu_k})e^{-\eta s}=\lim_{\eta\rightarrow 0^+} \sum_k |g_k|^2\left(2 \frac{\eta \overline{n}_k}{\eta^2+\nu_k^2} +\frac{\eta-i\nu_k}{\eta^2+\nu_k^2}\right).
\end{align}
Again, using $\pi \delta(x-a)=\lim_{\eta\rightarrow 0^+}\eta/(\eta^2+(x-a)^2)$, the real part reads
\begin{align}\label{eq:gamma_deph}
    \gamma=2{\rm Re}[\Gamma]=2\pi\sum_k |g_k|^2(2\overline{n}_k+1)\delta(\nu_k)=2\pi J(0)(2\overline{n}(0)+1),
\end{align}
using again the spectral density $J(\omega)=\sum_k |g_k|^2\delta(\omega-\nu_k)$. The imaginary part is irrelevant since the Lamb shift goes with $A_1^\dagger A_1=\sigma_z\sigma_z=\mathbb{I}$, so it does not shift the energy levels but rather introduce a constant energy shift. Hence, the resulting master equation reads as
\begin{align}\label{eq:MasterEq_SB}
    \frac{d}{dt}\rho_S(t)&=-i[H_S,\rho_S(t)]+\gamma\left(\sigma_z \rho_S(t)\sigma_z-\frac{1}{2}\{\sigma_z\sigma_z,\rho_S(t)\} \right)\nonumber\\&=-i[H_S,\rho_S(t)]+\gamma\left(\sigma_z \rho_S(t)\sigma_z-\rho_S(t) \right).
\end{align}
This was to be expected: The interaction $\sigma_z\sum_k(g_k b_k+g_k^* b_k^\dagger)$ produces pure dephasing noise. The dephasing rate $\gamma$ is proportional to the spectral density at zero frequency. However, in most physical cases $\lim_{\omega\rightarrow 0}J(\omega)\rightarrow 0$. At the same time, $\lim_{\omega\rightarrow 0}\overline{n}(\omega)\rightarrow \infty$, the number of excitations for any $T>0$ diverges as the frequency goes to $0$. The rate $\gamma$ may be finite considering a specific form of the environment, i.e. a functional dependence of the spectral density on $\omega$. 

As example, consider an Ohmic spectral density with exponential cutoff, $J(\omega)=\alpha \omega e^{-\omega/\omega_c}$. The Ohmic character means that $J(\omega)\propto \omega$, while sub- and super-Ohmic environment refer to $J(\omega)\propto \omega^s$ with $0\leq s<1$ and $s>1$, respectively. The cutoff frequency $\omega_c$  is typically introduced to ensure that  the system does not interact with arbitrarily large frequencies, $\lim_{\omega\rightarrow \infty} J(\omega)=0$. Note that $\alpha$ is just a constant factor that describes the overall coupling strength. In this manner, the dephasing rate for an environment at temperature $T$ given in Eq.~\eqref{eq:gamma_deph} becomes
\begin{align}\label{eq:gamma_deph_Ohm}
    \gamma=\lim_{\omega\rightarrow 0}2\pi J(\omega)(2\overline{n}(\omega)+1)=\lim_{\omega\rightarrow 0} 2\pi \alpha \omega e^{-\omega/\omega_c}\left(\coth\left(\frac{\omega}{2k_b T}\right)\right)=4\pi \alpha k_b T,
\end{align}
since $2\overline{n}(\omega)+1=2/(e^{\omega/(k_b T)}-1)+1=\coth(\omega/(2k_b T))$. 
In Eq.~\eqref{eq:gamma_deph_Ohm} we have used the L'H{\^o}pital rule, defining $f_1(\omega)=2\pi \alpha e^{-\omega/\omega_c}\omega$ and $f_2(\omega)=1/\coth(\omega/(2k_b T))$ so that  $\gamma=\lim_{\omega\rightarrow 0} f_1(\omega)/f_2(\omega)=f'_1(0)/f'_2(0)=4\pi\alpha k_b T$, where $f_2'(\omega)=1/(2k_b T) {\rm sech}^2(\omega/(2k_b T))$. This means that the off-diagonal elements of the density matrix (i.e. the $x$ and $y$ components of the Bloch vector) decay exponentially in time, $v_{x,y}(t)\propto e^{-2\gamma t}v_{x,y}(0)$.

\begin{figure}
    \centering
    \includegraphics[width=0.5\linewidth]{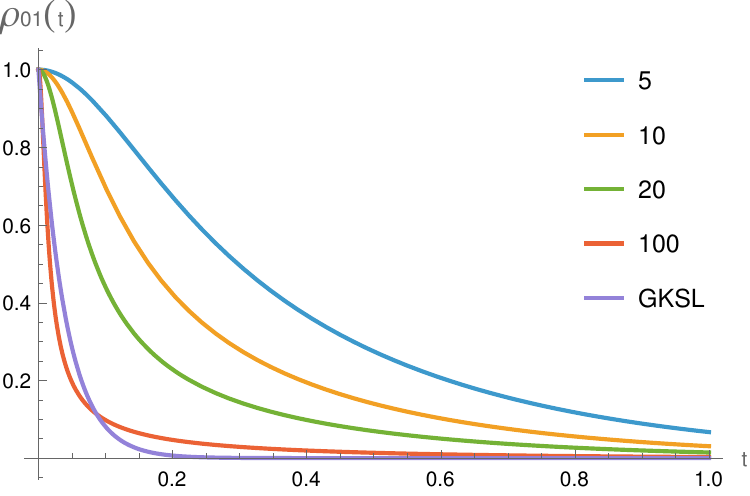}
    \caption{Decay of the off-diagonal term $\rho_{01}(t)$ as a function of time (assuming $\rho_{01}(0)=1$) for an Ohmic spectral density $J(\omega)=\omega e^{-\omega/\omega_c}$. The purple line shows the decay under the Born-Markov and rotating wave approximations (GKSL), while the others correspond to the exact expression derived in Eq.~\eqref{eq:deph_SB_exact}, which explicitly depends on the cut-off frequency ($\omega_c$) indicated in the legend (arbitrary units). }
    \label{fig:SB_comp}
\end{figure}

In any case, it is worth noting that the master equation is just an approximation to the exact dynamics. As mentioned above, this simple model can be worked out analytically to obtain the exact dynamics without any approximation (besides the typical uncorrelated initial state between system and environment). The derivation is however left as further material, included in Sec.~\ref{moremat:exact_SB} in case you are interested in the details.  In Fig.~\ref{fig:SB_comp} we illustrate the decay of the off-diagonal term $\rho_{01}(t)$ as a function of time for an Ohmic spectral density $(J(\omega)=\omega e^{-\omega/\omega_c}$ with $\alpha=1$, and $k_bT=1$ for the GKSL master equation (purple line), as well as for the exact solution for various cut-off frequencies $\omega_c$. Note that, in the GKSL approximated master equation, $\rho_{01}(t)=\rho_{01}(0)e^{-8\pi \alpha k_b T t}$, the cut-off frequency does not play any role. The exact expression does depend on $\omega_c$ (cf. Eq.~\eqref{eq:deph_SB_exact}).

\subsection{Atom in a lossy cavity}
\begin{figure}[t!]
    \centering
\includegraphics[width=0.9\linewidth]{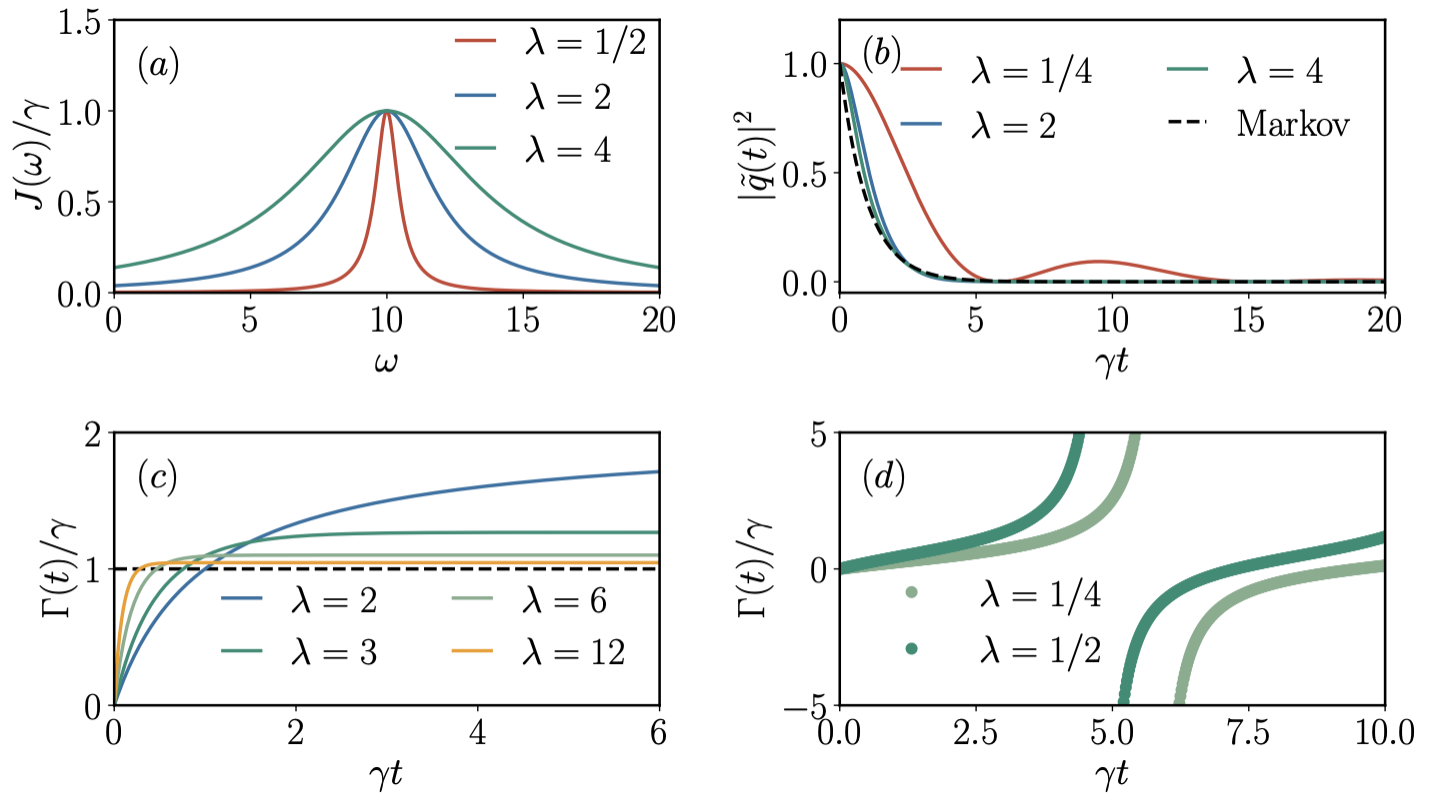}
    \caption{(a) Spectral density of the Lorentzian peaked environment at frequency $\omega_0$ (here set to $\omega_0=10$) for different frequency widths $\lambda$. (b) Comparison of the probability of the excited state of the atom $|\tilde{q}(t)|^2$ as a function of time for different $\lambda$ values. The dashed-black line corresponds to the Markovian approximation (GKSL master equation). Underdamped dynamics happen fro $\lambda<2$ ($\gamma=1$)  which reveal Rabi oscillations (coherent exchange of excitations between atom and bosonic environment), while for $\lambda\geq 2$ it is overdamped. (c) Instantaneous decay rate $\Gamma(t)$ in the overdamped regime. For $\lambda/\gamma \rightarrow \infty$ one recovers the Markov approximation, $\Gamma(t)\rightarrow \gamma$. (d) In the underdamped regime, the dynamics are clearly non-Markovian, where the instantaneous decay rate becomes negative during some time intervals.}
    \label{fig:atom_lossy}
\end{figure}

In the previous section we have studied a spin coupled to a large, virtually infinite, number of harmonic oscillators. We considered a specific coupling (longitudinal) such that the spin-boson model could be solved analytically. This allowed us to compare the Markovian approximation (i.e. the GKSL master equation) with the exact one. As a further demonstration that Markovian approximation may or not be a valid physical approximation we are going to see a final example:  Let us consider a two-level atom (i.e. a qubit) that interacts with a bosonic environment by exchanging excitations. In order to solve the system exactly, we take the spectral density to be a Lorentzian centered at the atom frequency and assume $T=0$ temperature. This is a common  scenario in several platforms, such as in cavity QED or superconducting circuits. The Hamiltonian of the system reads
\begin{align}
    H_S&=\omega_0\sigma^+\sigma^-,\\
    H_B&=\sum_k \omega_k b_k^\dagger b_k,\\
    H_I&=\sum_k (g_k \sigma^+ b_k+g_k^* \sigma^-b_k^\dagger).
\end{align}
so that the energy levels of the atom, $\ket{1}$ and $\ket{0}$ are separated by a frequency $\omega_0$. We assume that the environment is such that the spectral density takes the form of a Lorentzian centered at $\omega_0$ with width $\lambda$ (cf. Fig.~\ref{fig:atom_lossy}(a)), i.e.
\begin{align}
    J(\omega)=2\pi \sum_k |g_k|^2= \frac{\gamma \lambda^2}{(\omega-\omega_0)^2+\lambda^2},
\end{align}
where $\gamma$ gives account of an overall coupling strength. 
 As it should be clear by now, assuming the environment to be at $T=0$ temperature, the GKSL master equation will involve just a single operator $A=\sigma^-$ whose rate is $J(\omega_0)=\gamma$. This means that the atom spontaneously emits a photon into the environment at this rate $\gamma$. Neglecting possible Lamb shifts, in the interaction picture we have
 \begin{align}
     \frac{d}{dt}\rho_S(t)=\gamma \left(\sigma^-\rho_S(t) \sigma^+-\frac{1}{2}\{\sigma^+\sigma^-,\rho_S(t) \} \right).
 \end{align}
 
 The solution to this GKSL Markov master equation reads as (in terms of the Bloch vector)
 \begin{align}
     v_x(t)&=e^{-\gamma t/2}v_x(0),\\
     v_y(t)&=e^{-\gamma t/2}v_y(0),\\
     v_z(t)&=e^{-\gamma t}(1+v_z(0))-1.\label{eq:vz_atom_GKSL}
 \end{align}
 If the atom is initially in the excited state $\ket{1}$, then $\vec{v}(0)=(0,0,1)$, and the only term different from zero is $v_z(t)=2e^{-\gamma t}-1$. This means that the probability of finding the atom in the excited state is $p_1(t)=(1+v_z(t))/2=e^{-\gamma t}$. 
 
 The point we want address is the following: to what extent is this approximation valid? that is,  does the GKSL master equation accurately describe the dynamics of the atom? 

From the actual shape of the spectral density we can anticipate that, if $J(\omega)$ is very narrowed ($\lambda\rightarrow 0$), the Markovian approximation will break down. This consists to a well defined bosonic mode resonantly coupled to the atom, so that coherent exchange of excitations will play a significant role (i.e. backflow of information, thus non-Markovian).  Roughly speaking, the Markov approximation is akin to assume a flat spectral density. This corresponds to $\lambda\rightarrow \infty$. In that case, $J(\omega)\approx \gamma$. Another way to look at it: The inverse of the frequency width, $\lambda^{-1}$, establishes a characteristic time of the environment. The shorter this characteristic time, the better would be our Markov approximation that we assumed in the first place to derive the GKSL master equation. 

How do we compute the exact solution? Since the atom and modes exchange excitations, and the environment is initially empty (in the vacuum state), we can write the most general wavefunction containing just a single excitation. Note that the vacuum state for all elements does not change over time.  We will assume that at time $t=0$, the atom is in the excited state $\ket{1}$. Then,
\begin{align}
\ket{\psi(t)}_{SB}=\left[q(t)\sigma^++\sum_k c_k(t)b_k^\dagger\right]\ket{{\rm vac}},
\end{align}
where $\ket{{\rm vac}}=\ket{0}_S\bigotimes_k \ket{0}_k$ is the vacuum state for the whole system plus environment, and hence $q(t=0)=1$. Plugging this into the Schr\"odinger equation, we find
\begin{align}
    \dot{q}(t)&=-i\omega_0 q(t)-i \sum_k g_k c_k(t),\\
    \dot{c}_k(t)&=-i\omega_k c_k(t)-ig_k^* q(t).
\end{align}
If we define $\tilde{q}(t)=q(t)e^{-i\omega_0 t}$ and $\tilde{c}_k(t)=c_k(t)e^{-i\omega_k t}$ (equivalent to moving to an interaction picture with respect to $H_S+H_B$), we arrive to
\begin{align}
    \dot{\tilde{q}}(t)&=-i\sum_k g_k \tilde{c}_k(t)e^{-i(\omega_k-\omega_0)t},\\
    \dot{\tilde{c}}_k(t)&=-i g_k^* \tilde{q}(t)e^{i(\omega_k-\omega_0)t}.
\end{align}
 The last equation can be integrated formally,
 \begin{align}
     \tilde{c}_k(t)=-i\int_0^t ds \  g_k^* \tilde{q}(s)e^{i(\omega_k-\omega_0)s},
 \end{align}
and then it can be introduced in the equation for the atom amplitude leading to a integro-differential equation,
\begin{align}
    \dot{\tilde{q}}(t)&=-\sum_k |g_k|^2\int_0^t ds\  \tilde{q}(s) e^{i(\omega_k -\omega_0)s}e^{-i(\omega_k -\omega_0)t}\nonumber\\&=-\sum_k |g_k|^2e^{-i(\omega_k-\omega_0)t}\int_0^t ds \ \tilde{q}(s)e^{i(\omega_k-\omega_0)s}.
\end{align}
Taking the continuum limit (infinitely many bosonic modes), we have the following integro-differential equation,
\begin{align}\label{eq:qtilde_1}
 \dot{\tilde{q}}(t)=-\int d\omega \frac{J(\omega)}{2\pi}e^{-i(\omega-\omega_0)t}\int_0^t ds \ \tilde{q}(s)e^{i(\omega-\omega_0)s}.
\end{align}
If we define
\begin{align}
    f(t)=\int d\omega \frac{J(\omega)}{2\pi}e^{-i(\omega-\omega_0)t},
\end{align}
then Eq.~\eqref{eq:qtilde_1} can be expressed as
\begin{align}
    \dot{\tilde{q}}(t)=-\int_0^t ds \  f(t-s) \tilde{q}(s).
\end{align}
Indeed, the term $f(t-s)$ is nothing but the memory kernel of the system-environment interaction. In the Markov approximation, such memory kernel is replaced by a delta $f(t)\sim \gamma \delta(t)$. In the case of a Lorentzian spectral density, the exact expression can be computed as
\begin{align}
    f(t)=\int_0^\infty d\omega \ \frac{J(\omega)}{2\pi}e^{-i(\omega-\omega_0)t}=\int_0^\infty d\omega \ \frac{1}{2\pi}\frac{\gamma \lambda^2}{(\omega-\omega_0)^2+\lambda^2}e^{-i(\omega-\omega_0)t}.
\end{align}
Assuming that $\omega_0\gg 0$ so that the contribution for negative frequencies is negligible, we can approximate the previous function as
\begin{align}
    f(t)\approx \int_{-\infty}^\infty d\omega \ \frac{1}{2\pi}\frac{\gamma \lambda^2}{(\omega-\omega_0)^2+\lambda^2}e^{-i(\omega-\omega_0)t}=\frac{1}{2\pi}\int_{-\infty}^\infty d\Omega \frac{\gamma \lambda^2}{\Omega^2+\lambda^2}e^{-\Omega t}=\frac{\gamma \lambda}{2}e^{-\lambda t},
\end{align}
with $\lambda \geq 0$. From the previous result it is now clear why $\lambda^{-1}$ defines the characteristic time of the environment: it quantifies how the memory kernel decays in time. Therefore, the integro-differential equation for the atom amplitude $\tilde{q}(t)$ reads as
\begin{align}\label{eq:qtilde_2}
    \dot{\tilde{q}}(t)=-\int_0^t ds \frac{\gamma \lambda}{2}e^{-\lambda(t-s)}\tilde{q}(s).
\end{align}
The solution can be obtained using the Laplace transform (see~ Sec.\ref{moremat:Laplace_atom} for further details), leading to
\begin{align}\label{eq:qtilde_sol}
    \tilde{q}(t)=e^{-\lambda t/2}\left(\cosh(\Delta t/2)+\frac{\lambda}{\Delta}\sinh(\Delta t/2) \right),
\end{align}
where $\Delta=\sqrt{\lambda^2-2\gamma \lambda}$. Therefore, we can distinguish two cases: i) $\lambda\geq 2\gamma$, and ii) $\lambda<2\gamma$.

For the first case, $\lambda\geq 2\gamma$ the width of the spectral density is larger than the coupling, so we get closer to the Markovian limit (which takes place for $\lambda/\gamma\rightarrow \infty$). Here the amplitude of the qubit excitation shows an overdamped behavior,  that goes to zero in the long-time limit  following an exponential-like decay. Indeed, in the $\lambda\rightarrow \infty$ limit, we find $\lim_{\lambda\rightarrow \infty}\tilde{q}(t)=e^{-\gamma t/2}$. Since the probability of finding the qubit in the excited state is just $p_1(t)=|\tilde{q}(t)|^2$, we corroborate the Markovian result we got employing the GKSL master equation (see Eq.~\eqref{eq:vz_atom_GKSL}). 

For the second case, however, we get an underdamped behavior. Given that $\lambda<2\gamma$, the parameter $\Delta$ becomes purely complex, and so $\tilde{q}(t)=e^{-\lambda t/2}(\cos(|\Delta|t/2)+\lambda\sin(|\Delta|t/2)/\Delta)$. This case corresponds to a narrow spectral density compared to its coupling, so that there is a coherent exchange of excitations between the atom and the strongly-coupled bosonic mode of the environment (Rabi oscillations). Here we can clearly see that the behavior completely deviates from the Markovian approximation, even though the envelope still decays exponentially in time. See Fig.~\ref{fig:atom_lossy} for an illustration of the results comparing the probability $p_1(t)=|\tilde{q}(t)|^2$ with the Markovian approximation, $p_1(t)=e^{-\gamma t}$.

As a final remark, let us compute the instantaneous decay rate from the exact solution. In the Markovian approximation, a time-local GKSL master equation would read as
\begin{align}
    \frac{d}{dt}\rho_S(t)=\Gamma(t)\left(\sigma^-\rho_S(t)\sigma^+-\frac{1}{2}\{\sigma^+\sigma^-,\rho_S(t) \right).
\end{align}
Hence, the excited state population, $p_1(t)=|\tilde{q}(t)|^2$ follows the differential equation
\begin{align}
    \dot{p}_1(t)=-\Gamma(t)p_1(t),
\end{align}
which in terms of $\tilde{q}(t)$, can be expressed as
\begin{align}
    \frac{d}{dt}|\tilde{q}(t)|^2=\dot{\tilde{q}}(t)\tilde{q}^*(t)+\tilde{q}(t)\dot{\tilde{q}}^*(t)=2{\rm Re}[\dot{\tilde{q}}(t)\tilde{q}^*(t)] \Rightarrow 2{\rm Re}[\dot{\tilde{q}}(t)\tilde{q}^*(t)] =-\Gamma(t) \tilde{q}(t)\tilde{q}^*(t).
\end{align}
So, we find
\begin{align} 
\Gamma(t)=-2{\rm Re}\left[\frac{\dot{\tilde{q}}(t)}{\tilde{q}(t)} \right].
\end{align}
Plugging  $\tilde{q}(t)$ from Eq.~\eqref{eq:qtilde_sol} inside the previous expression, we obtain
\begin{align}
    \Gamma(t)=\frac{2\gamma \lambda}{\lambda+\Delta\coth(\Delta t/2)}.
\end{align}
If $\lambda/\gamma\rightarrow\infty$, we recover the Markovian result, $\Gamma(t)\rightarrow \gamma$. In general, however, the instantaneous decay rate will depend on time, deviating from just $\gamma$ and may be even negative during some intervals. This latter situation only happens when $\lambda<2\gamma$ (underdamped regime), reflecting what we already knew: the dynamics is non-Markovian and therefore, the dynamical map will fail to be CP-divisible during some time intervals (precisely when $\Gamma(t)<0$). For the overdamped regime ($\lambda\geq 2\gamma$), one can workout the long-time decay rate 
\begin{align}
    \lim_{t\rightarrow \infty}\Gamma(t)=\frac{2\gamma\lambda}{\lambda+\Delta}, \ {\rm if} \ \lambda\geq 2\gamma.
\end{align}
See Fig.~\ref{fig:atom_lossy} for the plots of these instantaneous decay rate $\Gamma(t)$ for different values of the parameters.

\subsection{Problems to practice}

\noindent\textbf{Problem 3.1}

\noindent Suppose a qubit $S$ that is weakly coupled with a collection of $N$ qubits that constitute the environment $B$. The Hamiltonian of the whole $S+B$ system can be written as
\begin{align}
    H_S=\frac{\omega}{2}\sigma_z,\quad
    H_B=\sum_{k=1}^N\frac{\nu_k}{2}\sigma_{z,k}, \quad 
    H_I=\sum_{k=1}^N\left(g_k \sigma^+\sigma^-_k+g_k^*\sigma^- \sigma^+_k\right),
    \nonumber
\end{align}
where the subscript $k$ refers to the Pauli operators of the environment qubits. Suppose an uncorrelated initial state, $\rho(0)=\rho_S(0)\otimes \rho_B$, where $\rho_B=\otimes_{k=1}^N \rho_{B,k}$. In addition, assume that each qubit of the environment is an equilibrium state of the form $\rho_{B,k}=p_k \ket{0}\bra{0}_k+(1-p_k)\ket{1}\bra{1}_k$, where $\sigma_{z,k}=\ket{0}\bra{0}_k-\ket{1}\bra{1}_k$, i.e. $\ket{1}_k$ is the ground state of the $k$th qubit. Answer:
\begin{itemize}
    \item[(i)] Under the usual approximations, derive the GKSL master equation and write it in the Schr\"odinger picture. Express the resulting rates and Lamb shift in the continuum limit ($N\rightarrow\infty$) in terms of the spectral density $J(\omega)=\sum_k |g_k|^2\delta(\omega-\nu_k)$.
    \item[(ii)] Explain (physically) what happens for the cases of $p_k=0,1$ and $1/2, \forall k$.
    \item[(iii)] Consider that the qubits are in a thermal state (at temperature $T$), and the qubit frequency is $\omega=2\pi\times 10$ GHz. Up to what temperature $T^*$ could we consider spontaneous absorption negligible? Assume that spontaneous absorption can be neglected when $\gamma_{s-abs}\leq 10^{-5}\gamma_{s-emi}$). Recall that $\hbar=1.054 \cdot 10^{-34} $ J$\cdot$s and $k_b=1.38\cdot 10^{-23}$ J/K. 
\end{itemize}

\vspace{0.5cm}
\noindent\textbf{Problem 3.2}
\vspace{0.2cm}

\noindent Consider a quantum harmonic oscillator interacting with many qubits that form the environment, whose total Hamiltonian is 
\begin{align}
    H_S=\omega a^\dagger a,\quad
    H_B=\sum_k \frac{\nu_k}{2}\sigma_{z,k},\quad
    H_I=a^\dagger a\sum_k g_k \sigma_{z,k}, 
    \nonumber
\end{align}
that is, the energy levels of the harmonic oscillator fluctuate depending on the state of the qubits. Assume $\rho_B$ to be an equilibrium/stationary state of $H_B$. 
Answer:
\begin{itemize}
    \item[(i)] Argue what would be the form of the GKSL master equation (assuming Born-Markov and rotating-wave approximations). No calculations needed.
    \item[(ii)] Describe qualitatively the impact of the dissipative part $D[\rho_S(t)]$ on a generic state $\rho_S(t)$. Show that if $\rho_S(t)$ is diagonal in the Fock basis $\{\ket{n}\}$, i.e. $\rho_S(t)=\sum_n p_n(t) \ket{n}\bra{n}$ then $D[\rho_S(t)]=0$. 
    \item[(iii)] Explain whether the Markov approximation remains valid in this context. 
\end{itemize}

\vspace{0.5cm}
\noindent\textbf{Problem 3.3}
\vspace{0.2cm}

\noindent Suppose an atom with three levels $\ket{g},\ket{e}$ and $\ket{f}$. The transition $\ket{g}\leftrightarrow \ket{f}$ can be driven by suitable laser radiation, while the other two are incoherent transitions as a result of the coupling of the atom with electromagnetic  environment $B$. In addition, selection rules forbid spontaneous transition between $\ket{f}$ and $\ket{g}$. The Hamiltonian can be written as (setting $\omega_g=0$) where $\omega_{e,g}=\omega_e-\omega_g$,
\begin{align}
    H_S=\omega_{e,g} \ket{e}\bra{e}+\omega_{f,g}\ket{f}\bra{f}, \quad H_B=\sum_k \nu_k a^\dagger_k a_k,\quad H_I= (\ket{e}\bra{f}+  \ket{g}\bra{e}) \sum_k g_k a_k +{\rm H.c.}.
    \nonumber
\end{align}

\begin{figure}[h!]
    \centering
    \includegraphics[width=0.4\linewidth]{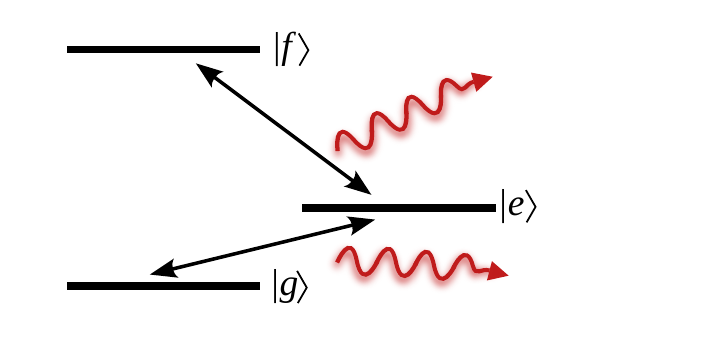}
\end{figure}

\begin{itemize}
    \item[(i)] Derive the GKSL master equation assuming thermal equilibrium states for the electromagnetic modes.
    \item[(ii)] Under what conditions would the incoherent transitions involve just spontaneous decay $\ket{f}\rightarrow\ket{e}\rightarrow\ket{g}$?
    \item[(iii)] If the atom has optical transitions, i.e. $\omega_f-\omega_g=2\pi \times 540 $ THz (green light), and $\omega_f-\omega_e=2\pi\times 400 $ THz (red light), up to what temperature $T$ could we consider that spontaneous absorption processes are negligible?  Consider the condition $\gamma_{\uparrow}=10^{-3}\gamma_{\downarrow}$ as the limit to neglect absorption $\gamma_{\uparrow}$. Recall that $\hbar=1.054\cdot 10^{-34}$ J$\cdot$s, and $k_b=1.38\cdot 10^{-23}$ J/K.
\end{itemize}

\newpage
\section{Quantum Thermodynamics}\label{c:thermo}

Thermodynamics stands as one of the most successful scientific theories, largely developed during the 19th century alongside the industrial revolution. The rapid progress in the design of thermal machines spurred theoretical investigations aimed at understanding their operating principles, limitations, and efficiency in converting heat into work—and vice versa. Pioneers such as Sadi Carnot, Rudolf Clausius, and William Thomson (better known as Lord Kelvin) laid the foundational concepts of this discipline. From these beginnings, thermodynamics evolved along multiple directions, spanning both applied engineering and fundamental physics, such as  statistical mechanics. In thermodynamics, we describe  the collective behavior of large ensembles of interacting particles through quantities such as pressure, work, heat, temperature, and entropy. The standard formulation of thermodynamics is inherently classical: it neglects the quantum nature of microscopic constituents and instead describes systems in terms of classical variables. This approach is remarkably accurate for macroscopic systems at normal conditions (e.g. at room temperature). However, as we scale down to smaller systems—those composed of a reduced number of particles, stochastic fluctuations become increasingly significant and can even be experimentally observed. Consider, for example, a heat engine whose working medium contains 
$N$ particles. For large 
$N$, the standard textbook case, quantities such as entropy, heat, and work behave deterministically, taking on well-defined values. As $N$ decreases, however, successive experimental runs may yield slightly different outcomes due to fluctuations. These quantities become stochastic, characterized by probability distributions. Although such fluctuations vanish in the macroscopic limit ($N\rightarrow\infty$), they play a crucial role in understanding small-scale machines, such in some biological processes. This area of study is known as stochastic thermodynamics, and is currently an active field of research.

The situation gets even more dramatic if we allow for classical and quantum fluctuations to enter into play. Quantum mechanics, as we will see, changes our understanding and predictions of thermodynamics. The key questions then are: How do quantum fluctuations modify the standard laws of thermodynamics? Can quantum resources be harnessed to enhance the performance of thermal machines? And can thermodynamic principles even be meaningfully applied to systems composed of only a few interacting particles? These questions are not merely of theoretical interest; they have  also profound practical implications. Much like in classical computation, as we keep on  miniaturizing machines, we will inevitably reach a scale where quantum effects become unavoidable, if not dominant. What will happen then, and what new phenomena might emerge? This section offers a brief overview on how thermodynamics can be reformulated and extended to the quantum domain \cite{Kosloff13,Vinjanampathy2016,Binder2018,Deffner2019,Landi2021}.

\subsection{Laws of thermodynamics in a quantum setting}
Before getting started, note that in classical thermodynamics, a state refers to possible configuration of classical variables that define the system (e.g. position and momentum). The possible configurations are known as microstates. The theory builds on the probability over a set of available microstates. In a similar fashion, quantum thermodynamics is developed by referring to the state of the system as the density matrix $\rho$ that accounts for the probabilities (but also coherence) between possible states.  

As in any standard textbook or course on thermodynamics, we commence by stating the pillars that support the theory: the four laws of thermodynamics. 
Let us recall them and discuss how they are translated in the quantum language.

\subsubsection{Zeroth law}

We start by the simplest, the zeroth law. Anecdotally, the zeroth law was introduced by R. H. Fowler in 1931, long after the first, second and third were well established. The statement is the following: 

\begin{enumerate} 
    \item[] \textbf{0th} If two systems $A$ and $B$ are in thermal equilibrium with another system $C$, then $A$ and $B$ are also in thermal equilibrium. This statement is sometimes restated by saying that if $A$ and $B$ are in thermal equilibrium, they both share the same temperature $T$.
\end{enumerate}

\noindent A quantum system in thermal equilibrium is described by its Gibbs state,
\begin{align}
    \rho_{eq}=\frac{e^{-\beta H}}{Z}\equiv \frac{1}{Z}\sum_n e^{-\beta E_n}\ket{n}\bra{n},
\end{align}
being $Z\equiv {\rm Tr}[e^{-\beta H}]$ the partition function, $\beta$ the inverse of the temperature, $\beta=(k_b T)^{-1}$ and $H$ the Hamiltonian of the system such that $H=\sum_n E_n \ket{n}\bra{n}$. Now, suppose a bipartite system $S+B$, then its combined thermal state is just
\begin{align}\label{eq:rho_eqSB}
    \rho_{eq;SB}=\frac{e^{-\beta (H_S+H_B+H_I)}}{Z_{SB}}.
\end{align}
Their reduced states, however, may differ from their local equilibrium state,
\begin{align}
    \rho_{eq;S}&={\rm Tr}_B[\rho_{eq;SB}]\neq \frac{e^{-\beta H_S}}{Z_S},\\
    \rho_{eq;B}&={\rm Tr}_S[\rho_{eq;SB}]\neq \frac{e^{-\beta H_B}}{Z_B}.
\end{align}
This is due to the interaction term that can shift or renormalize the energy of the other system, produce correlations or mix the eigenstates of the otherwise non-interacting subsystems. In general, for $S$ and $B$ interacting, $\rho_{eq;SB}\neq \rho_{eq;S}\otimes \rho_{eq;B}$. Is this a unique quantum feature? No, this also happens at the classical level. However, two remarks are in order: First, in a classical system the interaction only shifts local energies, i.e. it does not modify the basis of the equilibrium states. Second, in a classical macroscopic scenario the interaction term is typically much smaller than the energy of the local systems so it can be discarded, $H_I\rightarrow 0$. Intuitively, this can be understood since the interaction happens only at the boundary between $S$ and $B$, while $H_S$ and $H_B$ account for  energy of the whole macroscopic degrees of freedom. Yet, if such interaction is significant, the equilibrium state after switching on the interaction for a $S$ and $B$, initially in local thermal equilibrium states at $T$ with respect to $H_S$ and $H_B$, respectively,  would be of the form of Eq.~\eqref{eq:rho_eqSB} at a different temperature $T'$. In a quantum system this is often more problematic since we consider small number of constituents in $S$ and the interaction can be comparable to the energies of $H_S$. Hence, one must be careful with the meaning of equilibrium states. Moreover, there is a fundamental problem: if we consider $S+B$ as an isolated quantum system, is there any guarantee for it to thermalize? This is a long-standing problem in quantum statistical mechanics, which is outside the scope of this lecture (see for example Refs.~\cite{Deutsch1991,Rigol2008}).   

What about our master equation describing the thermalization of a quantum system? If you recall, the GKSL master equation was derived assuming a weak coupling. As we have mentioned in the previous chapter, a strong coupling renders the GKSL invalid and non-Markovian effects and backflow of information will likely arise. Moreover, we typically consider an infinite number of degrees of freedom in the environment, and therefore, the bath is unaffected by just a tiny interaction with $S$. To the contrary, $S$ is able to change considerably, i.e. to thermalize, by damping or absorbing excitations from the environment.  

However, we have avoided one key point: what does temperature even mean in a quantum system? Mathematically speaking, the temperature appears as a Lagrange multiplier that ensures that the resulting state maximizes the entropy constrained to an average energy $\langle U\rangle$ (more on this later). Physically, it just provides an imbalance of populations over the energy eigenstates of the system. It is not a direct observable quantity, it must be indirectly estimated from, for example,  populations over different states.

\subsubsection{First law}\label{ss:1st}
The first law of thermodynamics is just a reformulation of the energy conservation, mainly attributed to R. Clausius in 1850. The statement is the following:

\begin{enumerate} 
    \item[] \textbf{1st}  In any process, energy can never be created or destroyed; it can only be transferred from one object to another in the form of heat and/or work. That is,
    \begin{align}\label{eq:1st}
        \Delta \mathcal{U}=Q+W,
    \end{align}
    where $\Delta \mathcal{U}$ refers to the change in the internal energy of the system and $Q$ and $W$ to the heat and work that may flow in or out during the process. 
\end{enumerate}

The definition in Eq.~\eqref{eq:1st} keeps the convention that $Q>0$ and $W>0$ that correspond to heat supplied to and work done on the system. To the contrary, $Q<0$ means heat supplied by the system to the outside and $W<0$ work done by the system to an external agent. 

In addition, recall your classical thermodynamics lessons: The internal energy $\mathcal{U}$ is a state function. Therefore $\Delta \mathcal{U}$ depends only on the initial and final states, and not on how you got there. Work and heat, however, are path-dependent quantities: $W$ and $Q$ do depend on how you perform the transformation, that is, on the trajectory. If we consider classical fluctuations (stochastic thermodynamics), then $\mathcal \mathcal{U}$, $W$ and $Q$ become random or stochastic variables. This was to be expected since the stochastic trajectories reflect a distribution of possible values  for each of these standard thermodynamic variables.

In a quantum system, this scenario becomes more problematic. Clearly, due to its intrinsic probabilistic nature, the internal energy is a random variable, whose average value reads as
\begin{align}
    \langle \mathcal{U}(t)\rangle ={\rm Tr}[H(t)\rho(t)],
\end{align}
but it is still a state function. It is an observable quantity (to estimate it, just measure the energy $H(t)$ on your state $\rho(t)$). Thus, the change in the internal energy for a process taking place for $t=0$ (initial time) to $t=\tau$ (final time) is given by
\begin{align}
    \Delta \mathcal{U}={\rm Tr}[H(\tau)\rho(\tau)]-{\rm Tr}[H(0)\rho(0)].
\end{align}
Heat and work, however, are trajectory dependent. Yet, there is  analog of a classical trajectory in a quantum system. This is a major obstacle in quantum thermodynamics since $Q$ and $W$ cannot be defined in a universal and clear manner (cf. Sec.~\ref{s:qWQ}). The very notion of heat and work becomes blurred in a quantum system. A possible way forward is the following: Take the rate over time of $\langle \mathcal{U}(t)\rangle$, i.e.
\begin{align}
    \frac{d}{dt}\langle \mathcal{U}(t)\rangle=\frac{d}{dt}{\rm Tr}[H(t)\rho(t)]={\rm Tr}[H(t)\dot{\rho}(t)]+{\rm Tr}[\dot{H}(t)\rho(t)],
\end{align}
allowing the Hamiltonian to be time-dependent. 
In this manner, one could identify the heat current and power  as 
\begin{align}
    \langle \dot{Q}(t)\rangle &={\rm Tr}[H(t)\dot{\rho}(t)],\\
    \langle \dot{W}(t)\rangle&={\rm Tr}[\dot{H}(t)\rho(t)].
\end{align}
These definitions are reasonable  given that heat produces transitions between levels (e.g. relaxation), while work is tied to external agents changing some parameter of the system, and therefore changing the energy level structure. In this manner,  the total average heat and work for this process read as
\begin{align}\label{eq:Qdef}
    \langle Q\rangle &=\int_0^\tau dt \ {\rm Tr}[H(t)\dot{\rho}(t)],\\
    \langle W\rangle &=\int_0^\tau dt \ {\rm Tr}[\dot{H}(t)\rho(t)]. \label{eq:Wdef}
\end{align}
By construction, the first law is ensured: $\Delta \mathcal{U}=\langle Q\rangle +\langle W\rangle$. Again, note that, as in the classical case account for fluctuations, $\mathcal{U}$, $Q$ and $W$ are random variables whose values may differ from run to run. The first law is only satisfied in average, upon averaging the results over many experimental runs. Moreover, from the definitions of heat and work, it is clear that they are not observables in the conventional sense. They cannot be since there is no notion of a sure trajectory. Take the work $W$ for example. There is no observable whose expectation value provides $\langle W\rangle$. Instead, in order to compute $\langle W\rangle$, one needs to keep track of the average value of $\dot{H}(t)$ on the state $\rho(t)$ over time. However, any measurement to estimate ${\rm Tr}[\dot{H}(t)\rho(t)]$ unavoidably disturbs the system, forcing it to abandon the evolution or dynamics whose work we aimed at computing in the first place. This problem motivated another (useful) definition of work: the two-point measurement scheme. This will be discussed in Sec.~\ref{s:qWQ}

\subsubsection{Second law}
The second law of thermodynamics is perhaps the most profound among them. It has been reformulated several times over the years, and it was originally introduced in the 19th century by S. Carnot (early work in 1824), W. Thomson (1851) and R. Clausius (1854). Here we give the statement by M. Planck, which goes as follows:
\begin{enumerate} 
    \item[] \textbf{2nd} It is impossible for a cyclic machine to convert heat totally into work without producing change elsewhere in the universe.
\end{enumerate}
In essence, as you may already know, this statement tells us that the net change of entropy of a system $\Delta S$ must be larger or equal than the exchanged heat divided by the temperature $T$ of the environment,
\begin{align}\label{eq:2nd_class}
    \Delta S\geq \frac{Q}{T}.
\end{align}
This is known as the Clausius inequality, where the equality holds for reversible processes.

In order to translate the inequality for the change in entropy we must first define entropy for a quantum system. In a classical and isolated scenario, $S=-k_b \log \Omega$ where $k_b$ is the Boltzmann constant and $\Omega$ the number of available microstates. If the system is described instead by a probability distribution in the phase space, we have $S=-k_b \sum_i p_i \log p_i$. In a quantum scenario, we replace the Shannon by the von Neumann entropy,
\begin{align}
    S=-k_b {\rm Tr}[\rho \log \rho].
\end{align}

Under a series of conditions (Markovian dynamics, detailed balance to guarantee thermalization in the long-time limit), it can be shown that indeed Eq.~\eqref{eq:2nd_class} is fulfilled, but again replacing sure variables by their corresponding average values. For a general scenario, non-Markovian dynamics, non-detailed balance, etc. the relation is not as simple and similar to the classical case. Given the importance of entropy and the second law, this deserves a separate discussion;  we will come back to this in Sec.~\ref{s:entropy}.

\subsubsection{Third law}
The third law of thermodynamics is often referred to as the Nernst or unattainability principle, formulated by W. Nernst in the 1910s:
\begin{enumerate}
    \item[] \textbf{3rd} It is impossible for any process to reduce the entropy of a system to its absolute-zero value in a finite number of operations.
\end{enumerate}
An equivalent way to look at it: Given that the entropy of the $T=0$ state vanishes\footnote{A more general formulation should allow for non-zero entropy, as it is the case for systems with degenerate ground states.}, it is impossible to achieve a $T=0$ temperature state in a finite number of operations.

This is quite important in quantum technologies as we often assume the preparation of pure states: However, no matter how good your setup is, you will never be able to prepare the ground state of the system in an exact manner, i.e. with unit probability. Of course, your  initial state preparation may be sufficiently good, i.e. its fidelity may get very close to $1$ so that the associated error due to an imperfect state preparation becomes irrelevant (as typically other sources of errors dominate). If this is the case, the preparation can be treated as ideal, i.e. as if we could achieve a $T=0$ state.

This might sound reasonable, but you may wonder: What about if a system is measured, say, in the energy eigenbasis of the Hamiltonian?   If we write the Hamiltonian in its spectral form,
\begin{align}
    H=\sum_{n=0} e_n \ket{e_n}\bra{e_n},
\end{align}
 we can construct a PVM using the projectors $\Pi_n=\ket{e_n}\bra{e_n}$ associated to the energy outcomes $e_n$ (assuming no degeneracies). Assuming we can perform this measurement,  if we record outcome $e_n$, no matter the state of the system prior to the measurement, it collapses to $\ket{e_n}$, i.e.
\begin{align}
    \rho_n=\frac{\Pi_n \rho \Pi_n^\dagger}{{\rm Tr}[\Pi_n \rho \Pi_n^\dagger]}=\frac{\bra{e_n}\rho\ket{e_n}}{\bra{e_n}\rho\ket{e_n}}\ket{e_n}\bra{e_n}=\ket{e_n}\bra{e_n}.
\end{align}
Surely, if we observe $e_0$ our state is brought to the ground state, and this is nothing but a $T=0$ state. Although the measurement process is random, after a finite number of measurements we could retrieve $e_0$. In average, however, $\ket{e_0}$ is not prepared (it is not a deterministic operation). In any case, this seems to be at odds with the third law. Does this mean that the third law is violated in a quantum mechanical system? No, it is not! The third law of thermodynamics applies to quantum and classical systems alike. What is failing in our description then? The very notion of an ideal projective measurement. In reality, no matter how good the measurement device you have purchased, it will always come with uncertainty or finite error. In the context of measurement theory of quantum systems, the energy-basis measurement must be formulated as a POVM with a finite-energy resolution $\Delta E$, i.e.
\begin{align}
    M_e=\int de' f(e-e')\ket{e'_n}\bra{e'_n},
\end{align}
with $f(e-e')$ accounting for its energy resolution, such that $\int de M_e=\mathbb{I}$.  The measurement of $M_e$ would result in a collapse to a mixed state reflecting the uncertainty $\Delta E$. Moreover, since the measurement is the result of a physical interaction between systems (your system of interest to be measured and the measurement apparatus), pushing down the energy uncertainty  $\Delta E\rightarrow 0$ necessarily increases the interaction time (by virtue of the Heisenberg uncertainty principle for energy-time $\Delta E\Delta t\geq 1/2$ since $\hbar=1$). Hence, a perfect energy-projective measurement would require an infinite time (not to mention other technical impediments). While this is just an argument, not an actual demonstration, it suffices for the purpose of this course: The third law of thermodynamics holds in quantum mechanical systems.


\subsection{Quantum heat and work}\label{s:qWQ}
We have already introduced definitions for heat and work (cf. Eqs.~\eqref{eq:Qdef}-\eqref{eq:Wdef}). Let us test them under a generic Markov master equation in GKSL form (cf. Eq.~\eqref{eq:GKSL}). We start with heat $Q$. The average heat current or rate is given by
\begin{align}
    \langle \dot{Q}(t)\rangle &= {\rm Tr}[H_S(t)\dot{\rho}_S(t)]={\rm Tr}[H_S(t)\mathcal{L}\{\rho(t)\}]=-i{\rm Tr}[H_S(t)[H_S(t),\rho_S(t)]]+\nonumber\\&+\sum_k \gamma_k \left({\rm Tr}[H_S(t) A_k \rho_S(t)A_k^\dagger] -\frac{1}{2} {\rm Tr}[H_S(t)A_k^\dagger A_k \rho_S(t)]-\frac{1}{2}{\rm Tr}[H_S(t)\rho_S(t) A_k^\dagger A_k]\right).
\end{align}
For the first term, i.e. the trace with the commutator of $H_S(t)$ and $\rho_S(t)$, simply vanishes,
\begin{align}
    {\rm Tr}[H_S(t)H_S(t)\rho_S(t)-H_S(t)\rho_S(t)H_S(t)]={\rm Tr}[H_S(t)H_S(t)\rho_S(t)-H_S(t)H_S(t)\rho_S(t)]=0.
\end{align}
We can rearrange the other terms and get
\begin{align}
    \langle \dot{Q}(t)\rangle = \sum_k \gamma_k {\rm Tr}\left[\left(A_k^\dagger H_S(t) A_k-\frac{1}{2}\{ A^\dagger A_k,H_S(t)\} \right)\rho_S(t)\right],
\end{align}
so for each dissipative channel, we can define a heat rate operator $\dot{Q}_k(t)$,
\begin{align}\label{eq:Qkdot}
    \dot{Q}_k=\gamma_k \left( A_k^\dagger H_S(t) A_k-\frac{1}{2}\{ A^\dagger A_k,H_S(t)\} \right),
\end{align}
such that
\begin{align}
    \langle \dot{Q}(t)\rangle=\sum_k {\rm Tr}[\dot{Q}_k\rho_S(t)].
\end{align}
Note that each term $\dot{Q}_k$ in Eq.~\eqref{eq:Qkdot} is an Hermitian operator, $\dot{Q}_k^\dagger=\dot{Q}_k$, so it is an observable. The total heat, to the contrary, requires the integration of these rates over the whole process, from $t=0$ to $t=\tau$, 
\begin{align}
    \langle Q\rangle =\int_0^\tau dt \ \langle \dot{Q}(t)\rangle=\int_0^\tau dt \ \sum_k {\rm Tr}[\dot{Q}_k\rho_S(t)],
\end{align}
thus, it is not an observable in the traditional sense: There is no observable whose expectation value results in $\langle Q\rangle$.  In addition, note that this derivation tells us that heat can only be non-zero if the evolution is non-unitary (if there is some decay channel with operator $A_k$ such that $\gamma_k\neq 0$). That is why this definition is satisfying: Heat is associated to energy flow between system and environment, so if the system is closed, it must follow that $\langle Q\rangle=0$. 

Let us move now our attention to work $W$. If the system is closed, the average work is already defined: Given that $\langle Q\rangle=0$, it follows that $\Delta \mathcal{U}=\langle W\rangle={\rm Tr}[H(\tau)\rho(\tau)]-{\rm Tr}[H(0)\rho(0)]$, where we have used the first law, -i.e. Eq~\eqref{eq:1st}. In the closed system, the estimation of the work done on/by an external agent is \textit{easy}, it just involves the measurement of the initial and final energies of the system. In general, however, one needs to keep track of ${\rm Tr}[\dot{H}_S(t)\rho_S(t)]$ and integrate it over time. 

If the Hamiltonian is time independent, i.e. if there is no actual modification of the energy levels of the system by an external agent, then $\dot{H}(t)=0$, so that $\langle W\rangle=0$. Again, this is reasonable since any work is associated to external changes that modify the underlying Hamiltonian. Thus, in a non-driven case, $\dot{H}(t)=0$, and therefore any change in internal energy  comes solely from heat so $\Delta \mathcal{U}=\langle Q\rangle$. 

\subsubsection{The two-point measurement scheme}

Although the previous definitions for $Q$ and $W$ are reasonable, they are not very useful from a practical point of view. This motivated a new definition of work: Since for closed systems one can compute average work as $\langle W\rangle={\rm Tr}[H(\tau)\rho(\tau)]-{\rm Tr}[H(0)\rho(0)]$, one could define work relying on a two-point measurement scheme that precisely aims at estimating that quantity. Such scheme involves four steps:
\begin{itemize}
    \item[(i)] Prepare the initial state $\rho(0)$
    \item[(ii)] Perform an energy-projective measurement, in the eigenbasis of initial Hamiltonian, $H(0)=\sum_n E_n^0 \ket{n_0}\bra{n_0}$. This happens with probability \begin{align}
        p_n^0=\bra{n_0}\rho(0)\ket{n_0},
    \end{align} and the state collapses to
    \begin{align}
        \rho(0)\xrightarrow[]{ \textrm{outcome}\   E_n^0} \ket{n_0}\bra{n_0}.
    \end{align}
    \item[(iii)] Evolve the collapsed state under $H(t)$ and obtain $\rho_n(\tau)=V(\tau,0)\{ \ket{n_0}\bra{n_0}\}$, being $V(\tau,0)$ the dynamical map from start to end of the protocol.
    \item[iv)] Perform a final energy-projective measurement in the basis of the final Hamiltonian, $H(\tau)=\sum_m E_m^\tau\ket{m_\tau}\bra{m_\tau}$, such that the probability of recording $E_m^\tau$ reads as
    \begin{align}
p_{m|n}^\tau=\bra{m_\tau}\rho_n(\tau)\ket{m_\tau},
    \end{align}
    given that it is conditioned to the prior measurement that collapsed in $\ket{n_0}$.
\end{itemize}
The energy difference between these energies define a possible value for the work $W=E_{m}^\tau-E_n^0$. One can define a probability of work as $P(W)=\sum_{n,m}p_{m|n}^\tau p_n^0 \delta(W-(E_m^\tau-E_n^0))$. 
Performing the average over many runs, we obtain the average work of this two-point measurement scheme,
\begin{align}
    \langle W\rangle_{{\rm 2PM}}=\sum_{n,m}p_{m|n}^\tau p_n^0 (E_m^\tau-E_n^0).
\end{align}
Does this definition coincide with the one given in Eq.~\eqref{eq:Wdef}? Let us develop a bit more the expression,
\begin{align}
    \langle W\rangle_{{\rm 2PM}}&=\sum_{m}E_m^\tau \sum_n p_{m|n}^\tau p_n^0-\sum_n E_n^0\sum_m p_{m|n}^\tau p_n^0=\sum_m E_m^\tau p_m^\tau-\sum_n E_n^0 p_n^0\nonumber\\&=\langle E(\tau)\rangle -\langle E(0)\rangle,
\end{align}
where we have used the fact that $\sum_{n}p_{m|n}^\tau p_n^0=p_m^\tau$ since it is just the probability to get outcome $E_m^\tau$ in the second measurement, as well as $\sum_m p_{m|n}^\tau = 1$ since the conditioned probability must add up to $1$ when summing over all possibilities. Note that we have written $\langle E(\tau)\rangle$ and $\langle E(0)\rangle$ rather than ${\rm Tr}[H(\tau)\rho(\tau)]$ and ${\rm Tr}[H(0)\rho(0)]$. This is done on purpose because these quantities, in general, are different. If $[H(0),\rho(0)]\neq 0$, a measurement on $\{\ket{n_0}\}$ destroys the possible coherences that $\rho(0)$ might have had. Although the initial energies coincide, $\langle E(0)\rangle ={\rm Tr}[H(0)\rho(0)]$, the state that evolves until $t=\tau$ is different. This problem arises only due to quantum effects. Note that in Eq.~\eqref{eq:Wdef}, the state was arbitrary. In $\langle W\rangle_{{\rm 2PM}}$ the state that evolves is forced to be diagonal in $\{\ket{n_0}\}$ due to the measurement and subsequent collapse.  

As the initial state satisfies $[H(0),\rho(0)]=0$, then $\langle W\rangle_{2PM}={\rm Tr}[H(\tau)\rho(\tau)]-{\rm Tr}[H(0)\rho(0)]$. But this only coincides with $\langle W\rangle$ if the system is closed, i.e. $\langle Q\rangle=0$. That is, both definitions are equivalent if the system is closed and the initial state $\rho(0)$ commutes with $H(0)$, i.e. $\rho(0)$ diagonal in the $\{\ket{n_0}\}$ basis. This situation, however, is quite  restrictive. In general, these two definitions of work differ as they quantify distinct quantities. In any case, let us remark that the two-point measurement scheme provides a valuable (and efficient) tool to gain insight into the thermodynamics of the system. 


\subsubsection{Jarzynski equality} 
The Jarzynski equality \cite{Jarzynski97} is one of the most important and recent results in non-equilibrium statistical mechanics, that applies both to classical and quantum systems alike. It belongs to a larger family of fluctuation theorems that bridge dynamical with equilibrium properties. It was derived by C. Jarzynski in 1996. It provides a way to quantify equilibrium properties employing non-equilibrium observations, a key tool to study, for example, free energy differences in biological processes.

The derivation is simple, yet elegant. Let us make some assumptions first:
\begin{itemize}
    \item[(i)] The system $S$ is closed
    \item[(ii)] The state at initial time is a thermal equilibrium state (i.e. a Gibbs state) $\rho_{eq}$ given by
    \begin{align}
    \rho_{eq}=\frac{e^{-\beta H_0}}{Z_0}.
\end{align}
    \item[(iii)] The process changes the Hamiltonian from $H(0)$ to $H(\tau)$
\end{itemize}
    Then, as we have just seen above,  $\langle W\rangle=\langle W\rangle_{{\rm 2PM}}$. The Jarzynski equality follows then from computing the average value of $e^{-\beta W}$ under the two-point measurement scheme:
    \begin{align}
        \langle e^{-\beta W}\rangle&=\sum_{n,m} p_{m|n}^\tau p_n^0 e^{-\beta (E_m^\tau-E_n^0)}=\sum_{n,m} p_{m|n}^\tau \frac{e^{-\beta E_n^0}}{Z_0} e^{-\beta (E_m^\tau-E_n^0)}=\frac{1}{Z_0}\sum_{n,m} p_{m|n}^\tau e^{-\beta E_m^\tau}\nonumber\\&=\frac{1}{Z_0}\sum_{n,m} \bra{m}U(\tau,0)\ket{n}\bra{n}U^{\dagger}(\tau,0)\ket{m} e^{-\beta E_m^\tau}\nonumber\\&=\frac{1}{Z_0}\sum_m \bra{m}U(\tau,0)\left(\sum_n\ket{n}\bra{n}\right)U^\dagger(\tau,0)\ket{m} e^{-\beta E_m^\tau}=\frac{1}{Z_0}\sum_m e^{-\beta E_m^\tau}=\frac{Z_\tau}{Z_0}
    \end{align}
where $Z_\tau$ denotes the partition function of the Gibbs state at same temperature $\beta^{-1}$ but with Hamiltonian $H(\tau)$. Since the Helmholtz free energy is
\begin{align}
    F=\mathcal{U}-T S,
\end{align}
    which for a Gibbs state reads as
    \begin{align}
        F=-\frac{1}{\beta}\log Z,
    \end{align}
we can express the ratio of partition functions as
\begin{align}
    \frac{Z_\tau}{Z_0}=\frac{e^{-\beta F_\tau}}{e^{-\beta F_0}}=e^{-\beta (F_\tau-F_0)}=e^{-\beta \Delta F}.
\end{align}
    This leads to the Jarzynski equality,
    \begin{align}\label{eq:JE}
        \langle e^{-\beta W}\rangle=e^{-\beta \Delta F}.
    \end{align}
Note that the free energy is a purely equilibrium property: it quantifies the maximum amount of work that the system can perform in a thermodynamic process at constant temperature. The crucial step in the derivation is the micro-reversibility of the dynamics, i.e. the unitary evolution.  By virtue of the Jensen's inequality\footnote{The Jensen's inequality is widely used in statistics. It states that the expectation value of a random variable $X$ of a convex function $\varphi(X)$ obeys $\langle \varphi(X)\rangle \geq \varphi(\langle X\rangle)$.}, $\langle e^{X}\rangle\geq e^{\langle X\rangle}$ (because the exponential is a convex function), one can observe that
\begin{align}
    e^{-\beta \Delta F}=\langle e^{-\beta W}\rangle \geq e^{-\beta \langle W\rangle}\Rightarrow -\beta \Delta F \geq -\beta \langle W\rangle  \Rightarrow \langle W\rangle \geq \Delta F.
\end{align}
This is akin to a second law: the average work of a closed system when going from $H(0)$ to $H(\tau)$ must be larger than the free energy increment of its equilibrium states at $H(0)$ and $H(\tau)$. The equality holds for quasi-static (infinitely slow) evolutions. 

A natural question should be: Does Eq.~\eqref{eq:JE} remain valid if we allow for dissipative dynamics, i.e. beyond unitary evolution? The answer must be yes, since we can consider the global system $S+B$ as closed. Yet, if we  only have  access to the subsystem $S$, unfortunately Eq.~\eqref{eq:JE} does not hold in general (for example, the local Gibbs state for $S$ may already be a wrong assumption, see Sec.~\ref{ss:1st}). In any case, even if we assume local Gibbs states, and allow for a generic dynamical map, the Jarzynski equality must be modified: see Sec.~\ref{moremat:MJE} for further details.


\subsection{Entropy}\label{s:entropy}
The notion of entropy is a fundamental pillar of thermodynamics. In a quantum setting, the von Neumann entropy 
\begin{align}
    S(\rho)=-k_b {\rm Tr}[\rho \log \rho],
\end{align}
plays the role of the thermodynamical entropy~\cite{Landi2021}. It quantifies the uncertainty of the underlying probability distribution that $\rho$ represents (in its eigenbasis). Therefore, the minimum value takes place when $\rho$ is a pure state (no uncertainty), $S(\ket{\psi}\bra{\psi})=0$ since $\lim_{x\rightarrow 0}x\log x=0$. The maximum takes place in the opposite scenario, whenever $\rho$ is a completely mixed state, $\rho=\mathbb{I}/d$ being $d=\dim[\mathcal{H}]$. In that case, we have $S(\mathbb{I}/d)=-k_b \sum_{k=1}^d 1/d \log (1/d)=k_b\log d $. Therefore, for any $\rho\in\mathcal{D}(\mathcal{H})$, we have
\begin{align}
    0\leq S(\rho)\leq k_b \log d.
\end{align}
It is worth stressing that any unitary transformation does not modify the entropy of the state, $S(U\rho U^\dagger)=S(\rho)$ for $U^\dagger U =\mathbb{I}$. 

In addition, it is very useful to introduce another quantity: the quantum relative entropy, which is the extension to quantum states of the classical relative entropy (Kullback-Leibler divergence; see below). The quantum relative entropy is a measure of the distinguishability among two states $\rho, \sigma\in \mathcal{D}(\mathcal{H})$, and it is defined as
\begin{align}\label{eq:Srel}
    S(\rho \lVert \sigma)=k_b {\rm Tr}[\rho(\log\rho-\log \sigma)].
\end{align}
This expression can be rearranged in terms of the entropy $S(\rho)$ plus a relative term,
\begin{align}
    S(\rho \lVert \sigma)=-S(\rho)-k_b {\rm Tr}[\rho \log \sigma].
\end{align}
Note that the relative entropy is not symmetric ($S(\rho\lVert \sigma)\neq S(\sigma \lVert \rho)$) and is non-negative,
\begin{align}\label{eq:Klein}
    S(\rho\lVert \sigma)\geq 0,
\end{align}
being zero if and only if $\rho=\sigma$. The expression in Eq.~\eqref{eq:Klein} is known as Klein's inequality. In case $\rho$ and $\sigma$ commute, i.e. they can be diagonalized in the same basis $\rho=\sum_k p_k \ket{k}\bra{k}$ and $\sigma=\sum_k q_k \ket{k}\bra{k}$, then $S(\rho\lVert \sigma)$ reduces to the Kullback-Leibler divergence (a measure between different classical probability distributions), $S(\rho\lVert \sigma)=\sum_j p_j\log(p_j/q_j)$. In general, for $[\rho,\sigma]\neq 0$, one must be careful (see~\ref{examp:Srel}); indeed, the quantum relative entropy can diverge in some cases.

\begin{examp}[Quantum relative entropy]{examp:Srel}\label{examp:Srel}
Suppose two states $\rho,\sigma\in\mathcal{D}(\mathcal{H})$, such that
\begin{align}
    \rho=\sum_{k=1}^d p_k \ket{\phi_k}\bra{\phi_k}, \quad \sigma=\sum_{k=1}^d q_k\ket{\varphi_k}\bra{\varphi_k},
\end{align}
with $\{\ket{\phi_k}\}$ and $\{\ket{\varphi_k}\}$ two distinct orthonormal basis of $\mathcal{H}$.
Then, the quantum relative entropy can be written as
\begin{align}
    S(\rho\lVert \sigma)&=k_b \sum_{k=1}^d q_k \log q_k-k_b {\rm Tr}[\sum_{k,l=1}^d p_k \ket{\phi_k}\bra{\phi_k}\log q_l\ket{\varphi_l}\bra{\varphi_l} ]\nonumber\\&=k_b \sum_{k=1}^d q_k \log q_k -k_b \sum_{k,l=1}^{d} p_k \log q_l P_{k,l}
\end{align}
    where we have defined $P_{k,l}=|\langle \phi_k|\varphi_l\rangle|^2$. Thus, if for some $k,l$ one has  $q_l=0$ but $p_k>0$ with $P_{k,l}>0$, then $S(\rho\lVert\sigma)\rightarrow \infty$. This can be better formalized: $S(\rho\lVert \sigma)<\infty$ iff ${\rm ker}(\sigma)\subseteq {\rm ker}(\rho)$. 
\end{examp}

\subsubsection{System and environment as a closed system}
In the following we are going to derive the Clausius inequality, i.e. the second law of thermodynamics, for a quantum system. For that, let us consider the system interacting with an environment as a closed system. Furthermore, as we have considered in previous chapters, assume an uncorrelated initial state $\rho_{SB}(t_0)=\rho_S(t_0)\otimes \rho_B(t_0)$. In this way, we have
\begin{align}
    S(\rho_{SB}(t_0))=-k_b {\rm Tr}[\rho_{SB}(t_0)\log\rho_{SB}(t_0)]=S(\rho_S(t_0))+S(\rho_B(t_0)),
\end{align}
since the product state can be written as $\rho_{SB}(t_0)=\sum_k p_k q_l \ket{k,l}\bra{k,l}$, with $p_k$ and $q_l$ the corresponding eigenvalues of $\rho_{S}(t_0)$ and $\rho_B(t_0)$, respectively. 
Now, since $S+B$ is a closed system, it evolves under a unitary transformation,
\begin{align}
    \rho_{SB}(t)=U(t,t_0)\rho_{SB}(t_0)U^\dagger(t,t_0),
\end{align}
which does not change the entropy of the combined state, $S(\rho_{SB}(t))=S(\rho_{SB}(t_0))$, although it can definitely change the reduced states, $\rho_S(t)={\rm Tr}_B[\rho_{SB}(t)]$ and $\rho_B(t)={\rm Tr}_S[\rho_{SB}(t)]$, and thus their individual contributions to the total entropy. 

If we now assume that the initial state of the environment is a Gibbs state (thermal equilibrium state at inverse temperature $\beta$), and that the Hamiltonian of the environment remains constant $H_B$  we have
\begin{align}
    \rho_B(t_0)=\frac{e^{-\beta H_B}}{Z_B}, 
\end{align}
with $Z_B={\rm Tr}_B[e^{-\beta H_B}]$, so that
\begin{align}\label{eq:SBt0}
    S(\rho_B(t_0))=-k_b {\rm Tr}_B[\frac{e^{-\beta H_B}}{Z_B}(-\beta H_E-\log Z_B) ]=\frac{1}{T} {\rm Tr}_B[\rho_B(t_0)H_B] +k_b\log Z_B.
\end{align}

We can use now the quantum relative entropy between the state $\rho_{SB}(t)$ and the state $\rho_S(t)\otimes \rho_B(t_0)$. This a convenient way to group terms to arrive to the Clausius inequality. That is,
\begin{align}
    S(\rho_{SB}(t)\lVert \rho_S(t)\otimes \rho_B(t_0))=k_b {\rm Tr}[\rho_{SB}(t)\log \rho_{SB}(t)]-k_b {\rm Tr}[\rho_{SB}(t)\log (\rho_S(t)\otimes \rho_B(t_0)]
\end{align}
The logarithm of a product state can be split as the sum of logarithms for each part, i.e. $\log(\rho_1\otimes \rho_2)=\log \rho_1\otimes \mathbb{I}_2+\mathbb{I}_1\otimes \log \rho_2$. Therefore,
\begin{align}
    {\rm Tr}[\rho_{SB}(t)\log (\rho_S(t)\otimes \rho_B(t_0)]&={\rm Tr}[\rho_{SB}(t)\log \rho_S(t)]+{\rm Tr}[\rho_{SB}(t)\log \rho_B(t_0)]\nonumber\\&={\rm Tr}_S[\rho_S(t)\log \rho_S(t)]+{\rm Tr}_B[\rho_B(t)\log \rho_B(t_0)],
\end{align}
we have used ${\rm Tr}[ \cdot]={\rm Tr}_S[{\rm Tr}_B[\cdot]]$. The last term can be further manipulated because $\rho_B(t_0)$ is the Gibbs state, 
\begin{align}
    {\rm Tr}_B[\rho_B(t)\log \rho_B(t_0)]={\rm Tr}_B[\rho_B(t)(-\beta H_B-\log Z_B)]=-\beta {\rm Tr}[\rho_B(t)H_B]-\log Z_B.
\end{align}
Combining terms and identifying the entropies, we have
\begin{align}
    S(\rho_{SB}(t)\lVert \rho_S(t)\otimes \rho_B(t_0))=-S(\rho_{SB}(t))+S(\rho_S(t))+\frac{1}{T}{\rm Tr}_B[\rho_B(t)H_B]+k_b \log Z_B.
\end{align}
But as the total entropy remains constant,  $S(\rho_{SB}(t))=S(\rho_{SB}(t_0))=S(\rho_S(t_0))+S(\rho_B(t_0))$. Substituting the expression for the entropy of the initial state of the environment (Gibss state), cf. Eq.~\eqref{eq:SBt0}, we finally get
\begin{align}
    S(\rho_{SB}(t)\lVert \rho_S(t)\otimes \rho_B(t_0))=-S(\rho_S(t_0))-\frac{1}{T}{\rm Tr}_B[\rho_B(t_0)H_B]+S(\rho_S(t))+\frac{1}{T}{\rm Tr}_B[\rho_B(t)H_B].
\end{align}
Defining the increments, $\Delta S_S=S(\rho_S(t))-S(\rho_S(t_0)$ and $\Delta E_B={\rm Tr}_B[\rho_B(t)H_B]-{\rm Tr}_B[\rho_B(t_0)H_B]$, and because the quantum relative entropy is non-negative  $S(\rho_{SB}(t)\lVert \rho_S(t)\otimes \rho_B(t_0))\geq 0$ (Klein's inequality), we obtain
\begin{align}
    S(\rho_{SB}(t)\lVert \rho_S(t)\otimes \rho_B(t_0))=\Delta S_S +\frac{1}{T}\Delta E_B\geq 0.
\end{align}
The increment in energy of the environment happens only due to the exchange of heat ($H_B$ is time independent). Therefore, any increment in energy in $B$ is minus the heat that the system provides (recall that $Q<0$ means that heat flows out of the system). Hence,
\begin{align}\label{eq:AS_S}
    \Delta S_S-\frac{1}{T}\langle Q\rangle_S\geq 0 \Rightarrow \Delta S_{S}\geq \frac{\langle Q\rangle_S}{T}.
\end{align}
This is the Clausius inequality, i.e. second law, for a quantum system. The equality holds whenever the process is done in a quasi-static manner, so that $\rho_B(t)=\rho_B(t_0)$ at all times (environment in the same Gibbs state). In that case, the quantum relative entropy vanishes and so does $\langle Q\rangle_S=0$. 

Finally, we stress again that the second law is fulfilled in average. As commented above, this is not a unique feature of quantum systems.  If classical or quantum fluctuations are relevant, there might be a non-zero probability to violate the second law. This probability vanishes in the classical and thermodynamic limit of $N\rightarrow \infty$.

\subsubsection{Quantum data processing inequality}
As we have just seen, Eq.~\eqref{eq:AS_S} is the quantum counterpart of the second law, the Clausius inequality. This is formulated in terms of the change in entropy of a system and the heat exchanged with the environment. However, entropy in a quantum system is intimately linked with information carried by the states. In addition, as we have mentioned several times throughout this lecture notes, the time evolution or dynamics of a system is just a quantum channel that must be CPTP to ensure the physical validity of the evolved states. These two notions can be put together in the following question: How does information change under the action of a quantum channel? The answer to this question leads to an inequality, similar to the second law of thermodynamics, but only in terms of information. This relation is known as the quantum data processing inequality, and it is one of the most important results in quantum information theory. 

Suppose a CPTP quantum channel $\mathcal{E}(\rho): \mathcal{D}(\mathcal{H})\rightarrow \mathcal{D}(\mathcal{H})$. Then, it can be shown (cf. Sec.~\ref{moremat:QDPI}) that the action of $\mathcal{E}$ on two states $\rho,\sigma\in \mathcal{D}(\mathcal{H})$ cannot increase their distinguishability. That is, any physical operation on a system will decrease or keep the same degree of distinguishability among two possible initial states of the system $\rho,\sigma$: 
\begin{align}\label{eq:QDPI}
    S(\rho  \lVert \sigma)\geq S(\mathcal{E}(\rho)\lVert \mathcal{E}(\sigma)).
\end{align}
If you are interested in the details of the proof, see Sec.~\ref{moremat:QDPI}.

For unitary transformations (or reversible operations), the equality in Eq.~\eqref{eq:QDPI} holds. This is a fundamental result that may seem surprising at first, so let us examine its implications, and just as importantly, what it does not imply. Does this rule out preparation of orthogonal (perfectly distinguishable) states? No, it does not! The quantum data processing inequality is about the result of applying the very \textit{same} operation on two  states. If the operation is different, Eq.~\eqref{eq:QDPI} does not apply, and it is certainly possible to increase distinguishability under different operations. Another common issue that often arises regarding the quantum data processing inequality is how it can be reconciled with chaotic systems. As we know, two arbitrarily close initial states (say $x_A(0)$ and $x_B(0)$) evolving under the same (deterministic chaotic) dynamics will become completely far apart after a sufficiently long time. Since classical dynamics is just a special case of a quantum channel,  does Eq.~\eqref{eq:QDPI} rule this out? Again, it does not. Two classical arbitrarily close but different initial conditions $x_A$ and $x_B$ are, by definition, perfectly distinguishable in the quantum language. Therefore, $S(x_A(0)\lVert x_B(0))\geq S(x_A(t)\lVert x_B(t))$, where the strict inequality appears when there is some sort of dissipation in the system. If we allow for probability distributions of initial conditions $p_A$ and $p_B$, the same reasoning applies: For two overlapping distributions $p_A$ and $p_B$, they will continue to do so as the dynamics is deterministic (exactly equal initial condition evolve in the same way). The amount of overlap remains constant regardless of the evolution time. Recall that the operation $\mathcal{E}\{ \cdot\}$ must be exactly the same, as otherwise Eq.~\eqref{eq:QDPI} is not valid. Therefore, other characteristic features of chaotic dynamics (e.g. stochastic perturbations or non-deterministic chaotic systems) do not directly apply to the quantum data processing inequality.

Before moving on, let us consider another example but closer to open quantum system dynamics: think of a dynamical map $V(t,t_0)$ with a steady state $\rho_{ss}$. If the evolution time is long enough, the dynamical map $V\{\cdot\}$ brings any initial state into that particular steady state, $V\{\rho\}=\rho_{ss}$ and $V\{\sigma\}=\rho_{ss}$, thus erasing any possible information about the initial state. The irreversible character of the dynamical evolution is reflected in a decrease of distinguishability of quantum states under $V\{\cdot\}$.  This allows us to introduce the notion of entropy production $\Sigma$.

\subsubsection{Entropy production}
The entropy production refers to the rate at which entropy is being generated by a dynamical map. Suppose first the standard Markovian scenario: The generator of the dynamical map $V(t,t_0)\{\cdot\}$ is $\mathcal{L}\{\cdot\}$ such that $\frac{d}{dt}\rho(t)=\mathcal{L}\{\rho(t)\}$. Assume also that the dynamical map has a unique state state $\rho_{ss}$ (typically a Gibbs state at inverse temperature $\beta^{-1}$), i.e. $V(t,t_0)\{\rho_{ss}\}=\rho_{ss}$ or similarly, $\mathcal{L}\{\rho_{ss}\}=0$. We can then make use of the quantum data processing inequality (cf. Eq.~\eqref{eq:QDPI}) and find the inequality
\begin{align}
    S(\rho(t_0)\lVert \rho_{ss})\geq S( V(t,t_0)\{\rho(t_0)\}\lVert V(t,t_0)\{\rho_{ss}\})=S(\rho(t)\lVert \rho_{ss}).
\end{align}
If we take now the derivative with respect to time in the previous expression we get
\begin{align}
    \frac{d}{dt}S(\rho(t_0)\lVert \rho_{ss})\geq \frac{d}{dt}S(\rho(t)\lVert \rho_{ss}),
\end{align}
but because the left-hand side does not depend on time, we find
\begin{align}
    \frac{d}{dt}S(\rho(t)\lVert \rho_{ss})\leq 0.
\end{align}
This result is known as Spohn's inequality and it tells us that any initial state under a Markovian dynamical map gets closer (more indistinguishable) to the steady state as it evolves in time, i.e. the quantum relative entropy with respect to the steady state $\rho_{ss}$ decreases, and it does so in a monotonic manner. This is yet again a similar inequality than the second-law derived above (the Clausius inequality). 
This motivates the definition of entropy production as 
\begin{align}
    \Sigma=-\frac{d}{dt}S(\rho(t)\lVert \rho_{ss})\geq 0,
\end{align}
so that the system generates entropy at a non-negative rate at all times. This can developed a bit more to get
\begin{align}
    \Sigma=-k_b{\rm Tr}[\dot{\rho}\log \rho-\dot{\rho}-\dot{\rho}\log\rho_{ss}]=-k_b{\rm Tr}[\dot{\rho}(\log \rho-\log\rho_{ss})],
\end{align}
where we have used ${\rm Tr}[\dot{\rho}]=0$ ($\rho$ must be normalized). 
What is the physical meaning? We know that if we include the environment into consideration, entropy must remain constant. However, tracing out the environment reflects our lack of information of the total state: The system may get correlated with the environment, so that its initial information is spread into degrees of freedom that we are simply ignoring. In a Markovian evolution, information flow from the system to the environment happens at a positive rate, which is then reflected in a positive entropy production at all times.

If the steady state is not unique, the inequality still holds (replacing $\rho_{ss}$ by any of the possible steady states). However, if the dynamics fails to be Markovian, it is possible to have a negative entropy production during some time intervals, $\Sigma<0$. This was to be expected since non-Markovian dynamics takes place due to a backflow of information, i.e. information that reenters into the system from the environment, and therefore, the quantum relative entropy cannot not decrease monotonically. Hence, the Spohn's inequality does not hold under general dynamical maps.

\subsection{Quantum heat engines}
To conclude the section we are going to briefly discuss how heat engines can also be realized with quantum systems. This field has attracted considerable attention during the last decade or so, mainly due to three reasons. The primary motivation was to use these quantum heat engines as testbeds to find or inspect how the traditional laws of thermodynamics might bend under quantum mechanics \cite{Feldmann10, Torrontegui13}. The second reason consists in devising quantum processes that could circumvent the limitations of classical heat engines, or even surpass their performance constrained to the laws of classical physics \cite{Uzdin16}. The third reason goes to the heart of miniaturization: as we continue fabricating devices that are smaller and smaller in size, we may face new challenges where quantum phenomena will be relevant that need to be understood \cite{Uzdin19}.

Before we dive into the quantum setting, let us recall what is a classical heat engine and how it operates. A heat engine is a device made of a working medium that transforms heat into useful work. For that, the working medium is connected to a hot reservoir from where it takes heat. Then, the medium dumps heat into a cold reservoir by performing work on an external agent (e.g. by an expansion of a mechanical piston). These engines operate in a cyclic manner, i.e. they perform closed thermodynamical cycles involving different strokes between stages. Suppose a cycle of three stages $A$, $B$ and $C$, then the cycle consists of three strokes: $A\rightarrow B \rightarrow C\rightarrow A$. As a consequence, the change in internal energy for a full cycle is zero. 

Quantum heat engines \cite{Levy14} are not any different: they involve a working medium (now a quantum system of any sort) and cold and hot reservoirs. These reservoirs are two distinct environments to which the system can be coupled in a controllable manner, i.e. the system-environment interaction can be switched on and off. Quantum heat engines, however, can explore situations without classical analogue, such as squeezed environments and the impact of quantum coherence and/or entanglement in a working medium.

\subsubsection{Quantum Otto cycle}
Let us see how in detail the Otto cycle \cite{Geva92, Rezek17}, arguably one of the simplest thermodynamical cycles. The Otto cycle has four strokes between stages $A$, $B$, $C$ and $D$:
\begin{enumerate}
    \item Adiabatic compression stroke ($A\rightarrow B$): The working medium is disconnected from the reservoirs and evolves under a time-dependent Hamiltonian $H(t)$ in a quasi-static manner so that the energy difference between levels increases, keeping their populations constant. Since the system in this stroke is closed, no heat is exchanged, and the work done on the system reads as \begin{align}
    \langle W\rangle_{A\rightarrow B}={\rm Tr}[H(t_B)\rho(t_B)]-{\rm Tr}[H(t_A)\rho(t_A)], \quad \langle Q\rangle_{A\rightarrow B}=0.
    \end{align}
    \item Hot isochore ($B\rightarrow C$): The Hamiltonian remains constant at $H(t_B)=H(t_C)$ but the medium is connected to the hot reservoir. The system thermalizes (change in populations) to the Gibbs state at temperature $T_h$, $\rho(t_B)\rightarrow \rho(t_C)=\frac{e^{-\beta_h H(t_B)}}{Z_h}$. Since $\dot{H}=0$, no work is done on the system, and the average heat supplied to the system is given by 
    \begin{align}
        \langle W\rangle_{B\rightarrow C}=0, \quad \langle Q\rangle_{B\rightarrow C}={\rm Tr}[H(t_B)\rho(t_C)]-{\rm Tr}[H(t_B)\rho(t_B)].
    \end{align}
    \item Adiabatic expansion ($C\rightarrow D$): The system is again disconnected from the reservoirs and the Hamiltonian is changed lowering the energy difference among states. Hence,
    \begin{align}
        \langle W\rangle_{C\rightarrow D}={\rm Tr}[H(t_D)\rho(t_D)]-{\rm Tr}[H(t_C)\rho(t_C)], \quad \langle Q\rangle_{C\rightarrow D}=0.
    \end{align}
    \item Cold isochore ($D\rightarrow A$): The working medium is now connected to the cold reservoir, with a fixed Hamiltonian $H(t_D)=H(t_A)$. Thus, the state thermalizes to the Gibbs state $\rho(t_D)\rightarrow \rho(t_A)=\frac{e^{-\beta_c H(t_A)}}{Z_c}$. This brings us to
    \begin{align}
        \langle W\rangle_{D\rightarrow A}=0, \quad \langle Q\rangle_{D\rightarrow A}={\rm Tr}[H(t_A)\rho(t_A)]-{\rm Tr}[H(t_A)\rho(t_D)].
    \end{align}
\end{enumerate}
It is easy to see that $\Delta \mathcal{U}=0$ for the whole cycle. The efficiency of the Otto cycle is given by
\begin{align}
    \eta=\frac{|W_{net}|}{Q_{\rm supplied}}=1-\frac{|\langle Q\rangle_{D\rightarrow A}|}{\langle Q\rangle_{B\rightarrow C}},
\end{align}
and it is bounded by Carnot's efficiency (also in the quantum case): $\eta\leq \eta_{\rm Carnot}=1-T_C/T_H$ (see an example in~\ref{examp:Otto}). Although not explicitly shown here, note that as in the classical scenario, a slight modification can turn the heat engine into a refrigerator. The adiabatic strokes, if not performed infinitely slowly, will produce quantum excitations and coherence, reducing the efficiency, but yielding a finite power output. 
Actually, this cycle is not merely theoretical dream: The quantum Otto cycle has been implemented employing a single trapped ion as a working medium \cite{Singer}. The exploration of the interplay between quantum effects and thermodynamical quantities is currently an active area of research. 

\begin{examp}[Qubit as a working medium for Otto cycle]{examp:Otto}\label{examp:Otto}
Consider a quantum Otto cycle \cite{Feldmann02} that uses a single qubit as working medium connected to two reservoirs at $T_c$ (cold), and $T_h$ (hot). The compressions and expansion strokes change the frequency of the qubit from $\omega_s$ (small) to $\omega_l$ (large). 

Recall that after thermalizing, the qubit finds itself in the Gibbs state,
\begin{align}
    \rho=\frac{1}{Z}e^{-\beta H}=p\ket{0}\bra{0}+(1-p)\ket{1}\bra{1},
\end{align}
with $H=\omega\sigma_z/2$,  $Z=2\cosh(\beta\omega/2)$ and $p=(1+e^{\beta \omega})^{-1}$.

At the compression stroke the qubit is in the Gibbs state $\rho_c$ at $T_c$. Its frequency goes from $\omega_s$ to $\omega_l$, so that the work reads ($\rho_c$ and $\rho_h$ are cold and hot Gibbs states, while $H_{s,l}=\omega_{s,l}\sigma_z/2$ is the Hamiltonian at small or large frequencies),
\begin{align}
    \langle W\rangle_{A\rightarrow B}= {\rm Tr}[\rho_c H_l ]-{\rm Tr}[\rho_c H_s]=\frac{1}{2}(\omega_s-\omega_l)\tanh(\beta_c \omega_s/2).
\end{align}
The heat in the thermalization at $T_h$ reads
\begin{align}
    \langle Q\rangle_{B\rightarrow C}={\rm Tr}[\rho_h H_l ]-{\rm Tr}[\rho_c H_l]=\frac{1}{2}\omega_l\left[\tanh(\beta_c \omega_s/2)-\tanh(\beta_h\omega_l/2) \right],
\end{align}
the work in the expansion stroke,
\begin{align}
    \langle W\rangle_{C\rightarrow D}={\rm Tr}[\rho_h H_s]-{\rm Tr}[\rho_h H_l]=\frac{1}{2}(\omega_l-\omega_s)\tanh(\beta_h \omega_l/2),
\end{align}
and finally the heat in the cold thermalization,
\begin{align}
    \langle Q\rangle_{D\rightarrow A}={\rm Tr}[\rho_c H_s ]-{\rm Tr}[\rho_h H_s]=\frac{1}{2}\omega_s[\tanh(\beta_h \omega_l/2)-\tanh(\beta_c\omega_s/2)].
\end{align}
For this qubit to operate as a heat engine, we need $\langle W\rangle=\langle W\rangle_{A\rightarrow B} +\langle W\rangle_{C\rightarrow D} <0$ (work done by the system), so that $\beta_h\omega_l<\beta_c \omega_s$. Thus, the efficiency (work output divided by absorbed heat) reads
\begin{align}
   \eta_{\rm Otto}=\frac{|\langle W\rangle|}{\langle Q\rangle_{B\rightarrow C}}=\frac{\frac{1}{2}(\omega_l-\omega_s)[\tanh(\beta_c \omega_s/2)-\tanh(\beta_h \omega_l/2)]}{\frac{1}{2}\omega_l [\tanh(\beta_c \omega_s/2)-\tanh(\beta_h \omega_l/2)]} =1-\frac{\omega_s}{\omega_l},
\end{align}
which is clearly smaller than $1$ since $\omega_s<\omega_l$. Unit efficiency may appear to be reachable when  the small frequency vanishes, $\omega_s=0$. This, however, requires zero temperature of the cold reservoir ($\beta_c\rightarrow \infty$), such that the heat engine condition $\beta_h\omega_l<\beta_c\omega_s$ remains valid. In addition, note that $\beta_h \omega_l <\beta_c \omega_s$ implies  $\omega_s/\omega_l>\beta_h/\beta_c=T_c/T_h$, i.e.
\begin{align}
    \eta_{\rm Otto}=1-\frac{\omega_s}{\omega_l}< 1-\frac{T_c}{T_h}=\eta_{\rm Carnot},
\end{align}
with $\eta_{\rm Carnot}=1-T_c/T_h$ the Carnot efficiency. The equality $\eta_{\rm Otto}=\eta_{\rm Carnot}$ requires $\beta_h\omega_l=\beta_c\omega_s$  leading to zero work output. Note that, the Otto efficiency is achieved in the quasi-static limit, where the strokes are done infinitely slow (ensuring both a perfect thermalization, and constant populations throughout the work strokes). 
\end{examp}

\subsection{Problems to practice}

\noindent\textbf{Problem 4.1}

\noindent Consider two qubits with local Hamiltonians $H_i=\omega\sigma^z_i/2$ for $i=1,2$, that interact via 
\begin{align}
    H_I=g_z \sigma^z_1\otimes \sigma^z_2+g_x \sigma^x_1\otimes \sigma^x_2.
    \nonumber
\end{align}
Answer:
\begin{itemize}
    \item[(i)] Find the local and global equilibrium states for the qubits assuming a common temperature $\beta^{-1}$ and no interaction ($g_z=g_x=0$).
    \item[(ii)] If the interaction is relevant, find the global Gibbs state (at $\beta^{-1}$) and local equilibrium states (performing the partial trace) for $g_z=g$ and $g_x=0$ (classical case), as well as for $g_z=g_x=g$ (quantum interaction).  Discuss the differences
    \item[(iii)] Compute the quantum coherence $C(\rho)=\sum_{i\neq j}|\rho_{i,j}|$ of the global Gibbs state.
\end{itemize}
\textit{Hint:} For the second part of the exercise ($g_z=g_x=g$), you would need  to diagonalize the Hamiltonian ($4\times 4$ matrix), take its exponential and move back to the original basis. To ease the calculation, note that for $g_z=g_x=g$, it follows
\begin{align}
    e^{-\beta H}=e^{-\beta g}\begin{pmatrix} c_\Delta-\frac{\omega}{\Delta}s_\Delta & 0 & 0 & -\frac{g}{\Delta}s_\Delta\\
    0 & e^{2\beta g}c_g &-e^{2\beta g}s_g & 0 \\
    0 & -e^{2\beta g}s_g & e^{2\beta g}c_g & 0\\
    -\frac{g}{\Delta}s_\Delta & 0 & 0 & c_\Delta+\frac{\omega}{\Delta}s_\Delta
    \end{pmatrix},
    \nonumber
\end{align}
with $H=H_1+H_2+H_I$ in the computational basis $\{\ket{00},\ket{01},\ket{10},\ket{11}\}$, and the parameters $c_\Delta=\cosh(\beta\Delta), s_\Delta=\sinh(\beta\Delta), c_g=\cosh(\beta g), s_g=\sinh(\beta g)$, and $\Delta=\sqrt{g^2+\omega^2}$.

\vspace{0.5cm}
\noindent\textbf{Problem 4.2}
\vspace{0.2cm}

\noindent Show that the Helmholtz free energy $F=\mathcal{U}-TS$ for a Gibbs state $\rho_{eq}=e^{-\beta H}/Z$, can be written as 
\begin{align}
    F=-\frac{1}{\beta}\log Z.
    \nonumber
\end{align}
Here $\mathcal{U}$ denotes the internal energy, $T$ the temperature $k_b T=\beta^{-1}$, $S$ the entropy of the state $S=-k_b {\rm Tr}[\rho \log \rho]$ and $Z$ the partition function. 

\vspace{0.5cm}
\noindent\textbf{Problem 4.3}
\vspace{0.2cm}

\noindent Consider a single qubit undergoing Markovian dephasing well described by a GKSL master equation, with jump operator $\sigma_z$ and rate $\gamma> 0$. Answer the following:
\begin{itemize}
    \item[(i)] Compute the heat rate operator $\dot{Q}$ for $H_S=\omega \sigma^z_s/2$.
    \item[(ii)] Same as before but for $H_S=\omega \sigma^x_S/2$.
    \item[(iii)] If the qubit is initially in the $\ket{+}$ state and $H_S=\omega\sigma^x_S/2$, compute the total heat as $\langle Q\rangle_{0,\tau}=\int_0^\tau dt {\rm Tr}[\dot{Q}\rho_S(t)]$. 
    \item[(iv)] Compare the total heat $\langle Q\rangle_{0,\tau}$ to $\Delta \mathcal{U}={\rm Tr}[H_S \rho_S(\tau)]-{\rm Tr}[H_S \rho_S(0)]$, the internal energy variation.
    \item[(v)] Explain and discuss the results: Is heat absorbed or released by the qubit? What are the total heat and its rate for $\gamma \tau\rightarrow \infty$? 
\end{itemize}
\textit{Hint:} Find $\rho_S(t)$ from its Bloch vector decomposition under the corresponding master equation, i.e. $\rho_S(t)=\frac{1}{2}(\mathbb{I}+\vec{v}(t)\cdot \vec{\sigma})$. 

\vspace{0.5cm}
\noindent\textbf{Problem 4.4}
\vspace{0.2cm}

\noindent An isolated qubit is initially prepared in its Gibbs state for $H(0)=\omega_0 \sigma^z/2$ at temperature $\beta^{-1}$, and then driven until $H(\tau)=\omega_\tau \sigma^z/2$ by changing only $\omega$, that is, $H_S(t)\propto \sigma^z \ \forall t\in[0,\tau]$: 
\begin{itemize}
    \item[(i)] Find the work probability distribution under the two-point measurement scheme.
    \item[(ii] Verify that the Jarzynski equality is fulfilled.
\end{itemize}


\vspace{0.5cm}
\noindent\textbf{Problem 4.5}
\vspace{0.2cm}

\noindent One of the best examples to show that information and thermodynamics are deeply connected is the celebrated Landauer's principle: Erasing or forgetting information costs energy. Prove this principle for quantum information, i.e. that resetting a qubit, initially in any possible state, into a well-defined state (say $\ket{0}$) dumps at least an amount of heat $k_b T \log 2$ into the environment.

\newpage
\section{Further readings}
\label{c:fr}

\subsection{Relation between interaction picture frames}
\label{moremat:Int_Schro}

As shown in Eq.~\eqref{eq:psiI_dyn}, the state $\ket{\psi_I(t_0)}$ evolves according to $\frac{d}{dt}\ket{\psi_I(t)}=-i V_I(t)\ket{\psi_I(t)}$. Therefore, one can write
\begin{align}
    \ket{\psi_I(t)}=\mathcal{T}\left[e^{-i\int_{t_0}^tdt'V_I(t')} \right]\ket{\psi_I(t_0)}=U_I(t,t_0)\ket{\psi_I(t_0)}.
\end{align}
On the other hand, we know that
\begin{align}
    \ket{\psi(t)}=U(t,t_0)\ket{\psi(t_0)}, \qquad \ket{\psi_I(t)}=U_0^\dagger(t,t_0)\ket{\psi(t)}=U_0^\dagger(t,t_0)U(t,t_0)\ket{\psi(t_0)}.
\end{align}
Since $\ket{\psi_I(t_0)}=\ket{\psi(t_0)}$ (they coincide at initial time), the following relation must hold
\begin{align}\label{eq:Us_int}
    U_I(t,t_0)=U_0^\dagger(t,t_0)U(t,t_0).
\end{align}
The equivalence among both sides of the equation is not directly obvious. In order to verify Eq.~\eqref{eq:Us_int}, we note that from $\frac{d}{dt}\ket{\psi_I(t)}=-iV_I(t)\ket{\psi_I(t)}$ one finds
\begin{align}
    \frac{d}{dt}\left[U_I(t,t_0)\ket{\psi(t_0)}\right]=-i V_I(t)\ket{\psi_I(t)}=-iV_I(t)U_0^\dagger(t,t_0)U(t,t_0)\ket{\psi(t_0)},
\end{align}
which implies
\begin{align}\label{eq:UI_dot}
    \frac{d}{dt}U_I(t,t_0)=-i V_I(t)U_0^\dagger(t,t_0)U(t,t_0).
\end{align}
Now, if we compute $\frac{d}{dt}[U_0^\dagger(t,t_0)U(t,t_0)]$, we arrive to
\begin{align}
    \frac{d}{dt}\left[U_0^\dagger(t,t_0)U(t,t_0) \right]&=i U_0^\dagger(t,t_0)H_0(t)U(t,t_0)-iU_0^\dagger(t,t_0)(H_0+V(t))U(t,t_0)\nonumber\\&=-iU_0^\dagger(t,t_0)V(t)U(t,t_0),
\end{align}
using Eq.~\eqref{eq:ddtU},  $\frac{d}{dt}U_0(t,t_0)=-iH_0U_0(t,t_0)$, and $\frac{d}{dt}U(t,t_0)=-i(H_0+V(t))U(t,t_0)$. Making use of the definition of interaction picture for operators, $V_I(t)=U_0^\dagger(t,t_0) V(t)U_0(t,t_0)$, the previous equation can be written as
\begin{align}\label{eq:UI_dot2}
    \frac{d}{dt}\left[U_0^\dagger(t,t_0)U(t,t_0) \right]=-i V_I(t)U_0^\dagger(t,t_0)U(t,t_0).
\end{align}
Comparing Eqs.~\eqref{eq:UI_dot} and~\eqref{eq:UI_dot2}, and noting that the initial condition is the same for both differential equations, $U_I(t_0,t_0)=\mathbb{I}$ and $U_0^\dagger(t_0,t_0)U(t_0,t_0)=\mathbb{I}$, we corroborate that indeed Eq.~\eqref{eq:Us_int} holds, i.e $U_I(t,t_0)=U_0^\dagger(t,t_0)U(t,t_0)$.

\subsection{Proof of CPTP quantum channels}
\label{moremat:CPTP_Ep}
As mentioned, a quantum channel $\mathcal{E}(\rho)$ of the form given in Eq.~\eqref{eq:KrausQC} with $K_n^\dagger K_n\geq 0$ and the completeness relation in Eq.~\eqref{eq:Kraus_trace} fulfills all the three requirements to represent a physical operation. Complete positivity follows from the fact that $(\mathcal{E}\otimes \mathbb{I})(\rho)$ is also positive for any $\rho\in \mathcal{D}(\mathcal{H}_A\otimes \mathcal{H}_B)$. Let us see this more in detail. The transformed state $\rho'$ reads as 
\begin{align}
    \rho'=(\mathcal{E}\otimes \mathbb{I})(\rho)=\sum_{n} (K_n\otimes \mathbb{I})\rho(K_n^\dagger\otimes\mathbb{I}).
\end{align}
This state is positive if  $\bra{\phi}\rho'\ket{\phi}\geq 0, \forall \ket{\phi}\in\mathcal{H}\otimes \mathcal{H}$, i.e.
\begin{align}
    \bra{\phi}\sum_{n} (K_n\otimes \mathbb{I})\rho(K_n^\dagger\otimes\mathbb{I})\ket{\phi}=\sum_n \bra{\phi_n}\rho\ket{\phi_n}\geq 0.
\end{align}
Given that $\rho$ is also positive (physical state), one obtains that $\rho'\geq 0$, since it results in the sum of non-negative values.  Note that we have defined $\ket{\phi_n}\equiv (K_n^\dagger \otimes \mathbb{I})\ket{\phi}$. Hence, we conclude that $\mathcal{E}(\rho)=\sum_n K_n\rho K_n^\dagger \Rightarrow \mathcal{E}(\rho)$ is a CPTP quantum channel.

Now we face the proof for the other direction of the if and only if, that is, that any CPTP quantum channel can be written in the Kraus operator sum representation, Eq.~\eqref{eq:KrausQC}. We start assuming all three requirements for $\mathcal{E}(\rho): \mathcal{D}(\mathcal{H}_A)\rightarrow \mathcal{D}(\mathcal{H}_A)$, and the question is how to retrieve Eq.~\eqref{eq:KrausQC}. We start by considering an unnormalized maximally entangled state in $\mathcal{H}_A\otimes \mathcal{H}_B$,
\begin{align}
    \ket{\Omega}=\sum_{n=1}^{d}\ket{n}\otimes \ket{\tilde{n}},
\end{align}
where $\{\ket{n}\}$ form an orthonormal basis of $\mathcal{H}_A$, and similarly $\{\ket{\tilde{n}}\}$ for $\mathcal{H}_B$. For convenience and to make the distinction explicit, we denote $\ket{\psi}\in\mathcal{H}_A$, while $\ket{\tilde{\psi}}\in\mathcal{H}_B$. 
Then, since $\mathcal{E}(\rho)$ is CPTP, it follows that
\begin{align}
    C_\mathcal{E}=(\mathcal{E}\otimes \mathbb{I})(\ket{\Omega}\bra{\Omega})=\sum_{n,m}\mathcal{E}(\ket{n}\bra{m})\otimes (\ket{\tilde{n}}\bra{\tilde{m}})\geq 0,
\end{align}
the matrix $C_\mathcal{E}$ is positive semidefinite. This is the so-called Choi matrix. 
Now, we introduce a generic pure state for the system of interest with $\mathcal{H}_A$ (where $\mathcal{E}$ acts), and similarly for the other idle system with $\mathcal{H}_B$. That is, let $\ket{\psi}$ be a generic pure state of $\mathcal{H}_A$,
\begin{align}
    \ket{\psi}=\sum_k \alpha_k \ket{k},
\end{align}
and similarly we introduce a state for $\mathcal{H}_B$,
\begin{align}
    \ket{\tilde{\psi}}=\sum_k \alpha_k^*\ket{\tilde{k}}.
\end{align}
In this manner, we can see that
\begin{align}
    \bra{\tilde{\psi}}C_\mathcal{E} \ket{\tilde{\psi}}&=\bra{\tilde{\psi}}\left( \sum_{n,m}\mathcal{E}(\ket{n}\bra{m})\otimes (\ket{\tilde{n}}\bra{\tilde{m}})\right)\ket{\tilde{\psi}}=\sum_{n,m,k,k'} \mathcal{E}(\ket{n}\bra{m})\alpha_k \alpha_{k'}^*\langle \tilde{k}|\tilde{n}\rangle \langle \tilde{m}|\tilde{k}'\rangle\nonumber  \\& =\sum_{n,m} \alpha_n\alpha_m^* \mathcal{E}(\ket{n}\bra{m})=\mathcal{E}(\ket{\psi}\bra{\psi}),
\end{align}
where the last step follows from the linearity of the map. Since $C_\mathcal{E}$ is positive definite, it admits the spectral decomposition 
\begin{align}
    C_\mathcal{E}=\sum_{n}^r \lambda_n \ket{\psi_n}\bra{\psi_n},
\end{align}
with $r$ the rank of the Choi matrix $C_\mathcal{E}$, non-negative coefficients $\lambda_n\geq 0$ and $\ket{\psi_n}$ orthonormal vectors of $\mathcal{H}_A\otimes \mathcal{H}_B$. We can define a map on $\mathcal{H}_A$ as
\begin{align}
    K_n(\ket{\psi})=\sqrt{\lambda_n} \langle \tilde{\psi}|\psi_n\rangle,
\end{align}
so that
\begin{align}
    \sum_n K_n \ket{\psi}\bra{\psi}K_n^\dagger=\sum_{n} \lambda_n \langle \tilde{\psi}|\psi_n\rangle\langle \psi_n|\tilde{\psi}\rangle=\bra{\tilde{\psi}}C_\mathcal{E} \ket{\tilde{\psi}}=\mathcal{E}(\ket{\psi}\bra{\psi}).
\end{align}
Finally, by linearity of the map $\mathcal{E}$ we obtain the Kraus operator sum representation,
\begin{align}
    \mathcal{E}(\rho)=\sum_n K_n \rho K_n^\dagger.
\end{align}
So we conclude that $\mathcal{E}(\rho)$ is CPTP $\iff \mathcal{E}(\rho)=\sum_n^r K_n\rho K_n^\dagger$. 

We note that knowing how $\mathcal{E}\otimes \mathbb{I}$ acts on a maximally entangled state on the joint space $\mathcal{H}_A\otimes \mathcal{H}_B$ is sufficient to know how it acts on any state of the system of interest.

\subsection{Non-interacting environment}
\label{moremat:NonintB}
Let us proceed in a different way to show how one can recover the closed-system dynamics when $H_I=0$.
From Eq.~\eqref{eq:rhoS_exact} assuming $H_I=0$ we find
\begin{align}
    \rho_S(t)&={\rm Tr}_B[ U(t,t_0) \rho(t_0) U^\dagger(t,t_0)]\nonumber\\&={\rm Tr}_B[(U_S(t,t_0)\otimes U_B(t,t_0))\rho(t_0) (U_S^\dagger(t,t_0)\otimes U_B^\dagger(t,t_0) ],
\end{align}
since $H_S$ and $H_B$ act on different Hilbert spaces, they commute $[H_S,H_B]=H_S\otimes H_B- H_S\otimes H_B=0$ and therefore the time evolution operator can be written as $U(t,t_0)=U_S(t,t_0)\otimes U_B(t,t_0)$. 
Setting $t_0=0$ for convenience, and making use of the cyclic property of the trace (in this case only respect to the environment), we have
\begin{align}
{\rm Tr}_B[ (U_S(t)\otimes U_B(t))\rho(0) (U_S^\dagger(t)\otimes U_B^\dagger(t) ]&={\rm Tr}_B[(U_S(t)\otimes \mathbb{I}_B)\rho(0) (U_S^\dagger(t)\otimes \mathbb{I}_B)]\nonumber\\&=U_S(t) {\rm Tr}_B[\rho(0)]U_S^\dagger(t)=U_S(t)\rho_S(0)U_S^\dagger(t).
\end{align}
 This means that the system undergoes a purely unitary evolution,
\begin{align}
    \rho_S(t)=U_S(t,t_0)\rho_S(t_0) U_S^\dagger (t,t_0),
\end{align}
whose dynamical equation is simply that of a closed system $S$, cf. Eq.~\eqref{eq:rhoS_closed}.

\subsection{Dynamical map for closed dynamics}
\label{moremat:map_Hi0}
Given the dynamical map $\rho_S(t)=V(t,t_0)\{\rho_S(t_0)\}=\sum_{n,m}W_{n,m}(t,t_0)\rho_S(t_0)W_{n,m}^\dagger(t,t_0)$, let us show that, in the limit of non-interacting system environment, $H_I=0$, we recover the expected closed dynamics for $S$ alone. If $H_I=0$, it follows that $U(t,t_0)=U_S(t,t_0)\otimes U_B(t,t_0)$, and therefore
    \begin{align}
        W_{n,m}(t,t_0)&=\sqrt{\lambda_m}\bra{b_n}U(t,t_0)\ket{b_m}=\sqrt{\lambda_m}\bra{b_n}U_S(t,t_0)\otimes U_B(t,t_0)\ket{b_m}\nonumber\\&=U_S(t,t_0)\sqrt{\lambda_m}\bra{b_n}U_B(t,t_0)\ket{b_m}.
    \end{align}
Plugging this into the dynamical map, we obtain
    \begin{align}
        \rho_S(t)&=V(t,t_0)\{\rho_S(t_0)\}=\sum_{n,m}W_{n,m}(t,t_0)\rho_S(t_0)W_{n,m}^\dagger(t,t_0)\nonumber\nonumber\nonumber\\&=\sum_{n,m} U_S(t,t_0)\rho_S(t_0)U_S^\dagger(t,t_0) \lambda_m \bra{b_n}U_B(t,t_0)\ket{b_m}\bra{b_m}U_B^\dagger(t,t_0) \ket{b_n}\\&=U_S(t,t_0)\rho_S(t_0)U_S^\dagger(t,t_0)\sum_{n}\bra{b_n} U_B(t,t_0)\sum_m\lambda_m \ket{b_m}\bra{b_m}U_B^\dagger(t,t_0)\ket{b_n}\nonumber\\&=U_S(t,t_0)\rho_S(t_0)U_S^\dagger(t,t_0)\sum_{n}\bra{b_n}U_B(t,t_0)\rho_B(t_0)U_B^\dagger(t,t_0)\ket{b_n}\nonumber\\&=U_S(t,t_0)\rho_S(t_0)U_S^\dagger(t,t_0)\sum_{n}\bra{b_n}\rho_B(t)\ket{b_n}\nonumber\\&=U_S(t,t_0)\rho_S(t_0)U_S^\dagger(t,t_0){\rm Tr}_B[\rho_B(t)]=U_S(t,t_0)\rho_S(t_0)U_S^\dagger(t,t_0),
    \end{align}
where we have used the fact that the initial state for the environment evolves unitarily and it must be normalized at all times. 
So, indeed, we verify that the evolution of the system $S$ when it does not interact with $B$ follows the expected closed-dynamics of $S$ alone.

\subsection{Positive coefficient matrix}
\label{moremat:C_positive}
As mentioned in the main text, the coefficient matrix $c$, with elements defined in Eq.~\eqref{eq:c_ij} is positive. That is, given any complex vector $v$ of dimension $d^2$, $v\in\mathbb{C}^{d^2}$,  then $v^\dagger c v \geq 0$, i.e.
\begin{align}\label{eq:c_pos}
    \sum_{i,j}^{d^2}c_{i,j} v_i^*v_j\geq 0.
\end{align}
    In order to show this, we plug in the definition of $c_{i,j}$ and obtain
    \begin{align}
        \sum_{i,j} \sum_{n,m} (F_i,W_{n,m}(t))(F_j,W_{n,m}(t))^* v_i^* v_j=\sum_{n,m} \sum_{i,j} X_{i,n,m} X_{j,n,m}^*=\sum_{n,m}\left|\left(\sum_i X_{i,n,m} \right) \right|^2,
    \end{align}
where we define $X_{i,n,m}=v_i(F_i,W_{n,m})$. Given that the previous expression is the sum of positive numbers, $\sum_{n,m}\left|\left(\sum_i X_{i,n,m} \right) \right|^2\geq 0$, we prove that $c$ is positive, i.e. Eq.~\eqref{eq:c_pos} holds.

\subsection{Invariance of GKSL master equation}
\label{moremat:GKSL_invariance}
Let us see that the GKSL master equation is invariant under a transformation of a Lindblad operator $A_k$ into $A_k'=A_k+a_k \mathbb{I}_{d^2}$. Doing some algebra, one can see that the Lindblad operator is transformed according to
    \begin{align}
        A_k' \rho A_k'^\dagger-\frac{1}{2}\{ A_k'^\dagger A_k',\rho\}= A_k \rho A_k^\dagger-\frac{1}{2}\{ A_k^\dagger A_k,\rho\}+\frac{1}{2}[a^*A_k-aA_k^\dagger,\rho].
    \end{align}
The last term of the previous expression can be cast into a Hamiltonian form. Therefore,  if $A_k'=A_k+a \mathbb{I}_{d^2}$, then the Hamiltonian must be transformed accordingly to cancel the new extra contribution,
    \begin{align}
        H'=H-\frac{i\gamma_k}{2}(a^*A_k-aA_k^\dagger).
    \end{align}

\subsection{Hamiltonian evolution}
\label{moremat:H_vs_L}
The expression given in Eq.~\eqref{eq:rhot_vec} is similar to a pure state $\ket{\psi(t)}$ undergoing a fully-coherent (unitary)  Hamiltonian evolution. Let the Hamiltonian be $H$, then its spectral decomposition $H=\sum_k E_k \ket{\phi_k}\bra{\phi_k}$, with eigenstates fulfilling the completeness relation $\sum_k \ket{\phi_k}\bra{\phi_k}=\mathbb{I}$. This allows us to write the solution to the Schr\"odinger equation $\frac{d}{dt}\ket{\psi(t)}=-iH\ket{\psi(t)}$ as
\begin{align}
    \ket{\psi(t)}&=e^{-i(t-t_0)H}\ket{\psi(t_0)}=e^{-i(t-t_0)H} \left(\sum_{k}\ket{\phi_k}\bra{\phi_k} \right) \ket{\psi(t_0)}\nonumber\\ \label{eq:H_evol} &=\sum_k \langle \phi_k|\psi(t_0)\rangle e^{-i(t-t_0)H}\ket{\phi_k}= \sum_k c_k(t_0)e^{-i(t-t_0)E_k}\ket{\phi_k}.
\end{align}
where $c_k(t_0)=\langle \phi_k|\psi(t_0)\rangle$ are the expansion coefficients of the initial state in the eigenbasis of $H$. The main difference between Eq.~\eqref{eq:H_evol} and~\eqref{eq:rhot_vec} resides in: i) while $E_k\in\mathbb{R}$, the eigenvalues of $\mathbb{L}$  are in general complex, $\lambda_\alpha\in\mathbb{C}$, and ii) vectorized density matrices.

\subsection{Working with $A_\alpha(\omega)$}
\label{moremat:Aw_int}
The definition in Eq.~\eqref{eq:Aw_def} is very convenient as it allows us to express $A_I(t)$, i.e. the coupling terms in the interaction picture. For that, we first show that $[H_S,A_\alpha(\omega)]=-\omega A_\alpha(\omega)$, and then we use it to show the closed form in Eq.~\eqref{eq:HI_t}.   Using the definition of $A_\alpha(\omega)$, we can write
  \begin{align}
H_S A_\alpha(\omega)&=\sum_{\tilde{\varepsilon}}\tilde{\varepsilon} \ket{\tilde{\varepsilon}}\bra{\tilde{\varepsilon}} \sum_{\varepsilon'-\varepsilon=\omega} \ket{\varepsilon}\bra{\varepsilon} A_\alpha \ket{\varepsilon'}\bra{\varepsilon'} = \sum_{\varepsilon'-\varepsilon=\omega} \varepsilon \ket{\varepsilon}\bra{\varepsilon} A_\alpha \ket{\varepsilon'}\bra{\varepsilon'}.
  \end{align}
  Equivalently for $A_\alpha(\omega)H_S$,
  \begin{align}
A_\alpha(\omega)H_S&=\sum_{\varepsilon'-\varepsilon=\omega} \ket{\varepsilon}\bra{\varepsilon} A_\alpha \ket{\varepsilon'}\bra{\varepsilon'}\sum_{\tilde{\varepsilon}}\tilde{\varepsilon} \ket{\tilde{\varepsilon}}\bra{\tilde{\varepsilon}} = \sum_{\varepsilon'-\varepsilon=\omega} \varepsilon' \ket{\varepsilon}\bra{\varepsilon} A_\alpha \ket{\varepsilon'}\bra{\varepsilon'}.
\end{align}
  Combining both for the commutator, we arrive to
  \begin{align}\label{eq:HSAw}
[H_S,A_\alpha(\omega)]= \sum_{\varepsilon'-\varepsilon=\omega} (\varepsilon-\varepsilon') \ket{\varepsilon}\bra{\varepsilon} A_\alpha \ket{\varepsilon'}\bra{\varepsilon'}=-\omega \sum_{\varepsilon'-\varepsilon=\omega}\ket{\varepsilon}\bra{\varepsilon} A_\alpha \ket{\varepsilon'}\bra{\varepsilon'} =-\omega A_\alpha(\omega).
  \end{align}
  Now, we use this commutator to obtain the closed form in the interaction picture. From the Baker-Campbell-Hausdorff formula,
  \begin{align}
e^{i H_S t} A_\alpha(\omega) e^{-i H_S t} =  \sum_{n=0} \frac{(it)^n}{n!}[ H_S,A_\alpha(\omega)]_{n},
    \end{align}
  where $[X,Y]_n=[X_,[X,Y]_{n-1}]$ denotes the $n$th order nested commutator, with $[X,Y]_0=Y$, we obtain (using Eq.~\eqref{eq:HSAw})
  \begin{align}
      e^{i H_S t} A_\alpha(\omega) e^{-i H_S t}=\sum_{n=0} \frac{(it)^n}{n!} (-\omega)^n A_\alpha(\omega)=e^{-i\omega t} A_\alpha(\omega).
  \end{align}

\subsection{Splitting Fourier transforms}
\label{moremat:Gammas}
In order to understand why the Fourier transform of the environment correlation functions can be split as given in Eqs.~\eqref{eq:Sab} and~\eqref{eq:gab}, we first need to see what happens to $C_{\alpha,\beta}(s)$ upon conjugation, i.e.
    \begin{align}
        C_{\alpha,\beta}^*(s)=\left({\rm Tr}_B[B^\dagger_\alpha(s)B_\beta(0)\rho_B]\right)^\dagger={\rm Tr}_B[\rho_B B_\beta^\dagger(0) B_\alpha(s)]={\rm Tr}_B[B_\beta^\dagger(0)B_\alpha(s)\rho_B]=C_{\beta,\alpha}(-s).
    \end{align}
The last step is again a consequence of the stationary state $\rho_B$. In this manner, if we take the conjugate of $\Gamma_{\alpha,\beta}(\omega)$ we get
    \begin{align}
        \Gamma_{\alpha,\beta}^*(\omega)=\int_0^\infty ds e^{-i\omega s} C_{\alpha,\beta}^*(s)=\int_{0}^\infty ds e^{-i\omega s}C_{\beta,\alpha}(-s)=\int_{-\infty}^0ds e^{i\omega s}C_{\beta,\alpha}(s).
    \end{align}
Therefore, 
    \begin{align}
        \Gamma_{\alpha,\beta}(\omega)+\Gamma_{\beta,\alpha}^*(\omega)&=\int_{-\infty}^{\infty}ds e^{i\omega s} C_{\alpha,\beta}(s)\equiv \gamma_{\alpha,\beta}(\omega)\\
        \Gamma_{\alpha,\beta}(\omega)-\Gamma_{\beta,\alpha}^*(\omega)&=\int_{0}^\infty ds e^{i\omega s} C_{\alpha,\beta}(s)-\int_{-\infty}^0ds e^{i\omega s} C_{\alpha,\beta}(s)\equiv 2i S_{\alpha,\beta}(\omega),
    \end{align}
which account for the real and imaginary parts of the Fourier transform. Now, it is important to note that $\gamma_{\alpha,\beta}^*(\omega)=\gamma_{\beta,\alpha}(\omega)$ and $S_{\alpha,\beta}^*(\omega)=S_{\beta,\alpha}(\omega)$. This can be easily seen from the previous definitions,
    \begin{align}
        \gamma_{\alpha,\beta}^*(\omega)&=\int_{-\infty}^{\infty}ds e^{-i\omega s}C_{\alpha,\beta}^*(s)=\int_{-\infty}^{\infty}ds e^{-i\omega s}C_{\beta,\alpha}(-s)\nonumber\\&=\int_{-\infty}^{\infty}ds e^{i\omega s}C_{\beta,\alpha}(s)=\gamma_{\beta,\alpha}(\omega),\label{hermi}\\
        S_{\alpha,\beta}^*(\omega)&=\frac{-1}{2i}\left(\int_{0}^{\infty}ds e^{-i\omega s}C_{\alpha,\beta}^*(s)-\int_{-\infty}^0 ds e^{-i\omega s}C_{\alpha,\beta}^*(s) \right)\nonumber\\&=\frac{-1}{2i}\left(\int_{-\infty}^0ds e^{i\omega s}C_{\beta,\alpha}(s)-\int_{0}^{\infty}ds e^{i\omega s}C_{\beta,\alpha}(s) \right)=S_{\beta,\alpha}(\omega).
    \end{align}

\subsection{Positive semi-definite Hermitian $\gamma_{\alpha,\beta}(\omega)$}
\label{moremat:gab_pos}
The Hermitian property was already shown in Eq.~\eqref{hermi}, $\gamma_{\alpha,\beta}^*(\omega)=\gamma_{\beta,\alpha}(\omega)$. For the non-negativity suppose a generic observable of the environment $X_v(t)=\sum_\alpha v_\alpha B_\alpha(t)$ where $v_\alpha$ is a complex vector. The expectation value $\langle \int dt dt' f^*(t) f(t') X_v^\dagger (t)X_v(t')\rangle$ for any (square integrable) function $f(t)$ must be non-negative 
\begin{align}
    \int dt dt' \langle f^*(t) f(t') X_v^\dagger(t)X_v(t')\rangle \geq 0.
\end{align}
Using the stationary condition, $C_v(t-t')=\langle X_v^\dagger(t)X_v(t')\rangle$, and $f(t)=e^{i\omega t}$, we arrive to
\begin{align}\label{eq:corr_pos}
    \int dt dt' e^{-i\omega t'}e^{i\omega t}C(t-t')\propto \int d\tau e^{i\omega \tau} C_v(\tau)\geq 0.
\end{align}
Using the definition of $\gamma_{\alpha,\beta}(\omega)=\int_{-\infty}^{\infty}ds e^{i\omega s}C_{\alpha,\beta}(s)$, we can write
\begin{align}
     \int d\tau e^{i\omega \tau} C_v(\tau)=\int d\tau e^{i\omega \tau}\sum_{\alpha,\beta}v_\alpha^* v_\beta \langle B_\alpha^\dagger(s)B_\beta(0)\rangle=\sum_{\alpha,\beta} v_\alpha^* \gamma_{\alpha,\beta}(\omega)v_\beta.
\end{align}
Since Eq.~\eqref{eq:corr_pos} is non-negative, then it follows that $\sum_{\alpha,\beta} v_\alpha^* \gamma_{\alpha,\beta}(\omega)v_\beta\geq 0$, which is precisely the condition for $\gamma_{\alpha,\beta}(\omega)$ to be non-negative. Given that $\gamma_{\alpha,\beta}(\omega)$ is Hermitian and non-negative, its eigenvalues are non-negative and real, hereby demonstrating that Eq.~\eqref{eq:rhoI_9} is indeed a Markov GKSL master equation, which can be cast in diagonal form upon diagonalizing $\gamma_{\alpha,\beta}(\omega)$.

\subsection{Principal values and Lamb shift $H_{\rm LS}$}
\label{moremat:lamb}
The imaginary parts of the non-zero one-sided Fourier transforms of the two-time correlation functions of the environment are given in Eqs.~\eqref{eq:S11} and~\eqref{eq:S22}. Coming back to the master equation, the Lamb shift appears as
\begin{align}
    H_{\rm LS}=\sum_\omega \sum_{\alpha,\beta}S_{\alpha,\beta}(\omega)A_\alpha^\dagger (\omega) A_\beta(\omega).
\end{align}
In our case, we defined $A_1=a^\dagger$, and $A_2=a$, so that
\begin{align}
    H_{\rm LS}=S_{1,1}(\omega_0)a a^\dagger +S_{2,2}(\omega_0)a^\dagger a=S_{1,1}(\omega_0) \mathbb{I}+(S_{1,1}(\omega_0)+S_{2,2}(\omega_0))a^\dagger a.
\end{align}
The term $S_{1,1}(\omega_0)\mathbb{I}$ can be ignored (constant energy shift). However, $S_{1,1}(\omega_0)+S_{2,2}(\omega_0)$ modifies the frequency of the oscillator. Using Eqs.~\eqref{eq:S11} and~\eqref{eq:S22} we find
\begin{align}
    \Delta=S_{1,1}(\omega_0)+S_{2,2}(\omega_0)={\rm P.V.}\left(\sum_k |g_k|^2\frac{1}{\omega_0-\nu_k} \right).
\end{align}
The effect of this renormalization is best seen when dealing with a continuum of modes, characterized by a spectral density $J(\omega)$. This consists in replacing the sum over discrete modes into an integral over all frequencies $\omega$ of the environment,
\begin{align}
    \Delta={\rm P.V.} \int_0^\infty d\omega \frac{J(\omega)}{(\omega_0-\omega)}.
\end{align}
The actual frequency of the oscillator (called the dressed frequency due to the interactions, in contrast to the bare frequency $\omega_0$ when isolated) is then $\Delta+\omega_0$, i.e. $H_{S}+H_{\rm LS}=(\Delta +\omega_0)a^\dagger a$. Note that $\Delta$ only depends on the coupling terms and not on the specific state of the environment $\rho_B$ (it is independent of temperature).

\subsection{Stationary solution}
\label{moremat:steadystate_ho}

Suppose the Lindblad master equation given in Eq.~\eqref{eq:MasterEq_HO}. If we move to the interaction picture again, we simply have the dissipative part, with terms $\gamma_{\uparrow},\gamma_{\downarrow}$. We can project onto the diagonal terms, and find the populations for each of the Fock states, $P(n)\equiv \bra{n}\rho\ket{n}$. The master equation for the populations reads as (denoting $\overline{n}\equiv \overline{n}(\omega_0)$ as the average number of excitations in a thermal state at temperature $T$ and frequency $\omega_0$)
\begin{align}
    \frac{d}{dt}P(n)=\gamma_\downarrow\left[(n+1)P(n+1)-n P(n) \right]+\gamma_\uparrow \left[n P(n-1)-n(n+1)P(n) \right].
\end{align}
If we seek for the stationary state, then $\frac{d}{dt}P_s(n)\equiv 0, \forall n\geq 0$. The stationary solution follows then from the solution to
\begin{align}\label{eq:Psn}
    2\pi J(\omega_0) \left\{(\overline{n}+1)\left[(n+1)P_s(n+1)-n P_s(n) \right]+ \overline{n} \left[n P_s(n-1)-(n+1)P_s(n) \right]\right\}=0.
\end{align}
For that, one can use a generating function approach (more general), or simply propose a guess, i.e. an Ansatz of the form
\begin{align}
P_s(n)= C r^n.     
\end{align}
Inserting this into Eq.~\eqref{eq:Psn}, and simplifying the $2\pi J(\omega_0)$ factor, we find
\begin{align}
    (\overline{n}+1)\left[  (n+1) C r^{n+1}-n C r^n \right]+\overline{n} \left[n C r^{n-1}-(n+1)C r^n \right]=0.
\end{align}
Taking $C r^{n-1}$ a common factor, we arrive to
\begin{align}
    (\overline{n}+1)\left[ (n+1)r^2 -n r\right]+\overline{n}\left[ n-(n+1)r \right]=0.
\end{align}
Solving for $r$, we find two solutions, either $r=n/(n+1)$ or $r=\overline{n}/(\overline{n}+1)$. Since in our Ansatz $r$ does not depend on $n$, we obtain 
\begin{align}
    r=\frac{\overline{n}}{\overline{n}+1} \Rightarrow P_s(n)=C \left(\frac{\overline{n}}{\overline{n}+1}\right)^n.
\end{align}
The constant $C$ must ensure that the probabilities add up to one. Since $\sum_{n=0}^\infty C r^n=\frac{C}{1-r}$ for $|r|<1$, we identify the constant $C$ as
\begin{align}
    \sum_n P_s(n)=1 \Rightarrow \frac{C}{1-r}=1 \Rightarrow C=\frac{1}{\overline{n}+1},
\end{align}
and finally the stationary or steady populations are 
\begin{align}
    P_s(n)=\frac{1}{\overline{n}+1}\left(\frac{\overline{n}}{\overline{n}+1} \right)^n.
\end{align}
This is precisely the thermal or Gibbs state at temperature $T$, i.e. 
\begin{align}
    \rho_{s}\equiv \rho_{T}=\frac{e^{-H_S/(k_b T)}}{{\rm Tr}[e^{-H_S/(k_b T)}]}=\frac{1}{\overline{n}(\omega_0)+1}\sum_{n=0}\left(\frac{\overline{n}(\omega_0)}{\overline{n}(\omega_0)+1} \right)^n\ket{n}\bra{n}.
\end{align}
Therefore, we have verified that a harmonic oscillator interacting with a collection of bosonic modes at temperature $T$ relaxes to a thermal state at the same temperature.

\subsection{Exact solution to spin-boson model}
\label{moremat:exact_SB}
The spin-boson model with longitudinal coupling as considered here is exactly solvable. That is, the dynamics for the spin or qubit can be obtained under no approximation, besides the standard initial product state $\rho(0)=\rho_S(0)\otimes \rho_B$. Since our final goal is to compare the dynamics of the GKSL master equation (cf. Eq.~\eqref{eq:MasterEq_SB}) (under the Born-Markov and rotating-wave approximations) with the exact one, we consider again $\rho_B$ to be a thermal equilibrium state of $H_{B}$ at some temperature $T$. 

We start by computing the interaction term in the rotating frame with the local energy terms, $H_S+H_B$, i.e. $H_I(t)=e^{i t(H_S+H_B)}H_I e^{-it(H_S+H_B)}$, 
\begin{align}
    H_I(t)=\sigma_z \sum_k (g_k e^{-i\nu_k t}b_k+g_k^* e^{i\nu_k t}b_k^\dagger).
\end{align}
The commutator of $H_I(t)$ with $H_I(t')$ reads as
\begin{align}
    [H_I(t),H_I(t')]=&\mathbb{I}_S\otimes \sum_{k,k'}(g_k e^{-i\nu_k t}b_k+g_k^* e^{i\nu_k t}b_k^\dagger)(g_{k'} e^{-i\nu_{k'} t'}b_{k'}+g_{k'}^* e^{i\nu_{k'} t'}b_{k'}^\dagger)\nonumber\\-&\mathbb{I}_S\otimes\sum_{k,k'}(g_{k'} e^{-i\nu_{k'} t'}b_{k'}+g_{k'}^* e^{i\nu_{k'} t'}b_{k'}^\dagger)(g_k e^{-i\nu_k t}b_k+g_k^* e^{i\nu_k t}b_k^\dagger)\nonumber\\
    &=-2i \mathbb{I}_S\sum_k |g_k|^2\sin(\nu_k (t-t')).
\end{align}
This means that the commutator $[H_I(t),H_I(t')]$ commutes with $H_I(t'')$, so we can use this to write down the time-evolution operator
\begin{align}
    U_I(t)=\mathcal{T} e^{-\int_0^t ds' H_I(s)}=e^{\frac{1}{2}\int_0^tds_1 \int_0^s1 ds_2[H_I(s_1),H_I(s_2)]}e^{-i\int_0^t ds H_I(s)}.
\end{align}
Since $[H_I(t),H_I(t')]$ is just proportional to the identity, it provides a global phase to the qubit so it can be discarded. The dynamics is governed only by the second term,
\begin{align}
e^{-i\int_0^t ds H_I(s)}=e^{-i\int_0^t ds \sigma_z\otimes \sum_k g_k e^{-i\nu_k s}b_k +g_k^*e^{i\nu_k s}b_k^\dagger}=e^{\sigma_z\otimes Y(t)},    
\end{align}
    where the operator acting on the environment read as
    \begin{align}
        Y(t)=\sum_k \left(\alpha_k^*(t)b_k^\dagger - \alpha_k(t)b_k\right),
    \end{align}
    with $\alpha_k^*(t)=g_k^*(\frac{1-e^{i\nu_k t}}{\nu_k})$, so that $Y^\dagger(t)=-Y(t)$. 
The time-evolution operator can be further expanded (discarding already the global phase)
\begin{align}
    U_I(t)&=e^{\sigma_z \otimes Y(t)}=\sum_{n=0}^\infty \frac{(\sigma_z\otimes Y(t))^n}{n!}=\mathbb{I}_S \otimes\sum_{n=0}^\infty\frac{Y^{2n}(t)}{2n!}+\sigma_z\otimes \sum_{n=0}\frac{Y^{2n+1}(t)}{(2n+1)!}\nonumber\\
    &=\mathbb{I}_S\otimes \cosh(Y(t))+\sigma_z\otimes \sinh(Y(t)),
\end{align}
and similarly, $U_I^\dagger(t)=\mathbb{I}_S\otimes \cosh(Y(t))-\sigma_z\otimes \sinh(Y(t))$. 
Therefore, the state of the qubit at any time $t$ (in the interaction picture) follows from taking the partial trace over the environment $B$ of the evolved one $\rho(t)=U_I(t)\rho_S(0)\otimes \rho_BU_I^\dagger(t)$, 
\begin{align}
    \rho_{S}(t)={\rm Tr}_B[U_I(t)\rho_S(0)\otimes \rho_BU_I^\dagger(t)].
\end{align}
If we expand it, we get
\begin{align}
    \rho_S(t)=&\rho_S(0){\rm Tr}_B[\cosh^2(Y(t))\rho_B]-\rho_S(0)\sigma_z{\rm Tr}_B[\sinh(Y(t))\cosh(Y(t))\rho_B]\nonumber\\&+\sigma_z\rho_S(0){\rm Tr}_B[\cosh(Y(t))\sinh(Y(t))\rho_B]-\sigma_z\rho_S(0)\sigma_z{\rm Tr}_B[\sinh^2(Y(t))\rho_B].
\end{align}
Employing the relations for the hyperbolic cosine and sine, we obtain
\begin{align}
    \rho_S(t)=\begin{pmatrix}\rho_{00}(0) & \rho_{01}(0){\rm Tr}_B[e^{2Y(t)}\rho_B]\\
    \rho_{10}(0){\rm Tr}_B[e^{-2Y(t)}\rho_B] & \rho_{11}(0)
    \end{pmatrix}.
\end{align}
Note that $(\rho_{01}(0){\rm Tr}_B[e^{2Y(t)}\rho_B])^\dagger=\rho_{01}^*(0){\rm Tr}_B[e^{-2Y(t)}\rho_B]$ which shows that $\rho_{S}(t)$ is indeed Hermitian. The diagonal terms do not change (this corresponds to the interaction picture), while the off-diagonal are multiplied by a time-dependent factor. This was to be expected since the considered spin-boson model only produces pure dephasing noise. In the GKSL master equation, the off-diagonal terms are contracted by a factor $e^{-2\gamma t}$ (cf. Eq.~\eqref{eq:gamma_deph}). Let us compute the specific form of the exact dephasing term. Since $\rho_B$ is a thermal equilibrium state, only the diagonal terms $b_k^\dagger b_k$ provide a non-zero expectation value, i.e. ${\rm Tr}_B[Y(t)\rho_B]=0$.  In particular, it can be shown that in this case
\begin{align}
    \langle e^{2Y(t)}\rangle =e^{2\langle Y^2(t)\rangle},
\end{align}
where $\langle \dots\rangle$ denotes the expectation value over $\rho_B$. The expectation value $\langle Y^2(t)\rangle$ reads as
\begin{align}
    \langle Y^2(t)\rangle&\equiv {\rm Tr}_B[\sum_{k,k'}(\alpha_k^*(t)b_k^\dagger-\alpha_k(t)b_k)(\alpha_{k'}^*(t)b_{k'}^\dagger-\alpha_{k'}(t)b_k)\rho_B]\nonumber\\&=-\sum_{k}|\alpha_k(t)|^2(2\overline{n}_k(\nu_k)+1)=-\sum_k |\alpha_k(t)|^2 \coth(\nu_k/(2k_b T))
\end{align}
Using the spectral density $J(\omega)=\sum_{k}|g_k|^2\delta(\omega-\nu_k)$, and $\alpha_k(t)$, we can express the result as
\begin{align}
    \langle Y^2(t)\rangle=-2\int_0^\infty d\omega J(\omega)\coth\left(\frac{\omega}{2k_b T}\right)\frac{1-\cos(\omega t)}{\omega^2}.
\end{align}
As a consequence, the off-diagonal terms decay in time according to a dephasing rate $\Gamma(t)$, i.e. $\rho_{01}(t)=\rho_{01}(0)e^{-\Gamma(t)}$,
\begin{align}\label{eq:deph_SB_exact}
    \Gamma(t)=4\int_0^\infty d\omega J(\omega) \coth\left(\frac{\omega}{2k_b T}\right)\frac{1-\cos(\omega t)}{\omega^2}.
\end{align}
At initial time, $t=0$, we find $\Gamma(0)=0$ consistent with the initial condition. In general, $\Gamma(t)$ may exhibit a complicated time dependence due to $J(\omega)$, not necessarily linear on $t$ as in the GKSL master equation. See Fig.~\ref{fig:SB_comp} for a comparison of the distinct behavior of the pure dephasing noise resulting from a spin-boson model.

\subsection{Laplace transform for integro-differential equation}
\label{moremat:Laplace_atom}
The structure of Eq.~\eqref{eq:qtilde_2} recalls the Laplace transform, which is defined as 
    \begin{align}
        X(s)\equiv L\{x(t)\}=\int_0^\infty dt e^{-s t}x(t),
    \end{align}
for $s\in\mathbb{C}$. Indeed, the Laplace transform is often a key tool for solving integro-differential equations.  One of the key properties of the Laplace transform is that $L\{\dot{x}(t)\}=sX(s)-x(0)$. Applying it to Eq.~\eqref{eq:qtilde_2}, we have
    \begin{align}
        L\{\dot{\tilde{q}}(t)\}=s Q(s)-\tilde{q}(0)=sQ(s)-1,
    \end{align}
where we have used already the initial condition, $\tilde{q}(0)=1$. To continue, we must transform the right hand side of Eq.~\eqref{eq:qtilde_2}.  
For that, we recall another useful property of the Laplace transform, i.e. for convolutions we get the multiplication of the individual Laplace-transformed functions, 
    \begin{align}
        L\left\{ \int_0^t d\tau g_1(t)g_2(t-\tau)\right\}=G_1(s)G_2(s).
    \end{align}
Therefore, \begin{align}
        L\left\{-\int_0^t dt' f(t-t')\tilde{q}(t')\right\}=-F(s) Q(s),
    \end{align}
    where the Laplace transformed of the memory kernel reads as
    \begin{align}
        F(s)=L\{f(t)\}=\int_0^\infty dt \ e^{-s t}f(t)= \int_0^\infty dt \ e^{-s t}\frac{\gamma \lambda}{2}e^{-\lambda t}=\frac{\gamma\lambda}{2(\lambda+s)}.
    \end{align}
    In this manner, the integro-differential equation can now be solved algebraically for $Q(s)$, 
    \begin{align}
        sQ(s)-1=-F(s)Q(s)\Rightarrow Q(s)=\frac{1}{s+F(s)}=\frac{1}{s+\frac{\gamma \lambda}{2(\lambda+s)}}=\frac{\lambda+s}{s^2+\lambda s+\gamma \lambda/2}. 
    \end{align}
    The task now consists in reversing the Laplace transform to find $\tilde{q}(t)$. The previous equation for $Q(s)$ can be expressed as a sum of terms proportional to $1/(s+\alpha)$, whose inverse Laplace transform gives exponential functions, i.e. $L^{-1}\{\beta/(s+\alpha)\}=\beta e^{-\alpha t}$. Doing this last step, we arrive to Eq.~\eqref{eq:qtilde_sol}.

 \subsection{Modified Jarzynski for generic dynamical maps}
 \label{moremat:MJE}
In the derivation of the Jarzynski equality there is a key step: The conditional probability $p_{m|n}^\tau$ denotes the probability that the collapsed state $\ket{n}$ followed by its dynamics gives $\ket{m}$ in the second measurement. In general, this can be written as 
\begin{align}
    p_{m|n}^\tau=\bra{m}V(\tau,0)\{ \ket{n}\bra{n}\}\ket{m},
\end{align}
where $V(\tau,0)\{ \cdot\}$ is the dynamical map from $t=0$ to $\tau$. Plugging this into the calculation of $\langle e^{-\beta W}\rangle$ we obtain
\begin{align}
    \langle e^{-\beta W}\rangle&=\frac{1}{Z_0}\sum_{n,m}\bra{m}V(\tau,0)\{ \ket{n}\bra{n}\}\ket{m}e^{-\beta E_m^\tau}=\frac{1}{Z_0}\sum_m \bra{m}V(\tau,0)\{\sum_n \ket{n}\bra{n}\}\ket{m}e^{-\beta E_m^\tau}\nonumber\\&=\frac{1}{Z_0}\sum_m \bra{m}V(\tau,0)\{\mathbb{I}\}\ket{m}e^{-\beta E_m^\tau}.
\end{align}
This shows that, unless $\sum_m \bra{m}V(\tau,0)\{ \mathbb{I}\}\ket{m}=1$, the Jarzynski equality must be modified. The previous condition happens, as we know, if the map is unitary. In addition, if the map is unital, $V(\tau,0)\{\mathbb{I}\}=\mathbb{I}$, Eq.~\eqref{eq:JE} also holds. This can be further manipulated to write
\begin{align}
    \langle e^{-\beta W}\rangle=\frac{Z_\tau}{Z_0}{\rm Tr}[\rho_{eq;\tau}V\{\mathbb{I}\}].
\end{align}
Since $M={\rm Tr}[\rho_{eq;\tau}V\{\mathbb{I}\}]$ is just a positive number $M>0$, we have
\begin{align}
    \langle e^{-\beta W}\rangle = M e^{-\beta \Delta F}.
\end{align}
The left-hand side still refers to a purely dynamical process; yet, the right-hand side now contains the previous free energy difference and a sort of measure of how close the evolved identity overlaps with the final Gibbs state (explicitly dynamical).

\subsection{Proof of quantum data processing inequality}
\label{moremat:QDPI}
The proof of Eq.~\eqref{eq:QDPI} requires two steps: Monotonicity under partial trace and the fact that any map on a system $A$ can be written as a unitary transformation on an extended space, $A+B$, where $B$ is just an auxiliary system (this is just the Stinespring dilation). 

For the first step, we note that the quantum relative entropy can be written in a variational form as
\begin{align}
    S(\rho\lVert \sigma)=\sup_{X=X^\dagger}\left\{{\rm Tr}[\rho X]-\log {\rm Tr}[e^{X}\sigma] \right\},
\end{align}
    which is attained for $X=\log \rho -\log \sigma$. In that case, one falls back to Eq.~\eqref{eq:Srel}. If we consider now an extended system $A+B$, such that $\rho_A={\rm Tr}_B[\rho_{AB}]$ and $\sigma_A={\rm Tr}_B[\sigma_{AB}]$, the previous quantum relative entropy $S(\rho_{AB}\lVert \sigma_{AB})$ will be, in general, larger than $S(\rho_A\lVert \sigma_A)$ since the supremum for the later is restricted to operators $X=X_A\otimes \mathbb{I}_B$, while for the former it includes any $X\in\mathcal{H}_A\otimes \mathcal{H}_B$. In this manner, 
    \begin{align}\label{eq:QDPI_a}
        S(\rho_{AB}\lVert \sigma_{AB})\geq S(\rho_A\lVert \sigma_A).
    \end{align}
    That is, partial trace can only keep or decreases the degree of distinguishability among two states. 

Now, we use the fact that any map $\mathcal{E}\{\cdot\}$ on the system $A$ can be written as the result of a unitary transformation on an extended system $A+B$ after tracing $B$ (Stinespring dilation):
\begin{align}
    \mathcal{E}(\rho_A)={\rm Tr}_{B}\left[U_{AB}(\rho_A\otimes \nu_B)U_{AB}^\dagger \right],
\end{align}
and similarly for $\sigma_A$, 
\begin{align}
    \mathcal{E}(\sigma_A)={\rm Tr}_{B}\left[U_{AB}(\sigma_A\otimes \nu_B)U_{AB}^\dagger \right],
\end{align}
being $\nu_B$ just a fixed state of the auxiliary system $B$. Defining, $\tilde{\rho}=\rho_A\otimes \nu_B$ and $\tilde{\sigma}=\sigma_A\otimes \nu_B$, and using the fact that a unitary transformation on the states does not change their distinguishability (quantum relative entropy remains the same), we get
\begin{align}\label{eq:QDPI_b}
    S(\tilde{\rho}\lVert \tilde{\sigma})=S(\rho_A\otimes \nu_B \lVert \sigma_A\otimes \nu_B)=S(\rho_A\lVert \sigma_A),
\end{align}
where the last step can be shown as follows
\begin{align}
    &{\rm Tr}[(\rho_A\otimes \nu_B)\log(\rho_A\otimes \nu_B)-(\rho_A\otimes \nu_B)\log(\sigma_A\otimes \nu_B)]\nonumber\\&={\rm Tr}[(\rho_A\otimes \nu_B)(\log \rho_A-\log \nu_B)-(\rho_A\otimes \nu_B)(\log\sigma_A-\log\nu_B)]\nonumber\\
    &={\rm Tr}[(\rho_A\log \rho_A-\rho_A\log \sigma_A)\otimes \nu_B]={\rm Tr}_A[(\rho_A\log \rho_A-\rho_A\log \sigma_A)]{\rm Tr}_B[\nu_B]\nonumber\\&={\rm Tr}_A[(\rho_A\log \rho_A-\rho_A\log \sigma_A)]=S(\rho_A\lVert \sigma_A).
\end{align}
As expected, adding the same state of $B$ as a product to both $\rho_A$ and $\sigma_A$ does not change the distinguishability between $\rho_A$ and $\sigma_A$. We can combine now the results in Eq.~\eqref{eq:QDPI_a} and~\eqref{eq:QDPI_b}. That is, taking the trace over $B$ to $\tilde{\rho}$ and $\tilde{\sigma}$ (which is just the result of the quantum channel $\mathcal{E}\{\cdot\}$), we finally find the quantum data processing inequality:
\begin{align}
    S(\rho_A\lVert \sigma_A)=S(\tilde{\rho}\lVert\tilde{\sigma})\geq S({\rm Tr}_B[\tilde{\rho}]\lVert {\rm Tr}_B[\tilde{\sigma}])=S(\mathcal{E}(\rho_A)\lVert \mathcal{E}(\sigma_A)). 
\end{align}

\section{Conclusion}
The theory of open quantum systems occupies a central position in modern quantum physics. While the idealized framework of closed quantum systems remains indispensable for understanding the mathematical foundations of quantum mechanics, realistic quantum systems are inevitably coupled to external degrees of freedom. The interaction between a system and its environment gives rise to decoherence, dissipation, fluctuations, irreversibility, and thermalization, phenomena that cannot be captured within purely unitary dynamics of just the system of interest.

These lecture notes began by revisiting the formalism of closed quantum systems, including the Schr\"odinger, Heisenberg, and interaction pictures, together with the density matrix framework and the theory of quantum operations. These tools established the language required to describe subsystems and mixed states, thereby motivating the transition from isolated dynamics to reduced dynamics. The introduction of partial traces and quantum channels highlighted that the loss of information about environmental degrees of freedom naturally leads to non-unitary evolution.

The central objective of the notes was then to develop the theory of open quantum systems through the framework of quantum master equations. In particular, the Gorini–Kossakowski –Sudarshan–Lindblad (GKSL) equation emerged as the fundamental mathematical structure describing Markovian quantum dynamics under physically consistent assumptions. The Lindblad formalism provides a bridge between microscopic interactions and effective irreversible dynamics, while preserving complete positivity and trace preservation. Beyond its practical utility, it also clarifies the deep relation between quantum mechanics and statistical irreversibility.

The Liouvillian superoperator approach further demonstrated that dissipative dynamics possess a rich spectral structure. Concepts such as steady states, relaxation modes, spectral gaps, and Jordan decompositions reveal that dissipation is not merely a source of noise, but can itself become a resource for controlling and engineering quantum states. In contemporary quantum technologies, dissipation engineering has become an essential paradigm in quantum state preparation for quantum simulation, quantum computing, quantum sensing, and quantum communications.

The microscopic derivation of master equations illustrated that open-system dynamics are not fundamental postulates but emerge from approximations applied to larger closed systems. The Born-Markov and secular approximations provided physically transparent routes to effective reduced dynamics, while the discussion of their breakdown emphasized the limitations of Markovian descriptions. This naturally points toward ongoing developments in non-Markovian quantum dynamics, strong coupling regimes, structured reservoirs, and memory effects, active areas of current research.

The examples studied throughout the notes, including the damped harmonic oscillator, the spin-boson model, and cavity quantum electrodynamics, illustrated how open-system methods apply across atomic physics, condensed matter physics, quantum optics, and quantum information science. These models continue to serve as testing grounds for both analytical methods and emerging numerical techniques.

The final part of the lecture notes connected open quantum systems to thermodynamics. Quantum thermodynamics extends classical thermodynamic principles into regimes where coherence, entanglement, and quantum fluctuations become operationally relevant. The notions of quantum heat, work, entropy production, and fluctuation relations demonstrate that thermodynamics is fundamentally compatible with quantum mechanics, while simultaneously enriched by uniquely quantum phenomena. Quantum heat engines and nonequilibrium processes further stress how information-theoretic and thermodynamic concepts become deeply intertwined at the quantum scale.

Open quantum systems therefore stand at the intersection of multiple disciplines: quantum information, statistical mechanics, condensed matter physics, thermodynamics, and quantum technologies. As experimental quantum platforms continue to improve, the accurate description and control of environmental effects become increasingly important. In many contexts, understanding the environment is no longer simply a correction to ideal quantum dynamics, but rather an essential ingredient for the design of robust quantum devices.

Many important topics remain beyond the scope of this introductory text. Among them are non-Markovian dynamics, collision models, stochastic Schr\"odinger equations, tensor-network approaches to open systems, quantum trajectories, exceptional points in Liouvillian spectra, driven-dissipative phase transitions, and thermodynamics of strongly coupled systems. These directions illustrate that the field remains both mathematically rich and experimentally vibrant.

Ultimately, the theory of open quantum systems teaches a broader lesson about physics itself: no system is truly isolated. The emergence of irreversibility, decoherence, and thermodynamic behavior from underlying reversible quantum laws remains one of the deepest conceptual themes in theoretical physics. Understanding open quantum systems is therefore not only essential for modern applications, but also fundamental for understanding how the quantum world connects to the macroscopic phenomena observed in nature.

\section*{Acknowledgements}
We express sincere gratitude to all of our students who generously shared their time, insights, and feedback during the development of these lecture notes. Your thoughtful comments, questions, and suggestions helped us improve the clarity, accuracy, and relevance of the content. This work has greatly benefited from your engagement and perspective, and we truly appreciate your valuable contribution to making these notes more useful and accessible for future readers.

\paragraph{Funding information}
We acknowledge financial support from EQA under project ID 101298535 (EU-Digital) and the Spanish Government via projects PID2024-161371NB-C21 (MCIU/AEI/FEDER, EU) and TSI-069100-2023-8 (Perte Chip-NextGenerationEU). E. T. and R. P. acknowledge the Ram{\'o}n y Cajal (RYC2020-030060-I) and (RYC2023-044095-I) research fellowships.

\begin{appendix}
\numberwithin{equation}{section}

\section{Solutions to Introduction}

\noindent\textbf{Problem 1.1}

\noindent Using the Jacobi's formula for $U(t)=e^{tX}$, we have
    \begin{align}
        \frac{d}{dt}\det[ e^{t X}]=\det[e^{tX}] {\rm Tr}\left[e^{-t X}Xe^{t X} \right]=\det[e^{tX}] {\rm Tr}\left[X \right].
    \end{align}
    Therefore,
    \begin{align}
        \frac{d}{dt} \det [e^{t X}]=\det [e^{t X}] {\rm Tr}[X] \Rightarrow \det[e^{t X}]=C e^{t {\rm Tr}[X]},
    \end{align}
    with $C$ a constant. Since for $t=0$, we have $\det[e^{0 \cdot X}]=1$, it follows that $C=1$ and arrive to the requested identity,
    \begin{align}
        \det[e^{t X}]=e^{t {\rm Tr}[X]}.
    \end{align}

\vspace{0.5cm}
\noindent\textbf{Problem 1.2}
\vspace{0.2cm}
    
\noindent From the previous exercise, we know that
    \begin{align}
        \det[e^{t X}]=e^{t {\rm Tr}[X]},
    \end{align}
    and since $\det[U(t)]=1$ for $U\in\textit{SU}(d)$ (unitary transformation), then we have
    \begin{align}
        e^{t {\rm Tr}[X]}=1 \ \forall t
    \end{align}
    which leads to ${\rm Tr}[X]=0$ for $X\in\mathfrak{g}$. Since any $X\in\mathfrak{g}$ can be expressed as a linear combination of the $T_i$, we arrive to ${\rm Tr}[T_i]=0, \forall i=1,\ldots,d^2-1$.

\vspace{0.5cm}
\noindent\textbf{Problem 1.3}
\vspace{0.2cm}

\noindent The problem here is the trace operation on unbounded operators, which is not well defined. Indeed, ${\rm Tr}[a\adag]$ and ${\rm Tr}[\adaga]$ are both divergent sums, and their difference is also not well defined.

\vspace{0.5cm}
\noindent\textbf{Problem 1.4}
\vspace{0.2cm}

\noindent Alice and Bob qubits:
    \begin{itemize}
        \item[(i)] \begin{align}
            \rho_{AB}=\frac{1}{2+|\lambda|^2}\begin{pmatrix} |\lambda|^2 & 0 & \lambda & \lambda\\
            0 & 0 & 0 & 0\\
            \lambda^* & 0 & 1 & 1\\
            \lambda^* & 0 & 1 & 1
            \end{pmatrix}.
        \end{align}
        \begin{align}
            \rho_A={\rm Tr}_B[\rho_{AB}]=\frac{1}{2+|\lambda|^2}\begin{pmatrix} |\lambda|^2 & \lambda\\
                \lambda^* & 2
            \end{pmatrix}
        \end{align}
        \begin{align}
            \rho_B={\rm Tr}_A[\rho_{AB}]=\frac{1}{2+|\lambda|^2}\begin{pmatrix} |\lambda|^2+1 & 1\\
                1 & 1
            \end{pmatrix}
        \end{align}
        \item[(ii)] The purity is the trace of $\rho^2$, so that
        \begin{align}
            P={\rm Tr}[\rho_A^2]=\frac{4+2|\lambda|^2+|\lambda|^4}{(2+|\lambda|^2)^2}
        \end{align}
        \item[(iii)] As the combined state is pure, if the qubit reduced state of Alice's qubit is not in a pure state means that there was entanglement. From the purity, we look for the maxima and minima, $\frac{d}{d|\lambda|}P=0$, and find that $|\lambda|=0$ is the maximum, $P=1$, and $|\lambda|=\sqrt{2}$ gives the minimum value of the purity, $P=3/4$. Therefore, for any $|\lambda|\neq 0$, Alice and Bob share entangled qubits. 
        \item[(iv)] Since the measurement is done locally on Alice's qubit, and we are interested in the average state, Bob's qubit is simply the one we obtained above, $\rho_B$. This can be seen more formally as
        \begin{align}
            \rho_{av}=\sum_i(\Pi_i\otimes \mathbb{I}_B) \rho_{AB} (\Pi_i^\dagger\otimes \mathbb{I}_B)
        \end{align}
        so that 
        \begin{align}
            \rho_{B,av}={\rm Tr}_A[\rho_{av}]=\sum_{i'}\bra{i'} \sum_i(\Pi_i\otimes \mathbb{I}_B) \rho_{AB} (\Pi_i^\dagger\otimes \mathbb{I}_B)\ket{i'}=\sum_i (\bra{i}\otimes \mathbb{I}_B)\rho_{AB}(\ket{i}\otimes \mathbb{I}_B)=\rho_B.
        \end{align}
        Note that this is not equivalent to Bob having a state $\rho_B$ conditioned to a particular outcome of Alice's measurement! This can be seen, for example, if Alice obtains $\ket{0}$, then Bob's qubit collapses also to the $\ket{0}$ state, while $\ket{1}$ leads to $\ket{+}$ for Bob's qubit.
        \item[(v)] Take for example the reduce state of Alice, $\rho_A$. A projective measurement on $\ket{+}$ happens with probability $p_+={\rm Tr}[\rho_A \ket{+}\bra{+}]$, i.e.
        \begin{align}
            p_+=\frac{2+2{\rm Re}[\lambda]+|\lambda|^2}{4+2|\lambda|^2}.
        \end{align}
        If this event takes a probability of $13/18$, we can make $p_+=13/18$ and solve for $\lambda$. However, there an infinitely many solutions given that $\lambda\in\mathbb{C}$. One can rule out some values (for example, $\lambda=0$ is not valid), and also we require ${\rm Re}[\lambda]\neq 0$ as otherwise $p_+=1/2$. If $\lambda\in\mathbb{R}$, then this reduces to
        \begin{align}
            \frac{13}{18}=\frac{2+2\lambda +\lambda^2}{4+2\lambda^2} \Rightarrow 4-9\lambda+2\lambda^2=0 \Rightarrow \lambda=1/2,4
        \end{align}
        Even assuming a real parameter,   one cannot uniquely determine $\lambda$; there are two possibilities.
    \end{itemize}

\vspace{0.5cm}
\noindent\textbf{Problem 1.5}
\vspace{0.2cm}

\noindent POVMs and PVMs. 
    \begin{itemize}
    \item[(i)] Kraus operators for $A$ when sharp measurement on $B$. The state $\rho_A$ conditioned to having read the state $\ket{i}$ on $B$ reads as
    \begin{align}
        \rho_{A|i}&={\rm Tr}_B[(\mathbb{I}_A\otimes \Pi_i)U(\rho_A\otimes \ket{k}\bra{k})U^\dagger (\mathbb{I}_A\otimes \Pi_i)]=\sum_{i'}\bra{i'} \left(\ket{i}\bra{i} U(\rho_A\otimes \ket{k}\bra{k})U^\dagger \ket{i}\bra{i}\right)\ket{i'}\nonumber\\&=\bra{i}U(\rho_A\otimes \ket{k}\bra{k})U^\dagger \ket{i}=K_i \rho_A K_i^\dagger
    \end{align}
with $K_i=\bra{i}U\ket{k}$. One can easily check the properties of the POVM,
\begin{align}
    \sum_i K_i^\dagger K_i&=\sum_i \bra{k}U ^\dagger \ket{i}\bra{i}U\ket{k}=\bra{k} U^\dagger \sum_i \ket{i}\bra{i} U \ket{k}=\bra{k} U^\dagger \mathbb{I}_B U \ket{k}\nonumber\\
    &=\bra{k} \mathbb{I}_{AB}\ket{k}=\mathbb{I}_A \otimes \bra{k} \mathbb{I}_B\ket{k}=\mathbb{I}_A.
\end{align}
In addition (more general)
    \begin{align}
        p(i)&={\rm Tr}_{AB}[(\mathbb{I}_A\otimes P_i)U\rho_{AB}U^\dagger]={\rm Tr}_A[{\rm Tr}_B[(\mathbb{I}_A\otimes P_i)U(\rho_A\otimes \rho_B)U^\dagger]]\nonumber\\
        &={\rm Tr}_A[\rho_A {\rm Tr}_B[U^\dagger(\mathbb{I}_A\otimes P_i)U (\mathbb{I}_A\otimes \rho_B)]]={\rm Tr}_A[\rho_A M_i]
    \end{align}
    where the operator $M_i$ acts solely on qubit $A$ and is independent on $\rho_A$,
    \begin{align}
        M_i={\rm Tr}_B[U^\dagger(\mathbb{I}_A\otimes P_i)U (\mathbb{I}_A\otimes \rho_B)].
    \end{align}
    These operators $\{M_i\}$ form a POVM on $A$; they are Hermitian, and add up to the identity, $\sum_i M_i=\mathbb{I}_A$,
    \begin{align}
        \sum_i M_i&=\sum_i {\rm Tr}_B[U^\dagger(\mathbb{I}_A\otimes P_i)U (\mathbb{I}_A\otimes \rho_B)]={\rm Tr}_B[U^\dagger(\mathbb{I}_A\otimes \sum_i P_i)U (\mathbb{I}_A\otimes \rho_B)]\nonumber\\
        &={\rm Tr}_B[U^\dagger(\mathbb{I}_A\otimes \mathbb{I}_B)U (\mathbb{I}_A\otimes \rho_B)]={\rm Tr}_B[U^\dagger U (\mathbb{I}_A\otimes \rho_B)]\nonumber\\
        &={\rm Tr}_B[(\mathbb{I}_A\otimes \rho_B)]=\mathbb{I}_A {\rm Tr}_B[\rho_B]=\mathbb{I}_A.
    \end{align}
    In addition, $M_i$ is positive semi-definite. 
     \item[(ii)] If $\rho_B=\ket{0}\bra{0}$ and $U$ a controlled-$R_y(\theta)$ rotation, then we know that
     \begin{align}
         R_y(\theta)=e^{-i\theta/2 \sigma_y}=\begin{pmatrix} \cos(\theta/2) & -\sin(\theta/2) \\ \sin(\theta/2) & \cos(\theta/2)\end{pmatrix}
     \end{align}
     so that
     \begin{align}
         U=\ket{0}\bra{0}\otimes \mathbb{I}_B+\ket{1}\bra{1}\otimes e^{-i\theta/2 \sigma_y}=\begin{pmatrix} 1 & 0 & 0 &0 \\
         0 & 1 & 0 & 0 \\
         0 & 0 & \cos(\theta/2) & -\sin(\theta/2) \\ 0 & 0 & \sin(\theta/2) & \cos(\theta/2)\end{pmatrix}.
     \end{align}
     In this manner,
     \begin{align}
         K_0=\bra{0}_B U\ket{0}_B= \ket{0}\bra{0}_A+\cos(\theta/2)\ket{1}\bra{1}_A.
     \end{align}
     and
     \begin{align}
         K_1=\bra{1}_B U\ket{0}_B= \sin(\theta/2)\ket{1}\bra{1}_A.
     \end{align}
     One can verify that $K_0^\dagger K_0+K_1^\dagger K_1=\mathbb{I}_A$. 
     For a generic state $\rho_A$ we find
     \begin{align}
         \rho_A'=K_0\rho_A K_0^\dagger+K_1 \rho_A K_1^\dagger= \frac{1}{2}(1+\vec{v}'\cdot\vec{\sigma})
     \end{align}
     with $\vec{v}'=(v_x\cos\theta/2,v_y\cos\theta/2,v_z)$. In general, for $|\cos(\theta/2)|<1$, this corresponds to a dephasing channel for qubit $A$, as the off-diagonal elements decrease. Moreover, for $\cos(\theta/2)<0$, this also corresponds to a $\pi$ rotation around the $z$ axis. For $\theta=\pi$, the POVM reduces to a PVM (standard projective measurement). For $\theta=0$, this is just an identity operation. For $\theta=2\pi$, it corresponds to a $\pi$ rotation around the $Z$ axis.
     \end{itemize}

\section{Solutions to Open Quantum Systems}

\noindent\textbf{Problem 2.1}

\noindent For a product state form, we have
\begin{align}
    [H_S\otimes \mathbb{I}_B,\rho_S(t)\otimes \rho_B(t)]=H_S\rho_S(t)\otimes \rho_B(t)-\rho_S(t)H_S\otimes \rho_B(t)=[H_S,\rho_S(t)]\otimes \rho_B(t)
\end{align}
and
\begin{align}
    [\mathbb{I}_S\otimes H_B,\rho_S(t)\otimes \rho_B(t)]=\rho_S(t)\otimes [H_B,\rho_B(t)].
\end{align}
Now, taking the partial trace over the environment, we find
\begin{align}
    {\rm Tr}_B[\ [H_S,\rho_S(t)]\otimes \rho_B(t)]=[H_S,\rho_S(t)]{\rm Tr}_B[\rho_B(t)]=[H_S,\rho_S(t)]
\end{align}
given that $\rho_B(t)$ must be normalized at all times, ${\rm Tr}_B[\rho_B(t)]=1$. In addition,
\begin{align}
    {\rm Tr}_B[\ \rho_S(t)\otimes [H_B,\rho_B(t)]]=\rho_S {\rm Tr}_B[\ [H_B,\rho_B(t)]]=0
\end{align}
because of the cyclic property, ${\rm Tr}[A B]={\rm Tr}[B A]$, the trace of a commutator is zero. So we recover the von Neumann equation for the closed system,
\begin{align}
    \frac{d}{dt}\rho_S(t)=-i[H_S,\rho_S(t)]
\end{align}

\vspace{0.5cm}
\noindent\textbf{Problem 2.2}
\vspace{0.2cm}

\noindent Doing some algebra,
\begin{align}
    \sum_{n,m} W_{n,m}^\dagger(t,t_0)W_{n,m}(t,t_0)&=\sum_{n,m}\bra{b_m}U^\dagger(t,t_0) \sqrt{\lambda_m}\ket{b_n}\bra{b_n}U(t,t_0)\sqrt{\lambda_m}\ket{b_m}\nonumber\\&=\sum_m \bra{b_m}U^\dagger(t,t_0) \sqrt{\lambda_m}\sum_n \ket{b_n}\bra{b_n}U(t,t_0)\sqrt{\lambda_m}\ket{b_m}\nonumber\\&=\sum_m \bra{b_m}U^\dagger(t,t_0) \sqrt{\lambda_m}\mathbb{I}_B U(t,t_0)\sqrt{\lambda_m}\ket{b_m}\nonumber\\
    &=\sum_m \bra{b_m}\lambda_m U^\dagger(t,t_0)U(t,t_0)\ket{b_m}=\sum_m \lambda_m \bra{b_m} \mathbb{I}_S\otimes\mathbb{I}_B\ket{b_m}\nonumber\\&=\mathbb{I}_S\sum_m \lambda_m \langle b_m | b_m\rangle=\mathbb{I}_S
\end{align}

\vspace{0.5cm}
\noindent\textbf{Problem 2.3}
\vspace{0.2cm}

\begin{itemize}
    \item[(i)]  The state of the qubits $A$ and $B$ is given by
\begin{align}
    \ket{\psi(t)}_{AB}=U(t)\ket{\psi(0)}_{AB}=e^{-i Jt \sigma_x\otimes \sigma_x}\ket{00}=\cos Jt \ket{00}-i\sin Jt \ket{11}.
\end{align}
If we take the partial trace with respect to $B$, we obtain the state of $A$ at all times:
\begin{align}
    \rho_A(t)&={\rm Tr}_B[\ket{\psi(t)}\bra{\psi(t)}_AB]={\rm Tr}_B \left[ \begin{pmatrix}\cos^2(Jt) & 0 &0 &i\cos(Jt)\sin(Jt)\\
    0 & 0 &0 &0\\
    0 & 0 &0 &0\\
    -i\cos(Jt)\sin(Jt) & 0 &0 & \sin^2(Jt) \end{pmatrix} \right]\nonumber\\&=\begin{pmatrix} \cos^2(Jt) & 0 \\
    0 & \sin^2(Jt)       
    \end{pmatrix}
\end{align}
\item[(ii)] At times $Jt_s=\pi/4+n\pi/2$ with $n\in\mathbb{Z}$ we have $\cos^2(Jt_s)=\sin^2(Jt_s)=1/2$, so that the state of $A$ is completely mixed, $\rho_A(t_s)=\mathbb{I}_2/2$. At those times the combined state $\ket{\psi_{AB}(t_s)}$ is maximally entangled (a Bell state),
\begin{align}
    \ket{\psi_{AB}(t_s)}=\frac{1}{\sqrt{2}}(\ket{00}\pm i \ket{11}),
\end{align}
and therefore, tracing out the qubit $B$ renders the qubit $A$ in a maximally mixed state. 
\item[(iii)] Assuming a $\sigma_x$ Lindblad operator only with a time-dependent rate $\gamma_x(t)$ we have
\begin{align}
    \frac{1}{2}\begin{pmatrix} \dot{v}_z & \dot{v}_x-i\dot{v}_y\\
    \dot{v}_x+i\dot{v}_y & -\dot{v}_z        
    \end{pmatrix}=\gamma_x(t)\begin{pmatrix} -v_z & i v_y\\-i v_y & v_z        
    \end{pmatrix}.
\end{align}
Given that $v_x=v_y=0$, we simply reduce the problem to
\begin{align}\label{eq:dotv}
    \dot{v}_z=-2 \gamma_x(t) v_z,
\end{align}
with $v_z(0)=1$. Now, note that we know $\rho_A(t)$ at all times, from where we find
\begin{align}
    v_z(t)=2\cos^2(Jt)-1,
\end{align}
 so that 
\begin{align}
    \dot{v}_z=-4 J \cos(J t)\sin(J t).
\end{align}
Plugging this into Eq.~\eqref{eq:dotv}, we get 
\begin{align}
    \gamma_x(t)=-\frac{\dot{v}_z(t)}{2v_z(t)}=\frac{4 J \cos(Jt)\sin(Jt)}{4 \cos^2(Jt)-2}=J\tan(2J t).
\end{align}
The time dependent rate is not only divergent at times $Jt=\pi/4$, $3\pi/4$ (modulo $2\pi$), but it is negative for $Jt\in(\pi/4,\pi/2)\cup (3\pi/4,\pi)$, modulo $\pi$. This means that the master equation is not Markovian, it fails to be CP-divisible, at least for times $Jt\geq \pi/4$. Note that how a trivial example (completely solvable and simple) leads into a non-Markovian evolution when restricted to a single qubit. 
\begin{figure}
    \centering
    \includegraphics[width=0.4\linewidth]{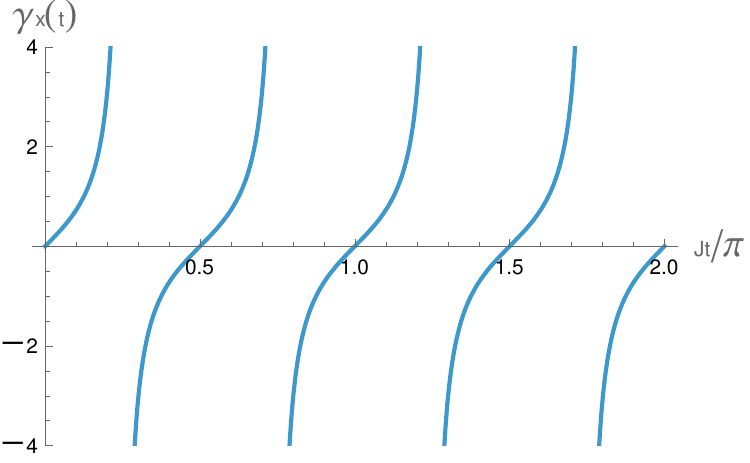}
    \caption{Time-dependent rate $\gamma_x(t)$ to describe the dynamics of the single qubit in exercise 3}
    \label{fig:gammax_rate}
\end{figure}

\end{itemize}

\vspace{0.5cm}
\noindent\textbf{Problem 2.4}
\vspace{0.2cm}

\begin{itemize}
    \item[(i)] The Lindblad equation reads
    \begin{align}
        \frac{d}{dt}\rho(t)&=-i[H,\rho(t)]+\gamma_z\left(\sigma_z \rho(t)\sigma_z-\frac{1}{2}\{\sigma_z\sigma_z,\rho \}\right)\nonumber\\&=-\frac{i\omega}{2}[\sigma_z,\rho(t)]+\gamma_z(\sigma_z\rho(t)\sigma_z-\rho(t))
    \end{align}
    Using the general form of the density matrix, we have $\rho(t)=\frac{1}{2}(\mathbb{I}+\vec{v}(t)\cdot\vec{\sigma})$, so that
    \begin{align}
        \frac{1}{2}\begin{pmatrix}\dot{v}_z(t) & \dot{v}_x(t)-i \dot{v}_y\\
            \dot{v}_x(t)+i \dot{v}_y & -\dot{v}_z(t)
    \end{pmatrix}&=\begin{pmatrix}0 & -\frac{i\omega}{2}(v_x(t)-i v_y)\\\frac{i\omega}{2}(v_x(t)+i v_y)
             & 0
        \end{pmatrix}\nonumber\\&+\begin{pmatrix}0 & \gamma_z(-v_x(t)+i v_y)\\\gamma_z(-v_x(t)-i v_y)
             & 0
        \end{pmatrix}.
    \end{align}
    Now, isolating for each component we have
    \begin{align}
        \dot{v}_x(t)&=-2\gamma_zv_x(t)-\omega v_y(t)\\
        \dot{v}_y(t)&=-2\gamma_z v_y(t)+\omega v_x(t)\\
        \dot{v}_z(t)&=0.
    \end{align}
    \item[(ii)] The solution to the differential equations reads as (take $u(t)=v_x(t)-iv_y(t)$ so that $\dot{u}(t)=(-2\gamma_z-i\omega)u(t)\Rightarrow u(t)=e^{(-2\gamma_z-i\omega)t}u(0)$)
    \begin{align}
        v_x(t)&=e^{-2\gamma_z t}\cos(\omega t)\\
        v_y(t)&=e^{-2\gamma_z t}\sin(\omega t)\\
        v_z(t)&=0.
    \end{align}
    This corresponds to a spiral in the $xy$ plane of the Bloch sphere, towards the center. See Fig.~\ref{fig:spiral} for a case with $\gamma_z=8\omega$. 
    \begin{figure}
        \centering
        \includegraphics[width=0.3\linewidth]{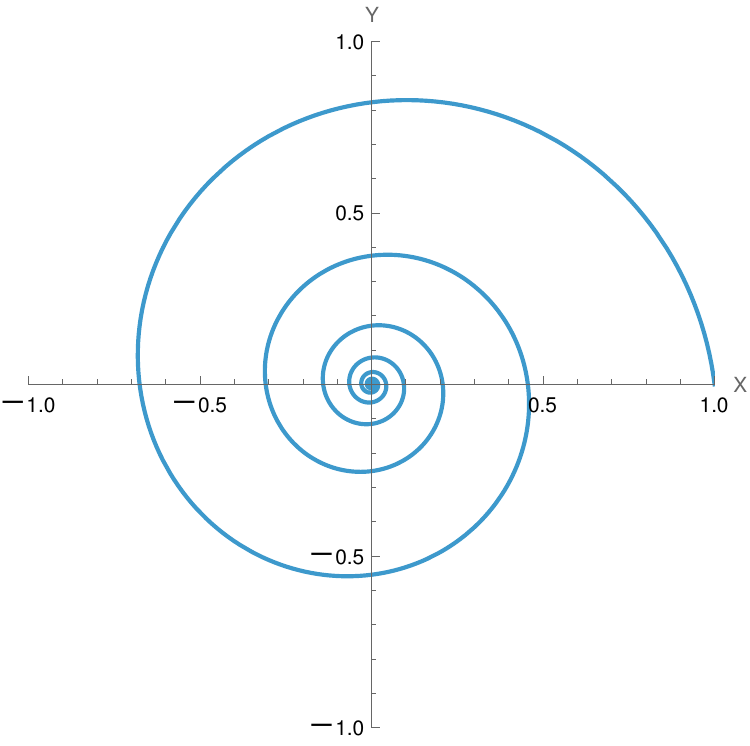}
        \caption{Trajectory of the qubit state in the $XY$ plane of the Bloch sphere for $\vec{v}(0)=(1,0,0)$ with $\gamma_z=8\omega$.}
        \label{fig:spiral}
    \end{figure}
    \item[(iii)] The characteristic time corresponds to the expectation value of $\langle \sigma_x\rangle$ decaying a factor of $1/e$. In this case, this corresponds to $T_2=\frac{1}{2\gamma_z}$. 
\end{itemize}

\vspace{0.5cm}
\noindent\textbf{Problem 2.5}
\vspace{0.2cm}

\noindent The steady state of the Lindblad equation can be obtained by setting $\frac{d}{dt}\rho_{ss}=0$. Since we look for the thermal equilibrium state, we can consider $\rho_{ss}=p_0\ket{0}\bra{0}+p_1\ket{1}\bra{1}$, with $p_0$ and $p_1$ given in the problem. Then, 
\begin{align}
    \mathcal{L}\{\rho_{ss}\}=0 \Rightarrow -i[H_S,\rho_{ss}]+\sum_{k=\pm}\gamma_k(\sigma_k \rho_{ss} \sigma_k^\dagger-\frac{1}{2}\{ \sigma_k^\dagger \sigma_k,\rho_{ss}\}).
\end{align}
Given that $H_{S}=\omega\sigma_z/2$, it commutes with $\rho_{ss}$. Using $\sigma^{+,-}$, we obtain the condition
\begin{align}
    \gamma_+ p_1-\gamma_- p_0=0 \Rightarrow \frac{\gamma_+}{\gamma_-}=\frac{p_0}{p_1}=e^{-\omega/kT}.
\end{align}
That is, provided the rates satisfy this ratio, the qubit will tend towards the thermal equilibrium state. This condition is known as detailed balance: $P(E_1\rightarrow E_2)/P(E_2\rightarrow E_1)=e^{(E_1-E_2)/kT}$. 

Now, if $\gamma_{\pm}=a\gamma_{\pm}$, the steady state remains unchanged, $\rho_{eq}$, provided $\gamma_{\pm}$ fulfill the relation found before. The constant $a$ determines how fast the qubit thermalizes, i.e. it fixes the time scale for equilibration.

\section{Solutions to Microscopic Models}

\noindent\textbf{Problem 3.1}
\vspace{0.2cm}

\noindent Qubit+Qubits:
    \begin{itemize}
        \item[(i)] GKSL master equation. First the correlation functions of the environment. Identify:
        \begin{align}
            A_1&=\sigma^+, B_1=\sum_k g_k \sigma_k^-\\
            A_2&=\sigma^-, B_2=\sum_k g_k^* \sigma_k^+
        \end{align}
        So that 
        \begin{align}
            C_{1,1}(s)={\rm Tr}_B[B_1^\dagger(s)B_1(0)\rho_B]={\rm Tr}_B[\left(\sum_k g_k^* e^{i\nu_k s}\sigma^+_k\right)\left(\sum_{k'} g_{k'} \sigma^-_{k'}\right)\rho_B]
        \end{align}
        Using ${\rm Tr}[\sigma_{k}^{\pm}\rho_B]=0$ and ${\rm Tr}[\sigma^{+}_k\sigma^-_k\rho_B]=p_k$ and ${\rm Tr}[\sigma^{-}_k\sigma^+_k\rho_B]=1-p_k$, we get
        \begin{align}
            C_{1,1}(s)=\sum_k |g_k|^2e^{i\nu_k s}p_k.
        \end{align}
        Similarly, $C_{1,2}(s)=C_{2,1}(s)=0$, and 
        \begin{align}
            C_{2,2}(s)={\rm Tr}_B[B_2^\dagger(s)B_2(0)\rho_B]=\sum_k |g_k|^2e^{-i\nu_k s}(1-p_k).
        \end{align}
        Now, since $A_1(t)=e^{i\omega t}\sigma^+$ and $A_2(t)=e^{-i\omega t}\sigma^-$, the one-sided Fourier transforms are
        \begin{align}
            \Gamma_{1,1}(\omega)&=\int_0^\infty ds e^{-i\omega s}C_{1,1}(s)=\int_0^\infty ds \sum_k e^{-i(\omega-\nu_k)s} |g_k|^2 p_k\\
            \Gamma_{2,2}(\omega)&=\int_0^\infty ds e^{i\omega s}C_{2,2}(s)=\int_0^\infty ds \sum_k e^{i(\omega-\nu_k)s} |g_k|^2 (1-p_k)
        \end{align}
        This results in
        \begin{align}
            \gamma_{1,1}(\omega)&=2{\rm Re}[\Gamma_{1,1}(\omega)]=2\pi \sum_k |g_k|^2p_k \delta(\omega-\nu_k)\\
            S_{1,1}(\omega)&=-{\rm P.V.}\left(\sum_k |g_k|^2p_k \frac{1}{\omega-\nu_k}\right)\\
            \gamma_{2,2}(\omega)&=2{\rm Re}[\Gamma_{2,2}(\omega)]=2\pi \sum_k |g_k|^2(1-p_k) \delta(\omega-\nu_k)\\
            S_{2,2}(\omega)&={\rm P.V.}\left(\sum_k |g_k|^2(1-p_k) \frac{1}{\omega-\nu_k}\right).
        \end{align}
        The Lamb shift is then $H_{\rm LS}=S_{1,1}A_1^\dagger A_1+S_{2,2}A_2^\dagger A_2=S_{1,1}(\omega)\sigma^-\sigma^++S_{2,2}\sigma^+\sigma^-$, shifting the energy levels. The GKSL master equation in the Schr\"odinger picture reads
        \begin{align}
            \frac{d}{dt}\rho_S(t)=-i[H_S+H_{\rm LS},\rho_S(t)]&+\gamma_{1,1}(\omega)\left(\sigma^+\rho_S \sigma^--\frac{1}{2}\{ \sigma^-\sigma^+,\rho_S(t)\} \right)\nonumber\\&+\gamma_{2,2}(\omega)\left(\sigma^-\rho_S \sigma^+-\frac{1}{2}\{ \sigma^+\sigma^-,\rho_S(t)\} \right).
        \end{align}
        The term $\gamma_{1,1}(\omega)$ corresponds to the rate of spontaneous absorption ($\ket{1}\rightarrow\ket{0}$), while $\gamma_{2,2}(\omega)$ for spontaneous decay. 
        \item[(ii)] In the continuum limit, we have
        \begin{align}
            \gamma_{1,1}(\omega)&=2\pi\int_0^\infty d\omega J(\omega)p(\omega),\\
            \gamma_{2,2}(\omega)&=2\pi\int_0^\infty d\omega J(\omega)(1-p(\omega)),\\
            S_{1,1}(\omega)&=-{\rm P.V.}\left(\int_0^\infty d\omega J(\omega)p(\omega)\frac{1}{\omega-\nu}\right),\\
            S_{2,2}(\omega)&={\rm P.V.}\left(\int_0^\infty d\omega J(\omega)(1-p(\omega))\frac{1}{\omega-\nu}\right),
        \end{align}
        where $p(\omega)$ denotes the population of the $\ket{0}$ state for the qubit at frequency $\omega$. 
        \item[(iii)] For $p_k=0$, the rate of spontaneous absorption vanishes (maximal for decay). Why? In this case, all the qubits of the environment are in the state $\ket{1}$ (ground state) and therefore, they cannot give an excitation to the system (only absorb, which corresponds to decay of $S$). Similarly, for $p_k=1$, spontaneous decay vanishes, and absorption is maximum. In the case of $p_k=1/2$, both rates are equivalent, so there is a balanced gain and loss. 

        \item[(iv)] The ratio between spontaneous absorption and emission 
        \begin{align}
            \frac{\gamma_{1,1}(\omega)}{\gamma_{2,2}(\omega)}=\frac{p(\omega)}{1-p(\omega)}
        \end{align}
        For a thermal state, we have $Z=2\cosh(\beta \omega/2)$ and $p(\omega)=e^{-\beta\omega/2}/Z$ so that
        \begin{align}
            \frac{\gamma_{1,1}(\omega)}{\gamma_{2,2}(\omega)}=\frac{p(\omega)}{1-p(\omega)}=\frac{e^{-\beta\omega/2}}{e^{\beta\omega/2}}=e^{-\beta\omega}=e^{-\omega/(k_b T)}
        \end{align}
        Setting $T^*$ for the condition of $10^{-5}$, we find 
        \begin{align}
            e^{-\omega/(k_b T^*)}=10^{-3} \Rightarrow T^*=\frac{\omega}{5k_b \log 10}
        \end{align}
        Taking into account the constants ($\hbar$ in particular), we have
        \begin{align}
            T^*=\frac{\hbar \omega }{5k_b \log 10}\approx 41.6 \ {\rm mK}
        \end{align}
        For temperatures smaller than $\approx  42$ mK, spontaneous absorption is negligible. 
    \end{itemize}

\vspace{0.5cm}
\noindent\textbf{Problem 3.2}
\vspace{0.2cm}

\noindent Quantum Harmonic oscillator with pure dephasing. 
    \begin{itemize}
        \item[(i)] From the form of the Hamiltonian we can say that (if the approximations hold), the master equation will be
        \begin{align}
            \frac{d}{dt}\rho_S(t)=-i[H_S+H_{\rm LS},\rho_S(t)]+\gamma \left(a^\dagger a\rho_S a^\dagger a-\frac{1}{2}\{ a^\dagger a a^\dagger a,\rho_S(t)\}\right)
        \end{align}
        where the Lamb shift will something of the type $H_{\rm LS}\propto a^\dagger a a^\dagger a$, i.e. a non-linear shift of the energy levels. It will slightly modify the quadratic potential. This term is known as self-Kerr interaction. 
        \item[(ii)] The dissipative part produces pure dephasing noise in the harmonic oscillator. That is, it does not change the energy but forces off-diagonal terms to decay exponential in time. For a generic state $\rho_S(t)$, only the diagonal part will survive in the long-time limit. 
        \item[(iii)] If $\rho_S(t)=\sum_n p_n(t)\ket{n}\bra{n}$, then
        \begin{align}
           D[\rho_S(t)]&=\gamma\left(a^\dagger a \sum_n p_n(t)\ket{n}\bra{n} a^\dagger a-\frac{1}{2}\{a^\dagger a a^\dagger a,\sum_n p_n(t)\ket{n}\bra{n}\}\right)\nonumber\\
            &=\gamma\sum_n \left(n^2p_n(t)\ket{n}\bra{n} -\frac{1}{2}n^2p_n(t)\ket{n}\bra{n}-\frac{1}{2}n^2p_n(t)\ket{n}\bra{n}\right)=0.
        \end{align}
        A diagonal state is a steady state of the dissipative part of the master equation. 
        \item[(iv)] The Markov approximation is not valid in this scenario. Therefore, the GKSL master equation breaks down. Why? This stems from the two-time correlation function of the environment. In our case, we have $B(t)=\sum_k g_k \sigma_{z,k}$, which is time independent. Therefore,
        \begin{align}
            C(s)={\rm Tr}_B[B^\dagger(s)B(0)\rho_B]={\rm Tr}_B[\sum_{k,k'}g_k g_{k'}\sigma_{z,k}\sigma_{z,k'}\rho_B].
        \end{align}
        Clearly, the two-time correlation function does not decay in time (it is a static quantity!). Hence, we cannot approximate the finite-time integration to $\infty$. The core of the Markov approximation resides in the fact that the two-time correlation functions decay in time sufficiently fast. However, $C(s)$ does not depend on time. 
    \end{itemize}

\vspace{0.5cm}
\noindent\textbf{Problem 3.3}
\vspace{0.2cm}

\noindent Atom. 
    \begin{itemize}
        \item[(i)] Define the operators as
        \begin{align}
            A_1=\ket{e}\bra{f}, A_2=\ket{f}\bra{e}, A_3=\ket{g}\bra{e}, A_4=\ket{e}\bra{g}.
        \end{align}
        So that $B_1=B_3=\sum_k g_k a_k$, and $B_2=B_4=\sum_k g_k^* a^\dagger_k$. 
        Again, the two-time correlation functions are
        \begin{align}
            C_{1}(s)=C_3(s)={\rm Tr}_B[B_1^{\dagger}(s)B_1(0)\rho_B]=\sum_k |g_k|^2 \overline{n}_k e^{i\nu_k s}
        \end{align}
        and 
        \begin{align}
            C_2(s)=C_4(s)=\sum_k |g_k|^2 (\overline{n}_k+1) e^{-i\nu_k s}
        \end{align}
        Doing the same calculations we find the rates
        \begin{align}
            \gamma_{f,e}(\omega_{f,e})&=2\pi \sum_k |g_k|^2 (\overline{n}_k+1)\delta(\omega_{f,e}-\nu_k)=2\pi J(\omega_{f,e})(\overline{n}(\omega_{f,e})+1)\\
            \gamma_{e,f}(\omega_{f,e})&=2\pi J(\omega_{f,e})\overline{n}(\omega_{f,e})
        \end{align}
         and similarly for $\gamma_{e,g}$ and $\gamma_{g,e}$. Here $\omega_{f,e}=\omega_f-\omega_e$ is the frequency of the transition between $\ket{f}$ and $\ket{e}$. Due to selection rules, we know that $\gamma_{f,g}=\gamma_{g,f}=0$. The rate $\gamma_{f,e}$ corresponds to spontaneous transition (incoherent) from $f$ to $e$, and similarly for the rest. Note that spontaneous decay is always larger than spontaneous absorption (due to the factor $\overline{n}(\omega)+1$).
         \item[(ii)] If the temperature is sufficiently low, then spontaneous absorption can be discarded, i.e. if $\overline{n}(\omega)\approx 0$. This depends on the interplay between frequency and temperature.
         \item[(iii)] For the given transition frequencies, we know that the limit will be set by the transition with smallest frequency. If we can discard spontaneous absorption for $\ket{g}\rightarrow\ket{e}$, then also for $\ket{e}\rightarrow\ket{f}$ since $\omega_{f,e}>\omega_{e,g}$. The limit condition in this case reads as
         \begin{align}
\gamma_{\uparrow}=10^{-3}\gamma_{\downarrow} \Rightarrow \frac{\gamma_{g,e}}{\gamma_{e,g}}=10^{-3}\Rightarrow \frac{\overline{n}(\omega_{e,g})}{\overline{n}(\omega_{e,g})+1}=10^{-3}\Rightarrow \overline{n}(\omega_{e,g})=1/999
         \end{align}
         Now, plugging numbers we have
         \begin{align}
             \overline{n}(\omega_{e,g})&=\frac{1}{999}\Rightarrow \frac{1}{(e^{\hbar (\omega_e-\omega_g)/(k_b T)}-1)}=\frac{1}{999}\nonumber\\&\Rightarrow T=\frac{\hbar(\omega_e-\omega_g)}{k_b\log 10^{3}}=\frac{1.054\cdot 10^{-34} 2\pi \cdot 140 \cdot 10^{12}} {1.38 \cdot 10^{-23} \cdot 6.908}\approx 972 {\rm K}\approx 1000 {\rm K}
         \end{align}
    \end{itemize}

\section{Solutions to Quantum Thermodynamics}

\noindent\textbf{Problem 4.1}
\vspace{0.2cm}

\noindent Local vs global equilibrium
    \begin{itemize}
        \item[(i)] For the first part, we have just
        \begin{align}
            e^{-\beta H_i}=e^{-\beta \omega/2}\ket{0}\bra{0}+e^{\beta \omega/2}\ket{1}\bra{1}
        \end{align}
        so that
        \begin{align}
            Z=e^{-\beta \omega/2}+e^{\beta\omega/2}=2\cosh(\beta\omega/2).
        \end{align}
        This results in
        \begin{align}
            \rho_{eq}=\frac{e^{\beta H}}{Z}=p_0\ket{0}\bra{0}+(1-p_0)\ket{1}\bra{1}
        \end{align}
        with $p_0=(1+e^{\beta\omega})^{-1}$. 
        \item[(ii)] For $g_z=g$ but still $g_x=0$ (classical), the Hamiltonian is diagonal
        \begin{align}
            H={\rm diag}(g+\omega,-g,,-g,g-\omega).
        \end{align}
        Hence, the global Gibbs state is
        \begin{align}
            \rho_{1,2;eq}=\frac{1}{2e^{\beta g}+e^{-\beta(g-\omega)}+e^{-\beta(g+\omega)}}{\rm diag}(e^{-\beta(g+\omega)},e^{\beta g},e^{\beta g},e^{-\beta(g-\omega)})
        \end{align}
        The local equilibrium states correspond to
        \begin{align}
            \rho=p_0\ket{0}\bra{0}+(1-p_0)\ket{1}\bra{1}
        \end{align}
        with 
        \begin{align}
            p_0=\frac{1}{2+\frac{e^{2\beta\omega}-1}{e^{\beta (2g+\omega)}+1}}
        \end{align}
        Clearly, if $g=0$, one recovers the non-interacting case. 

        For the quantum case, $g_z=g_x=g$, we use the hint, and compute $\rho_{1,2,eq}$. The important thing here is that the off-diagonal terms (correlations) are destroyed when describing locally the qubits. For the classical case, this does not matter (they are absent). 
        
        \item[(iii)] The quantum coherence of the latter case (for the global Gibbs state) in the computational basis reads as
        \begin{align}
            C(\rho_{1,2;eq})=\frac{g\sinh(\beta \Delta)+\Delta e^{\beta g}\sinh(\beta g)}{\Delta \cosh(\beta \Delta)+\Delta e^{\beta g}\cosh(\beta g)}
        \end{align}
        For fixed $g$ and $\omega$, this has a maximum as a function of $\beta$. For $\beta=0$, $C=0$. The larger $g$, the larger the coherence (obviously). 
    \end{itemize}

Note that $H=H_1+H_2+H_I$ can be written as $H=\sum_{k=1}^4 e_k \ket{\phi_k}\bra{\phi_k}$, with 
\begin{align}
    \ket{\phi_1}&=\frac{1}{\sqrt{2}}\left(\ket{01}+\ket{10} \right), \quad e_1=0\\
    \ket{\phi_2}&=\frac{1}{\sqrt{2}}\left(-\ket{01}+\ket{10} \right), \quad e_2=-2g\\
    \ket{\phi_3}&=-\frac{1}{\Delta_+}\ket{00}+\frac{1}{\Delta_-}\ket{11}, \quad e_3=g-\sqrt{g^2+\omega^2}\\
    \ket{\phi_4}&=\frac{1}{\Delta_-}\ket{00}+\frac{1}{\Delta_+}\ket{11},\quad e_4=g+\sqrt{g^2+\omega^2}
\end{align}
with $\Delta_{\pm}=\left(2+2\omega[\omega-\sqrt{g^2+\omega^2}]/g^2 \right)^{1/2}$.

\vspace{0.5cm}
\noindent\textbf{Problem 4.2}
\vspace{0.2cm}
    
\noindent The free energy for a Gibbs state reads
    \begin{align}
        F=\mathcal{U}-TS
    \end{align}
    Let us start from $\mathcal{U}$:
    \begin{align}
        \mathcal{U}={\rm Tr}[H e^{-\beta H}/Z]
    \end{align}
    while
    \begin{align}
        S=-k_b {\rm Tr}[e^{-\beta H}/Z (-\beta H-\log Z)]=\frac{1}{T}\mathcal{U}+k_b \log Z
    \end{align}
    Thus,
    \begin{align}
        F=\mathcal{U}-T\left( \frac{1}{T}\mathcal{U}+k_b \log Z\right)=-\frac{1}{\beta}\log Z.
    \end{align}

\vspace{0.5cm}
\noindent\textbf{Problem 4.3}
\vspace{0.2cm}

\noindent Heat in dephasing process
    \begin{itemize}
        \item[(i)] As we have seen, the heat rate reads as
        \begin{align}
            \dot{Q}_k=\gamma_k \left(A_k^\dagger H A_k-\frac{1}{2}\{A_k^\dagger A_k,H\}\right)
        \end{align}
        In this case, $A_k=\sigma_z$. So that $H\propto \sigma_z$ makes the heat rate zero.
        \item[(ii)] For $H=\omega \sigma_x/2$, we have
        \begin{align}
            \dot{Q}=-\gamma \omega \sigma_x
        \end{align}
        \item[(iii)] For this we need the state at all times. How? Using the Block vector:
        \begin{align}
            \dot{v}_x(t)&=-2 \gamma v_x(t)\\
            \dot{v}_y(t)&=-2\gamma v_y(t)-v_z(t)\omega\\
            \dot{v}_z(t)&=2v_y(t)\omega.
        \end{align}
        For $v_x(0)=1$, $v_{y,z}(0)=0$, it follows that 
        \begin{align}
            v_x(t)=e^{-2\gamma t},
        \end{align}
        while $v_y(t)=v_z(t)=0, \forall t$. 
        The total heat integrates the following quantity:
        \begin{align}
            {\rm Tr}[-\gamma \omega \sigma_x \frac{1}{2}(\mathbb{I}+v_x(t)\sigma_x)]=-\gamma \omega v_x(t)=-\gamma \omega e^{-2\gamma t},
        \end{align}
        that is,
        \begin{align}
            \langle Q\rangle_{0,\tau}=\int_0^\tau dt (-\gamma \omega e^{-2\gamma t})=-\frac{\gamma \omega}{-2\gamma}(e^{-2\gamma \tau}-1)=\frac{\omega}{2}(e^{-2\gamma \tau}-1). 
        \end{align}
        \item[(iv)] The total internal energy variation must be equal to the total heat:
        \begin{align}
            \Delta \mathcal{U}={\rm Tr}[\omega/2 \sigma_x \frac{1}{2}(\mathbb{I}+v_x(\tau)\sigma_x)]-{\rm Tr}[\omega/2 \sigma_x \frac{1}{2}(\mathbb{I}+v_x(0)\sigma_x)]&=\omega v_x(\tau)-\omega\nonumber\\&=\omega (e^{-2\gamma \tau}-1)
        \end{align}
        \item[(v)] Dephasing in this case corresponds to a negative heat, meaning that heat flows from the system to the environment. This was to be expected because initially, the qubit is in its excited state (of the Hamiltonian), and dephasing will bring it to an equal superposition of both states (with zero energy). If initially the system had $\omega/2$ of energy, it will release that amount to become in a $\frac{1}{2}\ket{+}\bra{+}+\frac{1}{2}\ket{-}\ket{-}$ state. For $\gamma\tau\rightarrow \infty$, the heat rate vanishes; the state is already (or arbitrarily close to) in the steady state, so heat current stops. 
    \end{itemize}

\vspace{0.5cm}
\noindent\textbf{Problem 4.4}
\vspace{0.2cm}
    
\noindent Qubit and Jarzynski: First, the Gibbs state reads as
    \begin{align}
        \rho_{eq}=\frac{1}{2\cosh(\beta \omega/2)}\left(e^{-\beta \omega/2}\ket{0}\bra{0}+e^{\beta \omega/2}\ket{1}\bra{1} \right)=p_0\ket{0}\bra{0}+(1-p_0)\ket{1}\bra{1}
    \end{align}
    with $p_0=(1+e^{\beta \omega})^{-1}$. Now, for the work probability distribution under the two-point measurement scheme, we need to take into account the projective measurement and then evolution. For this, is quite easy in this case, since the states $\ket{0}$ and $\ket{1}$ remain unaffected by the dynamics. Hence,
    \begin{align}
        P(W)= p_0 \delta(W-(\omega_1/2-\omega_0/2)+(1-p_0)\delta(W-(-\omega_1/2+\omega_0/2)
    \end{align}
    Therefore, if we compute the average of $e^{-\beta W}$, 
    \begin{align}
        \langle e^{-\beta W}\rangle= p_0 e^{-\beta(\omega_1-\omega_0)/2 }+(1-p_0) e^{-\beta (-\omega_1+\omega_0)/2 } =\cosh(\beta \omega_1/2){\rm sech}(\beta \omega_0/2).
    \end{align}
    On the other hand, 
    \begin{align}
        Z={\rm Tr}[e^{-\beta H}]\Rightarrow Z_k=2\cosh(\beta \omega_k/2)
    \end{align}
    Since $e^{-\beta \Delta F}=Z_1/Z_0$, corroborate the Jarzynski equality. 

    Note that the average work,
    \begin{align}
        \langle W\rangle = p_0 (\omega_1-\omega_0)/2+(1-p_0)(-\omega_1+\omega_0)=\frac{1}{2}(2p_0-1)(\omega_1-\omega_0)
    \end{align}
    while the increment in free energy 
    \begin{align}
        \Delta F=F_1-F_0=\frac{\omega_1}{2}-\frac{\log(1+e^{\beta w_1})}{\beta}-\left(\frac{\omega_0}{2}-\frac{\log(1+e^{\beta \omega_0})}{\beta} \right).
    \end{align}
    It can be shown that $\langle W\rangle-\Delta F\geq 0$. The equality holds if $\omega_1=\omega_0$, or for $\beta=0$. For $\beta\rightarrow \infty$, also in some cases depending on $\omega_1$ and $\omega_0$.

\vspace{0.5cm}
\noindent\textbf{Problem 4.5}
\vspace{0.2cm}
    
\noindent Landauer principle: From the second law, we know that
\begin{align}
    \Delta S\geq \frac{1}{T}\langle Q\rangle
\end{align}
Initially, the qubit may be in any state, $\rho_i=\mathbb{I}/2$, so that $S(\rho_i)=k_b \log 2$, while the final is a pure state $S(\ket{0}\bra{0})=0$. Thus, $\Delta S=-k_b \log 2$, so that
\begin{align}
    -k_b \log 2 \geq \frac{1}{T}\langle Q\rangle \Rightarrow -\langle Q\rangle \geq k_b T \log 2. 
\end{align}
but $-\langle Q\rangle$ is just the heat given by the system to the environment (recall that $Q>0$ is heat supplied to the system). Hence, 
\begin{align}
    \langle Q_{released}\rangle \geq k_b T \log 2
\end{align}

\end{appendix}







\end{document}